\documentclass[12pt, oneside, final]{uwthesis2}
\usepackage[hidelinks, pagebackref, bookmarksopen, bookmarksnumbered]{hyperref}
\hypersetup{pdfauthor={Antti Puurula}}
\usepackage{bookmark}

\usepackage{hyphenat}
\usepackage{bm}
\usepackage{epsf,amsfonts,amsmath}
\allowdisplaybreaks
\usepackage{setspace}
\usepackage{pdfpages}
\usepackage[titletoc]{appendix}
\usepackage{natbib}
\usepackage{slashbox}
\usepackage{tikz}
\usetikzlibrary{bayesnet}
\usepackage{subfigure}
\usepackage{algorithm}
\usepackage[noend]{algpseudocode}
\usepackage{afterpage}

\newcommand{\argmax}{\operatornamewithlimits{argmax}} 
\newcommand{\argmin}{\operatornamewithlimits{argmin}} 
\newcommand\numberthis{\addtocounter{equation}{1}\tag{\theequation}}
\newcommand{\ul}{\underline}

\usepackage[inner=1.0cm, outer=1.0cm,top=1cm,bottom=2cm,bindingoffset=0cm]{geometry}
\usepackage{multirow}
\usepackage{array}
\usepackage{fancyhdr}

\title{Scalable Text Mining with Sparse Generative Models}
\author{Antti Puurula}
\begin{document}
\maketitle
\frontmatter

\pagestyle{fancy}
\lhead{}
\rhead{}
\fancyhead[]{\raggedright{\leftmark}}
\pagenumbering{roman}
\setcounter{page}{2}

\chapter{Abstract}
The information age has brought a deluge of data. Much of this is in text form, insurmountable in scope for humans and incomprehensible in structure for computers. Text mining is an expanding field of research that seeks to utilize the information contained in vast document collections. General data mining methods based on machine learning face challenges with the scale of text data, posing a need for scalable text mining methods.\\

This thesis proposes a solution to scalable text mining: generative models combined with sparse computation. A unifying formalization for generative text models is defined, bringing together research traditions that have used formally equivalent models, but ignored parallel developments. This framework allows the use of methods developed in different processing tasks such as retrieval and classification, yielding effective solutions across different text mining tasks. Sparse computation using inverted indices is proposed for inference on probabilistic models. This reduces the computational complexity of the common text mining operations according to sparsity, yielding probabilistic models with the scalability of modern search engines.\\

The proposed combination provides sparse generative models: a solution for text mining that is general, effective, and scalable. Extensive experimentation on text classification and ranked retrieval datasets are conducted, showing that the proposed solution matches or outperforms the leading task-specific methods in effectiveness, with a order of magnitude decrease in classification times for Wikipedia article categorization with a million classes. The developed methods were further applied in two 2014 Kaggle data mining prize competitions with over a hundred competing teams, earning first and second places.



\tableofcontents
\newpage
\onehalfspacing


\chapter{List of Abbreviations and Acronyms}
\begin{center}
\renewcommand{\arraystretch}{0.75}
\begin{tabular}{p{1.75cm}p{10cm}p{1cm}}
BM25 & Best Match 25\\
BNB & Bernoulli Naive Bayes\\
DBN & Dynamic Bayes Network\\
DM  & data mining\\
EM & expectation maximization\\
HMM & Hidden Markov Model\\
IDF & inverse document frequency\\
IE & information extraction\\
IID & independent and identically distributed\\
IR & information retrieval\\
KDD  & knowledge discovery in databases\\
KDT & knowledge discovery in textual databases\\
LM & language model\\
LR & Logistic Regression\\
LSHTC & large-scale hierarchical text classification\\
MAP & mean average precision\\
Micro-F1 & micro-averaged F1-score\\
ML & machine learning\\
MNB & Multinomial Naive Bayes\\
NB & Naive Bayes\\
NDCG & normalized discounted cumulative gain\\
NLP & natural language processing\\
SVM & Support Vector Machine\\
TDM & Tied Document Mixture\\
TF & term frequency\\
TF-IDF & term frequency-inverse document frequency\\
TM & text mining\\
VSM & Vector Space Model\\
\end{tabular} 
\end{center}

\chapter{Notation and Nomenclature}
The notation used in the thesis follows closely the linear algebra notation used in statistical natural language processing and machine learning,
graphical models literature in particular \citep{Manning:99, Bishop:06}. Counts of words in a document are represented using a word vector $\bm w$,
and a sequence of words is represented using a word sequence $\ul{\bm w}$.  Mixture model notation is used to denote models 
conditional on variables, e.g. $p_l(n)= p(n|l)$ and $p_{li\ul{k}_j}(\ul{w}_j)= p(\ul{w}_j| l, i, \ul{k}_j)$ . Apostrophe is used to reduce notation, by indicating derived functions and variables explained in the context, e.g. 
$\sum{n'}$ indicates a sum over the same variable type as $n$ and $p'(n)$ indicates a function related to $p(n)$. Information retrieval and 
natural language processing terminology is reduced, e.g. "word" is used ambiguously to refer to word types and tokens.\\

\chapter{List of Notations}
\begin{center}
\renewcommand{\arraystretch}{0.75}
\begin{tabular}{p{1.75cm}p{10cm}p{1cm}}
$\bm w$ & vector of word counts $\bm w= [w_1, ... , w_n, ... ,  w_N]$, ordinarily integer counts $w_n\in \mathbb{N}^0$\\
$n$ & word variable, word vector index $1\le n \le N$\\
$N$ & number of distinct words, dimension of a word vector $N= |\bm w|$\\
$\ul{\bm w}$ & sequence of words $\ul{\bm w}= [\ul w_1, ... , \ul w_j, ... , \ul w_J]$\\
$j$ & word sequence index $1\le j \le J$, $1 \le \ul w_j \le N$\\
$J$ & length of document $J=|\bm w|_1= |\ul{\bm w}|$\\
$l$ & label index variable $1\le l \le L$\\
$\bm c$ & label vector variable $\bm c= [c_1, ... , c_l, ... , c_L]$, $c_l \in \{0, 1\}$\\
$L$ & number of distinct labels in a collection \\
$D$ & collection, a dataset of text documents\\
$i$ & document index variable $1\le i \le I$, $D^{(i)}$\\
$I$ & number of distinct documents\\
$D^{(i)}$ & document $i$ of dataset $D$, including possible meta-data, $D^{(i)}= \bm w^{(i)}$, $D^{(i)}= (\ul{\bm w}^{(i)}, l^{(i)})$\\
$m$ & mixture component index $1\le m \le M$, $\alpha_m$, $p(m)$, $p_m(n)$\\
$M$ & number of mixture components, maximum n-gram order\\
$\ul{\bm k}$ & sequence of mixture assignment indicators $\ul{\bm k}= [\ul k_0, … , \ul k_j, …, \ul k_J]$, $1\le \ul k_j \le M$\\
$\bm \theta$ & vector of model parameters, for a multi-class linear model $y(\bm \theta_l, \bm w)= \theta_{l0} + \sum_{n=1}^{N} \theta_{ln} w_n$, with bias $\theta_{l0}$\\
$C(l,n)$ & count of joint occurrences of variables in collection, $C(l,n)= \sum_{i:l^{(i)}= l} w_n^{(i)}$ \\
$D(l,n)$ & discount applied to a count $C(l,n)$\\
$p_l^u(n)$ & unsmoothed multinomial $p_l^u(n)= \frac{C(l,n)-D(l,n)} {\sum_n' C(l,n')-D(l,n')}$\\
$\alpha$ & back-off weight, determined by the smoothing method\\
$\ul{\bm r}$ & sequence of word weights, interpreted as probabilities of words occurring, $\bm r= [r_1, ... , r_j, ... , r_J]$, $r_j \in [0,1]$\\
\end{tabular} 
\end{center}

\mainmatter

\chapter{Introduction}
This chapter introduces the topic of the thesis and motivation for research. It presents a thesis statement 
based on the results of the research, lists the contributions of the thesis compared to the existing literature, references 
publications by the author related to the thesis, and describes the structure of the thesis.\\

\section{Motivation}
The information age has brought an information overflow. The Internet provides a highway where vast amounts of data can be instantly 
searched and retrieved. This data presents a source that can be mined for knowledge about the world 
and to improve decision making. A considerable, and perhaps the most valuable, portion of this data is in text form, such as newspapers, web pages, 
books, emails, blogs and chat messages.\\

The field of artificial intelligence known as \emph{text mining} has become an intersection between data mining, information retrieval, 
machine learning and natural language processing. Since its birth in the mid 90's, it has shown consistent growth in research publications, and has been
applied in numerous ways in both industry and academia. Companies use text mining to monitor opinions related to their brands,
while traditionally qualitative sciences such as the humanities use it as an empirical methodology for research. However, the 
fragmentation of text mining research has resulted in a variety of tools specialized for different applications.\\

Text mining applications are mapped into general statistical and machine learning tasks, such as classification, ranking, regression and clustering. 
The increase of data and computing power enables performing previously impossible tasks using statistical models. E-mail spam filters can be trained
with trillions of text documents, and cover billions of words. Documents can be classified into a Wikipedia article hierarchy with millions of categories.
A major limit on the possibilities of text mining is the \emph{scalability} in the dimensions of data. General data mining models have not been 
developed for the sparse and increasingly high dimensional data encountered in text processing tasks. Data mining models that can easily operate on thousands
of documents and words can be unusable on datasets with millions of documents and words. The main solution to scalability is to redesign algorithms that have 
no more than linear computational complexities in terms of documents, words, and class variables.\\

Given these problems of fragmentation and scalability, it would be useful to have models that are both \emph{versatile} and \emph{scalable}. 
A versatile model for text mining would have to applicable to different task types with high performance. A scalable model for text mining would have
to scale in all of the relevant dimensions of the applied task.\\

Following an overview of multinomial generative models of text, the thesis proposes extensions of the common Multinomial Naive 
Bayes (MNB) model as a family of versatile models for text mining. It is shown that when MNB is modified to correct
its modeling assumptions, it forms a versatile solution to text mining that is far from the ``punching bag of machine learning'' that basic Naive
Bayes models have been called \citep{Lewis:98, Rennie:03}. By using inverted index data structures, it is shown that many of the processing operations with 
MNB and its graphical model extensions can be solved as a function of sparsity, turning the ``curse of dimensionality'' \citep{Bellman:52, Feldman:06} with
sparse text data into a blessing. \emph{Sparse generative models} combining generative models with sparse inference offer a versatile and scalable 
solution to text mining.\\

\section{Thesis Statement}
The thesis statement is as follows: 

\emph{Generative models of text combined with inference using inverted indices provide sparse generative models for text mining that are both versatile and scalable, 
providing state-of-the-art effectiveness and high scalability for various text mining tasks.}\\

\section{Contributions of the Thesis}

The thesis presents a synthesis of the fragmented text mining literature, and describes the generative multinomial models of text used across various text 
mining tasks. Based on the literature, the thesis proposes a solution to text mining that is both scalable and applicable to diverse tasks. It is shown that 
many of the multinomial models of text can be formalized as instances and extensions of the MNB model, including the n-gram language models that are widely applied
across a variety of text processing applications. This common formalization enables the transfer of modeling innovations across different tasks and research 
literatures, and provides a high-performing baseline across different task types. Connections to linear and graphical models are shown, and extensions 
within these frameworks offer MNB even greater flexibility.\\

Lack of a proper formalization of a computational method means that the behaviour of a method is not explained by any existing theory, and it can
behave erratically. This thesis formalizes modifications to MNB such as smoothing, feature weighting and structural extensions, showing that for most uses 
the modified models are not heuristic, but well-defined graphical models that extend the basic MNB model, in some cases using approximate maximum 
likelihood estimates. Formalizing these modifications means that the models are well-defined in a probabilistic sense, and their 
behaviour can be understood from the underlying probability theory.\\

The proposed sparse inference builds on the use of inverted indices for information retrieval with multinomial models of documents. It is shown that this type
of inference is not limited to computing ranking scores for document retrieval, but can be used to compute exact posterior probabilities from any linear model for 
various uses, such as ranking, classification and clustering. Furthermore, sparse inference is extended to structural models, yielding the 
same improvements in scalability and computational complexity for many practical cases.\\

Finally, the thesis presents an extensive empirical evaluation of the proposed models and inference over tens of standard datasets used for text classification 
and ranked retrieval. The experiments give strong evidence in support of the thesis statement. In both types of tasks, modifications of MNB 
outperform or rival strong baseline methods commonly used for these tasks in effectiveness. Scalability experiments are conducted on a large Wikipedia
article classification task, showing that the proposed sparse inference improves scaling of the models into tasks with a million words, documents, 
and classes.\\

An extensive literature review has been conducted to verify the originality of the contributions. As expected, some of the contributions have prior and concurrent work.
The following list highlights the most fundamental contributions that are not found in the prior literature, to the best of the author's knowledge:
\begin{description}
\item[Synthesis of text mining methodology] Surveys of text mining have limited their perspectives to mostly a few of its influences, such as data 
mining \citep{Witten:04, Hotho:05, Feldman:06, Stavrianou:07, Weiss:12}, machine learning \citep{Weiss:12, Aggarwal:12}, information retrieval \citep{Aggarwal:12}, and
natural language processing \citep{Hearst:99, Black:06}. Chapter 2 gives a concise synthesis of current text mining methodology, incorporating the perspectives of various 
practitioners in a coherent framework. Text data is shown to be multiply structured data from a linguistic point of view, many of the core
algorithms used for text mining are shown to be cases of linear models, and possible views and solutions to the scalability problem are discussed.
\item[Formalization of smoothing with two-state Hidden Markov Models] Early, but discontinued research showed that the linearly smoothed n-gram language models used 
for information retrieval could be formalized with two-state Hidden Markov Models (HMM) \citep{Miller:99, Xu:00, Hiemstra:01}. Chapter 4 shows that all of the smoothing 
methods for multinomial generative models can be expressed as exact inference on a two-state HMM. The formalization further shows that most of the parameter estimates for the
smoothed multinomials derive from expected log-likelihood parameter estimation on a two-state HMM. Heuristically defined backoff models for 
Kneser-Ney \citep{Kneser:95} and power-law \citep{Huang:10} discounting are shown to derive from constraints applied to the two-state HMM.
\item[Formalization of feature weighting with probabilistic data] Concurrent research has derived the expected log-likelihood estimation of n-gram language models 
from probabilistically weighted data \citep{Zhang:14}. Chapter 4 extends this probabilistic data view to inference, showing that both estimation and inference
is well defined for probabilistically weighted word sequences and non-negative fractional word counts. This greatly extends the versatility of generative models of text,
as data can be weighted to correct modeling assumptions, without losing the probabilistic formalization.
\item[Sparse inference] Earlier research has applied inverted indices for reducing the classification times for K-nearest Neighbours \citep{Yang:94} and 
Centroid \citep{Shanks:03}. The same reductions are gained for computing posterior probabilities for linearly interpolated language models in information 
retrieval \citep{Hiemstra:98b, Zhai:01b}. Chapter 5 shows that inverted indices can be used to reduce the inference complexity according to sparsity for a variety 
of processing tasks, including linear models and structural extensions of linear models. Applied to Wikipedia classification with a million possible categories, 
an order of magnitude reduction of classification times is obtained for Tied Document Mixture, a novel structural extension of MNB.
\item[Evaluation of modified generative models] Evaluation of text mining methods has mostly been limited to experiments in one task type with a few possible
methods \citep{Sebastiani:02, Huston:14}. Chapter 6 presents a consistent framework for evaluation of text mining models applied for both text classification and
ranked retrieval tasks. Model parameters are searched using Gaussian random search optimization, avoiding the problem of local optima. Statistical testing is
conducted between datasets, measuring the strength of the discovered effects across text collections. Experiments are conducted over tens of text classification 
and retrieval datasets, comparing a large set of modified MNB variants with strong baseline methods for the tasks.
\end{description}

\section{Published Work}

The thesis includes and expands on earlier work by the author, most published in peer-reviewed conferences:
\begin{description}
\item[\citep{Puurula:11}] \emph{Mixture Models for Multi-label Text Classification}, New Zealand Computer Science Research Student Conference in Palmerston North, New Zealand, 2010. An analysis of generative mixture models for text modeling
\item[\citep{Puurula:11b}] \emph{Large Scale Text Classification with Multi-label Naive Bayes}, Second Smart Information Technology Applications Conference in Seoul, South Korea, 2011. A generative multi-label mixture model for modeling text 
\item[\citep{Puurula:12}] \emph{Scalable Text Classification with Sparse Generative Modeling}, Pacific Rim International Conference on Artificial Intelligence in Kuching, Malaysia, 2012. Earliest form of sparse posterior inference for MNB and a multi-label mixture model extension
\item[\citep{Puurula:12b}] \emph{Ensembles of Sparse Multinomial Classifiers for Scalable Text Classification}, ECML/PKDD - PASCAL Workshop on Large-Scale Hierarchical Classification in Bristol, United Kingdom, 2012. An ensemble of sparse generative models used successfully in a machine learning competition
\item[\citep{Puurula:12c}] \emph{Combining Modifications to Multinomial Naive Bayes for Text Classification}, Asian Information Retrieval Symposium in Tianjin, China, 2012. Combinations of modifications to Multinomial Naive Bayes examined
\item[\citep{Puurula:13}] \emph{Integrated Instance- and Class-based Generative Modeling for Text Classification}, Australasian Document Computing Symposium in Brisbane, Australia, 2013. Early version of the Tied Document Mixture model presented with sparse posterior inference
\item[\citep{Puurula:13b}] \emph{Cumulative Progress in Language Models for Information Retrieval}, Australasian Language Technology Association Workshop in Brisbane, Australia, 2013. Combinations of modifications to information retrieval language models examined
\item[\citep{Puurula:14}] \emph{Kaggle LSHTC4 Winning Solution}, Arxiv.org preprint, 2014. A description of the winning solution to the Kaggle Large Scale Hierarchical Text Classification competition, with an ensemble of sparse generative models
\item[\citep{Tsoumakas:14}] \emph{WISE 2014 Challenge: Multi-label Classification of Print Media Articles to Topics}, Web Information Systems Engineering in Thessaloniki, Greece, 2014. A report on the Kaggle WISE competition, where an ensemble using sparse generative models as components came second
\item[\citep{Trotman:14}] \emph{Improvements to BM25 and Language Models Examined}, Australasian Document Computing Symposium in Melbourne, Australia, 2014. Exploration of recent information retrieval ranking functions, including generative language models discussed in the thesis. \emph{Best paper award}\\
\end{description}

Open source code related to the thesis is distributed online, making the methods presented here available for wider use. The 
SGMWeka\footnote{http://sourceforge.net/projects/sgmweka/} open source toolkit for sparse generative modeling is available through SourceForge.net, 
as well as dataset preprocessing scripts required to reproduce the results shown in the thesis. The Kaggle LSHTC4 winning 
solution\footnote{http://www.kaggle.com/c/lshtc/forums/t/7980/winning-solution-description} is available via the Kaggle website, making it possible to replicate the winning 
methods. The competition description of the LSHTC4 solution is included in Appendix B of the thesis.\\

\section{Structure of the Thesis}

The thesis is written to be accessible to readers of differing backgrounds. The introductory chapters, as well as the experiments and conclusion are intended
to be readable by most. The chapters introducing novel mathematical ideas require extensive background knowledge in probability theory and statistical 
mathematics, and are recommended mainly for researchers. The rest of the thesis is structured as follows:
\begin{description}
\item[Chapter 2] introduces the topic of text mining, covering the terminology and methodology of text mining
that will be used in the following chapters. The emerging field of text mining is highly fragmented, and the used terms and methods differ widely. The 
chapter includes an extensive literature review of the topic, and presents the many facets
of text mining in an integrated framework that is accessible to readers without extensive mathematical background.\\
\item[Chapter 3] introduces the Multinomial Naive Bayes model for text mining and its extensions with generative graphical models. This chapter establishes 
the mathematical notation that will be used throughout the rest of the thesis, and defines concepts such as graphical models and dynamic programming.
This chapter is written to be accessible to readers with basic understanding of probability theory.\\
\item[Chapter 4] presents a more detailed analysis of the MNB model for text mining. It is shown that all of the commonly used smoothing methods for
correcting data sparsity with multinomial text models can be formalized as approximate maximum likelihood estimation on a constrained Hidden Markov Model. 
It is shown that feature weighting can be equally formalized for MNB models and its extensions. Furthermore, practical 
graphical model extensions of MNB are proposed that maintain the efficiency of the model, while providing greater effectiveness and modeling flexibility. This 
chapter is accessible to readers with experience in derivations for graphical models.\\
\item[Chapter 5] presents the idea of sparse inference for MNB, and more generally for linear models and structured extensions of linear models. The complexity 
of inference is reduced as a function of sparsity, by using inverted index representation of model parameters. This chapter is the most technically 
demanding in the thesis, and contains novel algorithms and derivations.\\
\item[Chapter 6] presents an extensive empirical evaluation of the modeling ideas presented in Chapters 4 and 5 in the context of text classification
and retrieval. In terms of effectiveness, it is demonstrated that the proposed extensions of MNB models greatly improve on the commonly used
generative models for these tasks, providing results competitive with strong baseline methods for both tasks. In terms of scalability, it is shown that MNB with
sparse inference easily scales to classification with a million features, documents and classes. Sparse inference on structured extensions of MNB scale with a 
similarly reduced time complexity, reducing inference times by an order of magnitude in the highest-dimensional cases examined.\\
\item[Chapter 7] concludes the thesis with a summary of a thesis, revisits the thesis statement, and discusses the implications of the 
findings, limitations of the thesis, and possible future work.

\end{description}

\chapter{Text Mining and Scalability}
This chapter presents a brief introduction to text mining, followed by a comprehensive overview of text mining methodology, and a discussion
on the scalability problem in text mining. The introduction discusses the variety of definitions for text mining, related fields preceding text mining,
and domains that apply text mining. The overview of text mining methodology provides a synthesis of viewpoints on text mining, starting
from the linguistic properties and representation of text data, followed by mapping of text mining problems into machine learning tasks, and
finally comparing text mining architectures to knowledge discovery processes. The discussion on scalability describes the scalability problem
in text mining with examples, implicit views on scalability taken by researchers and practitioners, and existing approaches to scalability.\\

\section{Introduction to Text Mining}
\subsection{Defining Text Mining}
Progress in information technology has brought an information overflow, with transformative societal implications that affect all aspects of 
human life. A considerable and possibly the most significant portion of this information is in the form of text data, such as books, news articles, 
microblogs and instant messages. These vast quantities of text data can only be accessed and utilized using computers, but
the automated processing of text is only possible using technology specialized for human language. Text mining (TM) in a broad sense refers to technology
that allows the utilization of large quantities of text data. In the following, this working definition will be amended by a more concise one.\\

\begin{figure*}
\centering
\includegraphics[scale=0.42, trim=10 10 10 160, clip=true]{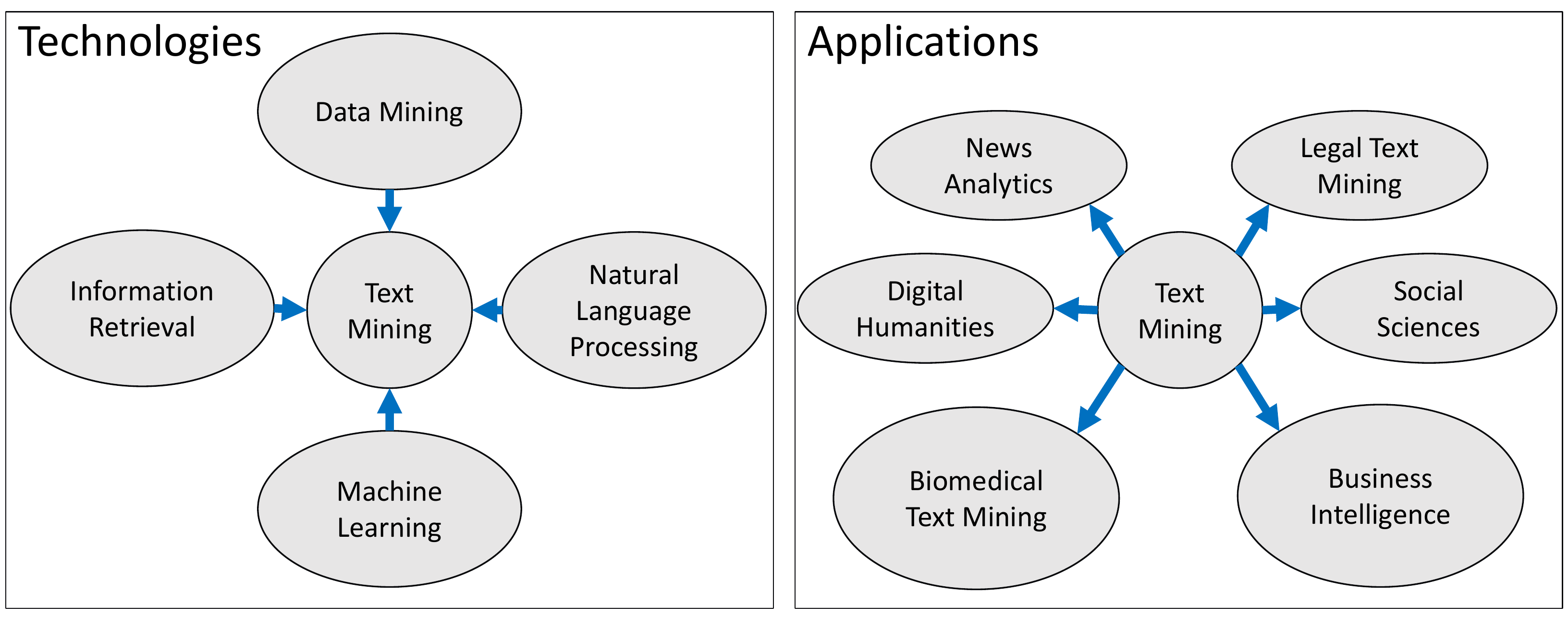}
\caption{Relationship of TM to the major related fields (technologies) and application domains (applications)}
\label{relations_figure}
\end{figure*}

Text mining originates from several earlier research fields, such as data mining, machine learning, information retrieval, natural language 
processing. Like these fields, TM has a foundation in computer science, with considerable influence from applied artificial 
intelligence \citep{Fayyad:96, Witten:04}. It is highly related and sometimes used interchangeably with terms such as information extraction, opinion mining and
text analytics. TM is used in a variety of application domains, such as biomedical TM and business intelligence. The related fields have influenced
TM in terminology and methodology, whereas the application domains have been influenced by TM. Figure \ref{relations_figure}
illustrates the relationship of TM to the major related fields and application domains. These relationships will be discussed in detail in Sections \ref{related_fields}
and \ref{application_domains}.\\

The term ``text mining'' originated in the data mining publications of mid 1990's. Feldman et al. wrote a series of publications starting from 1995 
under the term ``knowledge discovery in textual databases'' (KDT) \citep{Feldman:95, Feldman:97, Feldman:98, Feldman:06}, and by 1997 a number 
of authors used the term text mining
\citep{Ahonen:97, Rajman:97, Feldman:97, Tkach:97}. The early proponents of TM considered it to be an application of data mining to 
text data, and saw text data as ``unstructured data'' that needs to be structured for use in data mining: \emph{``before we can perform any kind of knowledge 
discovery in texts we must extract some structured information from them''} \citep{Feldman:95}. This KDT definition viewed TM as data mining,
with natural language processing and information retrieval as preprocessing and indexing steps in the mining 
process \citep{Feldman:95, Feldman:97, Feldman:98, Ahonen:97, Ahonen:97a, Albrecht:98, Dorre:99, Tkach:97, Liddy:00}. A slight variation of this 
KDT definition was TM as a different type of process from general data mining \citep{Rajman:97, Rajman:98, Merkl:98, Witten:99, Tan:99, Witten:00, Witten:00b}.\\

Within a few years TM started to interest other research communities. Natural language processing and computational linguistics researchers saw TM as a 
potential application \citep{Hearst:99}. Many of the leading machine learning methods of the next decade were developed and popularized in the context of 
modeling text \citep{Elkan:97, Joachims:98, Lewis:98, Lafferty:01, Hofmann:99, Blei:03}, providing TM practitioners an advanced toolkit. Information retrieval 
had extended earlier into information extraction, and its similarities to TM were discovered at this time as well \citep{Nahm:04, Mooney:05}. The definitions of 
TM gradually diversified away from Feldman's 
KDT definition of TM, as the term slowly started to be used in exceptionally diverse contexts in both academia and business. \cite{Witten:04} in his review discusses 
the problems of defining TM, and explicitly avoids providing a concise definition. Both \cite{Black:06} and \cite{Hotho:05} give a definition of TM close to the 
KDT definition, while noting the diversity of definitions. \cite{Cohen:05} in their survey of biomedical TM avoid providing an explicit definition. \cite{Stavrianou:07} 
in their survey give the KDT definition, while \cite{Weiss:12} appears to implicitly use the KDT definition.\\

Although the KDT definition is a very simple characterization of TM, it is not very descriptive in practice. Perhaps the biggest problem with the definition is that it 
does not capture what is unique about TM. TM overlaps many fields to the extent that any of its applications could equally be considered as problems of 
the related fields. What makes TM is unique is not the tasks and problems it shares with the related fields, but the \emph{interdisciplinarity} and 
\emph{integration of methods} for solving the problems.\\

An example of a TM problem could be a web monitoring system for analyzing sentiment related to a brand. A system of this type would require methods from 
information retrieval to search text data related to the brand, natural language processing and information extraction for extracting parts of the text that refer 
to the brand, and machine learning for predicting the sentiment. The system would further need text visualization and statistical tests for confirming the reliability of
the predictions. An integrated architecture for constructing such a system would most accurately be called a TM solution.\\

The view of TM as integration of artificial intelligence-based text processing technologies captures the main novelty of TM. Perhaps the first definition of TM 
from this point of view was reflected in the title of the KDD'2000 workshop on TM: \emph{``Text Mining as Integration of Several Related Research Areas''} 
\citep{Grobelnik:00}. \cite{Feinerer:08} avoids choosing a definition, but seems to support this view as well: \emph{``In general, text mining is an 
interdisciplinary field of activity amongst data mining, linguistics, computational statistics, and computer science''}. For the purpose of this thesis, a concise 
definition is proposed:\\

\emph{Text mining is an interdisciplinary field of research on the automatic processing of large quantities of text data for valuable information}.\\

\subsection{Related Fields}
\label{related_fields}
Text mining originates from earlier and well-established fields grounded in computer science and artificial intelligence, the four major ones being 
data mining (DM), information retrieval (IR), natural language processing (NLP) and machine learning (ML). All of these fields are interdisciplinary with 
a considerable amount of participation from a diverse range of academic subjects related to computer science. As information technology is becoming 
increasingly prevalent in the modern world, much of the research in these fields is becoming distributed and applied across every discipline in the academic 
world, including the ``soft sciences'' of the humanities that have previously relied on qualitative methodologies. Outside academia, these fields exist as 
viable industries, with a global market of start-up and large-cap companies alike. A comparison of TM and related fields is given in the following.\\

Data mining and knowledge discovery in databases (KDD) \citep{Fayyad:96, Chen:96b} deal with the discovery of useful patterns in large databases, and 
have origins in statistics, machine learning and databases. Within KDD, DM constitutes the algorithms used for discovery of patterns, whereas KDD refers to the 
overall interactive process, where the user explores a dataset \citep{Fayyad:96}. The term TM originated in DM research, and DM certainly remains one of the major 
influences in current TM. Many of the definitions used for KDD apply equally to TM today: \emph{``The KDD process can be viewed as a multidisciplinary activity that 
encompasses techniques beyond the scope of any one particular discipline such as machine learning.''} \citep{Fayyad:96} 
and \emph{``KDD also emphasizes scaling and robustness properties of modeling algorithms for large noisy data sets.''} \citep{Fayyad:96}. Multidisciplinarity and 
scalability are equally defining qualities of TM. To some extent, 
TM shares the idea of \emph{process models} that are applied in different tasks types \citep{Ahonen:97, Ahonen:97a, Liddy:00}. An early view was that TM 
is simply DM with a text-specific preprocessing phase and an additional document filtering phase \citep{Ahonen:97, Dorre:99}, but current TM systems are better 
described as architectures than a process \citep{Feldman:06}. A TM system neither requires a user discovering new patterns: TM can be used to automatically  
monitor existing well-known patterns in text, such as sentiment and topic. The DM characterization of text as ``unstructured data'' is also a broad generalization: 
text has shown to be a unique type of data structured in multiple ways, requiring specialized methods very different from general DM. The goals of TM and DM 
sometimes differ: the TM output is not necessarily hard facts or quantifiable values, but ``soft information'' in the form of text. Overall, there is surprisingly little
interaction between TM and DM today, although much of TM can be situated in the context of KDD.\\

Machine learning deals with systems that learn from data, and has origins in statistics, artificial intelligence and computer science. For a given learning task 
and performance 
measure, a learning system improves its performance using data \citep{Mitchell:97}. This contrasts with statistics, where the emphasis is on
finding the correct models for data, and not on directly optimizing performance \citep{Breiman:01}. The division in goals has led to a division between the 
``two cultures'' of traditional statistics and machine learning \citep{Breiman:01}. The success of ML has lead to it being adopted as a general 
framework in a variety of application domains requiring artificial intelligence. Much of ML has dealt with text data
 \citep{Joachims:98, Sebastiani:02, Lafferty:01, Blei:03}, and much of TM is based on the application of ML methods; text classification in particular. The division 
of TM into distinct task types \citep{Feldman:06, Aggarwal:12} also follows the general ML framework. Like with DM, the main reason 
for not considering TM as simply an application of ML is the uniqueness of text data. Techniques such as inverted indices have proven crucial for processing
text, yet these are virtually unknown in ML. Although the majority of TM methods originate in ML, TM systems also require tools that are specialized
for text, originating from a variety of disciplines, some requiring no learning, and some constructed with highly specialized human expertise.\\

Information retrieval deals with systems for retrieval and ranking of documents for a given information need. The ubiquitous case is web search, where 
the information need is expressed as a query consisting of words, and the ranked set of documents consists of webpage links. Modern research into 
IR started in the context of computerized library indexing systems \citep{Maron:61, Manning:08}, where the number of paper documents was 
increasing rapidly. This parallels the arrival of vast amounts of digital documents and the development of TM half a century later. Although IR is more
general and deals with different types of data, text is the main type of information used to index most forms of data, and text retrieval using word vectors and 
inverted indices constitutes the main methodology of current IR \citep{Manning:08}. These two text retrieval techniques also constitute key components for 
TM, since they enable scalable search of documents. IR has become possibly the main influence on TM, as it has expanded into more complex tasks 
after progress in ad-hoc text retrieval was considered to be stagnant at the end of the millennium \citep{Voorhees:99}.\\

Natural language processing or computational linguistics deals with the processing of text data using algorithms based on linguistic theory, and originated in 
the considerable efforts
to develop machine translation during the early years of the cold war \citep{Pierce:66}. Basic NLP tasks are segmenting, parsing, and annotating text according 
to the underlying linguistic structure of the data \citep{Manning:99}. Corpus linguistics refers to the goal of producing well-defined computational theories to 
understand natural language, whereas the term NLP is closer to the goal of developing practical language technology. TM was proposed early as 
an additional goal for computational linguistics \citep{Hearst:99}, and NLP has certainly had a large influence on TM. Statistical NLP in particular predated 
TM by integrating IR, statistics and NLP \citep{Manning:99}. Research in NLP has relied on annotated digital corpora such as the Brown 
corpus \citep{Kucera:67}. In contrast, TM often uses unannotated, large scale, and noisy datasets, with the goal of extracting information of 
value. NLP is not as interdisciplinary as TM; TM research exists across a wide variety of problem domains, whereas NLP research is 
seldom conducted in other fields.\\

Related to statistical NLP is another field at the intersection of NLP and IR deserving a mention. Information extraction (IE) research started in the 
context of the Message Understanding Conference series starting from 1987 \citep{Grishman:96, Sarawagi:08}. The goal of IE is the extraction of 
predetermined patterns from text collections, such as names, places and events. IE therefore has a more limited and concise goal than TM, but the two
fields overlap to the extent that sometimes little difference is seen between the two \citep{Sebastiani:02, Stavrianou:07}. One viewpoint is considering
IE as an intermediate step to TM, where the extracted patterns are used for discovery of more complex information \citep{Nahm:04, Mooney:05}. Text
summarization deals with the extraction of human-readable summaries of text, and is a type of TM task that falls outside the scope 
of IE \citep{Witten:04, Sarawagi:08, Das:07}.\\

\begin{figure*}
\centering
\subfigure[ScienceDirect]{
 \includegraphics[scale=0.98, trim=280 125 280 145, clip=true]{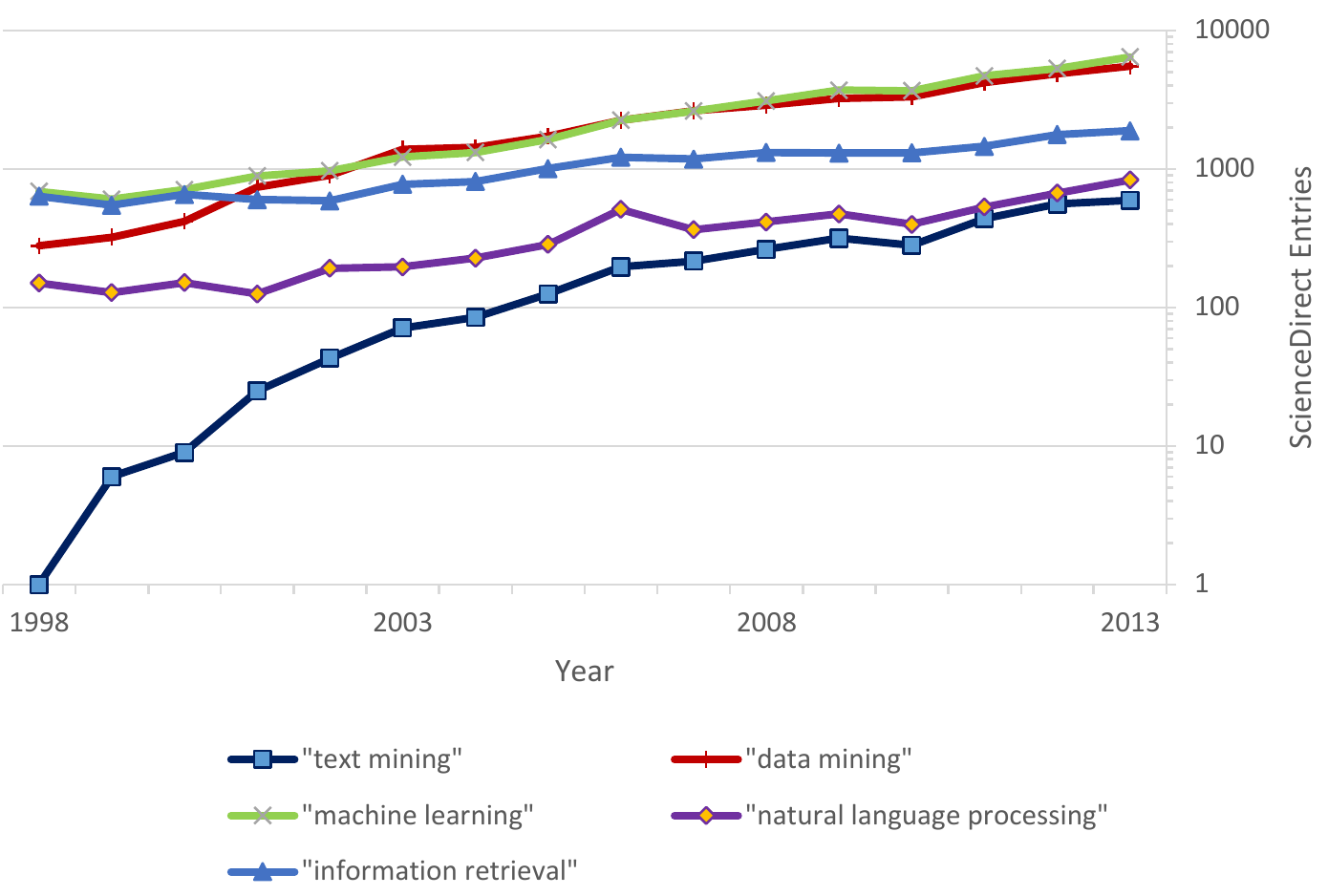}
}
\hspace{5pt}
\subfigure[Google Scholar]{
 \includegraphics[scale=0.98, trim=280 125 280 145, clip=true]{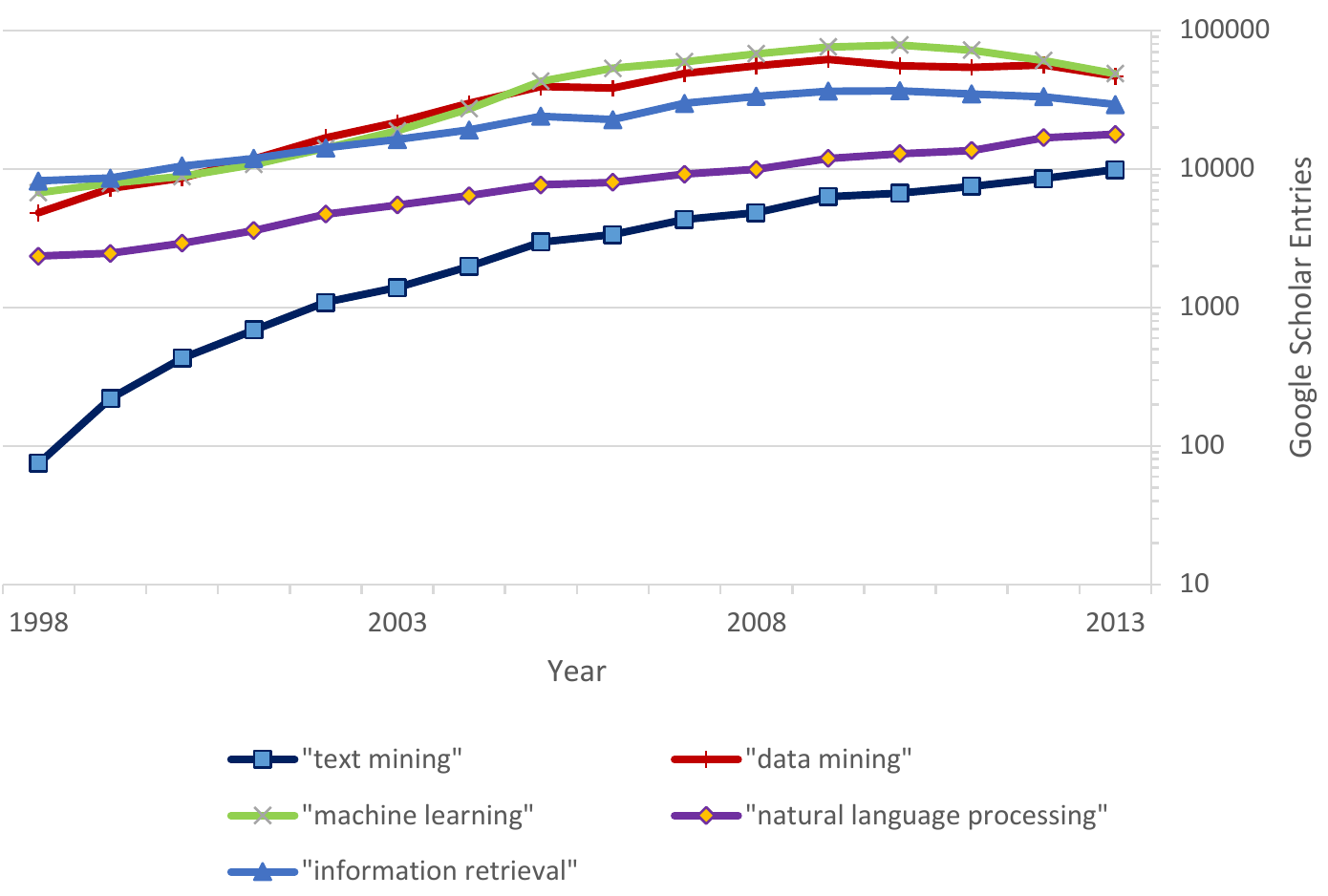}
}
\caption{Number of academic publications per year indexed by ScienceDirect and Google Scholar with the phrases ``text mining'', ``machine learning'', ``data mining'', ``natural language processing'', and ``information retrieval'', retrieved 2.8.2014. Both indices show TM research currently as active as the related 
fields were around the year 2000}
\label{publication_counts}
\end{figure*}

One way to measure the development of TM compared to the related fields is to count the number of publications that use the term over the years. 
Figure \ref{publication_counts} shows the number of publications for years 1998-2013 indexed by Google Scholar\footnote{scholar.google.com} and 
ScienceDirect\footnote{sciencedirect.com} containing the names of the fields. As can be seen, for a decade TM has grown at a steady rate of 800 more 
publications each year according to Google Scholar, and 50 according to ScienceDirect. Analyzing these graphs relies on a couple of assumptions: the quality of 
indexed publications is roughly similar, and that terminological ambiguity such as the use of ``computational linguistics'' for ``natural language processing'' does 
not distort the results. Since both indices agree on the main effect compared to the four other fields, this is most likely correct: published TM 
research has grown to a volume that the related fields had around year 2000.\\

\subsection{Application Domains}
\label{application_domains}

TM has propagated into a wide variety of domains both in academia and business. In academia, TM is enabling new forms of research by providing a 
methodology for meta-research of large quantities of publications \citep{Jensen:06}, as well as providing computational methods for fields that have lacked 
quantitative methodologies \citep{Schreibman:08}. Aside from the sheer volume shown in Figure \ref{publication_counts}, the diversity of TM applications 
is challenging and lacks a clear-cut categorization. Surveys in TM have suggested different application areas over the years. \cite{Black:06} cites business 
intelligence and biomedical TM as the main applications. \cite{Feldman:06} give corporate finance, patent research, and life sciences as the most 
successful applications. \cite{Fan:06} categorizes applications into medicine, business, 
government, and education. \cite{Weiss:12} gives a number of case studies: web market intelligence, digital libraries, help desk applications, news article
categorization, email filtering, search engines, named entity extraction, and customized newspapers. Overall, the general categories of biomedical TM and 
business intelligence capture the two most established application domains of TM.\\

Biomedical TM was proposed in the context of information retrieval as \emph{``the discovery of hidden connections in the scientific literature''} 
\citep{Swanson:88, Swanson:91, Hearst:99}. Biomedical TM has become an exceedingly popular application, since TM has provided a 
meta-analysis methodology to discover new facts by combining evidence 
from the vast biomedical research literature \citep{Cohen:05, Jensen:06, Rzhetsky:08, Zhou:10, Korhonen:12, VanLandeghem:13, Shemilt:13}. 
The overall output of scientific publications has increased exponentially for the last century, with some current estimates of annual growth at $4.73\%$ 
\citep{Larsen:10} and $8-9\%$ \citep{Bornmann:14}. In 2014, the biomedical article index PubMed\footnote{http://www.ncbi.nlm.nih.gov/pubmed} 
contains entries for over 24 million publications, and this exponentially growing literature can only be comprehended using new methods. The popularity
of biomedical TM has lead to a common perception of TM as simply biomedical literature mining.\\ 
 
The origins of TM in business intelligence can be traced to a 1958 IBM paper \citep{Luhn:58}, that proposed an automated system for managing 
information in documents. Business TM applications are diverse, including financial TM \citep{Kloptchenko:04, Gross-Klussmann:11}, 
marketing \citep{Decker:10, Pehlivan:11}, and
market intelligence \citep{Archak:07, Godbole:08, Baumgartner:09, Pehlivan:11, Ghose:11, Archak:11, Netzer:12}.
Perhaps the most common business use of TM is sentiment analysis, and more generally opinion mining \citep{Dave:03, Pang:08}, that seeks to 
analyze text data in order to monitor opinions related to companies, brands and products. Some business TM publications are starting to 
use the term text analytics as a synonym for TM \citep{Gruhl:04, Gross-Klussmann:11, Basole:13}, following the trend in the industry where the term
``analytics'' has become increasingly common over the past decade.\\

Outside these two major groups of applications, there are application domains that have recently adopted TM methodology. Domains
such as law \citep{Coscia:12}, political science \citep{Grimmer:13}, humanities \citep{Schreibman:08}, social science \citep{Brier:11} and
intelligence \citep{Maciolek:13} have combined TM methodology with traditional research methodologies. Many of these domains are applying TM to
online data sources such as blogs and micro-blogs, but some use TM on digitized publications. The ``soft sciences'' such as humanities and social science
in particular see TM as providing a new methodology for performing quantitative research on issues that previously relied only on qualitative methods  \citep{Schreibman:08}.\\

The extreme interdisciplinarity and variety makes assessing the overall scope of TM in great detail daunting, due to differences in: 1) terminologies, 
2) methodologies, and 3) traditions of publication. Firstly, a large portion of TM research occurs under related terms, such as text analytics, opinion mining, etc.
The different fields use their own terminology in addition to TM terms. This makes it difficult to find the relevant publications, and to synthesize 
a coherent picture from manuscripts written to address different issues using different terms. Secondly, TM is often mixed with the methodologies of the 
application domain. Comparing TM research often requires expertise of the theory and methodologies of both TM and the application domain. Thirdly, 
academic communities have varying traditions on publishing: computer scientists publish foremost in conferences, whereas humanities publish in the 
form of books, while most other disciplines prefer journal publication. A TM book written in the context of digital humanities could prove influential, 
yet lack the impact factors used in the natural sciences. This complicates assessing the quality of TM publications from external indicators such as
citation counts and impact factors. A further complication is grey literature: influential discourse on TM takes place not only in established and peer-reviewed 
academic contexts, but also in contexts such as non-reviewed articles\footnote{http://www.nature.com/news/trouble-at-the-text-mine-1.10184}, blogs\footnote{http://breakthroughanalysis.com/}, and white papers\footnote{http://hurwitz.com/index.php/component/content/article/394}.\\

\section{Text Mining Methodology}
\subsection{Text Documents as Multiply Structured Data}

Text data is commonly described as ``unstructured data''. This phrase originated in the earliest data mining enquiries into text, and has since been used in 
almost every description of TM. From a data mining point of view, raw text data is not organized into a database of numeric values, hence it can be considered
to be unstructured. The purpose of TM was to convert text into a structured form, where data mining could be applied \citep{Ahonen:97, Tan:99}.
The unstructured data description provides a simple introduction to TM from a data mining perspective, but is unfortunately misleading.\\

Text is more accurately called multiply structured data. The English word ``text'' comes from the Latin etymology ``textus'' meaning ``woven'', related 
to the words ``textile'' and ``texture''. Not only is language structured, but it occurs in \emph{documents} and \emph{collections} that have additional 
varying structure. Written text can be understood as sequential statements organized in hierarchies of structures such as sentences, passages, sections,
chapters, and so forth. Explicit structure in the form of metadata is virtually always available, whereas the ``unstructured data'' of human language has implicit 
structure, arguably among the most complex phenomena to have evolved in nature. The orthographic and typographic structure of written text is 
further complicated by the structured document mark-ups used in digital text, including hyperlinks that turn text data into hypertextual data better
understood in terms of graphs.\\

\begin{figure*}
\centering
\includegraphics[scale=0.5, trim=10 10 100 10, clip=true]{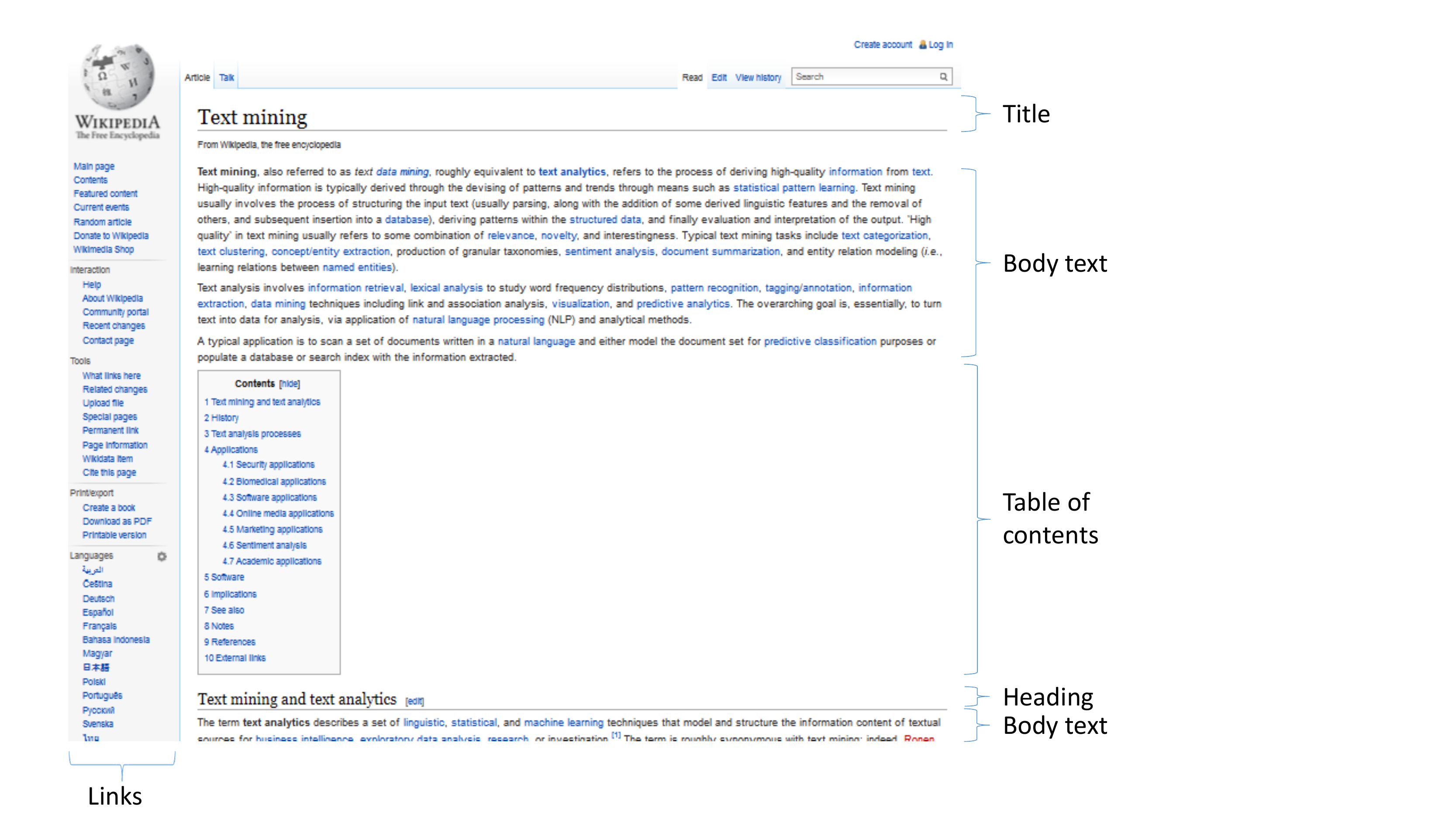}
\caption{Document structure in an example Wikipedia hypertext document}
\label{wikipedia_structure}
\end{figure*}

A collection of text data consists of documents that can number in millions or more. The collections can be static,
with no time component, or dynamic, with documents ordered by time. The collections can be fixed datasets, or streams that are not retained in
memory, but processed in an online manner. Collection and document metadata is typically very rich in TM applications, and can include internal and external 
hyperlink structures of the documents, locations and languages of the documents, author identities, years and dates of 
authorship, subcategories and ontologies of the documents, etc. The metadata can be unique to the dataset, or highly standardized \citep{Bargmeyer:00}.
Additional explicit metadata can be constructed by applying TM and ML methods on the dataset \citep{Pierre:02}. The document content is organized into fields 
such as titles, text sections and possibly link sections, and often contains non-text and partly textual media such as figures, illustrations, 
tables and multi-media. The layout, typography and mark-ups define the visual look of text and hyperlinks connect the text within 
the document, to other documents, and to resources outside the collection. Figure \ref{wikipedia_structure} illustrates document structure from the beginning
of a Wikipedia document.\\

A collection used for a specific task has an associated \emph{domain} of background knowledge \citep{Anand:95, Feldman:06}. \cite{Feldman:06} 
define domains in TM loosely as: \emph{``a specialized area of interest for which formal ontologies, lexicons, and taxonomies of information may be created''}.
The availability and usefulness of domain knowledge is one of the defining properties of TM \citep{Feldman:06}. 
The domains depend on the use of the collection, and knowledge from more than one domain can be beneficial. For example, a collection of Twitter 
microblog messages and newspaper articles used for monitoring a company's public image could benefit from having domain knowledge for spelling 
correction, normalization, and named entity recognition, for both types of data. The simplest form of domain knowledge is text collections of billions of words 
in the domain language that can be used to construct models for TM \citep{Napoles:12, Buck:14}. Natural language consists of languages, dialects, ethnolects and 
sociolects. It is common that TM collections and domains only cover a particular subset of one language.\\

\begin{figure*}
\centering
\includegraphics[scale=0.42, trim=10 10 10 10, clip=true]{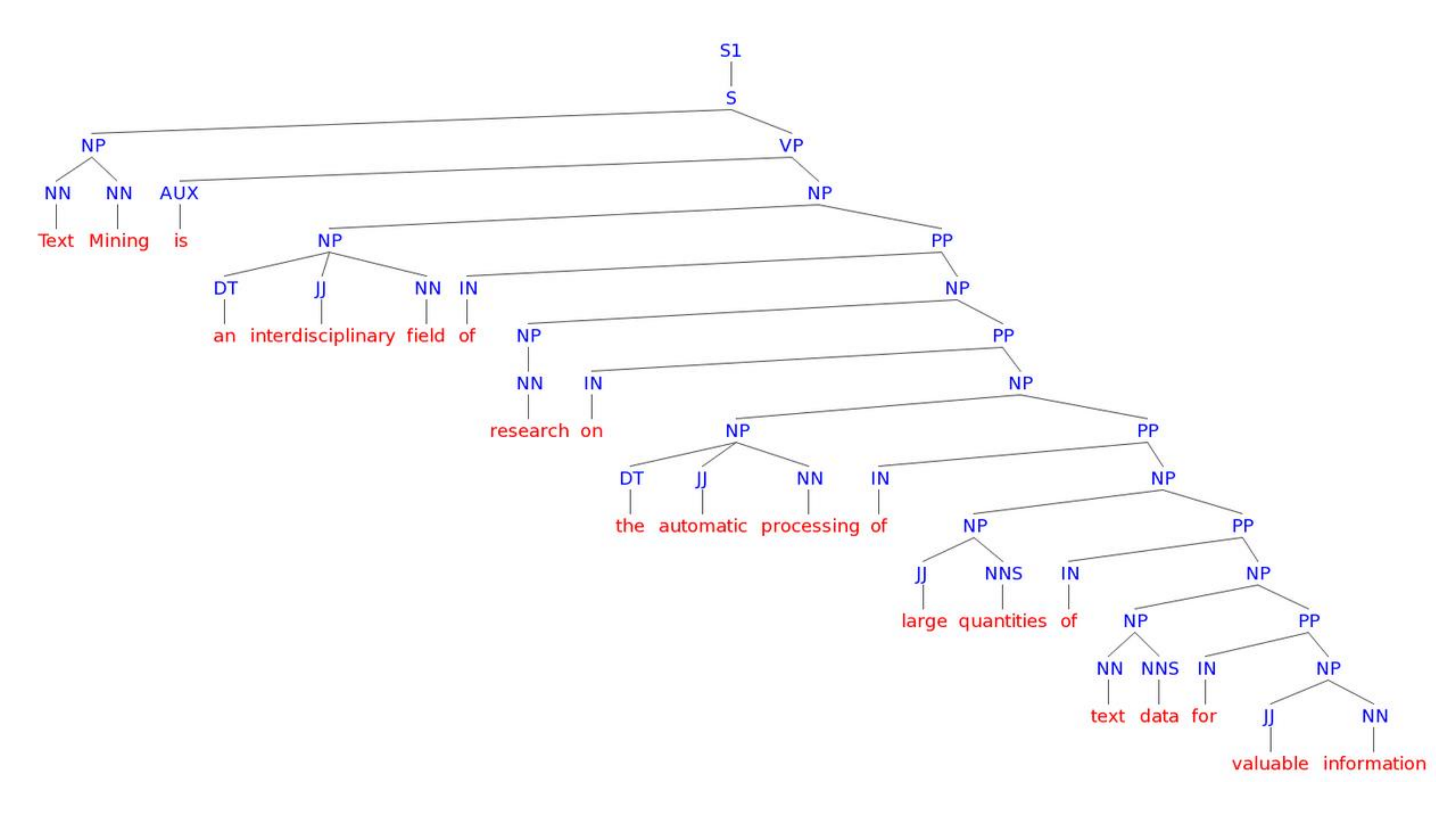}
\caption{Linguistic structure at the syntactic level for the TM definition used in this thesis, according to the NCLT wide-coverage parser \citep{Cahill:04}}
\label{syntactic_structure}
\end{figure*}

The actual text content of a text document consists of sequential information in the form of natural language. From the last century of linguistics research
into the structure of natural language, it has been established that language consists of structures at various levels. Starting from the highest level, these
are discourse, pragmatics, semantics, syntax, morphology, phonology and phonetics \citep{Jurafsky:08}. Linguistics itself has divided into subfields that 
each specialize in one of these levels. Discourse deals with topics and discussions, pragmatics with contextual meanings and interpretations,
and semantics with the meanings of linguistic constructions. Syntax deals with the generation of sentences from words, and morphology with the generation
of words from morphemes. Phonology and phonetics deal with phonemes and phones, the atomic units of speech. Figure \ref{syntactic_structure} shows
the structure at the syntactic level for the definition of TM used in this thesis, as provided by the NCLT wide-coverage parser\footnote{http://lfg-demo.computing.dcu.ie/lfgparser.html} \citep{Cahill:04}.\\

The lowest levels of language are easiest to describe formally, and syntax and the higher levels were considered too complex for formal descriptions until the 
advent of computational linguistics. While syntax and the lower levels have clearly defined elementary units such as words and morphemes, no consensus exists 
for elementary units in the higher levels of language. The levels are not strictly separated: morphosyntax, morphophonology, and
morphosemantics study some of the interactions between these levels. The levels are neither strictly hierarchical, but parallel. For example, within a word 
the morphological and syllabic segments commonly overlap: the word ``rated'' has the morphological boundaries ``rate + d'', and the
syllable boundaries ``ra - ted''. Text and speech analysis thus commonly annotates linguistic data with overlapping description tiers.\\

A further complication in natural language is ambiguity in the various levels. Ambiguous structures are very common in natural language. A common
example of ambiguity is the sentence \emph{``Time flies like an arrow; fruit flies like a banana''} \citep{Burck:65}. The words ``flies'' and ``like'' are both 
used ambiguously, ``flies'' as a verb in the first clause and as a noun the second, ``like'' as an adverb in the first and as a verb in the second. The sentence is 
also a ``garden path sentence'', since reading it forces the reader to disambiguate the word ``flies'' in the first clause, only to realize that ``time flies'' is not 
used as a noun phrase, and that the two clauses must be non-related. Resolving ambiguities requires understanding of the larger context, but generally 
provides higher efficiency to language as a form of communication \citep{Piantadosi:12}.\\


\subsection{Structured Representations for Text}

The view of text as unstructured data is misleading, but it considerably simplifies the overwhelming complexity of processing text documents. 
While human readers can easily understand most types of text documents, the simplest types are undecipherable for general computer algorithms. It is 
therefore necessary to use simplified representations of text to perform any processing of text that is natural for humans. TM uses different representations 
of text that depend on the use, also called intermediate forms \citep{Tan:99} or representational models \citep{Feldman:06}. In the following discussion, 
formal notation is introduced that will be extensively used in the later chapters of the thesis.\\

NLP commonly uses text processed into some type of \emph{normalized form}. This is a form of text with all non-linguistic elements removed or
separated from the text content, and the text is encoded using a standard such as ASCII or Unicode. The text can then be further normalized to remove 
unwanted variance \citep{Zhu:07, Puurula:09, Yang:13}. A common normalization is expansion of abbreviations and number words. Another common 
normalization is spelling correction. The normalized form depends on the intended use. For modeling text as word sequences using 
n-gram models, modeling effort can be reduced by removing sentence-initial capitalization and punctuation, and placing sentence boundary tokens.
Alternatively, the text can be normalized to recover the capitalization and punctuation instead, if the original text is missing these. A \emph{word sequence}
variable representing a normalized document of $J$ words can be formally expressed as $\ul{\bm w}$, where each integer variable $\ul{w}_{j}:1\le j \le J$ in 
the sequence indexes a word in a dictionary of $N$ possible words. Normally the dictionary size $N$ doesn't need to be defined. The dictionary can be
easily updated by maintaining a hash table, and mapping each previously unseen word to the integer value $N+1$.\\

The normalized word sequences can be further processed according to the intended use. For uses such as text classification, clustering and retrieval, there exists 
a set of common normalizations: stemming, stopwording and short word removal. Stemming removes word endings, so that for example the words
``connect, connected, connecting, connection, connections'' are all mapped to the same word ``connect'' \citep{Lovins:68, Porter:80}. This reduces variability
by performing a heuristic clustering of the words. Short words of less than three characters are removed, along with words that occur in a stop-word list: a
list of usually around 1000 common words that are not useful for the task. These linguistic processing methods depend on the language,
and in most languages advanced morphological processing is required \citep{Kurimo:10, Zhao:10}. The linguistic processing can also enrich the words
with tagging information, such as part-of-speech and dependency tags.\\

\begin{table*}
\centering
\caption{Examples of word sequence and word vector representations for two text snippets from Wikipedia as documents. For illustrating sparsity, zero counts are 
not shown for word vectors. Compared to word sequences, word vector variables grow in dimensionality and become increasingly sparse with more documents 
encoded by the dictionary}
\subtable[Normalized text form and corresponding word sequence representations with integer encoding]{
\begin{tabular}{|l|l|l|l|} \hline
document 1& $\ul{w}_j^{(1)}$ & document 2 & $\ul{w}_j^{(2)}$\\ \hline 
the & 1 & the & 1\\
book & 2 & bachman & 11\\
was & 3 & books & 12\\
released & 4 & is & 13\\
in & 5 & still & 14\\
1985 & 6 & in & 5\\
after & 7 & print & 15\\
the & 1 & in & 5\\
publication & 8 & the & 1\\
of & 9 & united & 16\\
the & 1 & kingdom & 17\\
first & 10 & although & 18\\
\hline
\end{tabular}
}
\hspace{5pt}
\subtable[Dictionary of integer encodings and word vector representations]{
\begin{tabular}{|l|l|l|l|} \hline
$n$ & word & $w_n^{(1)}$ & $w_n^{(2)}$\\
\hline
1 & the & 3 & 2 \\
2 & book & 1 &  \\
3 & was & 1 & \\
4 & released & 1 & \\
5 & in & 1 & 1 \\
6 & 1985 & 1 & \\
7 & after & 1 & \\
8 & publication & 1 & \\
9 & of & 1 & \\
10 & first & 1 &  \\ 
11 & bachman &  & 1\\
12 & books &  & 1\\
13 & is &  & 1\\
14 & still &  & 1\\
15 & print & & 1\\
16 & united & & 1\\
17 & kingdom &  & 1\\
18 & although &  & 1\\
\hline
\end{tabular}
}
\label{sequences_vectors}
\end{table*}

The normalized sequence forms used in NLP are insufficient for the common ML and DM methods that rely on data organized into vector forms. The majority of TM 
applications use a \emph{feature vector} representation of text documents originating from information retrieval \citep{Salton:63, Salton:75}. The most
basic type of a feature vector for a document is the ``bag-of-words'', where a feature vector $\bm w$ consists of the counts of each word in the document $w_n$,
where $n$ indexes the dictionary of $N$ words. The norms of a vector are used to measure document length. L1-norm is the length of the word sequence: 
$|\bm w^{(i)}|_1= J^{(i)} = \sum_n w_n^{(i)}$, whereas the "L0-norm" is the number of non-zero counts:  $|\bm w^{(i)}|_0= \sum_n \min(1, w_n^{(i)})$.
Sparsity of the vector equals the proportion of non-zero counts: $|\bm w^{(i)}|_0/N$. More complex feature vectors differ from the bag-of-words in how the 
features are chosen, and how the counts or weights of each feature are computed. Table \ref{sequences_vectors} shows a comparison of word sequences and 
word vector representations using integer counts.\\

Earlier text mining research believed that simple weighted words are not easily outperformed for most tasks \citep{Salton:88, Sebastiani:02}. Possible 
alternative features include word pairs \citep{Lesk:69}, linguistically tagged words \citep{Dave:03, Gamon:04}, factor concepts and topics \citep{Borko:63, 
Borko:64, Deerwester:88, Hofmann:99, Blei:03}, phrases \citep{Salton:88b} and parse trees \citep{Chubak:12}. Some applications such as authorship detection 
and essay scoring rely on non-typical features such as document length \citep{Larkey:98, Madigan:05}. More recently, word sequence 
features\footnote{called n-grams in publications, but these are word sequence features} have become a crucial part in some text classification tasks 
\citep{Dave:03, Gamon:04, Xia:11, Lui:12, Tsoumakas:13}. The recent results finding considerable improvements from combining other features with 
word vectors are due to the availability of more data, and advanced models for combining the feature sets. \\

For some uses binary weights $\forall n: w_n \in {0, 1}$ are sufficient. For most uses it is beneficial to weight the features so that the words relevant
for the modeling purpose are weighted higher. This results in non-negative fractional counts $\forall n: w_n \in \mathbb{R} \land w_n \geq 0$. 
With word features, the weighting functions typically dampen high count values, normalize the counts for varying document lengths, and weight the 
words according to rarity in the collection. A variety of possible weighting functions exist for choosing the weights, the most common being 
Term Frequency - Inverse Document Frequency (TF-IDF) \citep{Salton:88}. When weighting functions are used, the transformed counts can be denoted 
$\bm w$, whereas the original counts can be denoted $\bm w'$.\\ 

A collection of $I$ documents can be formalized as a set $D$, where the document variable $D^{(i)}$ for each document identifier $i$ consists of the structured 
variables used to represent the document. With word vectors for representation and no label information or other metadata, the document variable consists of 
the word vector: $D^{(i)}= (\bm w^{(i)})$. With word sequences and label variables $l^{(i)}: 1 \le l^{(i)} \le L $ of $L$ possible labels, the document variables 
would be $D^{(i)}= (l^{(i)}, \ul{\bm w}^{(i)})$.\\

The dictionary size $N$ for word vectors in a collection of millions of documents could typically be in the hundreds of thousands. Out of the possible 
words, typically only some tens or hundreds of words occur in a document. This means that the word vectors are extremely sparse, and both the dimensionality 
and sparsity increases as larger collections are processed. If counts are accumulated from all documents corresponding to a label, the label-conditional counts 
are almost as sparse. Like word vectors, most useful representations of text are \emph{high-dimensional sparse data}.\\

\begin{table*}
\centering
\begin{tabular}{|l|l|l|} \hline
word & $n$ & postings list\\
\hline
the & 1 & (1, 3) (2, 2)\\
book & 2 & (1,1)\\
was & 3 & (1, 1)\\
released & 4 & (1, 1) \\
in & 5 & (1, 1) (2, 1)\\
1985 & 6 & (1, 1)\\
after & 7 &(1, 1)\\
publication & 8 & (1, 1)\\
of & 9 & (1, 1)\\
first & 10 & (1, 1)\\ 
bachman & 11 & (2, 1)\\
books & 12 & (2, 1)\\
is & 13 & (2, 1)\\
still & 14 & (2, 1)\\
print & 15 & (2, 1)\\
united & 16 & (2, 1)\\
kingdom & 17 & (2, 1)\\
although & 18 & (2, 1)\\
\hline
\end{tabular}
\label{inverted_index}
\caption{Inverted index representation for the documents shown in Table \ref{sequences_vectors}. The postings lists are non-positional and
unweighted, containing only the document identifiers and word counts contained in the document word vectors}
\end{table*}

Representing a collection of high-dimensional sparse data can be done with an inverted index \citep{Zobel:06}, enabling scalable retrieval of documents as well as other types
of inference \citep{Yang:94, Shanks:03, Kudo:03, Puurula:12}. The scalability of modern web search engines is largely due to the representation of web pages 
using inverted indices \citep{Witten:94, Zobel:06}. An inverted index stores a document collection as a table of dictionary words or \emph{terms},
and a \emph{postings list} of document occurrences of the term $n$. Table \ref{inverted_index} illustrates an inverted index representation
for the example documents shown in Table \ref{sequences_vectors}. The inverted index representation is highly efficient, since the term occurrences are sparse 
and zero counts do not need to be considered when constructing, storing or using the index. Normally a posting contains a document identifier and the number 
of occurrences of the term in the document. Position information is sometimes included in the postings, for ranking functions that benefit from proximity 
information. The postings lists are commonly compressed for additional storage savings, and methods for further improving the efficiency of indices constitute 
an extensive literature \citep{Witten:94, Zobel:06}. Use of inverted indices can be described as a type of sparse matrix computation applied to text, 
although this view is not ordinarily taken in IR.\\

\subsection{Text Mining Applications as Machine Learning Tasks}

TM is applied in numerous ways across application domains. One possible way to categorize the applications is to use ML terminology and
consider the underlying learning problem that is solved in each application \citep{Feldman:06, Aggarwal:12}. This makes it possible to compare solutions used 
in different applications, and attempt solutions used for non-text data of the same task type. The basic framework of ML is described next, followed by a mapping 
of many common TM tasks into ML problems.\\

A commonly accepted definition of ML is: \emph{``A computer program is said to learn from experience $E$ with respect to some class of tasks 
$T$ and performance measure $P$, if its performance at tasks in $T$, as measured by $P$, improves with experience $E$.''} \citep{Mitchell:97}. For an example 
application of e-mail spam filtering, the experience $E$ could be a training dataset of spam/non-spam e-mail examples, the task $T$ could be the binary
classification of new examples into spam/non-spam, and the measure $P$ could be the percentage of correctly classified e-mails.\\

We can consider document collections $D$ a type of dataset for ML algorithms. The first division in learning problems is between inductive and transductive 
learning \citep{Gammerman:98, Joachims:99}. Given a training dataset $D$, inductive learning attempts to learn a function for general prediction, 
usually for making predictions on unseen data. Transductive learning does not attempt 
to learn a general function, but rather attempts to transfer properties found in a training dataset $D$ to a test dataset $D^\ast$. The transduced properties
are normally the label information available for dataset $D$, but not for $D^\ast$. Transduction improves prediction quality,
but the solutions will be optimized only for the used test set.\\

A second division in learning problems is between supervised, semisupervised and unsupervised learning. In supervised learning, a training dataset
is provided with label variables, whereas in unsupervised learning no label variables are provided. Supervised learning is considerably easier than unsupervised 
learning, as the label variables are usually reliable information for learning the function of interest. Unsupervised learning has to constrain the learning problem
to compensate for the lack of label information. For example, in a text clustering task the number of clusters is often fixed and prior distributions can be used to 
guide the learning towards a more plausible clustering. The corresponding supervised task of text classification would have the label variables provided, so that
the number of labels and their assignments to documents would not need to be learned, unlike in the unsupervised case. In semisupervised learning some portion 
of the label variables are provided. Semisupervised learning is very common in TM tasks, for example in many text classification applications there are 
a small number of labeled documents compared to large quantities of unlabeled documents.\\

The type of variables to be predicted divides ML problems further. Classification deals with the prediction of discrete label variables. Ranking deals with the ordering 
of discrete label variables. Regression deals with prediction of continuous label variables. The predicted variables can be structured. Binary classification deals 
with binary label variables $l\in {0, 1}$, multi-class classification with categorical label variables $l: 1\le l \le L$, multi-label classification with label vectors 
$\bm c= [c_1, ...,  c_L]$ of binary variables \citep{Tsoumakas:10} and multi-dimensional classification with label vectors of categorical variables \citep{Bielza:11}. 
Corresponding divisions exist for ranking and regression problems, as well as multi-output or multi-target prediction problems that contain mixed output variables.\\

More complex structured prediction problems arise when the input variable is not a simple vector. Sequence labeling classifies sequence variables into
corresponding structured variables. For example, syntactic parsers map word sequences into parse trees, and speech recognition maps sequences of acoustic
feature vectors into word sequences. Solutions for structurally complex problems often have a number of uses, and produce models that provide information 
about the learning problem. Multi-task learning attempts to solve different tasks at the same time, taking advantage of the related optimization problems 
to find better solutions \citep{Caruana:93}.\\

\begin{table*}
\centering
\caption{Mappings of TM applications to ML tasks}
\begin{tabular}{|l|l|l|l|} \hline
Application & Task & Publication \\ \hline 
sentiment analysis & binary classification & \citep{Pang:02}\\ 
spam filtering & binary classification & \citep{Medlock:06}\\ 
email categorization & multi-class classification & \citep{Joachims:97}\\
price prediction & multi-class classification & \citep{Lee:14}\\
news categorization & multi-label classification & \citep{Lewis:04}\\
document retrieval & ranking & \citep{Metzler:07}\\
sales prediction & regression & \citep{Archak:07}\\ 
essay grading & regression & \citep{Larkey:98}\\
patent mapping & clustering & \citep{Fattori:03}\\
event detection & clustering & \citep{Allan:98}\\
entity recognition & sequence labeling & \citep{McCallum:03}\\
\hline
\end{tabular}
\label{applications}
\end{table*}

Table \ref{applications} shows a sample of common TM applications and their mapping into ML tasks. An application can be mapped into a ML task in several 
different ways. For example text regression problems \citep{Archak:07, Joshi:10, Wang:10b, Archak:11, Ghose:11, Cao:11, Higgins:14, Lee:14} such as stock 
price prediction can often be solved using regression or classification. Commonly, the main improvements in ML come from defining the task 
well and choosing the features useful for that task, rather than the choice of learning
algorithms. Multiple ways of approaching a problem can work, and the combination of different types of solutions is highly beneficial. Complex learning
approaches are not necessary for applying the ML framework: classifiers such as K-nearest Neighbours \citep{Cover:67} and Naive Bayes (NB) \citep{Maron:61} 
can operate on the basis of counted training dataset statistics, without the use of iteratively learned parameters.\\

The performance measures $P$ depend on the application and task. For general classification tasks, accuracy is defined as the percentage of correctly
classified instances for a given test set. This might be ill-suited, if the use of the classification system is to find some relevant documents for each possible label.
Extensively studied tasks have highly specialized measures that attempt to quantify the usefulness of the ML system in the applications, such as NDCG that 
is used to measure ranking performance in web search engines \citep{Jarvelin:02}.\\

The ML framework involves segmentation of datasets into training and test portions, so that the performance is not measured on the same data that is used
to learn parameters. The most typical split is between a training portion for learning parameters, a development portion for calibrating meta-parameters of the 
algorithms, and a final test portion that is used for evaluation. Alternatively, cross-validation segments a dataset into a small number of exclusive training 
and development portions, and the performance measure can be averaged across the folds. More complex solutions can have a number of nested dataset 
segmentations, reserving unused testing data to optimize each layer in a system.\\

\subsection{Linear Models as Methods for Text Mining}

Mapping TM applications into established tasks enables the use of existing methods for solving problems. Earlier methods for TM relied on linguistic
methods for performing text preprocessing, and algebraic methods for performing text retrieval. The more recent methods for TM are algorithmic, often
model-based, and predominantly originate in statistics and ML. The new algorithmic methods based on statistics and learning have brought a paradigm shift 
in the field of artificial intelligence, and there remain few areas of TM where solutions based solely on domain expert-knowledge are preferred. Commonly,
domain knowledge such as stemmers and sentiment lexicons are used as additional information for learning algorithms.\\ 

Most algorithms on word vectors and related representations are applications of linear models, that perform predictions using linear combinations of feature 
values weighted by learned parameters. The tasks that are commonly solved using linear models include regression, classification, ranking and clustering.
In text regression regularized linear regression can be applied \citep{Archak:07, Joshi:10, Higgins:14}. In text classification classifiers such as 
Centroid \citep{Rocchio:71, Han:00}, Bernoulli Naive Bayes (BNB) \citep{Maron:61}, Logistic Regression (LR) and Support Vector Machines (SVM) \citep{Joachims:98, Fan:08} 
are linear models. In text ranking and text retrieval, all of the common scoring functions are linear models, including the Vector Space Model (VSM) \citep{Salton:75}, 
language models \citep{Kalt:96, Hiemstra:98}, as well as more recent discriminative ranking models \citep{Metzler:07}. Text clustering commonly uses linear
models, such as multinomial and Cosine distances \citep{Pavlov:04, Zhong:05, Banerjee:05, Rigouste:07}. The following presents a succinct overview of the 
linear model framework for TM.\\

A basic type of statistical model for solving modeling problems is the linear regression model, as used for solving text regression problems. Let us assume 
word vector features $\bm w$, with a dictionary of $N$ words. Let $\bm \theta$ denote a parameter vector of weights for the regression model, where $\theta_0$ 
is called the ``bias'' parameter and $\theta_n$ for values $1\le n \le N$ are the regression weights for each word feature $n$\footnote{Using $\bm w$ for the 
weight vector is the common notation with regression models \citep{Bishop:06}. The different notation $\bm \theta$ is used here to keep the notation consistent 
throughout the thesis.}. A linear regression predicting \emph{scores} $y(\bm \theta, \bm w)$ of a predicted continuous variable $y$ takes the form:

\begin{align*}
y(\bm \theta, \bm w)= \theta_0 + \sum_{n=1}^{N} \theta_n w_n
\numberthis \label{regression}\\
\end{align*}

The weights $\theta_n$ decide how much the predictors $w_n$ explain the observed variations of $y$ seen in a training dataset, where $y$ and
$\bm w$ are available, and $\bm \theta$ needs to be estimated. A basic way to estimate the parameters $\bm \theta$ is the method of least squares, so that
the sum-of-squares error function over the dataset $E_D(\bm \theta)=1/2 \sum_i (y^{(i)}- \bm \theta^T \bm w^{(i)})^2$ is minimized. This is equivalent
to maximum likelihood estimation assuming that the errors are generated by a Gaussian noise model.\\

Most applications of linear models use regularization to control overfitting \citep{Frank:93}. This adds a regularization term $R(\bm \theta)$ to the error function, 
so that the total error function becomes $E'_D(\bm \theta)= E_D(\bm \theta) + \Lambda R(\bm \theta)$, where $\Lambda$ is a weight for the 
regularization. Regularization often takes the form $1 /2 \sum_n |\theta_n|^q$, where $q$ is the L-norm of the regularizer. A common case $q=1$ is called the 
Lasso regularizer, and the case $q=2$ is the ridge or Tikhonov regularization. Regularization causes parameter estimates to shrink towards more 
conservative values, to zero in the case of the sparsity-inducing Lasso regularizer.\\

\begin{figure*}
\centering
\subfigure[Linear classifier: L1-regularized Logistic Regression]{
 \includegraphics[scale=0.5, trim=310 120 310 120, clip=true]{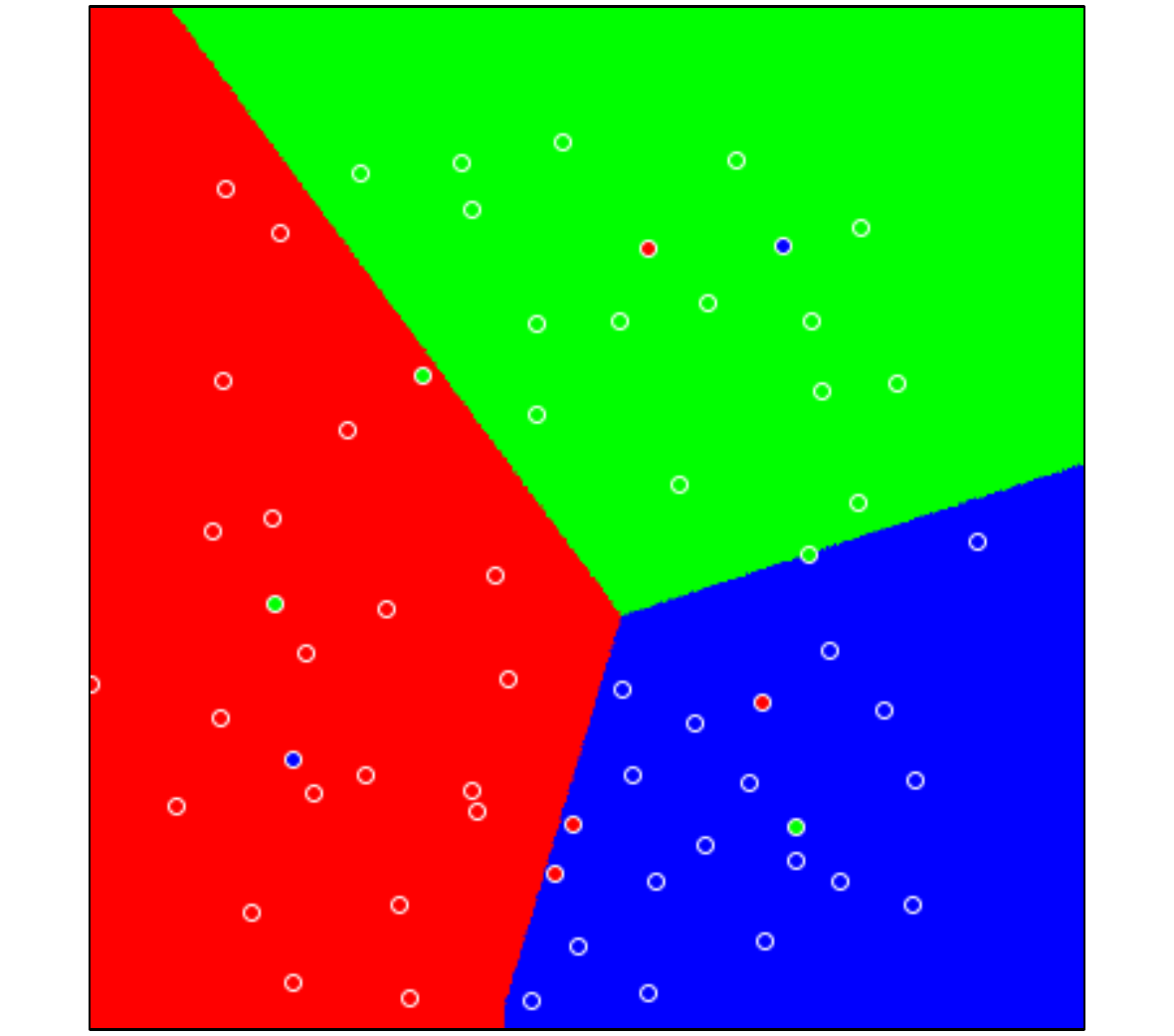}
}
\hspace{5pt}
\subfigure[Non-linear classifier: k=1 Nearest Neighbour]{
 \includegraphics[scale=0.5, trim=310 120 310 120, clip=true]{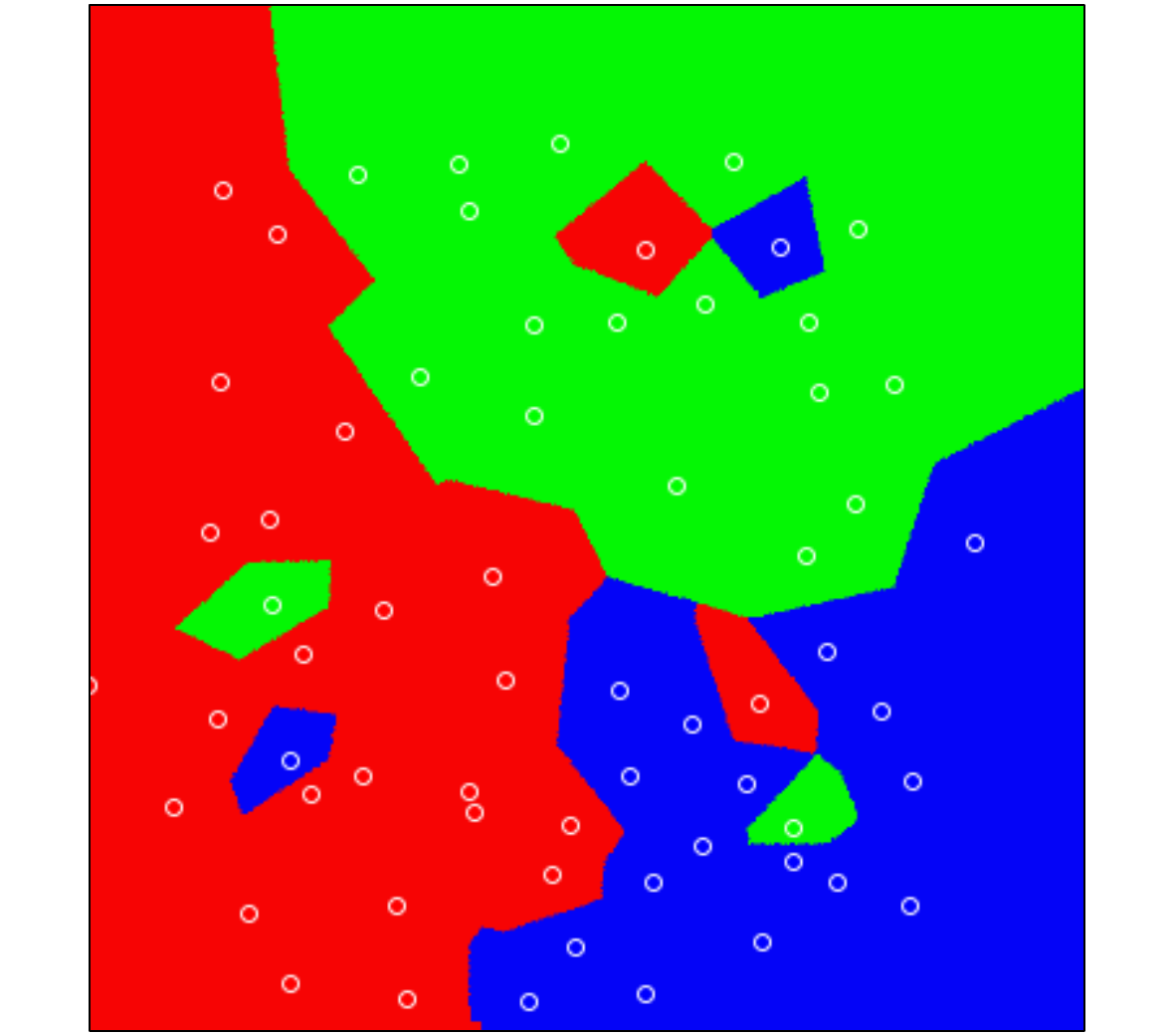}
}
\caption{Visualization of decision boundaries of a linear and non-linear classifier on 2-dimensional 3-class data. Logistic Regression forms linear decision boundaries,
K-Nearest Neighbours forms non-linear boundaries}
\label{linear_nonlinear}
\end{figure*}

Linear models for classification and ranking apply a further decision rule on the scores $y(\bm \theta, \bm w)$ to map the scores into categories and rankings. Binary
classification maps the scores based on the sign of the score: with a classification threshold of $0$, if $y(\bm \theta, \bm w) \ge 0$, $l=1$, else $l=2$. Multi-class 
classification involves label-dependent parameter vectors $\bm \theta_l$, and maps the scores by maximizing the score: $\argmax_l (y(\bm \theta_l, \bm w))$. 
Ranking sorts the scores for the parameter vectors, and maps the order of labels into ranks \citep{Metzler:07}. The multi-class linear scoring function can be 
expressed as:\\

\begin{align*}
y(\bm \theta_l, \bm w)= \theta_{l0} + \sum_{n=1}^{N} \theta_{ln} w_n
\numberthis \label{linear_score}\\
\end{align*}

This elementary function covers a swath of modeling approaches, with highly different semantics for the bias parameters $\theta_{l0}$ and the 
label-dependent parameters $\theta_{ln}$. In classification, models that can be expressed in the form of Equation \ref{linear_score} are called
linear classifiers, since they form linear decision boundaries as a function of the feature vectors. Figure \ref{linear_nonlinear} illustrates the decision boundaries
of a linear classifier compared to a non-linear classifier. With probabilistic approaches, the scores are related to the 
posterior probability of the label given the data through a link function. For example, with 
exponential-family models such as LR, BNB and Multinomial Naive Bayes (MNB), the posterior 
probabilities are $p(l| \bm w)= \exp(y(\bm \theta_l, \bm w))$.
With non-probabilistic approaches such as SVM, the scores are optimized entirely for classification and do not represent probabilities.\\

Estimating the linear model parameters depends on the models and the strategies used to reduce overfitting. With algebraic methods such as the Centroid Classifier 
and generative probabilistic models such as MNB, the label-dependent parameters $\theta_{ln}$ are estimated for each label independently,
while the bias parameters $\theta_{l0}$ are assumed uniform or estimated separately. With generative models, $\theta_{ln}= \log(p_l(n))$ are label-conditional 
log-probabilities, $\theta_{l0}= \log(p(l))$ are label prior log-probabilities, and both types of parameters can be smoothed and scaled to correct for overfitting. With
discriminative classifiers such as LR and SVM, the parameters are estimated by minimizing a regularized error function
\citep{Fan:08}, similarly to learning the regularized linear regression.\\

\begin{table*}
\centering
\caption{Parameter estimates for the common linear models used in TM. VSM, BMB and MNB are used for ranking, classification, and clustering,
BM25 is used for ranking and regularized LR/SVM for classification. BNB and MNB show unsmoothed parameter estimates. For BM25 a unique
label exists for each document: $l^{(i)}= i$, and for LR/SVM the labels are binary: $l^{(i)} \in (-1, 1)$.
$IDF(n)$ and $LN(i)$ for BM25 refer to the chosen IDF and length normalization functions, respectively. $R(\bm \theta)$
and $L(\bm \theta, D^{(i)})$ for LR/SVM refer to the chosen regularization and loss functions, respectively.}
\begin{tabular}{|l|l|l|} \hline
Model & Parameters $\bm \theta$ & Test $w_n$ \\ \hline
VSM \citep{Rocchio:71} & $\theta_{ln}= \frac{\sum_{i:l^{i}= l} w_n^{(i)}} {\sqrt{\sum_{n'}^N (\sum_{i:l^{i}= l} w_{n'}^{(i)})^2}}$ & $\frac{w'_n}{|\bm w'|_2}$\\
BNB \citep{Maron:61} & $\theta_{ln}= \log  \sum_{i:l^{i}= l} \frac{\min(1, w_n^{(i)})}{I}$ & $\min(1, w_n'^{(i)})$\\
MNB \citep{Kalt:96} & $\theta_{ln}= \log \frac{\sum_{i:l^{i}= l} w_n^{(i)}}{\sum_i \sum_n w_n^{(i)}}$ & $w_n'$\\
BM25 \citep{Manning:08} & $\theta_{ln}= IDF(n) \frac{(k_1 +1) LN(i)} {(k_1) LN(i)} $ & $\frac{(k_3+1)w_n'}{k_3+w_n'}$\\
LR/SVM \citep{Fan:08} & $\min_{\bm \theta} R(\bm \theta) + C \sum_i L(\bm \theta, D^{(i)})$ & $w_n'$ \\
\hline
\end{tabular}
\label{linear_models}
\end{table*}

Table \ref{linear_models} summarizes the parameter estimates for the commonly used linear models in TM. For BM25, the Croft-Harper IDF function is 
often used: $IDF(n)= \log \frac {I - I_n + 0.5}{I_n + 0.5}$ \citep{Manning:08}, where $I_n$ is the number of documents where the word $n$ occurs. Another 
common IDF function with
BM25 is $IDF(n)=  \log \frac {I +1}{I_n + 0.5}$ \citep{Fang:04}. The soft length normalization for BM25 is given by 
$LN(i)= w^{(i=l)}_n / (1- b+ b(|\bm w^{(i=l)}|_1/ A))$
, where the average document length $A$ is $\sum_i |\bm w^{(i)}|_1/I$. The loss function $L(\bm \theta, D^{(i)})$ for SVM/LR is 
$\log(1+ \epsilon^{-l^{(i)}\bm \theta^T \bm w^{(i)}})$ for LR, $\max(0, 1 - l^{(i)}\bm \theta^T \bm w^{(i)})$ for L1-loss SVM and 
$\max(0, 1 - l^{(i)}\bm \theta^T \bm w^{(i)})^2$ for L2-loss SVM \citep{Fan:08}. The regularization $R(\bm \theta)$ for SVM/LR is 
$\frac{1}{2} |\bm \theta|_2^2$ for L2 regularization and $|\bm \theta|_1$ for L1 regularization. BM25 requires the meta-parameters 
$k_1$, $k_3$ and $b$, LR/SVM requires the meta-parameter
$C$ for regularization. Use of feature transforms and smoothing for VSM, BNB, and MNB modifies the equations in Table \ref{linear_models}, and 
introduces additional meta-parameters.\\

Inference for different uses with parameters in the form of Equation \ref{linear_score} is a trivial summation and maximization for classification, and
summation and sorting for ranking. The algorithms used for estimation differ widely. For the Centroid Classifier and MNB, the estimation is 
a simple closed-form linear-complexity procedure of summing, normalizing and possibly smoothing the word count statistics. For LR and SVM, the 
algorithms depend on the error function and regularization \citep{Yuan:12}. Many of the practical models have the property of 
a convex error function, so that the estimation can be performed using efficient Gradient Descent algorithms, including the online version Stochastic Gradient 
Descent \citep{Bottou:10, Duchi:11, Bottou:12}. Stochastic Gradient Descent is applied on a large class of models, including L2 and L1-regularized 
LR \citep{Carpenter:08, Tsuruoka:09}, and often outperforms methods tailored for the particular problem.\\

Linear models can be extended in a number of ways to represent non-linear decision surfaces \citep{Keysers:03, Bishop:06, Chang:10}. The simplest 
way is mapping the original feature vectors to transformed ones, examples of which are factor decompositions \citep{Borko:64, Blei:03},
word pair features \citep{Lesk:69}, and explicit polynomial mappings \citep{Chang:10}. Generalized linear models apply an implicit link function to transform 
different types of prediction tasks into linear regression modeling problems, LR using the logit function being one example. Other types of
generalized linear models are not necessarily linear models in the sense of linear classifiers and the definition of Equation \ref{linear_score}. Replacing
the multinomial event model in MNB with a Gaussian model would likewise result in non-linear boundaries.\\

A second type of extension into non-linear decision boundaries is utilizing information present in individual documents of the collection. K-Nearest Neighbours
\citep{Cover:67}, Kernel Density Classifiers \citep{Parzen:62} and Mixture Models \citep{Li:97} are models that maintain parameters for a set of prototypes
for each class, and combine scores for each class from the prototype scores. These models can capture properties of local neighbourhoods in the documents
that would be lost with the representation of a single parameter vector. Kernel learning methods \citep{Boser:92, Joachims:98} use feature 
transformations called kernels into arbitrary spaces that are not explicitly computed, but rather evaluated implicitly by the learning algorithm. This gives kernel
learning a great deal of flexibility in learning decision surfaces, but with a computational cost that is not always preferable over a linear kernel maintaining
the original feature space \citep{Fan:08, Yuan:12}.\\

A third type of non-linear extension is multi-layer methods, such as tree-based learning \citep{Breiman:84, Quinlan:86}, neural networks 
\citep{Rosenblatt:58, Widrow:60}, 
and ensemble learning \citep{Breiman:01b, Friedman:02, Sill:09}. All of these combine layers of elementary base-learner algorithms, often dividing the 
original documents and feature vectors into different subsets for the base-learners. Tree-based methods combine component learners similar to mixture
models, but combine the components using hard decisions based on rules that best segment the data, rather than performing soft combination with fixed mixture 
weights assigned to each component. Neural networks extend simple base-learners such as LR with hidden layers of learners, with higher
modeling flexibility, but also introducing a difficult learning problem that is commonly approached using Stochastic Gradient Descent \citep{Widrow:60, Bottou:12} 
combined with heuristics. Ensemble methods combine a set of diverse base-learners to optimize a performance measure, commonly selecting the optimal
set of base-learners for the task and learning the best possible combination \citep{Sill:09, Puurula:12b}.\\

A further extension of linear models in TM is prediction and modeling in tasks that require structured variables, some of which cannot be accurately solved by 
decomposing the problems into simple linear problems. Examples of such tasks are entity and event detection performed
in information extraction, and syntactic tree generation in parsing sentences. However, a majority of the methods used for solving these problems are 
extensions of basic linear models into structured prediction: Conditional Random Fields \citep{Lafferty:01} extend LR, Hidden Markov 
Models \citep{Baum:70, Kupiec:92} extend Naive Bayes, and Max-margin Markov Networks \citep{Taskar:03} extend SVM. Structured prediction
models extending Naive Bayes are described in Chapter 3, and the methods developed in Chapters 4 and 5 can be equally extended into structured modeling.\\

\subsection{Text Mining Architectures as KDD Processes}

Preprocessing data to structured forms and applying ML to solve tasks forms the basic building blocks of TM. Combining these into complex solutions for TM
applications often requires integration of the available components into an \emph{architecture} for the TM application \citep{Feldman:06, Villalon:13, Maciolek:13}.
The concept of a TM architecture originates from viewing TM as a case of the KDD process \citep{Feldman:95, Feldman:97, Ahonen:97}.\\

A basic KDD process is defined as consisting of five steps \citep{Fayyad:96}: 1) selection, 2) preprocessing, 3) transformation, 4) data mining and 
5) interpretation/evaluation. Selection chooses documents and variables of interest for further analysis. Preprocessing consists of modeling noise and missing 
data. Transformation reduces and transforms the number of considered variables to a form more suited for analysis. Data mining applies algorithms to
find interesting patterns. Interpretation/evaluation performs interpretation of the discoveries, possibly visualizing the models or the data using the models. 
The number of five steps is not fixed, but an example of a possible basic process. All of the steps are interactive, with the user iteratively modifying the steps
and cycling through the process to discover more knowledge from the database.\\

TM was proposed in its earliest forms as KDD with an additional text-specific preprocessing step \citep{Ahonen:97, Dorre:99}. This text preprocessing
step consisted of preprocessing each document into a feature vector, and filtering of the feature vector to a form more easily processed by 
standard DM algorithms. It was also suggested that the filtering step was needed for scalability, as the resulting feature vectors would be
exceedingly high dimensional for the usual DM algorithms. Text datasets have since grown thousands of times in all relevant dimensions, 
and architectural decisions have been suggested to maintain scalability \citep{Villalon:13, Maciolek:13}.\\

Current TM architectures can perform the selection step by applying a search engine \citep{Villalon:13}, or applying a web crawler \citep{Maciolek:13} to 
retrieve documents related to the TM application. Unlike in typical KDD, the whole collection is therefore not necessarily known or available, but a sample of the 
vast amount of possible data is gathered in the first step. The preprocessing step can use extensive linguistic processing \citep{Villalon:13}, such as tagging 
words and phrases according to syntactic roles, identifying named entities and events, and categorizing documents into ontologies. The remaining basic
steps of transformation, data mining and interpretation/evaluation largely follow the general KDD process, but with some text specific solutions: transformation
can be done with topic modeling \citep{Hofmann:99, Blei:03} instead of general matrix factorization methods, data mining is done with algorithms that operate
well on high-dimensional sparse data such as Naive Bayes, and visualization is done with tools such as word clouds \citep{Silic:10}. \cite{Feldman:06} considers
domain knowledge sources as universally important for TM applications and presents extensive use of domain knowledge throughout TM architectures.\\

\begin{figure*}
\centering
\includegraphics[scale=0.5, trim=100 150 100 100, clip=true]{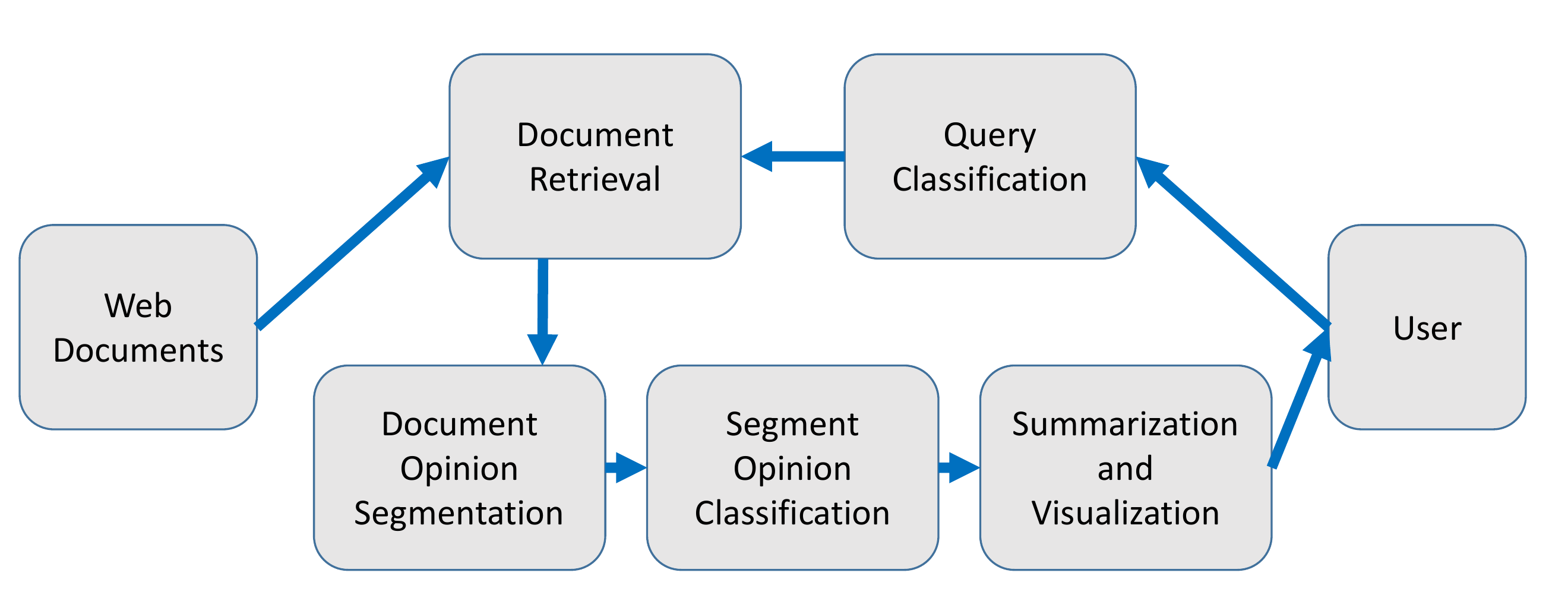}
\caption{Possible TM architecture of an opinion search engine}
\label{architecture}
\end{figure*}

As an example of a TM architecture for an application, a framework presented in a survey of opinion mining \citep{Pang:08} can be illustrated. This 
divides the construction of an opinion search engine system into four problems: 1) classification of queries into opinion/non-opinion related queries, 2)
finding matching documents and segmenting parts of opinionated content, 3) classifying the opinion related to the query in the relevant parts of documents, 
and 4) presenting the gathered opinion information with summarization and visualization. Figure \ref{architecture} shows a basic TM architecture for
this application, decomposed into ML tasks.\\

The KDD process appears to encompass everything contained in TM, but it can also be too general to describe some TM applications. Many applications of
TM involve systems that use the methods of TM such as inverted indices and domain knowledge resources, but do not aim at discovering new types
of knowledge, or require an interactive user. Similarly to information extraction, the outcome for TM can be one of known patterns, produced by a fully
automatic system. A typical case is text classification applications such as email spam and newspaper topic classification, that produce 
expected results automatically.\\


\section{The Scalability Problem}
\subsection{Scale of Text Data}

The information explosion has brought an overwhelming amount of data available to ordinary people and large institutions alike,
much of it in the form of text. A common truism originating from business intelligence research 
is that 80\%-90\% of data in corporate databases is in the form of unstructured text \citep{Rajman:97, Dorre:99, Godbole:08}.
Certainly a large majority of consumed data is in the form of written text such as newspapers, books, blogs, micro-blogs 
and chat messages. There is more textual data produced electronically in a single day than any individual person can digest in a 
lifetime, and the rate of production is increasing rapidly.\\

What has changed is not only the \emph{scale} of data, but its \emph{availability} and \emph{accessability}. For example, the largest 
library to date is the The British Library, containing 170 million items, closely followed by the Library of Congress with 152 million 
items\footnote{http://en.wikipedia.org/wiki/List\_of\_largest\_libraries}. Although considerable in scope, these repositories of 
data are not so readily accessed as databases existing digitally. The information explosion is not only making vastly larger 
amounts of data available, but making them rapidly accessible through technologies such as IR and TM.\\

The available text data is continuously expanding and of vast scale in several ways:
\begin{description}
\item[Number of Documents.]
The number of documents in many datasets
and streams is measured in millions, and in some cases billions. The online encyclopedia Wikipedia has 4.5 million articles in English
as of 2014\footnote{http://en.wikipedia.org/wiki/Wikipedia:Size\_of\_Wikipedia}. Google Books had digitized 30 million books by 
2013\footnote{http://en.wikipedia.org/wiki/Google\_Books}. The micro-blog provider Twitter announced in 2013 
that its 200 million users were sending 400 million tweets per day\footnote{https://blog.twitter.com/2013/celebrating-twitter7},
more than the global SMS mobile text message traffic combined. The popular social messaging app developer 
WhatsApp\footnote{http://techcrunch.com/2014/01/20/whatsapp-dld/} announced in 2014 that its 430 million users send 50 billion messages per day. 
The scale of data is such that the nascent field of stream mining has emerged
as a possible solution, because in many cases merely storing all this data is not practical or even feasible.\\

\item[Structural Metadata.]
Aside from the sheer number of documents available in databases and streams, text data in most cases comes with \emph{implicit} and \emph{explicit}
metadata. Implicit metadata is unstructured information such as topics, sentiment and named entities that can be discovered using text 
mining methods and incorporated into the document. Explicit metadata is information such as document hierarchy categorization, link 
data and various forms of tags that are attached to the document. For example, a Wikipedia article has category
labels and different types of external and internal links. The internal links alone can point to any of the millions of documents 
in Wikipedia, whereas the number of categories is close to half a million in the English Wikipedia. Various types of metadata are practically always present with text data, and the dimensionality of each type can reach millions or more.\\

\item[Representation Dimensionality and Sparsity.]
Once extracted from documents, unstructured text needs to be represented with a structured representation such as the word vector for most types of processing. With most useful structured representations, the dimensionality of the representation grows with the number of documents. Word vectors can grow to millions 
of words. At the same time, most useful representations are inherently sparse: while word vectors grow to millions, each individual document typically 
contains only some tens of unique words and the corresponding word vector is almost entirely empty.\\
\end{description}

Vast datasets have become common in TM research as well. Google N-grams\footnote{http://catalog.ldc.upenn.edu/LDC2006T13} 
released in 2006 contains 5-gram models estimated from 1 trillion words of web text \citep{Brants:07}, and has been widely used
for a great variety of TM tasks. The Annotated Gigaword\footnote{http://catalog.ldc.upenn.edu/LDC2012T21} is a collection of over 4 billion words of English text news, enriched with sentence segmentation, parse trees, dependency trees, named 
entities and in-document coreference chains \citep{Napoles:12}. Less annotated gigaword corpora have been produced for 
Arabic, Mandarin, Spanish and French. The 1B Word Language Modeling Benchmark\footnote{http://code.google.com/p/1-billion-word-language-modeling-benchmark/} started at the end of 2013 is a freely available billion word dataset for comparing 
progress in language modeling \citep{Chelba:13}. Text classification has started to tackle online ontology classifications such 
as Wikipedia article categorization\footnote{http://lshtc.iit.demokritos.gr}, where the number of categories reaches hundreds 
of thousands, and combinations of categories reach millions \citep{Puurula:12b}. Classification and retrieval tasks for 
web ontologies can have hundreds of millions of documents, accessed via cloud-based storage 
systems\footnote{http://trec-kba.org/} \citep{Gross:13}.\\

\subsection{Views on Scalability}
By definition TM seeks methods that discover new information by examining large quantities of data, as opposed to 
small-scale analysis that can be done by trained human specialists. There are three ways to view scalability in TM. 
Scalability can as an \emph{avoidable} challenge, as a \emph{beneficial} factor in building TM systems, or as 
\emph{necessary} for many TM tasks. This thesis argues that scalability is both beneficial and necessary. These 
three views are discussed next.\\

Scalability can be seen be seen as something to be avoided in the simplest way possible. It affects 
both system-level architectural and component-level algorithmic design. At the component level, most methods 
that originate in the related technical fields have problems adapting to the TM domain. IR has mostly worked with 
scalable methods for accessing text, but most methods originating within NLP, ML and DM are less scalable.\\

Although text is the natural domain of NLP, most NLP methods have been designed on much smaller corpora and emphasize 
theory, not practical large scale processing. Some methods such as shallow parsers and taggers can operate with linear 
complexity, but methods such as deep parsers and discriminative models are mostly non-linear and scale poorly to vast 
numbers of documents \citep{Lafferty:01, Blei:03, Collobert:11}. Non-NLP ML methods have mostly been defined for tasks on 
much lower dimensions. SVM excel in text classification effectiveness, and are scalable in both documents and 
features, but less so in the number of classes. Instance-based learning methods such as K-nearest Neighbours scale well to any amount of 
training data and classes, but are not generally competitive in effectiveness with very sparse high-dimensional vectors, 
and can become inefficient in inference with vast numbers of training documents. Decision trees in general do not scale to large 
numbers of classes or documents with high effectiveness. In brief, the majority of ML and DM methods 
are ill-suited to the scalability required in the text domain.\\

Some of the problems with unscalable components can be corrected by well designed system-level architectures. If a simple 
directional processing pipeline is used, the scalability of the TM system equals the scalability of the weakest component in the pipeline. 
Good decisions at the system level can reduce the bottlenecks in processing. Architectural design 
solutions can improve scalability, but designing better architectural solutions can be much harder than using better components. This is evidenced
in web-scale search engines, where answers to many fundamental questions in architectural design are still largely unknown, very difficult to 
exhaustively answer, and subject to change as the web evolves \citep{Cambazoglu:13, Asadi:13}. Before embarking on the difficult task of 
developing techniques for scalable TM, it is therefore necessary to ask whether scalability is warranted, or whether TM should be confined
to smaller problems using less scalable tools from the related technical fields. Certainly for most practitioners experienced in the related
fields, the obvious solution for TM would be to work on smaller problems with well-understood and widely adopted tools, coupled with 
data and feature selection, and the affordable large-scale parallelization enabled by cloud computing.\\

Scalability can also be seen as desirable. NLP has seen an increasing appreciation of scalable processing methods in the last decade. Motivated by 
the success of large-scale language models used in speech recognition and machine translation \citep{Buck:14}, the usefulness of large datasets for other 
NLP tasks has become widely acknowledged in a variety of supervised tasks, including disambiguation \citep{Banko:01}, parsing \citep{Pitler:10}, spelling 
correction \citep{Bergsma:10}, segmentation \citep{Huang:10b} and punctuation recovery \citep{Lui:13}. A general observation has been that 
effectiveness in tasks improves log-linearly with the amount of data \citep{Pitler:10}, so that each multiplication of the amount of data produces a 
constant relative improvement. This is in line with the improvement observed in large scale n-gram language models; there seems to be no limit to 
how much improvement comes from more data, as long as the models have sufficient complexity. For example, bigrams seem to 
saturate practically at some hundreds of millions of words, whereas trigrams should saturate at some billions of words \citep{Rosenfeld:00}.
Higher-order n-grams and other sufficiently complex models can learn from text data without saturating in the same manner 
\citep{Brants:07, Huang:10b, Buck:14}.\\

Many TM tasks have small amounts of labeled data available compared to large quantities of unlabeled training data.
Text classification tasks often require this type of semi-supervised learning. Both estimation based on the EM-algorithm \citep{Nigam:06}
and semi-supervised estimation \citep{Su:11} enable the utilization of unlabeled data. Large-scale data can also be used in unsupervised learning 
for tasks that do not require label information, such as the numerous uses of n-gram model counts \citep{Lapata:05}. Alternatively, 
large-scale unlabeled datasets can be automatically labeled, producing reusable machine-annotated resources for supervised 
tasks \citep{Napoles:12}.\\

The most recent view is that scalability is necessary for TM \citep{Rehurek:11}. Traditional IR methods have been developed to scale as
much data as possible, since IR systems are intended to retrieve as many relevant documents as possible. Unlike in NLP, scalability
has always been a requirement of IR. Similarly, some recent text classification tasks require classification into potentially millions of
classes, as in the case of Wikipedia categorization \citep{Puurula:12}. Tasks such as these require scaling to large
dimensions, and compromising dimensionality would reduce effectiveness of the methods considerably. It can
be further argued that some technologies, such as those utilizing machine translation, only became popular with the arrival of 
n-gram models trained on trillions of words \citep{Brants:07}.\\

This view of scalability as necessary stems from the definition of TM. In contrast to text analysis, TM
connotes a process of sifting through large quantities of less valuable material to find material of interest. Without data of vast scale,
TM reduces to a mere intersection of NLP and ML. Many new text processing tasks 
are starting to require processing at vast scale. The resulting processing systems are best called TM systems, as they incorporate 
methods from a variety of disciplines, and do not fit neatly into any of the related technical fields.\\

\subsection{Approaches to Scalable Text Mining}

A common solution to TM is to treat it as any large-scale processing problem: using general data processing components for processing,
and managing scalability with architectural decisions. This can be called a generic approach to scalable TM, and it has a couple of benefits. 
Components used in other types of data processing can be used in the text domain, and scalability can likewise be managed using well-known solutions. 
Components of this type include models such as Decision Trees and SVM, while architectural solutions include parallelization solutions such 
as Map-Reduce. Using generic solutions for both the components and architecture means that the solutions are to a large degree better understood, and both 
the required software and expertise is more widely available.\\ 

A number of generic techniques for scalability can be used at the architectural level:
\begin{description}
\item[Selection] can be applied to documents, features, classes in other
possible dimensions. For the example of feature selection, selection removes from the data the least important features under some measure of 
importance \citep{Lewis:92, Yang:97, Forman:03, Elsayed:08}. Documents can be selected by removing documents classified as spam or noise from the collection \citep{Chekuri:97, Manning:08}.
\item[Transformation] can be likewise applied to variables to reduce dimensionality, as well as to the processing problems
themselves. For example, a multi-label classification problem can be approximated by transforming it into a sequence of binary-label classification
problems, a multiclass classification problem, or a label ranking problem \citep{Tsoumakas:10}. Features can be combined or transformed using
topic modeling and matrix decomposition methods \citep{Hofmann:99}. 
\item[Precomputing] reduces inference time processing by computing and storing as much as possible of the processing offline \citep{Zhang:94, Mohri:04}. 
\item[Caching] stores and reuses solutions computed earlier \citep{Skobeltsyn:08}. Multiple caches can be used to store different types of subsolutions.
\item[Pruning and regularization] can be imposed on most generic modeling components, reducing the number of stored parameters 
and required computation \citep{Zhang:94, Skobeltsyn:08}.
\item[Streaming] processes data as a sequence of instances, rather than storing a full dataset in memory as a single batch. Mini-batch processing
stores smaller parts of the data, trading memory requirements for model performance. Many algorithms have online versions for stream training,
such as Stochastic Gradient Descent \citep{Bottou:10}, Online EM \citep{Liang:09}, and approximations for topic models \citep{Rehurek:11}.
\item[Parallelization] solves computing problems by using several processors simultaneously to solve sub-problems. Two basic types of 
parallelization should be distinguished: parallel computing and concurrent computing. Parallel computing solves problems as an array of smaller 
identical subproblems, and combines the results. Concurrent computing solves problems as a pipeline, forwarding data from one component to 
another continuously. Multi-core processors, computing clusters, grid computing and cloud computing are possible configurations for utilizing parallelization.\\
\end{description}

Recent general scientific literature commonly equates scalability with parallelization \citep{Hill:90,Kargupta:97,Dean:13}. This has become especially
prominent due to the popularity of Big Data and cloud computing as topics over the last decade. Parallelization gives reductions in processing time, 
up to the number of processors used. For parallel computing, an upper bound of this reduction is known as Amdahl's law \citep{Amdahl:67}, 
stating that the maximum speedup gained from parallel computing is dependent on how much of the problem can be solved by parallel computations. 
The statement can be extended for concurrent computing. Interestingly, Amdahl's 1967 paper is one of the most cited publications in parallel computing, 
but is very critical of parallel computing. The original reason for Amdahl's law was to present a rigid proof that parallel computing is not a 
panacea for scalability \citep{Amdahl:67}.\\

Much of the TM literature advocates general architectural methods for scalability, rather than ones taking the properties of text data 
into account. In particular, parallelization and cloud computing are considered by many to be solutions to TM: \citet{Baumgartner:09, Chard:11, Rehurek:11} 
and \citet{Tablan:13} present TM systems relying exclusively on cloud computing for scalability, \citet{Dunlavy:10} presents a system relying 
on parallelization and feature selection, while \citet{Villalon:13} presents a TM software library advocating parallelization, but implicitly using 
concurrency and document selection. For most TM tasks, selection is implicitly used, as document selection is commonly considered one of the 
main stages in TM \citep{Ahonen:97a}, and in data mining generally \citep{Fayyad:96}. \citet{Agichtein:05} reviews solutions that have been used in 
scalable information extraction, identifying four main approaches for scalability: 1) scanning the collection using rules, 2) selecting documents 
using search engines, 3) using customized indexes and 4) distributed processing. The first two are cases of selection and 
the last one refers to parallelization. The use of customized indexes is more specific to information extraction, and is one type of non-generic solution.\\ 

The downside of this generic approach to scalability is that it does not always scale well to TM, 
and simplifying a problem to suit generic components cannot always be done without unacceptable approximations. As an example, 
consider categorization of large-scale multi-label ontologies. \citet{Tsoumakas:13}
shows a leading solution to BioASQ\footnote{http://bioasq.org} biomedical article categorization of documents into combinations of 26k 
possible labels. A majority voting ensemble of four SVMs was trained for ranking the labels, and one for predicting the number of labels per 
document. The scalability was managed by document selection from 10.8M to 3.9M documents, feature selection by removing words occurring less than
six times in the collection, and parallelization with a cluster of 40 processors. Training the SVM classifiers took one and a half days on the 40 
processors. While this solution is still state-of-the-art for 26k labels, real-world ontologies are becoming larger in all relevant dimensions. 
A similar solution based on regularized hierarchical SVMs \citep{Gopal:13} was proposed for the LSHTC4\footnote{http://lshtc.iit.demokritos.gr/}  
datasets for Wikipedia categorization with 325k labels. Using a few approximations in the optimization, this resulted in 
training times of 37 hours with 512 processors, multiplying estimation times from a smaller Wikipedia dataset one tenth in size by a factor
of 25. The real-world datasets and streams mentioned earlier in this chapter are on a totally different scale. Clearly, there is a limit to how much 
can be accomplished by generic data processing techniques. More fundamental innovations for text processing must be applied, if the emerging 
vast datasets are to be fully utilized.\\


A second approach to scalability in TM is to use algorithms intended to scale well with text data by taking the properties
of text into account. This can be called the specialized approach to TM. The advantage of using specialized domain expertise is 
that highly developed solutions for the text domain can be utilized across different types of processing tasks. For example,
statistical n-gram language models offer practical and efficient solutions across a variety of text processing tasks.\\

To some extent many of the applied generic algorithms have already taken properties of text data into account, as the development of the applied 
algorithms and TM are intertwined in many places. For example, SVMs were proposed for text classification explicitly due to 
features of word count vectors: high dimensionality, sparsity, high proportion of irrelevant to relevant features, and linear separability in the text 
classification datasets available at that time \citep{Joachims:98}. Likewise, the first use of the Map-Reduce parallelization framework was the 
computation of web-scale n-grams used at Google \citep{Dean:08}. The vector space model was developed in the context of word vectors, and afterwards
applied to other tasks \citep{Salton:63}. Many of the related technical fields were developed in the context of processing 
text data, IR and IE in particular.\\

Taking the properties of text data into account does not warrant scalable processing. NLP has dealt exclusively with text
data, but much of the research has focused on finding computational models that work with small sets of text data, while possibly testing 
associated linguistic theories, such as formal frameworks for describing the grammar of natural languages. The emergence of vast text datasets has
made scalable processing methods more common, in particular shallow parsing methods in combination with machine learning \citep{Neumann:02, Lui:13}.\\

Specialized solutions for scalable TM can be broadly categorized into \emph{shallow processing}, \emph{hierarchical inference} and
\emph{inverted indices}, explained next in detail.\\

Shallow processing refers to NLP methods that attempt to approximate theoretically grounded deep grammatical processing methods. The typical case
is syntactic parsing, where the complexity of the correct model is still unknown, but is at least context free \citep{Chomsky:56}. Finite
state models provide an approximation to parsing with lower complexity \citep{Koskenniemi:90}. Many processing tasks, such as part of speech
tagging, phrase chunking and named entity recognition, can be done with low-complexity algorithms such as Hidden Markov Models \citep{Church:88}.
From a NLP perspective, shallow processing can be seen as an extension of text normalization \citep{Neumann:02}, whereas from the IE 
perspective it can be seen as a form of data enrichment \citep{Stajner:10}. Shallow processing components themselves do not necessarily need to scale 
in training, since they can be estimated from smaller amounts of data \citep{Tandon:10, Collobert:11, Lui:13}. Some linguistic theories argue that language
operates on gradually enriched semantic representations \citep{Neumann:02, Daum:03, Sagae:07}. By constraining 
deeper processing methods, shallow processing both improves efficiency and effectiveness of further processing stages.\\ 

Hierarchical modeling and inference can be applied in numerous ways to utilize the rich structure of text collections. As discussed earlier, text data
is embedded with multiple types of implicit and explicit structure. Hierarchical text classification organizes classes into a hierarchy, and classifies documents 
by traversing the hierarchy from the root towards the leaf classes. This reduces the number of classes
that need to be considered to the logarithm of the number of classes \citep{Koller:97, Tsoumakas:10, Gopal:13}. Hierarchical clustering can be used for clustering
documents, reducing the complexity of clustering to comparisons within each subcluster in the hierarchy \citep{McCallum:00}. Hierarchical ranking 
in text retrieval uses a cascade of inverted indices with increasing degrees of granularity \citep{Wang:11b}. Scalability in parsing and entity 
recognition can be improved by using hierarchical representations \citep{Petrov:07, Kiddon:10, Singh:11}. All of these cases of text processing use the 
same idea of coarse-to-fine processing, improving modeling effectiveness and scalability by organizing variables into a hierarchy. Graph-structured 
variables can be approximated using hierarchies, and unsupervised modeling can discover many types of variables implicitly present in text documents 
that can be used for hierarchical modeling.\\

Inverted indices have constituted the main data structure for efficient text retrieval for several decades \citep{Zobel:06}. A more recent development is the 
use of inverted indices in IE, starting from an IBM TM system 
called WebFountain \citep{Gruhl:04}. Indices can be enriched with a variety of information obtained through shallow processing of the documents. 
Words can be indexed after enrichment with person identification, location \citep{Gruhl:04}, part-of-speech and dependency information 
\citep{Cafarella:05, Cafarella:05b}, relation tuples \citep{Banko:07}, entity types, predicate-argument relationships, semantic frames, frame 
roles, frame-denoting elements, events, attributes and relations \citep{Hickl:07}. Virtually any type of information can be included in an 
enriched inverted index \citep{Gruhl:04, Agichtein:05}. This increases the index size, but lowers the complexity of retrieving documents matching an indexed 
annotation. For example, looking up documents that refer to a certain entity can be done simply by going through a postings list that contains documents 
classified to refer to that entity. This lowers the complexity of some types of processing, and has been considered the most promising method 
for making information extraction scalable \citep{Agichtein:05}.\\

These three strategies improve scalability of text processing by utilizing properties of text data. Shallow processing utilizes the structural nature of
implicit variables in text data. Hierarchical processing utilizes hierarchical representations of both implicit and explicit variables associated with text.
Enriched inverted indices and sparse processing utilize the sparsity of common representations of text, such as word vectors. Chapter 5 shows
how hierarchical processing and enriched inverted indices can be combined to perform scalable probabilistic inference for a variety of TM tasks.\\

\chapter{Multinomial Naive Bayes for Text Mining}
This chapter gives an overview of the Multinomial Naive Bayes (MNB) model for text mining, and its generative and graphical model
extensions. A basic broad definition of MNB is first given. Generative models related to MNB are described, including
mixture models, topic models, n-gram models, Hidden Markov Models, and Dynamic Bayes Networks. The notation of graphical models is 
introduced, and the connection of directed generative graphical models to the more general factor graphs is discussed. Dynamic programming
algorithms for operating with directed graphical models are described, including Viterbi, forward, and expectation maximization.\\

\section{Multinomial Naive Bayes}
\subsection{Introduction}

Multinomial Naive Bayes (MNB) is a probabilistic model of count data commonly used for various tasks in text mining. 
MNB originates in text classification research, but the model goes under different names in fields 
related to text mining. In text clustering, the equivalent
model is called a generative multinomial model \citep{Zhong:05}. In information retrieval, a special case of MNB is the
query likelihood language model (LM) \citep{Kalt:96, Hiemstra:98, Ponte:98, Zhai:01}. Many other methods can 
be related to MNB, either as extensions or modifications. MNB and related methods can be said to 
form one of the core statistical models of text mining.\\

The MNB model originates from the Naive Bayes (NB) model for text classification, which is a Bayes model 
that simplifies model estimation by making strong independence assumptions on features. NB was 
suggested in a 1961 paper by Maron \citep{Maron:61}. This paper was pioneering and even visionary
in multiple ways. The paper introduced the idea of automatic text classification, the Bernoulli NB model for text
classification, a correction to the zero-frequency problem of NB models, evaluation using held-out test
data, and used modern terminology as well as vector notation for describing the model. Maron's work received
limited continuation until the early 90s, when text classification started to become a major topic in the 
machine learning (ML) and data mining fields. Text classification and models related to NB were extensively 
researched for a decade, until learned linear classifiers such as Support Vector Machines (SVM) became 
popular due to their superior accuracy \citep{Joachims:98}.\\

By the end of the 90s, MNB was identified to be considerably better than Bernoulli NB for most text classification uses  
\citep{Lewis:98, McCallum:98b, Rennie:03}, but generally less accurate than discriminative classifiers such as 
Logistic Regression (LR) and SVMs \citep{Joachims:98, Rennie:03}. It was also noted that the strong 
modeling assumptions in MNB reduced performance, and modifying MNB could bring its performance closer to 
the discriminative classifiers \citep{Rennie:03, Schneider:05, Frank:06}. During the next decade, MNB and 
related methods spread to various other text mining tasks, in many cases becoming baseline methods, while 
research interest in generative models for text started to diversify into extensions such as mixture models 
\citep{Li:97, Monti:99, Toutanova:01, Novovicova:03} and topic models \citep{Hofmann:99, Blei:03}.\\

MNB and generative models of text have remained popular due to several advantages, specifically:

\begin{description}

\item[Simplicity]
NB models are very simple to describe and implement. They are among the first models taught to students 
in ML, prior to learned linear models such as SVM and LR. Simplicity also means that the estimated model parameters 
can be intuitively understood. More complex ensemble methods can be used to combine a set of NB classifiers, providing 
a high performance solution that is not a black-box \citep{Elkan:97}.

\item[Probabilistic Formulation]
The probabilistic formulation of MNB confers several advantages. The parameters and posterior probabilities of MNB can be 
easily visualized and interpreted. Text mining is sometimes used as a component for general data mining and statistical analysis. 
Probabilistic models can be better integrated into complex modeling than non-probabilistic components.

\item[Versatility]
Text mining applications vary greatly, and most text mining tools are specialized into solving particular problems.
MNB can be directly used in text classification, ranking and clustering, among other uses. Graphical model extensions such 
as HMMs, mixture models, and conditional N-gram models can be used for modeling structured data. Specialized extensions 
can be made for handling different types of structured data, such as multi-label outputs \citep{McCallum:99} and multi-field documents
\citep{Wang:10}.

\item[Robustness]
The variability in text mining applications causes different types of modeling challenges, such as limited training data 
and mismatch between training and test datasets. Despite making strong assumptions, NB models seem to perform 
well empirically \citep{Domingos:97} and MNB is commonly used as a baseline model in text mining 
applications.

\item[Scalability]
The amount of available data is growing at an exponential rate. Text datasets as word vectors are high-dimensional
in the number of documents, words and labels. MNB has a 
simple form that results in linear time and space complexity for both estimation and inference. Unlike many methods
for text mining, NB models scale linearly in the number of words, documents and labels. This means that MNB can be used 
on vast datasets, where anything exceeding linear scaling is intractable.

\item[On-line training]
The vast text datasets that have become available can no longer be stored in the memory of a single computer.
This is causing a shift from batch processing to on-line processing, where the data is not kept in memory, but processed as 
a stream. Many data streams are time-ordered, so that older documents are less useful for building predictive models. 
Examples of data streams are news stories, microblog messages and other forms of media used for 
rapid communication. Models working on data streams should 
support on-line training and down-weighting of older data points. These are trivially implemented for NB models.

\item[Parallelization]
Large-scale data processing can be tackled with parallelization across multiple processors.
Parallel computing frameworks such as MapReduce and Hadoop have become popular solutions for dealing with 
scalability. One of the main original uses for the MapReduce framework was training large-scale language models by 
parallelized count accumulation and combination \citep{Dean:08}. Other types of NB models can be parallelized in the same 
fashion in both estimation and inference.
\end{description}

The disadvantage of MNB is the effectiveness compared to more complex models. Even in earlier text classification 
research, NB was used as a ``straw man'' baseline for comparing more complex models \citep{Domingos:97, Lewis:98, Rennie:03}. 
The introduction of SVM \citep{Joachims:98} brought about a gradual decline of interest in MNB and other generative 
models for text classification, following a general trend in ML towards discriminative classifiers. Currently, discriminative classifiers such 
as SVMs, LR and Maximum Entropy models are considered the most effective solution for text classification uses 
such as spam classification, sentiment analysis and document categorization. 

\subsection{Definition}
The MNB model considers word count vectors $\bm w$ as being generated by underlying multinomial distributions of words
$n$ associated with label variables $l$. Usually the instances of text are documents and the label variables are classes, 
document identifiers, or clusters, depending on the task. The label-dependent multinomial distributions $p_l(n)$ are called 
conditional distributions, and are combined with a categorical prior distribution $p(l)$ of the label variables.
A common intuitive explanation is a ``generative process'', where documents are generated by first sampling a 
label from the categorical distribution, and then sampling words from the multinomial associated with the label.\\

A generative model in ML terminology is a model of the joint distribution $p(\bm w, l)$ of input and output 
variables \citep{Bishop:06,Klinger:07,Sutton:07}. For MNB the inputs are word vectors $\bm w$ and the outputs are document labels $l$. 
A generative Bayes model factorizes the joint distribution as $p(\bm w,l|\bm \theta)= p_l(\bm w|\bm \lambda) p(l|\bm \pi)$, 
so that the parameters $\bm \theta$ are assumed to factorize into parameters $\bm \lambda$ for 
the \emph{conditionals} $p_l(\bm w|\bm \lambda)$ and $\bm \pi$ for the \emph{prior} $p(l|\bm \pi)$. Posterior inference 
can be done by applying Bayes theorem: $p(l | \bm w)= p(\bm w|l) p(l) / p(\bm w)$, where $p(l | \bm w)$ is called the 
\emph{posterior} and $p(\bm w)=\sum_l p(\bm w, l)$ the \emph{marginal}. For ranking and classification, the marginal
can be omitted and the inference done by maximizing the \emph{joint} $p(\bm w, l)$ instead. In classification, the 
optimization becomes $\argmax_l p_l(\bm w) p(l)$. Regression can be performed with continuous variables 
for labels \citep{Frank:98}.\\

A problem with Bayes classifiers is estimating the dependencies of the input variables or features $n$ for computing 
$p_l(\bm w)$. By making independence assumptions, computing $p_l(\bm w)$ can be simplified. A common 
assumption is ``naive independence', which assumes that the conditional probabilities $p_l(n, w_n)$ are 
independent, that is, $p_l(\bm w) = \prod_n p_l(n, w_n)$. Models of this type are commonly called NB models. The 
parameterization of $p_l(\bm w)$ can take many forms. A common type of NB model is the multivariate Bernoulli NB 
\citep{Maron:61}, where the counts $w_n$ are restricted to binary values $\forall_n: w_n\in \{0,1\}$ and each class conditional
probability $p_l(n, w_n)$ is modeled by a Bernoulli distribution. The Bernoulli distribution models biased ``coin-flip'' outcomes
of a variable by using a single parameter describing how biased the coin flips are. For example, with parameter 
$\lambda_{ln}= \log(0.1)$, the probability of the word $n$ occurring in a document for label $l$ would be $0.1$. The Bernoulli
parameters can be estimated simply by counting the training documents $\bm w^{(i)}$ for label $l$ with the word $n$, 
and dividing by the number of documents for that label. \\

The multivariate Bernoulli model for NB 
\citep{Maron:61, Robertson:76, Domingos:97, Lewis:98, McCallum:98b, Craven:00} is also known as the Binary Independence 
Model and Bernoulli NB, and was the first type of NB model suggested for text mining \citep{Maron:61}. Other possible 
models for $p_l(\bm w)$ in text mining include Gaussian \citep{Domingos:97}, multinomial \citep{Lewis:98, McCallum:98b, Craven:00, 
Rennie:03, Schneider:05, Frank:06, Puurula:12c}, Poisson \citep{Church:95, Kim:06, Li:06}, Von Mises-Fisher 
\citep{Banerjee:05}, asymmetric distributions \citep{Bennett:03}, kernel densities \citep{Ciarelli:09}, and finite mixtures of 
distributions \citep{Church:95, Banerjee:05, Li:06}.\\

For most text mining uses the multivariate Bernoulli model has been replaced by the multinomial model of text and
its extensions. A multinomial distribution can be seen as a generalization of the Bernoulli distribution into multiple coin-flip 
outcomes and multiple coin tosses, much like multiple rolls of a biased dice with $N$ sides. A multinomial models the 
sums of $n$ possible outcomes from $J= \sum_n w_n$ dice rolls. The exact order of the dice rolls in a sequence of $J$ 
rolls $\ul{w}_j$ is not needed for counting the sums of outcomes $w_n$. The probability mass function
for a label-conditional multinomial distribution of word vectors becomes 
$p_l(\bm w)= Z(\bm w) \prod_n p_l(n)^{w_n}$. The normalizer $Z(\bm w)= \frac{(\sum_n w_n)!}{\prod_n w_n!}$ takes into account that
word vectors can correspond to a number of different word sequences. As it is constant for a given word vector, it 
can be omitted in most uses. A common special case of the multinomial 
is the binomial distribution $N=2$. Another special case of note is the categorical distribution
$\sum_n w_n= 1$.\\

Using multinomials for Bayes model conditional distributions, we get the joint probability distribution for MNB:
\begin{align*}
p(\bm w, l)=& p(l) p_l(\bm w)\\
			  =& p(l) Z(\bm w) \prod_n p_l(\bm w)^{w_n},
\numberthis \label{MNB1}
\end{align*}
where $p(l)$ and $p_l(\bm w)$ are parameterized with a categorical and a multinomial, respectively.\\

Since document lengths $J$ vary, models are defined over all possible lengths, and the multinomials for different lengths have shared 
parameters. The shared parameters are ``tied'', since they are constrained to be equal regardless of length. In addition, a distribution 
such as Poisson must be assumed for generating different document lengths, so that the model can generate the joint distribution over 
documents of all lengths. The length factor has no practical effect in most applications, and is omitted for 
posterior inference uses such as clustering, ranking and classification. Therefore both the length factor and 
the multinomial parameter tying are commonly omitted in the MNB literature \citep{McCallum:98b}.\\

A problem with varying document lengths is that the posterior probabilities for MNB models get increasingly
close to either $0$ or $1$ as the document length increases, since the conditional probability $p_l(\bm w)$ is
computed by multiplying the $N$ probabilities $p_l(\bm w)$ independently \citep{Frank:98, Monti:99, Bennett:00, Craven:00}.
This scale distortion of the posteriors does not affect classification, ranking or hard clustering, since the rank-order of 
probabilities for different labels is preserved. For uses such as soft clustering the posterior probabilities can be improved 
with feature transforms \citep{Pavlov:04}, feature selection \citep{Rigouste:07, Pinto:09}, and using KL-divergence instead of posterior
probabilities to correct for the document lengths \citep{Craven:00, Schneider:05, Pinto:09}. Nevertheless, for some
applications the indirectly estimated posterior probabilities from MNB can be insufficient, and a model directly 
optimizing the posterior probabilities is preferred, such as LR.\\

\subsection{Estimation}

Estimation of MNB parameters is commonly done by applying maximum likelihood estimation 
\citep{McCallum:99, Rennie:01, Juan:02, Vilar:04, Madsen:05, Frank:06}. Due to the data sparsity problem with
text data, various smoothing methods are commonly used to correct maximum likelihood parameter estimates.
In some cases the smoothed estimation is presented as maximum a posteriori estimation \citep{Rennie:01, 
Schneider:05, Smucker:07}. Despite the name ``Bayes'', NB models are commonly not Bayesian in the sense
of Bayesian estimation, where a distribution over parameters is maintained instead of a point estimate of parameters.
A fully Bayesian version of MNB has been proposed, but shown to be less suitable than maximum likelihood point estimates \citep{Rennie:01}.
The most common type of estimation is supervised estimation, where a training dataset $D$ consists of $I$ pairs 
$D^{(i)}= (\bm w^{(i)}, l^{(i)})$ of word vectors and labels, assumed to be independent and identically distributed 
(IID). The unsmoothed maximum likelihood estimation approach for the supervised case is described next.\\

The maximum likelihood method of statistical estimation selects a vector of parameters for a model that maximizes the 
likelihood of the parameters given the data $\mathcal{L}({\bm \theta}| D)$. This equals the probability of the data given the 
parameters $p(D| {\bm \theta})$. A conceptual difference between these two is that likelihood is a function of parameters for given 
training data, whereas probability assumes a model with parameters and can refer to both seen and future data. The MNB likelihood 
function can be derived:
\begin{align*}
\mathcal{L}({\bm \theta}|D)&=p(D| {\bm \theta}) \\
 				&=\prod_i p(l^{(i)}|{\bm \theta})  p_{l^{(i)}}(\bm w^{(i)}|{\bm \theta}) &\text{IID data assumption}\\
 				&=\prod_i p(l^{(i)}|\bm \lambda)  p_{l^{(i)}}(\bm w^{(i)}|\bm \pi) &\text{Bayes model}\\
				&=\prod_i p(l^{(i)}|\bm \pi) \prod_n p_{l^{(i)}}(n|\bm \lambda)^{w^{(i)}_n} \frac{(\sum_nw^{(i)}_n)!}{\prod_n w^{(i)}_n!} &\text{Multinomial conditional}
\numberthis \label{MNB1_likelihood}
\end{align*}

Maximizing the likelihood can be simplified by noting that the log of the likelihood has the same maximum, but is easier to 
handle computationally.  The MNB log-likelihood decomposes into terms that can be separately optimized. The 
maximization of the log-likelihood function can be derived:
\begin{align*}
&\argmax_{\bm \theta}(\mathcal{L}(\bm \theta|D))= \argmax_{\bm \theta}(\log \mathcal{L}({\bm \theta}|D))
							= \argmax_{\bm \theta}(\log p(D| {\bm \theta}))\\
						      &= \argmax_{(\bm \lambda, \bm \pi)}(\log(\prod_i p(l^{(i)}|\bm \pi) \prod_n p_{l^{(i)}}(n|\bm \lambda)^{w^{(i)}_n} \frac{(\sum_n w^{(i)}_n)!}{\prod_n w^{(i)}_n!}))\\
						      &= \argmax_{(\bm \lambda, \bm \pi)}(\sum_i (\log p(l^{(i)}|\bm \pi) + \sum_n w^{(i)}_n \log p_{l^{(i)}}(n|\bm \lambda) +\log(\frac{(\sum_n w^{(i)}_n)!}{\prod_n w^{(i)}_n!})))\\
						      &= \argmax_{(\bm \lambda, \bm \pi)}(\sum_i \log p(l^{(i)}|\bm \pi) + \sum_i \sum_n w^{(i)}_n \log p_{l^{(i)}}(n|\bm \lambda))\\
						      &= \argmax_{(\bm \lambda, \bm \pi)}(\sum_l C(l) \log p(l |\bm \pi) + \sum_l \sum_n C(l, n)  \log p_l(n|\bm \lambda)),
\numberthis \label{loglikelihood}
\end{align*}
where $C(l)$ and $C(l, n)$ refer to the accumulated counts of the variables in the training data.\\

Optimizing the parameters for the prior and conditionals can now be done separately, and for both cases this consists of choosing 
the vector of parameters that most likely generated the accumulated vector of counts. The maximum likelihood solution to this estimation of
a categorical distribution $p(l|\bm \pi)$ is the relative frequency estimate $C(l)/\sum_{l'} C(l')$: $\pi_l = \log(C(l)/\sum_{l'} C(l'))$ and 
$\lambda_{ln} = \log(C(l, n)/\sum_{n'} C(l, n'))$. This is a standard result 
in statistics and commonly proven using the method of Lagrange multipliers \citep{Bilmes:98, Juan:02}.\\

In practice, estimating the MNB model consists of accumulating the counts in training data and normalizing by the sums of 
counts, with time complexity $O(IN)$ and space complexity $O(LN)$. Taking sparsity into account, the space complexity 
of estimation is reduced to $O(L+\sum_l \sum_{n: \exp(\lambda_{ln})>0} 1)$ and time complexity to $O(\sum_i |\bm w^{(i)}|_0)$. 
The parameters can be represented efficiently using sparse matrix representations. A hash table with $(l, n)$ pairs as keys and 
parameters as values is one popular choice, with amortized constant time complexities for updating the counts. Another
common choice is sparse vectors of word counts and periodic merging of the accumulated vectors with list merge 
operations. With large-scale datasets this can be done using the map-reduce framework \citep{Dean:08}.\\ 

Given labeled training data, maximum likelihood estimation of parameters for MNB can be done 1) exactly, 2) with a closed form solution, 
3) as online learning, and 4) in linear time complexity in terms of features, documents and classes. These four 
advantages make MNB highly useful in practical applications. The maximum likelihood estimates are exact, so there are no approximations
required, and any system using MNB does not have to consider possible errors from approximations. The closed form solution
means that the estimates can be computed by applying elementary mathematical 
operators, such as summation and division. The estimation can be done as online learning from streams of text documents, 
and the effect of older documents can be removed from the estimates trivially. Lastly, the maximum likelihood estimation has linear
complexities in all relevant dimensions. Among the common text mining methods, only Centroid and K-nearest Neighbours classifiers have 
the same scalability in training.\\

Extending MNB estimation to weighted data can be done by weighting the accumulated counts from each document. 
In some cases documents are softly labeled, by using a distribution of weights over labels instead of a single label. Extending the 
estimation to soft labeling can be done similarly, by weighting the counts by the label weights. In case of unsupervised and 
semi-supervised estimation, the expectation maximization (EM) algorithm \citep{Dempster:77, Bailey:94, Bilmes:98, Zhong:05, Nigam:06, 
Gupta:11} can be used to estimate parameters. This consists of initialization of parameters followed by EM-iterations of computing the soft labeling
$p(l|\bm w^{(i)})$ (expectation step) and re-estimating the model from the soft labeling (maximization step). Each iteration
improves the log-likelihood until a stationary point of the likelihood function is reached. Combining EM 
with multiple random initializations reduces the probability of not reaching a global optimum of the likelihood. Improvements of EM such 
as online EM \citep{Liang:09} can be used to estimate parameters from stream data and reduce the required amount of EM-iterations.\\

\section{Generative Models Extending MNB}
\subsection{Mixture Models}
Mixture modeling techniques are commonly used to extend MNB and multinomial models of text, providing both improved
modeling precision and a means for incorporating structure into the models. The use of mixtures enables
the modeling of multi-modal distributions, with accuracy increasing as a function of the number of added components and the
amount of available data to estimate the components.\\

A basic type of mixture is the Finite Mixture Model \citep{Pearson:94}. 
This models data as being generated by a normalized linear combination of $L$ component distributions, so
that for each component $l$ a weight $p(l)$ and a component-conditional distribution $p_l()$ is estimated. The
component weights are constrained $0\le p(l) \le 1$ and $\sum_l p(l)=1$. For example, a finite mixture of multinomials
takes the form:
\begin{align*}
p(\bm w)= \sum_{l} p(l) Z(\bm w)\prod_n p_{l}(n)^{w_n}
\numberthis\\
\end{align*}

By replacing the component variable with the label variable in the MNB model of Equation \ref{MNB1}, 
supervised Bayes models can be viewed as mixture models with known component probabilities $p(m|\bm w^{(i)})$ for each 
training document \citep{McCallum:98b, Novovicova:03, Nigam:06}. Commonly the components are unknown variables which 
are estimated in training using approximate algorithms such as EM and optimization 
on held-out data. Depending on the application and the type of mixture modeling, different optimization algorithms and 
constraints on the parameters can be used to simplify the estimation problem.\\

Multinomial models of text can be extended by adding mixtures over both documents and words. Document-clustering
mixtures treat documents as being generated by a mixture, with each component corresponding to a prototypical
document. Word-clustering mixtures treat words similarly, with each component corresponding to a prototypical
word. The document-clustering components can be called \emph{themes}, while word-clustering components
can be called \emph{topics} \citep{Keller:04}. Combination and extension of these two basic types of mixture
for text data result in the various mixture and topic models of text.\\

The earliest proposed document-clustering mixture extension of MNB conditions the label variables 
on the components \citep{Kontkanen:96, Monti:99}. In 
this use the mixture components cluster the training data into soft partitions:
\begin{align*}
p(\bm w, l)= \sum_m p(m) p_m(l) Z(\bm w)\prod_n p_{ml}(n)^{w_n}
\numberthis\\
\end{align*}

Replacing each conditional multinomial $p_l(\bm w)$ in MNB by a finite mixture of $M$ multinomial components 
produces the more common Multinomial Mixture Bayes Model \citep{Monti:99,Novovicova:03,Nigam:06}:
\begin{align*}
p(\bm w, l)= p(l) \sum_{m} p_l(m)  Z(\bm w)\prod_n p_{lm}(n)^{w_n}
\numberthis \label{doc_mix}\\
\end{align*}

Word-clustering mixtures are used in topic models and multi-label mixture models \citep{Li:97, Hofmann:99, 
Li:00, Blei:03, McCallum:99,Ueda:02}. A basic model of this type extends the multinomial:
\begin{align*}
p(\bm w) = Z(\bm w) \prod_n (\sum_m p(m) p_{m}(n))^{w_n}
\numberthis \label{topic_model} \\
\end{align*}

The earliest proposed topic model of this form \citep{Li:97} replaced the multinomial in MNB, using 
hard clustering of words to form shared components $p_m(n)$, with separate distributions $p_l(m)$.
A related model called the Stochastic Topic Model \citep{Li:00} used this type of modeling without label variables, 
performing inference using the components directly for topic segmentation and analysis. Multi-label classification
models use this type of topic model as well, but learn the components from multi-label data. The Multi-label Mixture 
Model \citep{McCallum:99} uses Equation \ref{topic_model}, but performs inference by greedily adding label 
components $l$ and estimates $p_l(m)$ for each document, using a prior $p(\bm c)$ over the labelsets $\bm c= [1,...,l,...,L]$ 
instead of labels. Parametric Mixture Model \citep{Ueda:02} 
is similar, but uses a uniform distributions for $p_l(m)$ and the labelset prior $p(\bm c)$. Further multi-label
mixture models have built on these two models \citep{Ueda:02b, Kaneda:04, Sato:07, Wang:08, Ramage:09}.\\

Probabilistic topic models became more widely known with Probabilistic Latent Semantic Analysis \citep{Hofmann:99}.
This uses a unique label variable for each document, so that the joint probability becomes:
\begin{align*}
p(\bm w, l) \propto p(l) \prod_n (\sum_{m} p_l(m) p_{m}(n))^{w_n}
\numberthis\\
\end{align*}

Although popular, the Probabilistic Latent Semantic Analysis model is not a fully generative model of documents \citep{Blei:03, Keller:04}, since
it does not generate new document variables $l$. The number of components $M$ needs to be optimized, and 
$L$ is tied to the number of training set documents. Thus the model is not very scalable and is prone to overfitting
\citep{Blei:03}. To address these issues, a model called Latent Dirichlet Allocation \citep{Blei:03, Minka:02} 
was proposed by replacing the document variables in Probabilistic Latent Semantic Analysis with a Dirichlet distribution $p(\bm \omega)$ of
the component weights $p(m |\bm \omega) =\omega_m$:

\begin{align*}
p(\bm w) = \int p(\bm \omega|\bm \tau) Z(\bm w) \prod_n(\sum_m p(m|\bm \omega) p_{m}(n))^{w_n} d \bm \omega ,
\numberthis\\
\end{align*}
where $p(\bm \omega| \bm \tau)$ is modeled by a Dirichlet distribution, given by:

\begin{align*}
p(\bm \omega| \bm \tau) \propto \frac {\Gamma(\sum_m \tau_{m})} {\prod_m \Gamma(\tau_{m})} \prod_m \omega^{\tau_{m}}_m ,
\numberthis
\end{align*}
where $\bm \tau$ are the parameters for the Dirichlet distribution and $\Gamma$ is the gamma function.\\

The integration over possible component weight vectors has no closed form solution, and must be approximated
using algorithms such as variational Bayes, Gibbs sampling, and expectation propagation \citep{Minka:02, Asuncion:09}.
The Latent Dirichlet Allocation model proved exceptionally popular and turned probabilistic topic modeling into an active field of research
\citep{Griffiths:04, Blei:06, Li:06b, Asuncion:09, Blei:12}.\\

A substantial literature exists on various models based on Latent Dirichlet Allocation. One conceptually useful model is the Theme Topic Mixture 
Model \citep{Keller:04}. This presents a discretized version of Latent Dirichlet Allocation that does not require approximate inference. The 
Dirichlet over component weights is replaced by a document-clustering mixture: 

\begin{align*}
p(\bm w) = \sum_l p(l) Z(\bm w) \prod_n (\sum_m p_l(m) p_{m}(n))^{w_n}
\numberthis\\
\end{align*}

Theme Topic Mixture Model presents a direct combination 
of the document clustering and word clustering finite mixture models. Combinations and extensions of these two 
types of mixture modeling are used throughout text mining applications, both with multinomial models and 
distributions other than the multinomial.\\
 
\subsection{N-grams and Hidden Markov Models}
The multinomial model of text considers words within documents to be distributed independent of their context. 
Consequently all phrase- and sentence-level information present in the documents is left unmodeled. This sequence 
information is crucial for many applications of generative models of text, including machine translation, speech recognition,
optical character recognition and text compression. Even in tasks where multinomial text models are considered 
sufficient, sequence modeling has been shown to be beneficial \citep{Song:99, Miller:99, Peng:03, Medlock:06}. 
Incorporating sequence information is most commonly 
done using higher order sequential models called n-grams, also known as Markov chain models. These models
originate from the early days of computer science, and aside from the many practical uses they have been instrumental
in the development of computer science and information theory \citep{Shannon:48, Chomsky:56, Markov:71}.\\

An n-gram model generalizes the multinomial model to take the preceding sequence of $M-1$ words into account,
where $M$ is the order of the n-gram. Each \emph{history} of preceding $M-1$ words models a 
separate categorical. The models of the first three orders are commonly called unigram, bigram and trigram models, 
corresponding to zeroth, first and second order Markov chain models, respectively. Over 
the last decades higher order models such as 4-grams and 5-grams have become standard, with the availability of web-scale 
text datasets and the development of computer processors and memory. A full n-gram model of order $M$
and vocabulary size of $N$ requires $N^M$ parameters, i.e. counts, for the $N^{M-1}$ categorical distributions.
Due to the Zipf-law distribution of text data, the models will be extremely sparse, and only a fraction of these 
counts will be seen in any amount of training data. For these reasons the main foci in n-gram research 
have been efficiency of implementation \citep{Siivola:07, Brants:07, Watanabe:09, Pauls:11} and methods for 
smoothing high order n-grams with lower order estimates \citep{Jelinek:80, Ney:94, Chen:96, Chen:99, Goodman:00, Huang:10, Schutze:11}.\\

Let $\ul{\bm w}$ denote any word sequence that corresponds to the counts in the word vector $\bm w$: $w_n= \sum_{j:\ul w_j= n} 1$. 
Let $\ul{w}_{j-M+1}...\ul{w}_{j}$ denote a subsequence of $M$ words ending at word $j$. The history or
context of an n-gram is the sequence 
$\ul{w}_{j-M+1}...\ul{w}_{j-1}$ of preceding $M-1$ words, a sequence of 0 words in the unigram case $M=1$. 
The probability of a word sequence is given by:
\begin{align*}
p(\ul{\bm w}) &= \prod_j p_M(w_j | j, \ul{\bm w})\\
                      &= \prod_j p_M(w_j | \ul{w}_{1}...\ul{w}_{j-1}) &\text{Markov chain}\\
                      &= \prod_j p_M(w_j | \ul{w}_{j-M+1}...\ul{w}_{j-1}) &\text{finite history}
\numberthis\\
\end{align*}

The probabilities for the first $M-1$ n-grams are undefined, since their histories would span over the
word sequence boundaries. To correct this, the sequence can be ``padded'' by adding start symbols 
``\textless s\textgreater'' at the beginning of the sequence. Sequence end symbols 
 ``\textless /s\textgreater'' can be added, and both of these improve modeling accuracy at the boundaries.\\

There is a considerable literature on methods for smoothing the n-gram language models. Virtually all of these
interpolate n-grams hierarchically with lower order n-grams \citep{Chen:99}.
Let $p_m()$ denote the smoothed $m$-th order model in the hierarchy and $p_m^u()$ denote the 
unsmoothed model of the same order. The interpolation smoothed n-gram probabilities can be expressed as:
\begin{align*}
p_m(w_j | \ul{w}_{j-m+1}...\ul{w}_{j-1}) =& (1-\alpha_m) p_m^u(\ul{w}_j | \ul{w}_{j-m+1}...\ul{w}_{j-1}) \\
&+\alpha_m \: p_{m-1}(\ul{w}_j | \ul{w}_{j-m+2}...\ul{w}_{j-1}),
\numberthis
\end{align*}
where $\alpha_m$ are backoff weights for order $m$ chosen by the smoothing method. With Jelinek-Mercer 
smoothing \citep{Jelinek:80, Chen:99}, $\alpha_m$ are simply fixed parameters estimated on held-out data.\\

Often a uniform zerogram model is included to end the recursion. Without a zerogram, the different n-grams in 
hierarchical interpolation methods form a hierarchy of smoothing with $M$ levels. The smoothing weights can be expanded:
\begin{align*}
p(\ul{\bm w}) &= \prod_j p_M(\ul{w}_j | \ul{w}_{j-M+1}...\ul{w}_{j-1})\\
	&= \prod_j \sum_m p(m) p^u_m(\ul{w}_j | \ul{w}_{j-m+1}...\ul{w}_{j-1})\\
	&= \prod_j \sum_m (\prod_{m'=m+1}^{M} \alpha_{m'} -\prod_{m'=m}^{M} \alpha_{m'}) p^u_m(\ul{w}_j | \ul{w}_{j-m+1}...\ul{w}_{j-1}),
\numberthis \label{ngram_expand}\\
\end{align*}

In this form it is seen that n-gram smoothing methods utilize a word-level mixture, generating
each word in a sequence as a finite mixture model of the different order models. Hierarchical smoothing
means that the mixture component weights are generated dynamically as a product of the higher order back-off 
weights: $p(m)= \prod_{m'=m+1}^{M} \alpha_{m'} -\prod_{m'=m}^{M} \alpha_{m'}$ \citep{Bell:89}. Writing a word-clustering 
mixture of Equation \ref{topic_model} in the sequence form shows a further surprising connection:
\begin{align*}
p(\ul{\bm w}) &= \prod_j \sum_m p(m) p_m(\ul{w}_j)
\numberthis\\
\end{align*}

Comparing the word-clustering mixture in this form to the expanded n-gram model of Equation \ref{ngram_expand},
we can note the similarity between topic models and hierarchically smoothed n-grams. Both models generate 
words as a mixture of components. In topic modeling the components correspond to topics, and weights are generated 
by the chosen topic modeling method. In n-gram modeling the components correspond to the n-gram smoothing 
hierarchy, and weights are generated dynamically by the chosen smoothing method.\\

Unlike mixture models, n-gram models have no hidden (unknown) variables in estimation and are efficiently estimated by 
normalizing the known count statistics. Extending n-grams with hidden variables leads to a more powerful 
class of models called Hidden Markov Models (HMM) \citep{Baum:66, Rabiner:89, Bilmes:98, Miller:99}. 
A categorical HMM considers text to be generated as categorical outputs of a hidden Markov 
chain model, so that only the outputs $\ul w_j$ of the Markov chain are seen. A HMM can be seen
as an extension of a mixture model, so that the weights of each component become dependent on the history of
the last $M-1$ components that generated an output. In HMM terminology the outputs are called emissions
or observations, and the hidden variables are called states. In text modeling the outputs are commonly words, 
and the hidden variables are structural variables such as parts of speech, named entities, sections, topics and authors 
that are to be discovered using the HMM.\\

Due to the abundance of applications for HMMs, a number of variants exist that can be mentioned.  Arc-emission HMMs output
a variable on each transition to a state, whereas state-emission HMMs have the output variables attached to the states. Historically 
HMMs were defined more commonly as arc-emission HMMs, rather than state-emission HMMs. These two types of HMMs are 
equivalent, in the sense that either can be transformed to the other. In many applications the output distributions are 
Gaussian or mixture models, instead of categoricals. Epsilon or $\epsilon$-states can be used, that have no attached output distribution. 
HMMs of this type are called $\epsilon$-transition HMMs and are more powerful in the distributions that can be 
represented, since $\epsilon$-states can be used to model complex sequences of hidden variables behind each output.
But these also pose problems for inference, since each output word can be generated by arbitrarily long loops of 
$\epsilon$-transitions.\\

Let $\ul{\bm{k}}$ denote a hidden state sequence of states $m$ corresponding to a word sequence $\ul{\bm{w}}$. 
Let $M$ be the number of hidden states, $p_{\ul{k}_j}(\ul{w}_j)$ the categorical output distribution of state 
${\ul{k}_j}$, and 
$p({\ul{k}_j}|\ul{k}_{j-1})$ the transition probability to state $\ul{k}_j$ given the previous state $\ul{k}_{j-1}$. 
A first-order categorical state-emission HMM without $\epsilon$-transitions produces the joint probabilities:
\begin{align*}
p(\ul{\bm{w}},\ul{\bm{k}}) &= \prod_j p(\ul{k}_j | \ul{k}_{j-1})  p_{\ul{k}_j}(\ul{w}_j),
\label{hmm} \numberthis
\end{align*}

where the hidden $p(\ul{k}_1 | \ul{k}_0)= p(\ul{k}_1)$ is provided by a categorical distribution $p(\ul{k}_1)$
for the initial states $\ul{k}_1$. Alternatively, the sequence can be padded with boundary symbols the same way 
as with n-gram models. 

\subsection{Directed Graphical Models}
MNB and the discussed extensions can be described in the general framework of graphical models
\citep{Pearl:86, Lauritzen:88, Loeliger:04, Frey:05, Klinger:07, Parikh:07, Sutton:07} as generative directed graphical models. 
A graphical model is a model of a joint distribution of variables that factorizes according to an underlying
graph. Commonly this is an \emph{independency graph} that encodes the independence assumptions of the model.
Nodes in an independency graph represent the variables, and lack of an edge between two variables indicates an independence
assumption. Since computation is only required for related variables, factorizing a joint distribution according to 
the assumptions greatly simplifies modeling. Algorithms developed for the estimation and inference of graphical models
are applicable to new types of models, reducing the time required for research and development, while the graphical
representation reduces the time required for presentation of new models.\\

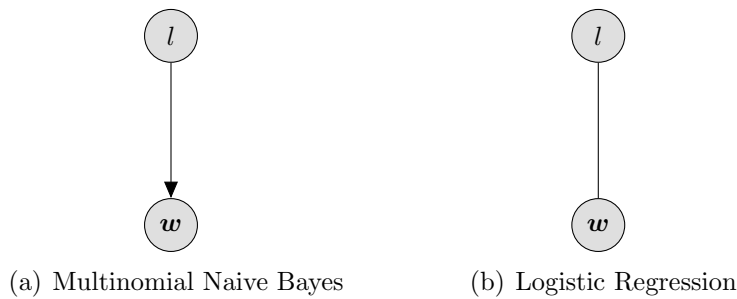
\begin{figure}[h]
\centering
\subfigure[Multinomial Naive Bayes]{
 \hspace{2.0cm}
%
%
%
%


\begin{tikzpicture}[x=1.7cm,y=1.8cm]


  \node[obs]                   (w)      {$\bm{w}$} ; %
  \node[obs, above=of w]    (l)      {$l$} ; %
  \edge {l} {w} ; %
  



\end{tikzpicture}

 \hspace{2.0cm}
} \hspace{0.0cm}
\subfigure[Logistic Regression]{
 \hspace{2.0cm}
%
%
%
%


\begin{tikzpicture}[x=1.7cm,y=1.8cm]


  \node[obs]                   (w)      {$\bm{w}$} ; %
  \node[obs, above=of w]    (l)      {$l$} ; %
  \draw (l) -- (w);
  \end{tikzpicture}

 \hspace{2.0cm}
} 
\caption{Independency graphs for a Multinomial Naive Bayes and Logistic Regression models. Multinomial Naive Bayes is a directed graphical model,
Logistic Regression is an undirected graphical model
}
\label{MNB_LR}
\end{figure}

The various types of graphical models use different factorizations and graphical notations.
Traditionally, graphical models come from two types, called Bayesian Networks \citep{Pearl:86} and 
Markov Random Fields \citep{Kindermann:80}. Both types visualize the variable nodes in the graph with a circle, and 
encode the independence assumptions in a model by omissions of edges. Both represent the assumptions
using an independency graph $G= (V, E)$ of variable nodes $V$ and edges $E$. Bayesian Networks are called 
\emph{directed graphical models}, as the edges, depicted as arrows, encode a directional conditionality of variables. 
Markov Random Fields are called \emph{undirected graphical models}, and the undirected edges imply dependency, but not directionality.
The two types of models are different in the distributions they can represent and the used algorithms. MNB forms
a basic directed graphical model, while LR is a corresponding undirected graphical model.
Figure \ref{MNB_LR} shows a comparison of MNB and LR using elementary independency graph notation, with the word vector $\bm w$ and
label variables $l$ forming the variable nodes $\bm x$ of the graph.\\

Graphical models factorize a joint distribution $p(\bm x)$ to $T$ factors $\Psi_t$: $p(\bm x)= \prod_t \Psi_t(\text{nd}(t))$,
where $\text{nd}(t)$ is the subset of the variable nodes $x_t$ connected to factor $\Psi_t$. The factors for undirected graphical
models are also called cliques, and are constrained to be arbitrary non-negative functions $\Psi_t \ge 0$, where
$\Psi_0= Z= 1/\sum \Psi_{t=1}^T(\text{nd}(t))$ is an additional normalization factor, also called the partition function.
The factors for a directed graphical model are conditional probability distributions of the form $\Psi_t(\text{nd}(t))= p(x_t| \text{nd}(t))$ for each
node $t$, with the special case $\text{nd}(t)= \emptyset$ producing marginal probability distributions $p(x_t)$. No
additional normalization factor is required for directed graphs, as the conditional distributions are normalized probabilities.
Since the factors for directed graphs are defined for each node $t$, the factorization can be read directly from the graph, whereas
the clique factors for undirected graphs are visualized less directly by the independency graph notation.\\

The notation for graphical models is ongoing constant evolution, and additional notation has been introduced 
\citep{Minka:08, Dietz:10, Andres:12}. Usually shaded nodes represent known variables, and 
clear nodes represent hidden ones. Additional plate notation is standard for repeating parts in graphical models, such 
as repeated segments in variable sequences. This represents ``unrolling'' of the graph, so that the 
segment is repeated a certain number of times. Nevertheless, in practice the plate notation is inconvenient for representing 
dependencies in sequences. Visualizations of sequence graphical models depict repeated fragments of the models 
instead \citep{Murphy:02, Deviren:04, Deviren:05, Frey:05, Wiggers:06, Sutton:07, Parikh:07, Klinger:07}, sometimes
indicating the repetition by use of ellipses, or combining the fragment notation with the plate notation \citep{Wang:07}.\\

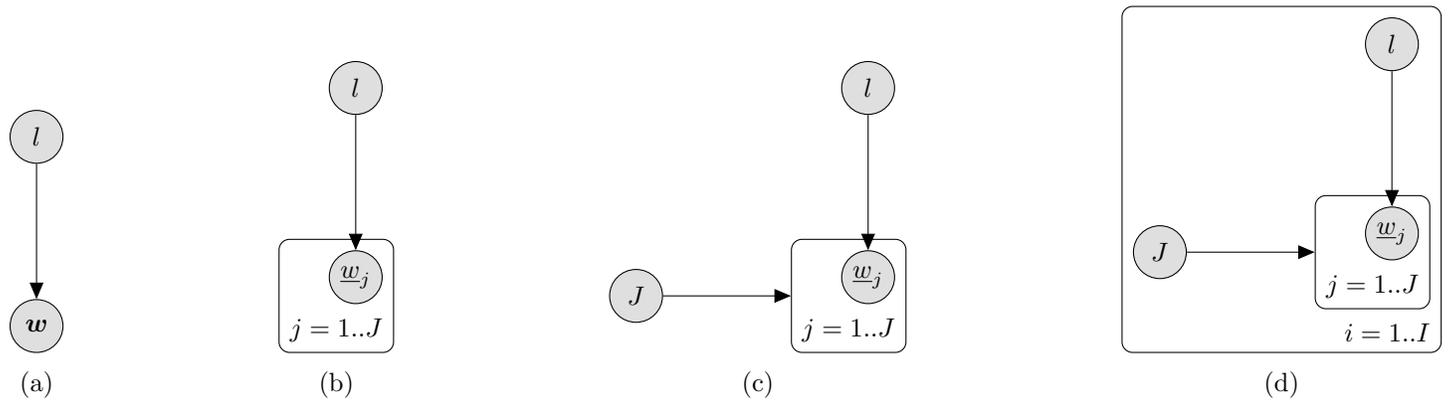
\begin{figure}[h]
\centering
\subfigure[]{
%
%
%
%


\begin{tikzpicture}[x=1.7cm,y=1.8cm]


  \node[obs]                   (w)      {$\bm{w}$} ; %
  \node[obs, above=of w]    (l)      {$l$} ; %
  \edge {l} {w} ; %
  



\end{tikzpicture}

}\hfill
\subfigure[]{
%
%
%
%


\begin{tikzpicture}[x=1.7cm,y=1.8cm]


  \node[obs]                   (w_j)      {$\ul{w}_j$} ; %
  \node[obs, above=of w_j]    (l)      {$l$} ; %
  \edge {l} {w_j} ; %
  \plate {doc} {(w_j)} {$j= 1..J$}; %
  



\end{tikzpicture}

}\hfill
\centering
\subfigure[]{
%
%
%
%


\begin{tikzpicture}[x=1.7cm,y=1.8cm]


  \node[obs]                   (w_j)      {$\ul{w}_j$} ; %
  \node[obs, above=of w_j]    (l)      {$l$} ; %
  \edge {l} {w_j} ; %
  \plate {doc} {(w_j)} {$j= 1..J$}; %
  
  \node[obs, left=of doc] (J)  {$J$};
  \edge {J} {doc} ; %



\end{tikzpicture}

}\hfill
\centering
\subfigure[]{
%
%
%
%


\begin{tikzpicture}[x=1.7cm,y=1.8cm]


  \node[obs]                   (w_j)      {$\ul{w}_j$} ; %
  \node[obs, above=of w_j]    (l)      {$l$} ; %
  \edge {l} {w_j} ; %
  \plate {doc} {(w_j)} {$j= 1..J$}; %
  
  \node[obs, left=of doc] (J)  {$J$};
  \edge {J} {doc} ; %
  \plate {dataset} {(w_j) (J) (doc) (l)} {$i= 1..I$}; %



\end{tikzpicture}

}
\caption{Independency graph notations for illustrating MNB, with an increasing degree of explicitness from left to right. The
label variable $l$ is considered known for estimation on training data and unknown for inference on test data. Edge from variable
$J$ to the plate indicates the plate is unrolled $J$ times}
\label{MNB_directedgraph}
\end{figure}

As an example of a directed graphical model, we can first take the MNB $p(l, \bm w)= \Psi_1(l) \Psi_2(l, \bm w)$. 
In this case the factors are categoricals $\Psi_1(l)= p(l)$ and multinomials $\Psi_2(l, \bm w)= p_l(\bm w)$. An
equivalent definition can be done using sequence variables and graph unrolling. In this case 
$p(l, \bm{\ul{w}}) = \prod_t \Psi_t(\text{nd}(t))$, where $\Psi_1(l)= p(l)$ is categorical and $\Psi_{t+1}(l, \ul{w}_t)=
 p_l(\ul{w}_t)$ for $2 \le t \le J+1$ are categorical draws from the multinomial, corresponding to the unrolled variables.
Figure \ref{MNB_directedgraph} shows directed independency graph notations for MNB, with varying degrees of 
explicitness.\\

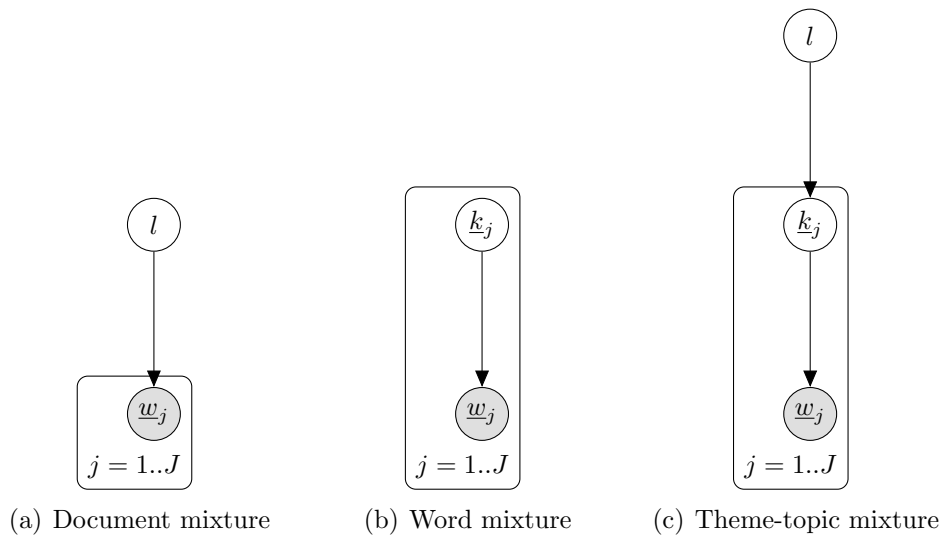
\begin{figure}[h]
\centering
\subfigure[Document mixture]{
 \hspace{1.0cm}
%
%
%
%


\begin{tikzpicture}[x=1.7cm,y=1.8cm]


  \node[obs]                   (w_j)      {$\ul{w}_j$} ; %
  \node[latent, above=of w_j]    (l)      {$l$} ; %
  \edge {l} {w_j} ; %
  \plate {doc} {(w_j)} {$j= 1..J$}; %
  



\end{tikzpicture}

 \hspace{1.0cm}
}
\subfigure[Word mixture]{
 \hspace{1.0cm}
%
%
%
%


\begin{tikzpicture}[x=1.7cm,y=1.8cm]


  \node[obs]                   (w_j)      {$\ul{w}_j$} ; %
  \node[latent, above=of w_j]    (k_j)      {$\ul{k}_j$} ; %
  \edge {k_j} {w_j} ; %
  \plate {doc} {(w_j) (k_j)} {$j= 1..J$}; %
  



\end{tikzpicture}

 \hspace{1.0cm}
}
\subfigure[Theme-topic mixture]{
 \hspace{1.0cm}
%
%
%
%


\begin{tikzpicture}[x=1.7cm,y=1.8cm]


  \node[obs]                   (w_j)      {$\ul{w}_j$} ; %
  \node[latent, above=of w_j]    (k_j)      {$\ul{k}_j$} ; %
  \node[latent, above=of k_j]    (l)      {$l$} ; %
  \edge {k_j} {w_j} ; %
  \edge {l} {k_j} ; %
  \plate {doc} {(w_j) (k_j)} {$j= 1..J$}; %
  



\end{tikzpicture}

 \hspace{1.0cm}
}
\caption{Comparison of the basic mixture model extensions of multinomials using directed independency graph notation}
\label{mix_directedgraph}
\end{figure}

Graphical model notations are most useful for expressing an overview of a statistical model, compared
to similar models. The mixture models discussed in this section are compared Figure \ref{mix_directedgraph}.
When some properties of the model are not crucial for presenting the main modeling ideas, 
they can be omitted from the illustration. For example, the models discussed in this chapter all share common 
modeling ideas, such as the use of categorical distributions and assumption of IID data. Inclusion of these types 
of properties in the visualization would cause the graphical notations to be less accessible. \\

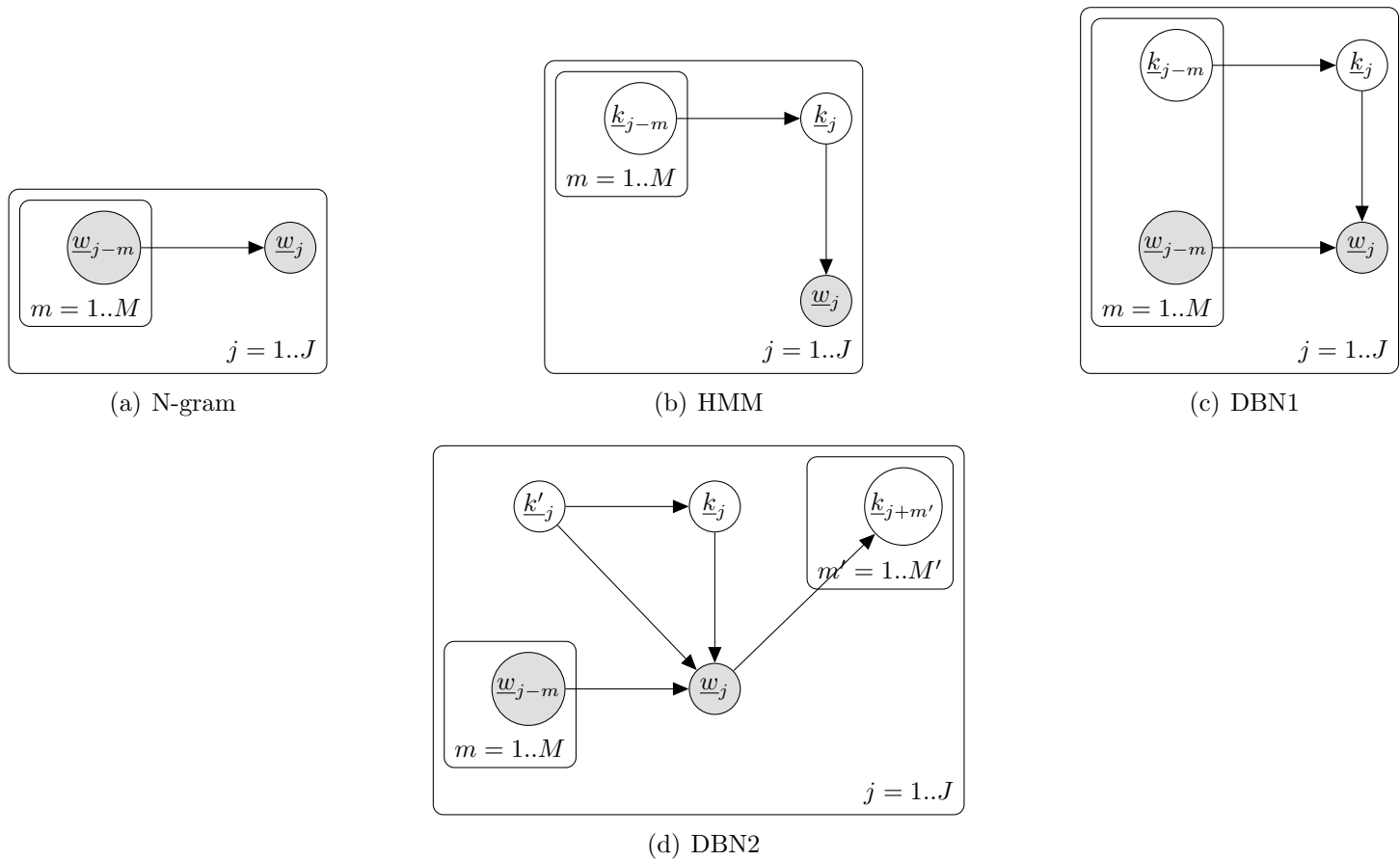
\begin{figure}[h]
\centering
\subfigure[N-gram]{
%
%
%
%


\begin{tikzpicture}[x=1.7cm,y=1.8cm]


  \node[obs]                   (w_j)      {$\ul{w}_j$} ; %
  \node[obs, left=of w_j]    (w_j-)      {$\ul{w}_{j-m}$} ; %
  \edge {w_j-} {w_j} ; %
  \plate {hist} {(w_j-)} {$m= 1..M$};
  \plate {doc} {(hist) (w_j) (w_j-)} {$j= 1..J$}; %
  



\end{tikzpicture}

}\hfill
\subfigure[HMM]{
%
%
%
%


\begin{tikzpicture}[x=1.7cm,y=1.8cm]


  \node[obs]                   (w_j)      {$\ul{w}_j$} ; %
  \node[latent, above=of w_j]    (k_j)      {$\ul{k}_j$} ; %
  \node[latent, left=of k_j]    (k_j-)      {$\ul{k}_{j-m}$} ; %
  \edge {k_j} {w_j} ; %
  \edge {k_j-} {k_j} ; %
  \plate {hist} {(k_j-)} {$m= 1..M$};
  \plate {doc} {(hist) (w_j) (k_j) (k_j-)} {$j= 1..J$}; %
  



\end{tikzpicture}

}\hfill
\subfigure[DBN1]{
%
%
%
%


\begin{tikzpicture}[x=1.7cm,y=1.8cm]


  \node[obs]                   (w_j)      {$\ul{w}_j$} ; %
  \node[obs, left=of w_j]    (w_j-)      {$\ul{w}_{j-m}$} ; %
  \node[latent, above=of w_j]    (k_j)      {$\ul{k}_j$} ; %
  \node[latent, left=of k_j]    (k_j-)      {$\ul{k}_{j-m}$} ; %
  \edge {k_j} {w_j} ; %
  \edge {k_j-} {k_j} ; %
  \edge {w_j-} {w_j} ; %
  \plate {hist} {(k_j-) (w_j-)} {$m= 1..M$};
  \plate {doc} {(hist) (w_j) (k_j) (k_j-) (w_j-)} {$j= 1..J$}; %
  



\end{tikzpicture}

}\hfill
\subfigure[DBN2]{
%
%
%
%


\begin{tikzpicture}[x=1.7cm,y=1.8cm]


  \node[obs]                   (w_j)      {$\ul{w}_j$} ; %
  \node[obs, left=of w_j]    (w_j-)      {$\ul{w}_{j-m}$} ; %
  \node[latent, above=of w_j]    (k_j)      {$\ul{k}_j$} ; %
  \node[latent, left=of k_j]    (k'_j)      {$\ul{k'}_{j}$} ; %
  \node[latent, right=of k_j]    (k_j+)      {$\ul{k}_{j+m'}$} ; %
  \edge {k_j} {w_j} ; %
  \edge {k'_j} {w_j} ; %
  \edge {k'_j} {k_j} ; %
  \edge {w_j-} {w_j} ; %
  \edge {w_j} {k_j+} ; %
  \plate {hist1} {(w_j-)} {$m= 1..M$};
  \plate {hist2} {(k_j+) } {$m'= 1..M'$};
  \plate {doc} {(hist1) (hist2) (w_j) (k_j) (k'_j) (w_j-) (k_j+)} {$j= 1..J$}; %
  



\end{tikzpicture}

}
\caption{Comparison of the sequence model extensions of multinomials using directed graphical model notation. N-gram
and HMM are $M$-order models. DBN1 combines the n-gram and HMM. DBN2 is a highly connected example of a DBN,
with a second hidden variable type $\ul{m'}_j$, and $\ul{w}_j$ conditioning $M'$ future hidden variables $\ul{k}_{j+m'}$.}
\label{seq_directedgraph}
\end{figure}

The use of unrolled Bayes Networks for sequence modeling has been called Dynamic Bayes Network 
models (DBN) \citep{Deviren:04, Deviren:05, Wiggers:06}. Graph unrolling enables describing sequence models 
such as topic models, HMMs, and n-grams in the notation of directed graphical models. 
Topic variables, HMM hidden states and n-gram order interpolation weights can be described as variables in 
the graph. For example, a categorical HMM can be expressed as $p(\ul{\bm{w}},\ul{\bm{k}})
= \prod_t \Psi(\text{nd}(t))$, where $\Psi_t(\text{nd}(t))= p_{\ul{k}_{t}}(\ul{w}_{t})$ for $1 \le t \le J$ and $\Psi_t(\text{nd}(t))= 
p(\ul{k}_{t\%M} | \ul{k}_{t\%M-1})$ for $J+1 \le t \le (J+1)M$, where $M$ is the number of HMM hidden states $m$.\\

DBNs extend this graphical model view of HMMs, so that a number of hidden variables can underlie an observed word output, 
instead of a single hidden state variable. For example, a model of text
can have topic variables, dialog types, word history, word clusters, parts of speech etc. as the hidden variables 
\citep{Deviren:04, Deviren:05, Wiggers:06}. Including some of these variable types can considerably improve 
over simple n-gram models of text. Interpolated n-gram models can be included in DBNs by linking to each word the different n-gram 
order nodes, and an interpolation variable node that outputs the n-gram order mixture weights. With DBNs any conceivable 
variables can be used to condition the sequence variables, as long as the conditioning does not form cycles in the graph.
Figure \ref{seq_directedgraph} shows a comparison of the sequence model extensions of multinomials.\\

\subsection{Factor Graphs and Gates}

Graphical models using both directed and undirected edges include chain graphs \citep{Frydenberg:90} and 
ancestral graph Markov models \citep{Richardson:02}. More recently, factor graphs \citep{Kschischang:01, Frey:03, Loeliger:04, Frey:05, Lazic:13} 
has been proposed as a superset of the earlier frameworks. The original formalism itself
has been followed by a number of extensions \citep{Loeliger:04, Frey:03, Minka:08, McCallum:09b, Dietz:10, Andres:12},
such as directed factors \citep{Frey:03, Dietz:10}, gates \citep{Frey:03, Minka:08, Dietz:10, Oberhoff:11} and factor templates \citep{McCallum:09b}.\\

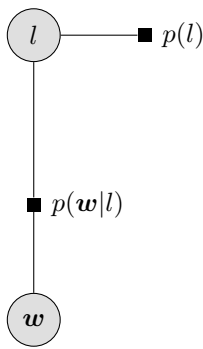
\begin{figure}[h]
\centering
%
%
%
%


\begin{tikzpicture}[x=1.7cm,y=1.8cm]
  \node[obs]                       (w)           {$\bm w$} ; %
  \factor[factor,above=of w]      {pwl}     {right:$p(\bm w| l)$}{}{}  ; %
  \node[obs,above=of pwl]          (l)           {$l$} ; %
  \factor[factor,right=of l]      {pl}     {right:$p(l)$}{}{}  ; %

  \factoredge {l} {pl} {} ; %
  \factoredge {w} {pwl} {} ; %
  \factoredge {pwl} {l} {} ; %

\end{tikzpicture}

\caption{Multinomial Naive Bayes illustrated using factor graph notation, with the factors $p(l, \bm w)= \Psi_1(l) \Psi_2(l, \bm w)$}
\label{mnb_factor}
\end{figure}

The factor graph notation visualizes any graphical model using a graph $G= (V, F, E)$ of variable nodes $V$, factor nodes $F$ and
edges $E$. The graph is bipartite, so that each edge connects a variable node to a factor node. The variable nodes are 
visualized using circles, the factor nodes using small squares and the edges using lines. Figure \ref{mnb_factor} shows 
MNB as a factor graph. Factor graphs are a strict superset of undirected and directed graphical models, since 
both types of models can be represented as factor graphs, while many factor graphs cannot be represented by either 
types of models \citep{Frey:03}.\\

Gates are a proposed extension of factor graph notation that allows more explicit visualization of mixture models and
context-dependent models \citep{Minka:08, Winn:12}. Gates notation uses a gate or switch function \citep{Frey:03}
to select the behaviour of a sub-graph based on a key variable, such as a mixture model component indicator.
For example, labels for MNB can be written as a vector of variables $\bm c$: $\forall l: 0 \le c_l \le 1$, 
and $\sum_l c_l= 1$. The joint distribution for MNB can then be rewritten as a gate:
\begin{align*}
p(\bm w, \bm c)= \prod_l (p(\bm w, l))^{c_l}
\numberthis \label{MNB_GATES}\\
\end{align*}

The label indicator variables $c_l$ in Equation \ref{MNB_GATES} performs the role of a switch, changing the output of the gate to $1$ 
for all labels $l$ with key value $c_l=0$. The gate formalism enables expression of factor graphs 
where more general functions are used for the key variables, such as context-dependent variables. For a given key value, only the 
parts of the gated sub-graph with key value $c_l>0$ need to be computed.\\

\begin{figure}[h]
\centering
\hspace{1.0cm}
\subfigure[Gates]{
%
%
%
%


\begin{tikzpicture}[x=1.7cm,y=1.8cm]
  \node[obs]                       (w)           {$\bm w$} ; %
  \factor[factor,above=of w]      {pwl}     {right:$p(\bm w| l)$}{}{}  ; %
  \node[obs,above=of pwl]          (l)           {$l$} ; %
  \factor[factor,below right=of l]      {pl}     {right:$p(l)$}{}{}  ; %

  \factoredge {l} {pl} {} ; %
  \factoredge {w} {pwl} {} ; %
  \factoredge {pwl} {l} {} ; %
  \gate {X-gate} {(pwl)(pl)(pwl-caption)(pl-caption)} {}
  \factoredge {l} {X-gate} {} ; %
\end{tikzpicture}

}
\hfill
\hspace{2.0cm}
\subfigure[Expanded gates, for a model with L=2]{
%
%
%
%


\begin{tikzpicture}[x=1.7cm,y=1.8cm]
  \node[obs]                       (w)           {$\bm w$} ; %
  \factor[factor,above=of w]      {pwl1}     {left:$p(\bm w| l=1)$}{}{}  ; %
  \factor[factor,above=of pwl1]      {pl1}     {left:$p(l=1)$}{}{}  ; %
  \factor[factor,right=of pwl1]      {pwl2}     {right:$p(\bm w| l=2)$}{}{}  ; %
  \factor[factor,above right=of pwl2]      {pl2}     {right:$p(l=2)$}{}{}  ; %
  \node[obs,above=of pl2]          (l)           {$l$} ; %
  \factoredge {l} {pl1} {} ; %
  \factoredge {w} {pwl1} {} ; %
  \factoredge {pwl1} {l} {} ; %
  \factoredge {l} {pl2} {} ; %
  \factoredge {w} {pwl2} {} ; %
  \factoredge {pwl2} {l} {} ; %
  \vgate {V-gate}
  {(pwl1)(pl1)(pwl1-caption)(pl1-caption)} {$l=1$}
  {(pwl2)(pl2)(pwl2-caption)(pl2-caption)} {$l=2$} {}
  \factoredge {l} {V-gate} {} ; %
\end{tikzpicture}

\hspace{1.0cm}
}
\caption{Multinomial Naive Bayes illustrated using gated factor graph notation, with the gate 
$p(\bm w, \bm c)= \prod_l (p(\bm w, l))^{c_l}$ for factors $p(l, \bm w)= \Psi_1(l) \Psi_2(l, \bm w)$}
\label{mnb_gates2}
\end{figure}
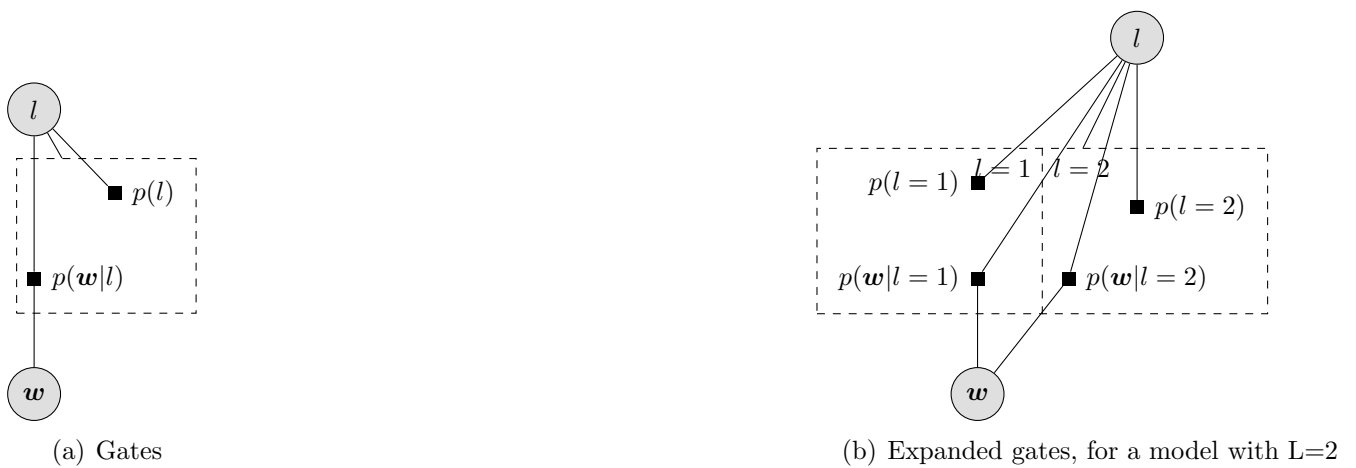

Gates are visualized using dashed rectangles, either as a single rectangle similar to a plate, or in an expanded form
with separate rectangles for the different values of the key. The latter is useful in complex cases, when the key values 
result in different types of computations within the gate. The key variable is shown connected to the gate rectangle by a line. 
Figure \ref{mnb_gates2} shows gated factor graphs for MNB with gates and expanded gates. The gate notation is unnecessary for
simple models such as MNB, but becomes useful for illustrating inference in more complex graphical models, as will be shown
in Chapter 5 of the thesis.\\

\subsection{Inference and Estimation with Directed Generative Models}
\subsubsection{Overview of Algorithms}
Directed generative models commonly have efficient algorithms for both inference and estimation. MNB has linear time and space 
complexity algorithms for both, as do constrained extensions of multinomials such as n-grams. Mixture extensions generally
complicate the estimation, and multiply the inference complexities by the number of components in each mixture. 
Dynamic programming \citep{Bellman:52, Viterbi:67} is an algorithmic innovation that can be used to reduce the complexities for HMM and 
DBN extensions. Latent Dirichlet Allocation and some of the more complex directed graphical models require other approximate algorithms for both inference and estimation,
such as Variational Bayes and Gibbs Sampling \citep{Asuncion:09, Blei:12}.\\

The joint $p(\bm w, l)$ for MNB is computed by the product $p(l) \prod_n p_l(n)^{w_n}$ over the prior and the label  
conditionals, and the naive inference algorithm for this is a trivial product over the terms. Inferring the marginal $p(\bm w)$ 
from the joint and posterior $p(l|\bm w)$ with the Bayes rule can be done with the closed form operations of sums and products.
The naive inference time and space complexities for the MNB posterior $p(l|\bm w)$ is $O(|\bm w|_0 L)$ \citep{Manning:08}.
Document-clustering mixtures introduce sums over documents, and word-clustering mixtures introduce sums over words;
both multiply the space and time complexities of inference by the number of components $J$. Replacing the multinomial in MNB 
with a HMM introduces a sum over all possible hidden state sequences. This can be reduced using dynamic 
programming techniques. Inference on more complex directed graphical models can require other algorithms, such as the approximate
algorithms for Latent Dirichlet Allocation \citep{Asuncion:09, Blei:12}.\\

MNB is estimated by gathering and normalizing the sufficient statistics of counts. When sparsity is utilized, this has the time complexity 
$O(\sum_i |\bm w^{(i)}|_0)$ and space complexity $O(L+\sum_l \sum_{n: \exp(\lambda_{ln})>0} 1)$. 
Extension with mixtures complicates the estimation, due to a sum in the likelihood function that does not decompose into 
separate optimizations. A common practical solution to this 
problem is the EM algorithm based on dynamic programming. This treats the mixture components as random variables, iteratively 
optimizing the expected conditional log-likelihood until a stationary point of the likelihood 
function is reached. As with inference, more complex graphical model extensions can require approximations such as Variational Bayes and
Gibbs Sampling. Aside from parameter estimation, sometimes the graphical model structure itself is unknown and learned using a variety 
of approximate methods \citep{Friedman:97, Lazic:13}.\\

Additional meta-optimization in estimation is commonly required, to choose the meta-parameters required by the models, 
and to avoid bad local minima of the EM-estimated likelihood function. Meta-parameters required by the models
can include the number of components in the mixtures and smoothing parameters. Basic solutions for setting these are 
heuristic values and grid searches. Local minima are encountered with complex multimodal likelihood functions, such as those
for mixture models. A basic solution is random restarts for lowering the probability of a low quality local optima. Chapter 6 of the thesis
describes a Gaussian random search algorithm that can be used to solve the meta-optimization problem in a principled manner.

\subsubsection{Dynamic Programming}
Dynamic programming \citep{Bellman:52} is an algorithmic innovation that can be used to solve problems
with overlapping subproblems efficiently. The application of dynamic programming makes HMMs practical, 
by lowering the complexity of the required computations. These
are next described in brief for the case of a first-order categorical HMM. The probability of a word sequence 
$\ul{\bm{w}}$ for the HMM of Equation \ref{hmm} can be marginalized:
\begin{align*}
p(\ul{\bm{w}}) &= \sum_{\ul{\bm{k}}} \prod_j p(\ul{k}_j | \ul{k}_{j-1})  p_{\ul{k}_j}(\ul{w}_j)
\numberthis\\
\end{align*}

Marginalizing $p(\ul{\bm{w}})$ by summation over the possible state sequences is not usually feasible,
as there are $M^J$ possible state sequences. Computing this observation state probability
efficiently is considered the first of the three main computational problems with HMMs \citep{Rabiner:89}.
The second problem is optimizing $\argmax_{\ul{\bm{k}}} p(\ul{\bm{k}} | \ul{\bm{w}})$, used for
segmenting data to the hidden state sequences. The third problem is estimating the model, given that
the hidden states are unknown in training data.\\

The summation over the state sequences 
$\sum_{\ul{\bm{k}}}$ in computing $p(\ul{\bm{w}})$ can be considered a brute-force solution to the 
problem. A dynamic programming solution is computing the probabilities $p(\ul{\bm{w}})$ over
the sequence indices $j$ instead, summing for each possible state $m$ the probability of all sequences 
leading to the state, called the forward probability $\xi_j(m)$. This is known as the forward algorithm.
Using the forward variables, the marginal probability for the HMM can be rewritten in a recursive form:
\begin{align*}
p(\ul{\bm{w}}) &= \sum_{\ul{\bm{k}}} \prod_j p(\ul{k}_j | \ul{k}_{j-1})  p_{\ul{k}_j}(\ul{w}_j) \\
					 &= \sum_m \xi_{J}(m)\\
\xi_j(m)&=\begin{cases}
         p(m) p_m(\ul{w}_j), & \text{if $j=1$}\\
         \sum_{m'} (\xi_{j-1}(m') p(m | m'))  p_m(\ul{w}_j), & \text{otherwise} ,\\
        \end{cases}
\numberthis
\end{align*}
where $p(m)= p(\ul{k}_j)$ and $p_m(\ul{w}_j)=p_{\ul{k}_j}(\ul{w}_j)$.\\

The forward algorithm solves $p(\ul{\bm{w}})$ recursively by computing $\xi_j(m)$ from $j=1$ to $j=J$, 
reducing the time complexity from $O(JM^J)$ to $O(JM^2)$. The space complexity is also reduced to $O(2K)$,
since only the forward variables for the current and previous sequence index need to be kept in memory.
Alternatively, the algorithm can be run in the reverse order.
This is called the backward algorithm and it produces exactly the same results. These algorithms
can be extended to virtually all types of HMMs. The zeroth order HMM can be 
shown to be a special case, requiring only $O(JM)$:
\begin{align*}
p(\ul{\bm{w}}) &= \sum_m \xi_{J}(m)\\
                        &= \sum_m \sum_{m'} ((\xi_{J-1}(m') p(m))  p_m(\ul{w}_J))\\
                        &= \sum_{m'} (\xi_{J-1}(m')) \sum_m (p(m) p_m(\ul{w}_J))\\
                        &= \prod_j \sum_m p(m) p_m(\ul{w}_j))
\numberthis \label{forward}\\
\end{align*}

Related dynamic programming algorithms are used for solving the other two problems with HMMs. Changing 
the sum over $m$ in Equation \ref{forward} to a max returns the probability of the 
word sequence by the single most likely sequence of states, solving the second problem. This is known as the Viterbi 
algorithm \citep{Viterbi:67} and its efficient implementations underlie many of the applications of HMMs into sequence 
classification. Lastly, the forward and backward algorithms can be combined to compute the posterior probabilities 
$p(\ul{k}_j | \bm{\ul{w}})$ of each state $\ul{k}_j$ for each sequence index $j$. This is 
known as the forward-backward algorithm, and the posteriors can be used for the Expectation step
in EM estimation \citep{Baum:70, Rabiner:89, Bilmes:98}.\\

In case a directed graphical model has no cycles of edges, efficient exact dynamic programming inference is possible using
extensions of the forward, Viterbi and forward-backward algorithms to general graphs \citep{Pearl:86, Kschischang:01, Loeliger:04}. 
For more complex graphical models a variety of less efficient exact and approximate inference algorithms exist. 
Most of these build on the idea of variable elimination
\citep{Zhang:94}, that works by marginalizing variables away to arrive at the inferred probability distribution.
The generalized case of forward algorithm to graphs is the sum-product algorithm or belief propagation. 
Analogously the generalized case of Viterbi algorithm is the max-product algorithm. These extend the use of the forward and
maximum variables $\xi_j(m)$, so that a message variable is computed for each node in the graph and the inference
is done by passing the messages in an efficient order. 

\subsubsection{Expectation Maximization}
The training data for the HMM of Equation \ref{hmm} consists of instances {$D^{(i)}= (\ul{\bm w}^{(i)}, \ul{\bm k}^{(i)})$ 
of known and hidden variables, where $\ul{\bm w}^{(i)}$ is the known sequence and $\ul{\bm k}^{(i)}$ is unknown variables for 
the $i$-th instance of training data. The likelihood function becomes:
\begin{align*}
\mathcal{L}(\bm \theta|D)&= p(D| {\bm \theta})\\
&=\prod_i \sum_{\ul{\bm k}^{(i)}}  p(\ul{\bm{w}}^{(i)},\ul{\bm{k}}^{(i)} |\bm \theta)\\
&= \prod_i \sum_{\ul{\bm k}^{(i)}} \prod_j (p(\ul{k}_j^{(i)}|\ul{k}_{(j-1)}^{(i)}, \bm \theta) \:  p_{\ul{k}_j^{(i)}}(\ul{w}^{(i)}_j| \bm \theta))
\numberthis \label{likelihood2}\\
\end{align*}

The summation over components causes a problem for optimizing the log-likelihood of the model, since
the log-likelihood no longer factorizes into parts that can be separately optimized. If the hidden states were
known and each word was generated by a single component, the log-likelihood function would factorize easily. 
In this case the likelihood function is:
\begin{align*}
\mathcal{L}(\bm \theta|D)&=\prod_i  p(\ul{\bm{w}}^{(i)},\ul{\bm{k}}^{(i)} |\bm \theta)\\
&= \prod_i \prod_j (p(\ul{k}_j^{(i)}|\ul{k}_{(j-1)}^{(i)}, \bm \theta) \: p_{\ul{k}_j^{(i)}}(\ul{w}^{(i)}_j| \bm \theta))
\numberthis\\
\end{align*}

In most cases the hidden states are not known. With mixture models and HMMs the main solution to this 
problem is using the EM algorithm to estimate the model. Instead of attempting to maximize the log-likelihood, the 
EM algorithm treats the components as random variables and iteratively optimizes the conditional expectation
of the log-likelihood $Q(\bm \theta|D, \bm{\hat{\theta}})$. Let $p(\ul{\bm k}^{(i)}| \ul{\bm w}^{(i)}, \bm{\hat{\theta}})$
indicate the expectation of a hidden variable sequence $\ul{\bm k}^{(i)}$ for word sequence $\ul{\bm w}^{(i)}$ according to some prior 
parameters $\bm{\hat{\theta}}$: $p(\ul{\bm k}^{(i)}| \ul{\bm w}^{(i)}, \bm{\hat{\theta}})=p(\ul{\bm k}^{(i)}, \ul{\bm w}^{(i)}| \bm{\hat{\theta}}) 
/ \sum_{\ul{\bm k}^{(i)}} p(\ul{\bm k}^{(i)}, \ul{\bm w}^{(i)}| \bm{\hat{\theta}})$ The Q-function becomes:
\begin{align*}
Q(\bm \theta|D, \bm{\hat{\theta}})&= E(\log(\mathcal{L}(\bm \theta|D, \bm{\hat{\theta}}))) \\
&=\sum_i \sum_{\ul{\bm k}^{(i)}} {p(\ul{\bm k}^{(i)}| \ul{\bm w}^{(i)}, \bm{\hat{\theta}})} \log(p(\ul{\bm{w}}^{(i)},\ul{\bm{k}}^{(i)} |\bm \theta))\\
&= \sum_i  \sum_{\ul{\bm k}^{(i)}} \sum_j p(\ul{\bm k}^{(i)}| \ul{\bm w}^{(i)}, \bm{\hat{\theta}}) \log(p(\ul{k}_j^{(i)}|\ul{k}_{(j-1)}^{(i)},\bm \theta) \: p_{\ul{k}_j^{(i)}}(\ul{w}^{(i)}_j| \bm \theta)))
\numberthis\\
\end{align*}

Since $p(\ul{\bm k}^{(i)}| \ul{\bm w}^{(i)}, \bm{\hat{\theta}})$ is treated as a random variable, $Q(\bm \theta|D, \bm{\hat{\theta}})$ 
becomes a random variable as well. With random variables the sum of expectations equals the expectation
of the sum, so that for computing $\mathcal{L}(\bm \theta|D, \bm{\hat{\theta}}))$ the expectation
can be pushed inside the sums, placing the expectation $p(\ul{\bm k}^{(i)}| \ul{\bm w}^{(i)}, \bm{\hat{\theta}})$ as a weight for 
each possible hidden state sequence. Let $\bm \lambda_m$ indicate the conditional parameters for 
component $m$, $\bm \lambda$ the conditional parameters for all components and $\bm \alpha$ the 
parameters for component weights. The conditional log-likelihood can be optimized:
\begin{align*}
&\argmax_{\bm \theta}(Q(\bm \theta|D, \bm{\hat{\theta}}))\\
&= \argmax_{(\bm \lambda, \bm \alpha)}(\sum_i \sum_{\ul{\bm k}^{(i)}} \sum_j p(\ul{\bm k}^{(i)}| \ul{\bm w}^{(i)}, \bm{\hat{\theta}}) \log(p(\ul{k}_j^{(i)}|\ul{k}_{(j-1)}^{(i)},\bm \alpha) \: p_{\ul{k}_j^{(i)}}(\ul{w}^{(i)}_j| \bm \lambda)))\\
&= \argmax_{(\bm \lambda, \bm \alpha)}(\sum_i \sum_{\ul{\bm k}^{(i)}} \sum_j p(\ul{\bm k}^{(i)}| \ul{\bm w}^{(i)}, \bm{\hat{\theta}}) \log(p(\ul{k}_j^{(i)}|\ul{k}_{(j-1)}^{(i)},\bm \alpha))\\
&+ \sum_i \sum_{\ul{\bm k}^{(i)}} \sum_j p(\ul{\bm k}^{(i)}| \ul{\bm w}^{(i)}, \bm{\hat{\theta}}) \log(p_{\ul{k}_j^{(i)}}(\ul{w}^{(i)}_j| \bm \lambda)))
\numberthis \label{loglikelihood3}\\
\end{align*}

Replacing the current set of parameters $\bm{\hat{\theta}}$ with the estimated parameters 
$\bm{\hat{\theta}}'= \argmax_{\bm \theta}(Q(\bm \theta|D, \bm{\hat{\theta}}))$
and repeating the optimization until
$Q(\bm \theta|D, \bm{\hat{\theta}})= Q(\bm \theta|D, \bm{\hat{\theta}}')$ produces the
EM-estimated stationary point of the likelihood function. Depending on the likelihood function and initialization,
the stationary point can remain far from a global optimum, and EM is often repeated with randomly sampled initial 
parameters. Other practical variants include online versions of EM \citep{Liang:09} and combination with genetic 
algorithms \citep{Martinez:00, Pernkopf:05,Puurula:10}.\\

For more scalable algorithmic implementation, EM is implemented using the forward-backward variables 
$p(\ul{k}^{(i)}_j | \bm{\ul{w}^{(i)}},\bm{\hat{\theta}})$
instead of the state sequence variables $p(\ul{\bm k}^{(i)}| \ul{\bm w}^{(i)}, \bm{\hat{\theta}})$. 
Summing these weighted posterior probabilities produces expected counts $E(C())$ in place of the counts $C()$ used in Equation \ref{loglikelihood}.
Using the forward-backward algorithm, the posteriors can be computed 
in the same time complexity as the forward algorithm, but require $O(JM)$ in space complexity, as the full $JM$ matrix needs
to be stored for estimating with the weights.\\

\chapter{Reformalizing Multinomial Naive Bayes}
This chapter presents a thorough reformalization of Multinomial Naive Bayes (MNB) as a probabilistic model. The issue of smoothing
is first discussed, noting that most of the multinomial smoothing methods necessary for MNB are not correctly formalized under maximum likelihood
estimation. A unifying framework for the methods is proposed, and a two-state Hidden Markov Model (HMM) formalization is shown
to derive the smoothed parameter estimates. Feature weighting is formalized
for both estimation and inference as well-defined approximation with expected log-probabilities, given probabilistically weighted word sequences.
A formalization of MNB is defined that takes these corrections into account, followed by a more general graphical model extension
that includes label-conditional document length modeling, and scaling the influence of label priors.\\

\section{Formalizing Smoothing}
\subsection{Smoothing Methods for Multinomials}
\label{smoothing_section}
The sparsity of text data causes problems for maximum likelihood estimation of multinomials. Words occurring zero times
for a label will cause the corresponding parameter estimates to be zero as well, resulting in 0-probabilities when computing
the probabilities with \emph{unsmoothed models} $p_l^u(\bm w| \bm \lambda^u)$. This complication is known as the 
zero-frequency problem in statistics, and smoothing methods for solving this problem have been extensively researched. However, the 
\emph{smoothed models} $p_l(\bm w| \bm \lambda)$ are no longer justified by maximum likelihood, and only a subset of the 
smoothing methods can be formalized under other principles, such as maximum a posteriori \citep{MacKay:95, Rennie:01}. This 
section proposes a unified framework for smoothing generative
models of text, and shows that practically all of the smoothing methods for multinomials can be formalized as approximate 
maximum likelihood estimation on a constrained Hidden Markov Model (HMM).\\

An early solution to the zero-frequency problem dates two centuries and is known as Laplace correction \citep{Laplace:14}. 
On discussing the probability of the sun rising tomorrow, Laplace argued that despite 
the event of the sun not rising has never been seen, it should still have a probability assigned to it. The proposed Laplace 
correction adds a single count to each possible event, thereby avoiding the zero frequency problem. In language modeling 
this correction is known as Laplace smoothing, and a parametric generalization adding a fractional count $\mu$ instead 
was proposed by Lidstone in 1920 \citep{Lidstone:20}. By further multiplying the added count by a prior background model 
$p^u(n)$, this correction becomes the Dirichlet prior method for smoothing.\\

Maximum a posteriori estimation of multinomials with a Dirichlet prior takes the form:
\begin{align*}
\lambda_{ln}= \log(\frac {C(l,n)+\mu \: p^u(n)}{\sum_{n'} C(l,n')+ \mu \: p^u(n')}), \numberthis
\label{Dir_smooth}\\
\end{align*}
where the special case $p^u(n)= 1/N$ is Lidstone smoothing and $p^u(n)= 1/N$, $\mu=N$ 
is Laplace correction \citep{Rennie:01, Smucker:07}.\\ 

Another basic way to avoid the zero-frequency problem is to linearly interpolate parameter estimates with a background model. 
This is known as Jelinek-Mercer smoothing \citep{Jelinek:80}, and takes the form:

\begin{align*}
\lambda_{ln}= \log((1-\beta) \frac {C(l,n)} {\sum_{n'} C(l,n')} + \beta \: p^u(n)) \numberthis
\label{JM_smooth}\\
\end{align*}

Dirichlet prior and Jelinek-Mercer differ only in how the weight for the background model is chosen. 
A general interpolation function covers both types of smoothing \citep{Johnson:32,Smucker:07}:

\begin{align*}
\lambda_{ln}= \log((1-\alpha_l) p^u_l(n) + \alpha_l p^u(n)), \numberthis
\label{int_smooth}
\end{align*}

where $p^u_l(n)$ is the unsmoothed label-conditional multinomial for label $l$. Fixing $\alpha_l$ to a pre-determined value 
$\alpha_l= \beta$ produces Jelinek-Mercer smoothing. Fixing $\alpha_l =\frac{\mu} {\mu+\sum_n C(l,n) }= 1- \frac{\sum_n C(l, n)}{\mu + \sum_n C(l, n)}$ 
produces Dirichlet prior smoothing. Combining these as $\alpha_l= 1- \frac {\sum_n (C(l, n) - \beta C(l, n))}{\mu+\sum_n C(l, n)}$ 
results in two-stage smoothing, a smoothing method suggested for information retrieval \citep{Zhai:01, Smucker:07}.\\

\label{background_models}
The background model is commonly a uniform distribution $p^u(n)=\frac{1}{N}$, or a label-independent collection model
$p^u(n)= \frac {\sum_l C(l, n)} {\sum_{n'} \sum_l C(l, n')}$. A uniform-smoothed collection model with smoothing weight $\Upsilon$ interpolates 
between a uniform and a collection model: $p^u(n)=(1-\Upsilon) \frac {\sum_l C(l, n)} {\sum_{n'} \sum_l C(l, n')}+\Upsilon \frac{1}{N}$. 
In n-gram language modeling literature the uniform background model
is called the zerogram, and the collection model is called the unigram background model \citep{Chen:99}. 
Alternatively, if several documents per label exist, the counts from each document can be normalized by length, or weighted according 
to usefulness. External datasets can likewise be used to estimate the background model, such as large text datasets of the same language. 
The choice of background model can also 
be motivated by the task, such as using a collection model to introduce relevance information in ranked retrieval \citep{Zhai:01}.\\

Jelinek-Mercer and Dirichlet prior are the two most common types of smoothing for MNB models. One view
of smoothing is that smoothing methods discount seen occurrences of words in order to redistribute the subtracted
probability mass $a_l$ to the background model. Under this view both of these discount the seen counts linearly 
by $\alpha_l$. A third basic type of smoothing is called is called absolute discounting 
\citep{Ney:94, Chen:96, Chen:99, Zhai:01b, Zhai:04}. This works similar to Jelinek-Mercer smoothing, but subtracts 
a parameter value $\delta: 0 \leq \delta \leq 1$ from all counts for a label, and uses the subtracted probability 
mass for choosing the smoothing coefficient. The discounted counts can be denoted $C'(l, n)= C(l,n) -\delta$. Using Equation 
\ref{int_smooth} and choosing $p^u_l(n)= \frac {C'(l,n)} {\sum_{n'} C'(l,n')}$, 
and $\alpha_l= 1- \frac{ \sum_n C'(l,n)} {\sum_n C(l,n)}$ produces absolute discounting.\\

A problem with absolute discounting is that usually separate discount values are optimal for different counts, 
with higher discounts for 
higher counts \citep{Ney:94}. Empirical analyses \citep{Chen:99, Durrett:11, Schutze:11, Neubig:12} have shown 
that optimal discount values seem to follow a power-law distribution, rather
than the constant ones in absolute discounting. A recent improvement over absolute 
discounting is power-law discounting \citep{Momtazi:10c, Momtazi:10b, Huang:10}. This method discounts according to a 
power function 
$C'(l, n)=C(l,n) - \delta C(l,n)^\delta$, with $0\le \delta \le 1$. Combining this with 
a Dirichlet prior and reorganizing terms produces Pitman-Yor smoothing, with 
$p^u_l(n)= \frac {C(l,n) -\delta C(l, n)^{\delta}} {\sum_{n'} (C(l,n') -\delta C(l, n')^{\delta})}$ and 
$\alpha_l= 1- \frac{ \sum_n (C(l, n)-\delta C(l, n)^{\delta})} {\mu+ \sum_n C(l, n)}$, that
approximates inference on a Pitman-Yor process  \citep{Momtazi:10c, Momtazi:10b, Huang:10}.\\

\begin{table}[ht]
\centering
\caption{Smoothing methods used with MNB models. Parameter estimation formulas for smoothing weights $\alpha_l$ and unsmoothed multinomials $p^u_l(n)$.}
\begin{tabular}{|l|c |c |} \hline
Smoothing method & Smoothing weight $\alpha_l$ & Multinomial $p_l^u(n)$ \\ \hline
Jelinek-Mercer & $\beta$ & $\frac{C(l, n)} {\sum_{n'} C(l, n')}$ \\ \hline
Dirichlet prior & $1- \frac{\sum_n C(l, n)}{\mu + \sum_n C(l, n)}$ & $\frac{C(l, n)} {\sum_{n'} C(l, n')}$ \\ \hline
Two-stage smoothing & $1- \frac {\sum_n (C(l, n) - \beta C(l, n))}{\mu + \sum_n C(l, n) }$  & $\frac{C(l, n)} {\sum_{n'} C(l, n')}$ \\ \hline
Absolute discounting & $1- \frac{ \sum_n (C(l, n)-\delta)} { \sum_n C(l, n)}$ & $\frac {C(l,n) -\delta} {\sum_{n'} (C(l,n') -\delta)}$ \\ \hline
Power-law discounting & $1- \frac{ \sum_n (C(l, n)-\delta C(l, n)^{\delta})} { \sum_n C(l, n)}$ & $\frac {C(l,n) -\delta C(l, n)^{\delta}} {\sum_{n'} (C(l,n') -\delta C(l, n')^{\delta})}$ \\ \hline
Pitman-Yor approx. & $1- \frac{ \sum_n (C(l, n)-\delta C(l, n)^{\delta})} {\mu+ \sum_n C(l, n)}$ & $\frac {C(l,n) -\delta C(l, n)^{\delta}} {\sum_{n'} (C(l,n') -\delta C(l, n')^{\delta})}$ \\ \hline
Generalized smoothing & $1- \frac{ \sum_n (C(l, n)-D(l, n))} {\mu+ \sum_n C(l, n)}$ & $\frac {C(l,n) -D(l, n)} {\sum_{n'} (C(l,n') -D(l, n'))}$ \\ \hline
\end{tabular}
\label{mnb_smoothing}
\end{table} 

The discussed smoothing methods can be covered by a general function that is called here generalized smoothing. By 
choosing $\alpha_l=1- \frac{ \sum_n (C(l, n)-D(l, n))} {\mu+ \sum_n C(l, n)}$ 
and $p^u_l(n)= \frac {C(l,n) -D(l, n)} {\sum_{n'} (C(l,n') -D(l, n'))}$, and $D(l,n)$ according to the chosen discounting, 
we recover all of the smoothing methods as special cases of Equation \ref{int_smooth}. Generalized smoothing 
with $\mu=0$ and $D(l, n)= \beta C(l, n)$ implements 
Jelinek-Mercer smoothing as linear discounting, $D(l,n)= C(l, n)-\delta$ implements absolute 
discounting and $D(l,n)= \delta C(l, n)^{\delta}$ implements power-law discounting. A discounting function combining
Jelinek-Mercer and power-law discounting can be defined as: $D(l,n)= \delta C(l, n)^{\delta}+ \beta C'(l, n)$, 
where $C'(l,n) = C(l,n)-  \delta C(l, n)^{\delta}$.
Chapter 6 of the thesis experiments with this combined discounting function. Table \ref{mnb_smoothing} summarizes the smoothing methods 
in terms of smoothing weights $a_l$ and unsmoothed multinomials $p^u_l(n)$.\\

The parameters can be chosen to maximize the likelihood of held-out data, or a performance measure
related to the task. Closed form approximations requiring no held-out data are possible for some parameters. 
The discounting parameter $\delta$ for absolute and power-law discounting can be approximated 
using a leave-one-out likelihood estimate \citep{Ney:94}. Denoting the frequency of 1-counts as 
$n_1= \sum_{n:(\sum_l C(l, n))=1}1$ and 2-counts as $n_2= \sum_{n:(\sum_l C(l, n))=2}1$, the discount 
parameters can be approximated as $\delta=n_1/(n_1+2n_2)$ \citep{Ney:94, Chen:99, Huang:10, Zhang:14}. This Kneser-Ney
estimate provides an approximate upper bound of the optimal discount value, and has been demonstrated to work well in 
practice \citep{Chen:99, Goodman:00, Vilar:04, Zhang:14}.\\

The smoothing methods for multinomial text models can be extended into the smoothing methods used 
for higher-order n-gram language models. A substantial literature exists for advanced n-gram smoothing techniques \citep{Chen:99, Rosenfeld:00}. 
These extend the multinomial smoothing techniques hierarchically, by placing the 
lower-order $m-1$ n-gram as the background model for each order $m$. For example, Witten-Bell smoothing \citep{Moffat:90}  
is hierarchical linear interpolation with a nonparametric estimate for the interpolation weights \citep{Chen:99}. Using 
the label-conditional model as a higher order model and the background distribution as a lower-order model, the Witten-Bell 
estimate for the smoothing weight is $\alpha_{l}= 1- \frac{\sum_n C(l, n)} {\sum_{n:C(l, n)>0} 1 + \sum_n C(l, n)}$. We 
can note that Witten-Bell smoothing is a case of Dirichlet prior smoothing with a heuristic estimate for the Dirichlet 
parameter: $\mu_l= \sum_{n:C(l, n)>0} 1 = \sum_n \min(1, C(l, n))$. Witten-Bell smoothing originates from text compression 
modelling, where linearly interpolated n-gram models are known as Prediction by Partial Matching (PPM) models \citep{Cleary:84}. The Witten-Bell
smoothed PPM is known as PPM-C and has been a baseline for text compression for over two decades. In general applications 
of n-grams more effective smoothing methods can be applied.\\

%
Interpolated Kneser-Ney smoothing \citep{Chen:99, 
James:00, Goodman:00, Siivola:05, Goldwater:06, Teh:06, Heafield:13} has been the standard LM smoothing method for 
n-grams for over a decade. This combines n-grams hierarchically using absolute discounting, but replaces the lower order $m<M$
estimates of counts by the number of contexts the count occurs in \citep{Kneser:95}. For a multinomial, the modified background model 
becomes $p^u(n)= \frac {\sum_{l:C(l,n)>0} 1} {\sum_{n'} \sum_{l:C(l,n')>0} 1}$. A common example for this method is the phrase
``San Francisco''. A unigram-model estimate for ``Francisco'' could be very high, but since the unigram is likely to occur
in only this one context, the bigram Kneser-Ney estimate for this lower-order n-gram count would likely be 1. The modified model
would correctly consider the unigram ``Francisco'' as very unlikely to occur outside this context. All but the highest-order unsmoothed 
models are replaced by the modified counts before discounting, 
providing a considerable improvement in modeling precision \citep{Chen:99, James:00}.\\

Some improvements over interpolated Kneser-Ney have been suggested over the years, with limited acceptance.
Modified Kneser-Ney smoothing \citep{Chen:99, Siivola:05, Heafield:13, Zhang:14} replaces the discount for each order with 
three different discounts for counts 1, 2 and 3+, optimized together on held-out data for perplexity 
\citep{Chen:99, Siivola:05}. For both interpolated 
and modified Kneser-Ney, the discount parameters can be estimated on held-out data, or approximated with heuristics such 
as the discussed discount estimate $\delta= n_1/(n_1+2n_2)$.
Power-law discounting LM \citep{Huang:10} replaces the
absolute discounting in interpolated Kneser-Ney with Pitman-Yor Process smoothing.\\

Smoothing the parameter estimates directly as in Equation \ref{int_smooth} would cause zero-value parameters to become 
non-zeros, resulting in loss of the parameter sparsity. The parameter estimates $p^u_l(n)=\lambda_{ln}^u$ and 
$p^u(n)=\lambda_n^u$ are often kept separate, so that the complexity of storing $p_l(n)$ is not increased. The space 
complexity of estimation for a smoothed MNB model is $O(L+N+\sum_l |\bm \lambda_l^u|_0)$ and the time complexity is 
$O(\sum_i |\bm w^{(i)}|_0)$.\\

\subsection{Formalizing Smoothing with Two-State Hidden Markov Models}
\label{smoothing_hmm}
Parameter interpolation is the standard formalization of parameter smoothing for multinomial and n-gram language 
models \citep{Jelinek:80, Chen:96, Chen:99, Zhai:01b, Zhai:04, Smucker:07}. Although parameter interpolation provides a
principled method for estimating multinomial parameters, the estimated parameters are not strictly speaking maximum 
likelihood estimates of the multinomial model, but rather ad-hoc estimates \citep{Hiemstra:04}.
The parameter interpolation introduced in Equation \ref{int_smooth} and Table \ref{mnb_smoothing} shows that all smoothing 
methods can be expressed as a normalized mixture of an unsmoothed multinomial and a background distribution. This analysis 
is extended next by showing that the smoothing methods can be formulated as maximum expected log-likelihood estimation on a 
constrained generative model. The proof proceeds by showing that the 
interpolated parameters used in the smoothed multinomials can be implemented by 
a categorical HMM, and that constraining the HMM appropriately in parameter estimation reproduces a model with the same 
joint probabilities as the smoothed multinomial. Compared to formalization of smoothing methods as approximate inference on a 
Pitman-Yor process and other Bayesian models \citep{MacKay:95, Teh:06, Goldwater:06, Neubig:12}, the proposed HMM 
formalization has the advantage that inference using the estimated models is exact.\\

An early formulation of smoothed document models for IR used a two-state HMM for
formalizing Jelinek-Mercer smoothing \citep{Miller:99}. This model used a HMM with one
hidden state for the document distribution and one for collection smoothing \citep{Miller:99, Xu:00}. In addition, this work 
showed how to integrate bigram models, feature weighting, translation models and relevance feedback using 
the HMM model. However, the parameter estimation for this model was not well defined, the model was considered to 
be very different and unrelated to multinomial LMs, and model smoothing was limited to  
Jelinek-Mercer smoothing. The connection to LMs was discovered shortly afterwards \citep{Hiemstra:01}, as
was the need to use constraints such as parameter tying and fixed parameters for the formalization 
\citep{Hiemstra:01}. Using parameter tying, Jelinek-Mercer smoothed higher-order LMs could be implemented
as HMMs \citep{Manning:99}. Despite considerable interest at the time, the two-state HMM model never became 
popular in IR, nor was its connection to LM smoothing methods explored. We show in the following that this model 
can be used to formalize all of the discussed smoothing methods in the maximum likelihood framework.\\

%

Let $l$ be a label variable indicating one of the $L$ multinomials sharing a background model. Let 
$\ul{\bm w}$ be any sequence of $J$ words, and $\ul{w}_j: 1 \le \ul{w}_j \le N$ correspond to the words
counted in $\bm w$, so that $w_n= \sum_{j: \ul{w}_j= n}1$. Let $\ul{\bm k}$ be an unknown sequence of $M=2$
hidden states $\ul{k}_j: 1 \le \ul{k}_j \le M$ generating the sequence $\ul{\bm w}$. A probability model 
over the joint probabilities $(l, \ul{\bm w}, \ul{\bm k})$ can be defined:
\begin{align*}
p(l, \ul{\bm{w}},\ul{\bm{k}}) &= p(l) \prod_j p_l(\ul{k}_j)  p^u_{l\ul{k}_j}(\ul{w}_j), 
\numberthis \label{model3}
\end{align*}
where $p_l(\ul{k}_j)$ is a categorical and $p^u_{l\ul{k}_j}(\ul{w}_j)$ is a categorical conditional on the label $l$ and the hidden state $\ul{k}_j$.\\

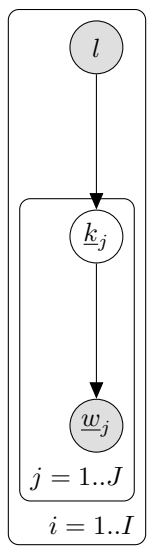
\begin{figure}[h]
\centering
%
%
%
%


\begin{tikzpicture}[x=1.7cm,y=1.8cm]
  \node[obs]                   (w_j)      {$\ul{w}_j$} ; %
  \node[latent, above=of w_j]    (k_j)      {$\ul{k}_j$} ; %
  \node[obs, above=of k_j]    (l)      {$l$} ; %
  \edge {l} {k_j} ; %
  \edge {k_j} {w_j} ; %
  \plate {doc} {(w_j) (k_j)} {$j= 1..J$}; %
  \plate {dataset} {(w_j) (doc) (l)} {$i= 1..I$}; %
\end{tikzpicture}

\caption{Graphical model for the two-state HMM. The hidden variables $\ul{k}_j$ for smoothing components are unknown
and the context label variables $l$ are known in estimation}
\label{HMM_smooth}
\end{figure}

The model of Equation \ref{model3} is a special case of a 0th order categorical HMM, where $p(l)$ correspond to initial state
probabilities, $p_l(\ul{k}_j)$ to HMM state transition probabilities, and $p^u_{l\ul{k}_j}(\ul{w}_j)$ to state emission 
probabilities. Transitions between the label-conditional states are not allowed, so that each word sequence is produced by a single label. 
Figure \ref{HMM_smooth} shows the graphical model for the two-state HMM.\\ 

The conditional probabilities $p_l(\bm w)$ for the model can be derived:
\begin{align*}
p_l(\bm w) &= Z(\bm w) \sum_{\ul{\bm k}} \prod_j p_l(\ul{k}_j) p^u_{l\ul{k}_j}(\ul{w}_j)\\
&= Z(\bm w)  \prod_j \sum_m (p_l(m) p^u_{lm}(\ul{w}_j)) \\
&= Z(\bm w)  \prod_n (\sum_m p_l(m) p^u_{lm}(n))^{w_n},
\numberthis \label{marginal}
\end{align*}
where $p_l(m)=p_l(\ul{k}_j)$, $p^u_{lm}(n)= p^u_{l\ul{k}_j}(\ul{w}_j)$, and $Z(\bm w)$ is the multinomial normalizer.\\

The multinomial normalizer $Z(\bm w)=\frac{(\sum_n w_n)!}{\prod_n w_n!}$ accounts for the fact that a count vector 
$\bm w$ can be generated by the number $Z(\bm w)$ of permutations of the sequence 
$\ul{\bm w}$. As discussed in Chapter 3, computing the product of sums $\prod_j \sum_m$ corresponds to 
the forward algorithm, whereas computing the equivalent sum of products $\sum_{\ul{\bm k}} \prod_j$ gives 
the brute-force solution to the same problem \citep{Rabiner:89}.\\ 

The conditional probabilities $p_l(\bm w) = Z(\bm w) \prod_n (\sum_m p_l(m) p^u_{lm}(n))^{w_n}$ can be implemented by 
a multinomial $p_l(\bm w| \bm \lambda) = Z(\bm w) \prod_n \exp(\lambda_{ln})^{w_n}$, with parameter vector 
$\bm \lambda_l$: $\lambda_{ln}= \log(\sum_m p_l(m) p^u_{lm}(n))$. From this form it can readily be seen that all of the smoothing methods 
can be implemented with the two-state HMM of Equation \ref{model3}, by choosing $p^u_{lm=1}(n)= p^u_{l}(n)$, $p^u_{lm=2}(n)= p^u(n)$, and 
$p_l(m=2)= \alpha_l$ as the smoothing weight.\\ 

The training data for model of Equation \ref{model3} consists of documents {$D^{(i)}= (\ul{\bm w}^{(i)}, l^{(i)}, \ul{\bm k}^{(i)})$, 
where $\ul{\bm w}^{(i)}$ are known word sequences, $l^{(i)}$ are label indicators, and $\ul{\bm k}^{(i)}$ are unknown component assignments for 
the $i$-th document of training data. The likelihood function becomes:
\begin{align*}
\mathcal{L}(\bm \theta|D)&= p(D| {\bm \theta})\\
&=\prod_i p(l^{(i)}|\bm \theta) \sum_{\ul{\bm k}^{(i)}}  p_{l^{(i)}}(\ul{\bm{w}}^{(i)},\ul{\bm{k}}^{(i)} |\bm \theta)\\
&= \prod_i p(l^{(i)}|\bm \theta) \sum_{\ul{\bm k}^{(i)}} \prod_j p_{l^{(i)}}(\ul{k}_j^{(i)}|\bm \theta) \:  p^u_{l^{(i)}\ul{k}_j^{(i)}}(\ul{w}^{(i)}_j| \bm \theta)\\
&= \prod_i p(l^{(i)}|\bm \theta) \prod_j \sum_m p_{l^{(i)}}(m|\bm \theta) \: p^u_{l^{(i)}m}(\ul{w}^{(i)}_j| \bm \theta)
\numberthis \label{likelihood2}
\end{align*}

This is difficult to optimize, due to the sum within the product. Similarly to the EM-algorithm, $\ul{\bm k}$ can be treated as random variables.
Given an initial distribution over component assignments $p_{l^{(i)}}(\ul{\bm k}^{(i)}|\ul{\bm{w}}^{(i)},\bm{\hat{\theta}})$, the conditional expectation of the 
log-likelihood becomes:
\begin{align*}
Q(\bm \theta|D, \bm{\hat{\theta}}) =& E(\log(\mathcal{L}(\bm \theta|D, \bm{\hat{\theta}})))\\
&=\sum_i \log(p(l^{(i)}|\bm \theta)) + \sum_{\ul{\bm k}^{(i)}} {p_{l^{(i)}}(\ul{\bm k}^{(i)}|\ul{\bm{w}}^{(i)},\bm{\hat{\theta}})} \log(p_{l^{(i)}}(\ul{\bm{w}}^{(i)},\ul{\bm{k}}^{(i)} |\bm \theta))\\
&= \sum_i \log(p(l^{(i)}|\bm \theta)) \\
& + \sum_{\ul{\bm k}^{(i)}} \sum_j p_{l^{(i)}}(\ul{\bm k}^{(i)}|\ul{\bm{w}}^{(i)},\bm{\hat{\theta}}) \log(p_{l^{(i)}}(\ul{k}_j^{(i)}|\bm \theta) \: p^u_{l^{(i)}\ul{k}_j^{(i)}}(\ul{w}^{(i)}_j| \bm \theta))
\numberthis\\
\end{align*}

Maximizing the conditional expected likelihood parameters for the unsmoothed multinomial $p^u_{lm=1}(n)$, 
background model $p^u_{lm=2}(n)$, and weights $p_l(m)$ decouples into
separate optimizations, given the distribution over component assignments $p_{l^{(i)}}(\ul{\bm k}^{(i)}|\ul{\bm{w}}^{(i)},\bm{\hat{\theta}})$.
Let $\bm \lambda_{lm}=[\lambda_{l1}, ... , \lambda_{lN}]$ indicate the conditional parameters for component $m$ of label $l$, 
$\bm \alpha_{lm}= [a_{l1}, ..., a_{lM}]$ the parameters for component weights of label $l$, and $\bm \pi= [\pi_1, ..., \pi_L]$ parameters for the label priors. 
Let $\bm \lambda$ and $\bm \alpha$ indicate the combined parameters for the label-conditionals and component weights.
Given $p_{l^{(i)}}(\ul{\bm k}^{(i)}|\ul{\bm{w}}^{(i)},\bm{\hat{\theta}})$, the maximization decouples:
\begin{align*}
&\argmax_{\bm \theta}(Q(\bm \theta|D, \bm{\hat{\theta}}))\\
=& \argmax_{(\bm \lambda, \bm \alpha, \bm \pi)}(\sum_i \log(p(l^{(i)}|\bm \pi)) \\
& + \sum_i \sum_{\ul{\bm k}^{(i)}} \sum_j p_{l^{(i)}}(\ul{\bm k}^{(i)}|\ul{\bm{w}}^{(i)},\bm{\hat{\theta}}) \log(p_{l^{(i)}}(\ul{k}_j^{(i)}|\bm \alpha) \: p^u_{l^{(i)}\ul{k}_j^{(i)}}(\ul{w}^{(i)}_j| \bm \lambda))\\
=& \argmax_{(\bm \lambda, \bm \alpha, \bm \pi)}(\sum_i \log(p(l^{(i)}|\bm \pi)) \\
& + \sum_i \sum_{\ul{\bm k}^{(i)}} \sum_j p_{l^{(i)}}(\ul{\bm k}^{(i)}|\ul{\bm{w}}^{(i)},\bm{\hat{\theta}}) \log(p_{l^{(i)}}(\ul{k}_j^{(i)}|\bm \alpha))\\
& + \sum_i \sum_{\ul{\bm k}^{(i)}} \sum_j p_{l^{(i)}}(\ul{\bm k}^{(i)}|\ul{\bm{w}}^{(i)},\bm{\hat{\theta}}) \log(p^u_{l^{(i)}\ul{k}_j^{(i)}}(\ul{w}^{(i)}_j| \bm \lambda)))
%
\numberthis  \label{Q2state}\\
\end{align*}


The parameter estimates for the smoothing methods derive from different assumed 
distributions \\ $p_{l^{(i)}}(\ul{\bm k}^{(i)}|\ul{\bm{w}}^{(i)},\bm{\hat{\theta}})$, except the background model for some smoothing methods.
Both discounting and linear interpolation imply distributing the expected count mass from a label-conditional model to a background model. 
Estimation of a shared background model requires further tying of parameters, removing the 
label-conditional dependency for the second component: $\forall l:p^u_{lm=2}(n)= p^u(n)$. 
The label-conditional unsmoothed models $p^u_{lm=1}(n)$ and the shared background model $p^u(n)$ become estimated from the
count mass distributed by $p_{l^{(i)}}(\ul{\bm k}^{(i)}|\ul{\bm{w}}^{(i)},\bm{\hat{\theta}})$, while the smoothing weight $p_l(m=2)$ 
equals the proportion of count mass distributed to the background model for the label $l$.\\

\begin{table}[ht]
\centering
\caption{Two-state HMM parameter estimates with different assumptions for $p_{l^{(i)}}(\ul{\bm k}^{(i)}|\ul{\bm{w}}^{(i)},\bm{\hat{\theta}})$,
with the unsmoothed label-conditional probabilities $p^u_{lm=1}(n)$, smoothing weights $p_l(m=2)$, and shared background model probabilities
$p^u_{lm=2}(n)= p^u(n)$
}
\begin{tabular}{|l|c |c |c|} \hline
$p_{l^{(i)}}(\ul{\bm k}^{(i)}|\ul{\bm{w}}^{(i)},\bm{\hat{\theta}})$ & $p^u_{lm=1}(n)$ & $p_l(m=2)$ & $p^u(n)$ \\ \hline
Jelinek-Mercer & $\frac{C(l, n)} {\sum_{n'} C(l, n')}$ & $\beta$ & $\frac{\sum_{l'}  C(l', n)} {\sum_{n'} \sum_{l'} C(l', n')}$ \\ \hline
Absolute disc. & $\frac {C(l,n) -\delta} {\sum_{n'} (C(l,n') -\delta)}$ & $1- \frac{ \sum_n (C(l, n)-\delta)} { \sum_n C(l, n)}$ & $\frac {\sum_{l:C(l,n)>0} 1} {\sum_{n'} \sum_{l:C(l,n')>0} 1}$ \\ \hline
Power-law disc. & $\frac {C(l,n) -\delta C(l, n)^{\delta}} {\sum_{n'} (C(l,n') -\delta C(l, n')^{\delta})}$ & $1- \frac{ \sum_n (C(l, n)-\delta C(l, n)^{\delta})} { \sum_n C(l, n)}$ & $\frac {\sum_{l'} \delta C(l',n)^{\delta}}{\sum_{n'} \sum_{l'} \delta C(l',n')^{\delta}}$ \\ \hline
\end{tabular}
\label{twostate_estimates}
\end{table} 

For some common smoothing methods, the background model is correctly estimated as derived from the two-state HMM formalization. 
Table \ref{twostate_estimates} shows parameter estimates, when $p_{l^{(i)}}(\ul{\bm k}^{(i)}|\ul{\bm{w}}^{(i)},\bm{\hat{\theta}})$ distributes to form the
parameter estimates for Jelinek-Mercer smoothing, absolute discounting and power-law discounting. With Jelinek-Mercer, the collection model
estimates derive as the background model. With absolute discounting and power-law discounting, the derived background model estimates equal
the modified background models \citep{Kneser:95, Huang:10} proposed for these methods. These modified background models were proposed
to satisfy marginal constraints of interpolating a single background model to label-conditional models, but are commonly used to form all $M-1$ order background
models in hierarchical n-gram smoothing \citep{Goodman:00, Zhang:14}. According to the two-state HMM derivation, the models for $M-2$ orders should 
be estimated from the expected count mass recursively distributed from the higher order models, instead of observed counts \citep{Goodman:00}, or expected 
counts \citep{Zhang:14} of each order independently.\\

Smoothing methods including a Dirichlet prior will result in untypical background models, since a Dirichlet prior gives weight to the background
model in proportion of the sum of counts: $\alpha_l=1- \frac{\sum_n C(l, n)}{\mu + \sum_n C(l, n)}$. This distributes the residual count mass for
estimating $p^u(n)$ unevenly, so that labels with more accumulated counts contribute less to the background model: $p^u(n)= 
\frac{\sum_{l} \alpha_l C(l, n)} {\sum_{n'} \sum_{l} \alpha_{l} C(l, n')}$. Using a uniform background model will analogously
imply untypical label-conditional models. A uniform background model would distribute the same count mass for each word type $n$ from the 
label-conditional models. Discounting a fraction $\delta$ from the first observation of a word would result in a uniform background model, as would 
discounting uniformly from each observation of a word: $D(l,n)= \delta/\sum_{l'} C(l', n)$. A uniform background model constrains the 
choices for the label-conditional and smoothing weight. For example, if a single smoothing weight is used for all labels, the weight will be upper bound by
the proportion of discounted count mass: $p_l(m=2) \le \argmin_n (\sum_{l'} D(l',n))/ (\sum_{l'} C(l',n))$. Common practice chooses the background model 
as either a collection or uniform distribution, regardless of the label-conditional models and smoothing weights. Alternatively, the background model and 
the smoothing method hyperparameters can be optimized together for a performance measure of the task \citep{Puurula:12c}.\\

Maximizing the expected conditional likelihood corresponds to a single iteration of the 
EM algorithm on the HMM. Full EM estimation of the parameters would replace $\bm{\hat{\theta}}$
by $\bm{\theta}$, and iterate the expectation step of computing $p_{l^{(i)}}(\ul{\bm k}^{(i)}|\ul{\bm{w}}^{(i)},\bm{\hat{\theta}})$ and the 
maximization step of solving Equation \ref{Q2state} with the updated expectation \citep{Hiemstra:04}. Nevertheless, 
the EM parameter estimates would not necessarily reach a global optimum of the model likelihood
\citep{Baum:70, Rabiner:89, Bilmes:98}. The Q-function parameters estimates can be exact maximum likelihood estimates in some cases, 
and exact closed form estimates can exist for some sets of constraints. For example, if segmentations of words into states $\ul{\bm k}^{(i)}$ are 
provided, the parameters are exact maximum likelihood estimates. The case of linear interpolation with
the background model and smoothing parameters fixed has a closed form exact maximum likelihood solution \citep{Zhang:08b}.\\

The HMM framework for formalizing smoothed multinomials can be extended for formalizing smoothing in more structured
generative models. This involves constraining the transitions according to the model \citep{Miller:99, Manning:99, Hiemstra:01},
but the approximate maximum likelihood estimation remains the same. 
Due to tied parameters, the resulting HMM topologies can become complicated, 
especially if the model structure and parameter tying is complex. Visualizing the HMMs can be simplified by presenting 
label-dependent parts of the HMM topology separately \citep{Miller:99} and illustrating fragments of the 
model \citep{Manning:99, Hiemstra:01}.\\

\section{Extending MNB for Fractional Counts}
\subsection{TF-IDF and Feature Transforms with MNB}
Parameter smoothing constitutes the primary means for correcting assumptions in models for 
text such as MNB. In some fields of text mining this is seen as sufficient. For example, in 
information retrieval the common view is that collection smoothing performs the same task as Inverse 
Document Frequency (IDF) weighting, and as such integration of IDF to LMs is not necessary 
\citep{Zhai:01, Hiemstra:04, Zhai:08}. This has been challenged by both experimental results 
\citep{Smucker:06, Momtazi:10, Puurula:13b} and analyses of IDF compared to collection smoothing \citep{Robertson:04}. In 
text classification and clustering, MNB has been combined with Term Frequency (TF) and IDF transforms (TF-IDF)
\citep{Rennie:03, Kibriya:04, Pavlov:04, Frank:06, Puurula:12c}.\\

Various versions of both IDF and TF-IDF have been proposed for text data \citep{Robertson:76, Salton:88}. 
TF-IDF functions comprise three transforms: term count normalization, document length normalization and 
document frequency weighting. The first two are commonly performed with a combined function and
form the TF-part of feature weighting. Document frequency weighting is usually considered an 
independent factor, and forms the IDF-part of TF-IDF. A combined TF-IDF feature transform is 
given by:
\begin{equation}
w_n = \log(1+\frac{w'_n} {|\bm w'|_0}) \log \frac {I} {I_n},
\label{tfidf}
\end{equation}
where $\bm w'$ is the original unweighted word vector, and $I_n$ is the number of collection documents having 
the word $n$.\\

The choice of log1+ transform for TF normalization can be justified as a correction to multinomials for better 
modeling the power-law distributions seen in text data \citep{Rennie:03}. The use of ``L0 norm'' or the number 
of non-zeros in the word vector is called unique length normalization, and has been shown to be robust across
datasets \citep{Singhal:96}. The IDF factor $\log \frac {I} {I_n}$ is called Robertson-Walker IDF 
\citep{Robertson:04}, and forms the most commonly used version of IDF.\\

A large number of variations exist for each of the three components and for how they are combined. Term count
normalization can be omitted, or use stronger damping \citep{Singhal:98}. Length normalization can use L1 
norm or L2 norm, and can be applied before or after term count normalization. Document frequency weighting by IDF
can take a number of forms, one common variant being Croft-Harper IDF \citep{Croft:79}, which downweights common
words more severely. Parameterized versions also exist, adding versatility to the transforms \citep{Lee:07}. A 
generalized version of Equation \ref{tfidf} can be defined \citep{Puurula:12c}:

\begin{equation}
w_n = \frac{\log (1+\frac{w'_n} {|\bm w'|_0^{\phi}})} {|\bm w|_0^{1-\phi}} \log(\max(1, \upsilon + \frac {I} {I_n})),
\label{tfidf2}
\end{equation}
where $\phi$ controls length scaling and $\upsilon$ IDF lifting. $\phi=1$ performs length normalization before term
count normalization, $\phi=0$ performs it after. Values $0<\phi<1$ produce smooth combinations of term count and length
normalization, while $\phi>1$ and $\phi<0$ produce more extreme normalizations. $\upsilon=0$ produces Robertson-Walker
IDF, while $\upsilon=-1$ produces unsmoothed Croft-Harper IDF.  Values $\upsilon>1$ produce weaker IDF normalizations,
while $\upsilon<-1$ produces stronger IDF normalizations.\\

\subsection{Methods for Fractional Counts with Multinomial Models}
Transforming data is standard practice for correcting model assumptions in statistical modeling. This enables the
use of well-understood simple models with complex data. Common transforms include flooring and ceiling values to 
accepted bounds, binning, log transforms, and standard score normalization. 
Feature transforms are less commonly used in text modeling since the 
models are defined on count data, and most normalizations of count data would produce fractional
counts that are undefined for multinomial and categorical models \citep{Juan:02, Vilar:04, Tam:08, Bisani:08, Zhang:14}. This means 
that the common use of TF-IDF with MNB produces models that are not well-defined in a probabilistic sense. However, there
are a few methods in common use that allow fractional counts in restricted uses.\\

For inference uses, a method commonly used to weight words in ranking is the 
Kullback-Leibler (KL) divergence \citep{Zhai:01c}. In ranking, KL-divergence is mostly used to incorporate feedback 
information into test documents, but it has also been used to integrate IDF weighting of words 
\citep{Smucker:06, Momtazi:10, Momtazi:10b}. In soft classification it has been used to correct differences in 
scores caused by varying document lengths \citep{Craven:00, Schneider:05}. 
KL-divergence is a measure between two probability functions. For the case of two multinomial distributions $p_l(\bm w)$ and 
$p'(\bm w)$, the negative KL-divergence is:
\begin{align*}
-D(p_l(\bm w) || p'(\bm w))&= - \sum_n p'(n) \log \frac{p'(n)} {p_l(n)} \nonumber \\
&=  \sum_n p'(n) \log p_l(n) - \sum_n p'(n) \log p'(n) 
\numberthis \label{KL}\\
\end{align*}

The second term $-\sum_n p'(n) \log p'(n)$ is the entropy for model $p'(\bm w)$. When KL-divergence 
is used for ranking or posterior scoring, the model $p'(\bm w)$ is the test document or query model, and its entropy 
can be omitted since it has constant effect on each label model $p_l(\bm w)$. If the model $p'(\bm w)$ is estimated 
as the unsmoothed estimate $p'(\bm w)= \frac {w_n} {\sum_n w_n}$ for a test document, then scoring by negative 
KL-divergence gives rank-equivalent scores to the posterior log-probabilities 
$-D(p_l(\bm w) || p'(\bm w)) \overset{rank}{=} \log(p(l)) \sum_n w_n \log(p_l(n))$.\\

KL-divergence thus provides a framework for generalizing posterior inferences $p(l| \bm w)$ by replacing the counts for a 
test document by model parameters. A common use is to incorporate pseudo-feedback information from the top ranked labels to 
the document \citep{Zhai:01c}, in order to rerank the document with the updated model. A more recent use is transforming
features \citep{Smucker:06, Momtazi:10}, so that parameters are weighted and renormalized, according to 
a weighting such as Inverse Collection Frequency \citep{Smucker:06, Momtazi:10} or IDF \citep{Momtazi:10}. For example,
using IDF would replace the test document model as $p'(n)= Z \: w_n \: IDF(n)$, where $Z$ is a normalization term and
$IDF(n)$ the IDF-weight of word $n$. A problem with the KL-divergence framework is that it is not probabilistic in a 
strict sense, since the KL-divergence scores are not probabilities. In addition, it cannot be used to incorporate feature 
transforms or normalizations in model estimation.\\

A fully probabilistic alternative to KL-divergence is to define a model that directly uses feature weights. The query term 
weighting model has been proposed \citep{Momtazi:10b} for weighting MNB conditional probabilities, so that $p(\ul{\bm w}|l, \ul{\bm r})= 
Z_{\ul{\bm r}} \prod_j p_l(\ul{w}_j)^{\ul{r}_j}$, $Z_{\ul{\bm r}}$ is a document-dependent normalizer and $\ul{r}_j$ is an arbitrary non-negative
weight, for example $\ul{r}_j= IDF(\ul{w}_j)$. This can be seen as a log-linear model \citep{Darroch:72}, where the
weights for each feature in the sequence are fixed. The same method was demonstrated earlier for the two-state categorical 
HMM models in IR, for the special case of weighting query sections, without considering normalization \citep{Miller:99}. This method 
presents a simple modification that provides well-defined posterior probabilities for infererence. Like KL-divergence inference, 
this method cannot be used for model estimation.\\

For model estimation, static model interpolation \citep{Stolcke:02} enables weighting of training data for n-gram LMs. This works identically
to Jelinek-Mercer method used for smoothing, but combines weighted components from different training data sources. For example,
using $K$ component datasets with multinomial parameters $p_k(n)$ and weights $p(k)$, the interpolated parameters would be $p(n)= \sum_k p(k) p_k(n)$.
Basic linear interpolation can be equally implemented by weighting and storing fractional counts 
from each dataset. Details of the interpolation can vary, and some smoothing heuristics such as the Kneser-Ney discounting estimate require integer 
count information \citep{Ney:94, Zhang:14}. In general this method allows integration of fractional counts in estimation, while maintaining the 
probabilistic framework.\\

\subsection{Formalizing Feature Transforms and Fractional Counts with Probabilistic Data}
\label{weighted_hmms}

A method that enables transformed features for both estimation and inference is the formalization of fractional counts as probabilistic data.
Concurrent research in estimation of n-grams for machine translation has formalized fractional counts as expectations of counts, given a probability 
distribution over possible word sequences \citep{Zhang:14}. This method can be extended for estimation and inference of generative models in a variety
of applications, as discussed next in detail.\\

A weight sequence $\ul{\bm r}= [\ul{r}_1, ..., \ul{r}_J]$ matching a word sequence $\ul{\bm w}$ can be interpreted as probabilities of words occurring, 
similar to the distribution over hidden components provided by the EM-algorithm. Each weight $\ul{r}_j$ indicates the probability
of the corresponding word $\ul{w}_j$ to have occurred in the data, so that the weights define a distribution over possible word sequences. 
A possible word sequence $\ul{\bm{\hat{w}}}$ given $\ul{\bm w}$ and $\ul{\bm r}$ can be called a \emph{realization}, a special case being 
a realization with no 
words $\ul{\bm{\hat{w}}}= \epsilon$. A sequence of binary indicator variables $\ul{\bm{\hat{r}}}$ called an \emph{occurrence sequence} 
indicates a draw from distribution defined by the weight variables $\ul{\bm r}$. A realization can be generated by different occurrence sequences, 
and the mapping can be denoted $\ul{\bm{\hat{w}}} = d(\ul{\bm{\hat{r}}}, \ul{\bm{w}})$.\\

\begin{table*}
\centering
\caption{Statistics for a word sequence weighted by probabilities}
\subtable[Document word sequence $\ul{\bm{w}}$ with weights $\ul{\bm{r}}$]{
\hspace{10pt}
\begin{tabular}{|l|l|l|} \hline
j &	$\ul{w}_j$ & $\ul{r}_j$ \\ \hline
$1$ & $1$ & $0.7$ \\
$2$ & $2$ & $0.8$ \\ 
$3$ & $1$ & $0.9$ \\
\hline
\end{tabular}
\hspace{10pt}
}
\hspace{10pt}
\subtable[Realizations $\ul{\bm{\hat{w}}}$ and probabilities $p(\ul{\bm{\hat{w}}})$]{
\hspace{10pt}
\begin{tabular}{|l|l|} \hline
$\ul{\bm{\hat{w}}}$ & $p(\ul{\bm{\hat{w}}})$ \\ \hline
$\epsilon$ & $0.006$ \\
$1$ & $0.068$ \\
$2$ & $0.024$ \\
$12$ & $0.056$\\
$11$ & $0.126$ \\
$21$ & $0.216$ \\
$121$ & $0.504$ \\ 
\hline
\end{tabular}
\hspace{10pt}
}
\hspace{10pt}
\subtable[Probabilities of counts $p(w_n=c)$ and expectations of count frequencies $E(\sum_{n:w_n=c} 1)$]{
\hspace{10pt}
\begin{tabular}{|l|l|l|l|} \hline
& $c=1$ & $c=2$ & $c>0$\\ \hline
$p(w_1=c)$ & $0.34$ & $0.63$ & $0.97$ \\
$p(w_2=c)$ & $0.80$ & $0.00$ & $0.80$ \\
$E(\sum_{n:w_n=c} 1)$ & $1.14$ & $0.63$ &\\
\hline
\end{tabular}
\hspace{10pt}
}
\hspace{10pt}
\subtable[Expected fractional counts $E(\bm w_n)$]{
\hspace{10pt}
\begin{tabular}{|l|l|} \hline
$E(w_1)$ & $1.6$\\
$E(w_2)$ & $0.8$\\
\hline
\end{tabular}
\hspace{10pt}
}
\label{count_distr}
\end{table*}

The probability of an occurrence sequence $\ul{\bm{\hat{r}}}$ can be computed by multiplying the weights: 
$p(\ul{\bm{\hat{r}}}| \ul{\bm w}, \ul{\bm r})= \prod_j \ul r_j^{\ul{\hat{r}}_j} (1-\ul r_j)^{|\ul{\hat{r}}_j-1|}$.
The probability of each realization $p(\ul{\bm{\hat{w}}})$ can be computed by summing its occurrence sequences: 
$p(\ul{\bm{\hat{w}}}| \ul{\bm w}, \ul{\bm r})= \sum_{\ul{\bm{\hat{w}}}=d(\ul{\bm{\hat{r}}}, \ul{\bm{w}})} p(\ul{\bm{\hat{r}}}| \ul{\bm w}, \ul{\bm r})$.
Assuming the words occur independently, the probabilities of counts $p(w_n=c)$ are distributed according to a Poisson-binomial distribution: 
$p(w_n=c)= \sum_{j:\ul{w}_j =n} \ul{r}_j$. The expectations of count frequencies $E(\sum_{n:w_n=c} 1)$ can then be computed with a recursive
algorithm \citep{Zhang:14}. Table \ref{count_distr} summarizes these basic statistics that can be computed by treating weights for words as probabilities.\\

The method proposed by \cite{Zhang:14} uses the expectations of count frequencies $E(\sum_{n:w_n=c} 1)$ in place of the $n_1$ and $n_2$ statistics for 
the Kneser-Ney estimate, and the expected counts $E(w_n)$ in place of the counts $w_n$. The resulting LMs maximize the expected conditional likelihood given the 
distribution over realizations, similarly to the two-state HMMs in Section \ref{smoothing_hmm}. Let training data $D$ for a multinomial distribution consist of 
word and weight sequences: $D^{(i)}= (\ul{\bm w}^{(i)}, \ul{\bm r}^{(i)})$. The expected conditional log-likelihood can be written:
\begin{align*}
Q(\bm \theta|D, \bm{\hat{\theta}}) &= E(\log(\mathcal{L}(\bm \theta|D, \bm{\hat{\theta}}))) \\
&=\sum_i \sum_{\ul{\bm{\hat{w}}}^{(i)}} p(\ul{\bm{\hat{w}}}^{(i)}| \ul{\bm{w}}^{(i)}, \ul{\bm{r}}^{(i)}) \log(p(\ul{\bm{\hat{w}}}^{(i)}))\\
&=\sum_i \sum_{\ul{\bm{\hat{w}}}^{(i)}} \sum_{\ul{\hat{w}}_j^{(i)}} p(\ul{\bm{\hat{w}}}^{(i)}| \ul{\bm{w}}^{(i)}, \ul{\bm{r}}^{(i)}) \log(p(\ul{\hat{w}}_j^{(i)}))\\
&=\sum_n E(C(n)| D) \log(p(n))
\numberthis \\
\label{expected_Q}
\end{align*}

This method formalizes the use of fractional counts for estimation, but contains one flaw. With higher order n-grams different realizations have 
different word histories, since omission of a word causes an n-gram history to skip a word. For example, with a realization $\ul{\bm{\hat{w}}}= [9 3]$ 
for a weighted word 
sequence $\ul{\bm w}= [9 2 4 5 3]$, the subsequence $[2 4 5]$ would not be realized, and an n-gram history for the last word $\ul w_5= 3$ in the 
sequence would have to start with $\ul w_1= 9$. Different realizations will yield different sets of n-gram histories, 
and assuming fixed histories would become increasingly incorrect with long word sequences and low weights $\ul r_j$. 
Correct estimation should take the differing histories into account. The method presented by \cite{Zhang:14} sidesteps this issue
by allowing weighting only at the level of sentences. Nevertheless, experimental improvements are demonstrated from the expected 
Kneser-Ney smoothing \citep{Zhang:14}.\\

Probabilistic data can be applied equally for inference. The expectation of log-probability $E(\log(p(\ul{\bm{w}})))$ equals the Q-function for a single document:
\begin{align*}
E(\log(p(\ul{\bm{w}})))&= \sum_{\ul{\bm{\hat{w}}}} p(\ul{\bm{\hat{w}}}| \ul{\bm{w}}, \ul{\bm{r}}) \log(p(\ul{\bm{\hat{w}}}))\\
&= \sum_{\ul{\bm{\hat{w}}}} \sum_{j=1}^{|\ul{\bm{\hat{w}}}|} p(\ul{\bm{\hat{w}}}| \ul{\bm{w}}, \ul{\bm{r}}) \log(p(\ul{{\hat{w}}}_{j}))\\
&= \sum_j \ul{r}_j \log(p(\ul{w}_j))
\numberthis \\
\end{align*}

The expectation of probability $E(p(\ul{\bm{w}}))$ takes the form:
\begin{align*}
E(p(\ul{\bm{w}}))&= \sum_{\ul{\bm{\hat{w}}}} p(\ul{\bm{\hat{w}}}| \ul{\bm{w}}, \ul{\bm{r}}) p(\ul{\bm{\hat{w}}})\\
&= \sum_{\ul{\bm{\hat{w}}}}  p(\ul{\bm{\hat{w}}}| \ul{\bm{w}}, \ul{\bm{r}}) \prod_{j=1}^{|\ul{\bm{\hat{w}}}|} p(\ul{{\hat{w}}}_j)\\
&= \sum_{\ul{\bm{\hat{r}}}} \prod_j p(\ul{\hat{r}}_j) \prod_j p(\ul{w}_j)^{\ul{\hat{r}}_j}\\
&= \sum_{\ul{\bm{\hat{r}}}} \prod_j p(\ul{\hat{r}}_j) p(\ul{w}_j)^{\ul{\hat{r}}_j}
\numberthis\\ 
\end{align*}

These two are not necessarily equivalent, as demonstrated by Jensen's inequality: $E(\log(p(\ul{\bm{w}}))) \le \log(E(p(\ul{\bm{w}})))$. It is not clear which
one to use for inference. $E(p(\ul{\bm{w}}))$ seems to be the natural choice, since it equals the mean of the probability over the distribution.
If a model is estimated using the Q-function of equation \ref{expected_Q}, then $E(\log(p(\ul{\bm{w}})))$ is consistent with the estimation.
$E(\log(p(\ul{\bm{w}})))$ is trivially computed from the weighted counts, whereas $E(p(\ul{\bm{w}}))$ is more complicated, but
can be computed using the forward algorithm from the word and weight sequences in time linear to the sequence length. 
For this thesis, we will use $E(\log(p(\ul{\bm{w}})))$, since it gives results that equal the use of fractional counts in existing literature, while no
results using $E(p(\ul{\bm{w}}))$ exist.\\

\section{Formalizing MNB as a Generative Directed Graphical Model}
\label{graphical_mnb}
The present chapter has formalized the smoothing and feature weighting commonly used with MNB, but there are several  
omissions in the standard descriptions of MNB. It can be argued that the MNB model is not multinomial, naive, or Bayesian:

\begin{description}
\item[1)] MNB is usually not a Bayesian model since no distribution over parameters is kept, but rather a generative 
model using the Bayes rule for posterior inference.
\item[2)] MNB is not a Naive Bayes model, since the conditional distribution is modeled by a single multinomial, not by 
conditionally independent models for each feature. 
\item[3)] The conditional multinomials in MNB are in fact tied multinomials with parameters shared 
for all possible documents lengths, combined with a length-generating distribution such as Poisson.
\item[4)] The conditional multinomials themselves have not been formalized correctly, but using a variety of 
smoothing methods with no connection to maximum likelihood.
\item[5)] Feature weighting such as TF-IDF has not been generally formalized for either the training or test set 
documents, despite being very commonly used with MNB.\\
\end{description}

Given the common misconceptions about MNB, it is useful to formalize it with a precise definition. We can redefine 
MNB as a generative model over sequences that factorizes into distributions for labels, 
word sequence lengths and 2-state HMMs for label-conditionals. Feature weighting can be formalized as estimation
and inference using expectations of log likelihoods and log probabilities over probabilistic data. The joint probability for 
the model factorizes as:
\begin{align*}
p(\ul{\bm{w}}, l, \ul{\bm k}) &= p(l) p(J) p(\ul{\bm{w}}, \ul{\bm k}| \:  l, J)\\
&= p(l) p(J) \prod_j p_l(\ul{k}_j) p^u_{l\ul{k}_j}(\ul{w}_j),
\numberthis \label{hmm_mnb}\\
\end{align*}
where the prior $p(l)$ is categorical, the length generation factor $p(J)$ is Poisson and 
label-conditionals $p(\ul{\bm{w}}, \ul{\bm k}| \: l, J)$ are modeled by the 2-state HMMs with $M=2$ and categoricals
$p_l(\ul{k}_j)$ for each $l$, and $p_{l\ul{k}_j}(\ul{w}_j)$ for each $l$ and $m$.\\

\begin{figure}[h]
\centering
%
%
%
%


\begin{tikzpicture}[x=1.7cm,y=1.8cm]


  \node[obs]                   (w_j)      {$\ul{w}_j$} ; %
  \node[latent, above=of w_j]    (k_j)      {$\ul{k}_j$} ; %
  \node[obs, above=of k_j]    (l)      {$l$} ; %
  \edge {l} {k_j} ; %
  \edge {k_j} {w_j} ; %
  \plate {doc} {(w_j) (k_j)} {$j= 1..J$}; %
  
  \node[obs, left=of doc] (J)  {$J$};
  \edge {J} {doc} ; %
  \plate {dataset} {(w_j) (J) (doc) (l)} {$i= 1..I$}; %



\end{tikzpicture}

\caption{Graphical model for Multinomial Naive Bayes, formalizing multinomial smoothing as a 2-state Hidden Markov Model}
\label{MNB2}
\end{figure}
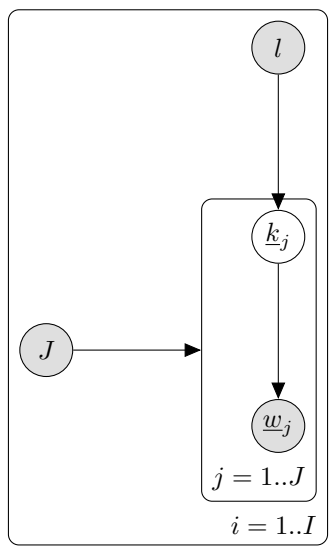

The two-state HMM terms correspond to the original MNB terms: $p^u_{lm=1}(n)= p^u_l(n)$ for the label-conditional models, $p^u_{lm=2}(n)= p^u(n)$
for the shared background model, and $p(m=2)= \alpha_l$ for the smoothing weight. Figure \ref{MNB2} shows the 
graphical model notation for the correctly formalized MNB.\\

The hidden states can be marginalized away:
\begin{align*}
p(\ul{\bm{w}}, l) &= p(l) p(J) \sum_{\ul{\bm k}} \prod_j (p_l(\ul{k}_j) p^u_{l\ul{k}_j}(\ul{w}_j))\\
&= p(l) p(J)  \prod_j \sum_m (p_l(m) p^u_{lm}(\ul{w}_j))\\
&= p(l) p(J)  \prod_n ((1-\alpha_l) p^u_l(n) + \alpha_l \: p^u(n))^{w_n}
\numberthis \label{reform_seq_joint} \\
\end{align*}

The label posterior $p(l | \ul{\bm{w}})$ given a word sequence becomes:
\begin{align*}
p(l | \ul{\bm{w}}) &= \frac{p(\ul{\bm{w}}, l)} {\sum_{l'} p(\ul{\bm{w}}, l')}\\
&\propto p(l) \prod_n  ((1-\alpha_l) p^u_l(n) + \alpha_l \: p^u(n))^{w_n}
\numberthis \label{reform_seq_post} \\
\end{align*}

The label posterior $p(l | \bm{w})$  given a word vector becomes:
\begin{align*}
p(l|\bm w) &= \frac{p(l) p(J) Z(\bm w) \prod_j \sum_m (p_l(m) p^u_{lm}(\ul{w}_j))} {\sum_{l'} p(l') p(J) Z(\bm w) \prod_j \sum_m (p_{l'}(m) p^u_{l'k}(\ul{w}_j))}\\
&\propto p(l) \prod_n ((1-\alpha_l) p^u_l(n) + \alpha_l \: p^u(n))^{w_n},
\numberthis \label{reform_vector_post}
\end{align*}
where $Z(\bm w)$ is the multinomial normalizer.\\

The word vectors can be used to compute exactly the same label posteriors, meaning that for posterior inference
only the word vectors are required, and not the word sequence information. The joint can be marginalized to produce
the probability of a word sequence: 

\begin{align*}
p(\ul{\bm{w}})&= \sum_l \sum_{\ul{\bm{k}}} p(\ul{\bm{w}}, l,\ul{\bm{k}})\\
&= p(J)  \sum_l p(l) \sum_{\ul{\bm{k}}} p_l(\ul{\bm{w}},\ul{\bm{k}})\\
&= p(J)  \sum_l p(l) \prod_n ((1-\alpha_l) p^u_l(n) + \alpha_l \: p^u(n))^{w_n}
\numberthis \label{reform_seq_joint2} \\
\end{align*}

From the marginal probabilities MNB can be seen as a mixture of label-conditional models 
\citep{McCallum:98b, Novovicova:03, Nigam:06}, where the label variables are known for each document in a training
dataset. The MNB factorization is simplified by a number of independence assumptions. The length factor $p(J)$ is assumed to 
be mutually independent with the other factors \citep{Juan:02}.\\ 

The training data consists of documents $D^{(i)}= (l^{(i)}, \ul{\bm w}^{(i)}, \ul{\bm k}^{(i)})$, where $l^{(i)}$ and 
$\ul{\bm w}^{(i)}$ are known, and $\ul{\bm k}^{(i)}$ is unknown. The likelihood function is:
\begin{align*}
\mathcal{L}({\bm \theta}|D)&=p(D| {\bm \theta}) \\
&=\prod_i p(l^{(i)}|\bm \pi) p(|\ul{\bm{w}}|_0^{(i)}|\bm \chi) \sum_{\ul{\bm k}^{(i)}} \prod_j p_{l^{(i)}}(\ul{k}_j^{(i)}|\bm \alpha) \: p^u_{l^{(i)}\ul{k}_j^{(i)}}(\ul{w}^{(i)}_j| \bm \lambda)
\numberthis \label{}\\
\end{align*}

Treating $\ul{\bm k}$ as a random variable distributed according to a prior distribution $p_{l^{(i)}}(\ul{\bm k}^{(i)}|\ul{\bm{w}}^{(i)},\bm{\hat{\theta}})$, 
the conditional expectation of the log-likelihood becomes:
\begin{align*}
Q(\bm \theta|D, \bm{\hat{\theta}}) &= E(\log(\mathcal{L}(\bm \theta|D, \bm{\hat{\theta}})))\\
&= \sum_i (\log(p(l^{(i)}|\bm \theta)) + \log(p(|\ul{\bm{w}}|_0^{(i)}|\bm \theta)) \\
&+ \sum_{\ul{\bm k}^{(i)}} \sum_j p_{l^{(i)}}(\ul{\bm k}^{(i)}|\ul{\bm{w}}^{(i)},\bm{\hat{\theta}}) \log(p_{l^{(i)}}(\ul{k}_j^{(i)}|\bm \theta) \: p^u_{l^{(i)}\ul{k}_j^{(i)}}(\ul{w}^{(i)}_j| \bm \theta)))
\numberthis \label{loglikelihood3} \\
\end{align*}

The maximization decouples into separate optimizations:
\begin{align*}
\argmax_{\bm \theta}(Q(\bm \theta|D, \bm{\hat{\theta}})))&= \argmax_{(\bm \pi, \bm \lambda, \bm \alpha, \bm \chi)}(\sum_i \log (p(l^{(i)}|\bm \pi)) \\
&+\sum_i \log(p(|\ul{\bm{w}}|_0^{(i)} |\bm \chi))\\
&+\sum_i \sum_{\ul{\bm k}^{(i)}} \sum_j p_{l^{(i)}}(\ul{\bm k}^{(i)}|\ul{\bm{w}}^{(i)},\bm{\hat{\theta}}) \log(p_{l^{(i)}}(\ul{k}_j^{(i)}|\bm \alpha))\\
&+ \sum_i \sum_{\ul{\bm k}^{(i)}} \sum_j p_{l^{(i)}}(\ul{\bm k}^{(i)}|\ul{\bm{w}}^{(i)},\bm{\hat{\theta}}) \log(p^u_{l^{(i)}\ul{k}_j^{(i)}}(\ul{w}^{(i)}_j| \bm \lambda)))
\numberthis \label{loglikelihood4} \\
\end{align*}

The parameters $\bm \pi$ and $\bm \chi$ are invariant to the
expectation over the hidden states, and are therefore exact maximum likelihood estimates. For many uses the parameters 
$\bm \chi$ for word sequence lengths can be omitted, since these have no effect on the posterior probabilities.
The parameters $\bm \lambda$ and $\bm \alpha$ are expected log-likelihood estimates given 
$p_{l^{(i)}}(\ul{\bm k}^{(i)}|\ul{\bm{w}}^{(i)},\bm{\hat{\theta}})$, as discussed in
Section \ref{smoothing_hmm}. Choosing $p_{l^{(i)}}(\ul{\bm k}^{(i)}|\ul{\bm{w}}^{(i)},\bm{\hat{\theta}})$ and a form for the shared background 
model $p^u(n)$ implements smoothing.\\

Feature weighting is incorporated by performing estimation and inference over probabilistic data, as described in Section \ref{weighted_hmms}. 
Given a probabilistic weight sequence $\ul{\bm r}$ matching the word sequence $\ul{\bm w}$, probabilities can be approximated by the expectations
of log-probabilities given $\ul{\bm r}$. For both inference and estimation, this reproduces the results that come from simply using fractional counts in 
algorithms instead of integer counts, as has been done with applications of MNB using weighted words. In Equations \ref{reform_seq_joint}, \ref{reform_seq_post}, 
\ref{reform_vector_post}, and \ref{reform_seq_joint2}, approximation with expected log-probabilities replaces the integer vector $\bm w$ by the 
fractional expected counts $E(\bm w| \ul{\bm r})$, that are provided by any chosen feature weighting function. Maximum expected log-likelihood estimation introduces the weight terms $r_j$ to Equations \ref{loglikelihood3} and \ref{loglikelihood4}. The multinomials for the two-state HMM become weighted by 
both the distribution of occurring/non-occurring terms defined by $\ul{\bm r}$, and the assumed distribution over 
the HMM component assignments $p_{l^{(i)}}(\ul{\bm k}^{(i)}|\ul{\bm{w}}^{(i)},\bm{\hat{\theta}})$. The length-modeling 
factor $p(|\ul{\bm{w}}|_0^{(i)} |\bm \chi))$ should also take the weighted distribution over occurring word sequences into account, unless the length model 
is approximated by ignoring weights. Further derivation of Poisson length model estimation on expected sequence lengths is omitted, as the experiments 
conducted in the thesis evaluate length modeling separately to feature weighting.\\

\section{Extending MNB with Prior Scaling and Document Length Modeling}

The extension of MNB so far has formalized smoothing methods as a directed graphical model. Further useful extensions to MNB can be
defined by modifying the graphical model factorization. Two such extensions are label-conditional document length modeling and scaling 
of the label prior.\\

Document lengths in MNB are assumed to be generated by a shared distribution such as Poisson \citep{McCallum:98b, Blei:03}, that can be 
omitted for most uses. Use of label-conditional distributions has been suggested in the literature \citep{McCallum:98b}, but has not experimented.
Since document lengths are known to be among the strongest features for some tasks, such as automatic essay scoring \citep{Larkey:98}, explicit
length modeling could prove useful.\\

Prior scaling is applied to LMs in uses such as speech recognition, where a LM models the prior distribution of possible word sequences in a
Naive Bayes framework \citep{Gales:07}. In these uses the prior has a very different scale than the conditional distribution, and a scaling factor is applied to
match the contributions of the prior and conditional optimally. Prior scaling can be applied to TM tasks equally, but this has not been attempted.\\

An Extended MNB incorporating prior scaling and document length modeling can be defined as:
\begin{align*}
p(\ul{\bm{w}}, l, \ul{\bm k}) &= p(l) p(J|l) p(\ul{\bm{w}}, \ul{\bm k}| \:  l, J)\\
&= p(l) p(J|l) \prod_j p_l(\ul{k}_j) p^u_{l\ul{k}_j}(\ul{w}_j),
\numberthis \label{extended_MNB}
\end{align*}

where the label prior $p(l)$ and length model $p(J|l)$ are scaled and renormalized versions of the original distributions $p'(l)$ and $p'(J|l)$ : $p(l)=
Z(\vartheta) \: p'(l)^\vartheta$ and $p(J|l)= Z(\varsigma) \: p'(J|l)^\varsigma$. $Z(\vartheta)$ and $Z(\varsigma)$ normalize the factors to be probability distributions, and $\vartheta$ and $\varsigma$ are meta-parameters to be estimated on held-out data.\\
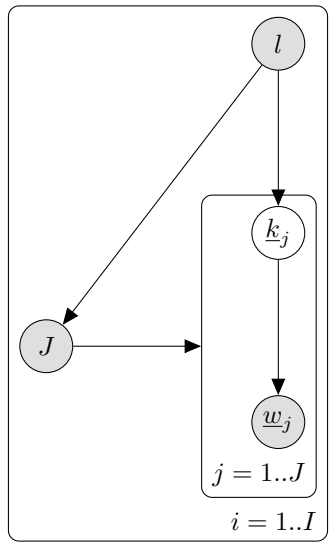
\begin{figure}[h]
\centering
%
%
%
%


\begin{tikzpicture}[x=1.7cm,y=1.8cm]


  \node[obs]                   (w_j)      {$\ul{w}_j$} ; %
  \node[latent, above=of w_j]    (k_j)      {$\ul{k}_j$} ; %
  \node[obs, above=of k_j]    (l)      {$l$} ; %
  \edge {l} {k_j} ; %
  \edge {k_j} {w_j} ; %
  \plate {doc} {(w_j) (k_j)} {$j= 1..J$}; %
  
  \node[obs, left=of doc] (J)  {$J$};
  
  \edge {l} {J} ; %
  \edge {J} {doc} ; %
  \plate {dataset} {(w_j) (J) (doc) (l)} {$i= 1..I$}; %



\end{tikzpicture}

\caption{Graphical model for the Extended MNB, with label-conditional document length modeling, but without scaling of the factors for $p(l)$ and $p(J|l)$.}
\label{MNB3}
\end{figure}

The Extended MNB model with fixed scaling factors $\vartheta=1$ and $\varsigma= 1$ constitutes a directed generative model, and the
corresponding graphical model is illustrated in Figure \ref{MNB3}. With scaling weights other than 1 and 0, the parameter estimates of the scaled factors no 
longer derive from a directed factorization. One solution to formalize the model is to consider the combination of factors as a type of log-linear model combination
\citep{Klakow:98, Hinton:02, Bouchard:04, Suzuki:07}:
\begin{align*}
p(\ul{\bm{w}}, l, \ul{\bm k}) &= Z \: p'(l)^\vartheta \: p'(J|l)^\varsigma \: p(\ul{\bm{w}}, \ul{\bm k}| \:  l, J)
\numberthis \label{model_mnb4}\\
\end{align*}

With the log-linear model of Equation \ref{model_mnb4}, the factors for label priors, document lengths and label-conditional word distributions 
become feature functions combined using the log-linear weights $\vartheta$ and $\varsigma$. Maximum likelihood estimation of the model becomes complicated.
One approximate solution is to keep the maximum likelihood estimates for the directed model, and directly optimize the new parameters $\vartheta$ and 
$\varsigma$ for a performance measure on held-out development data \citep{Metzler:05}. This approximation is used for the experiments conducted in 
Chapter 6. It maintains the simplicity of directed models, while allowing the additional parameters to improve model performance.\\

\chapter{Sparse Inference}
This chapter proposes computation based on sparse model representations for scalable inference. This reduces the time and space 
complexity of inference for a variety of linear models and structural extensions. First a basic case of sparse posterior
inference is derived for uses such as Multinomial Naive Bayes (MNB) classification, ranking and clustering. This is extended into a more general case of
joint inference on hierarchically smoothed sequence models, and further into joint inference on mixtures of such models.
Additional efficiency improvements for the inference are discussed, and a structural extension of MNB benefiting from sparse inference
is proposed, called Tied Document Mixture.\\

\section{Basic Case: Sparse Posterior Inference}
\label{posterior_inference}
Once estimated, a generative model such as Multinomial Naive Bayes (MNB) can be used for several types of inference, depending on the task and application. 
The most common types are classification in text categorization, ranking in information retrieval, soft classification in 
document clustering, and model combination. 
All of these perform inference related to the posterior $p(l| \bm w)$.
In classification, the most likely label to generate a document is selected: 
$\argmax_l p(l|\bm w)$. In ranking, scores are computed for each label: $y(\bm \theta_l, \bm w)\overset{rank}{=} p(l|\bm w)$. In 
soft classification and model combination, the posterior is used directly.\\

The time and space complexity of the inference is crucial in many applications of MNB, since this determines the scalability of the model
to large-scale tasks. The numbers of labels, words and training documents can exceed millions in many practical tasks 
involving web-scale text datasets. The required computation can be reduced depending on the type of inference. For example, for 
posterior inference the multinomial normalizer can be omitted, since $p(l| \bm w)= \frac {Z \: p(l) \prod_n p(w_n |l)}{\sum_l Z\: p(l) \prod_n p(w_n |l)}
= \frac {\: p(l) \prod_n p(w_n |l)}{\sum_l\: p(l) \prod_n p(w_n |l)}$. \\

A naive algorithm for performing posterior inference with MNB computes 
$p(l|\bm w)= \frac {p(l, \bm w)} {\sum_{l'} p(l', \bm w)} = Z\: p(l) \prod_{n:w_n\ne 0} p_l(n)^{w_n}$
for each label $l$ by computing the joint probabilities $p(l, \bm w)$ as a product over terms $n$ for each label $l$, and normalizing
by the sum over the joints $Z= \sum_{l'} p(l', \bm w)$. This has complexity $O(|\bm w|_0 L)$, and this is widely considered 
to be optimal for linear models \citep{Manning:08}. However, the time complexity for posterior inference can be substantially 
reduced by taking into account sparsity in the parameters. Earlier work with inverted indices has shown that classifier scores for 
Centroid classifier \citep{Shanks:03} and K-nearest Neighbours \citep{Yang:94} can be done as a function of sparsity.
The posteriors for MNB with uniform priors can be computed similarly \citep{Hiemstra:98b, Zhai:01b}, for both Jelinek-Mercer \citep{Hiemstra:98b} 
and other basic smoothing methods \citep{Zhai:01b}. The \emph{sparse inference} proposed here generalizes this inference to MNB with 
categorical priors, a variety of linear models and structural extensions. A basic sparse posterior inference algorithm for MNB is described next. It 
combines three techniques for efficient computation:\\

\begin{description}
\item[Log-domain computation] refers to use of log probability values, a standard practice with large-scale statistical 
models. This changes products into sums and exponents into products in the log-domain, which are much 
cheaper to compute with modern processors. For example, $\log(p(l)p_l(\bm w))= \log(p(l))+\log(p_l(\bm w))$ and 
$\log( p_l(n)^{w_n})= \log(p_l(n)) w_n$. Another benefit of log-computation is reducing numerical inaccuracy that 
comes from computing with small probabilities. Alternatives such as scaling \citep{Rabiner:89} for correcting 
inaccuracy with large-scale models are less useful in general.\\

\item[Precomputing] organizes an algorithm and necessary data structures so that less computation is necessary when the
algorithm is used. A basic sparse posterior inference algorithm computes first the smoothing distribution probabilities 
$p^u(n)$, and then updates these for each $l$ by multiplying with $\frac{(1-\alpha_l) p_l^u(n)+\alpha_l p^u(n)} {p^u(n)}$. 
The original parameter values $p^u_l(n)$ can be replaced by precomputed ones, since for computing $p(l, \bm w)$, $p^u_l(n)$
are only needed for updating $p^u(n)$ for each $l$. The parameters for $p(l)$, $p^u(n)$, 
$\alpha_l$ and $p^u_l(n)$ can be replaced by the precomputed log values 
$p'(l)= \log p(l)$, $p'^u(n)=\log (p^u(n))$, $\alpha'_l= \log(\alpha_l)$, and $p'^u_l(n)= \log((1-\alpha_l) p_l^u(n)+\alpha_l p^u(n)) - \alpha'_l - p'^u(n)$
for all $p^u_l(n)>0$.\\

\item[Inverted index] data structures have formed the core technique of information retrieval for the last decades \citep{Zobel:06}. 
An inverted index uses a vector $\bm \zeta$ of \emph{postings lists} for each word, so that for 
each word a list $\bm \zeta_n$ of occurrence information can be accessed in constant time. 
For sparse inference with MNB, it is sufficient to maintain the label and parameter in the 
postings, so that each posting is a pair $(l, p'^u_l(n))$. 
$p_l(n)$ can first be computed by $p^u(n)$, and 
updated for each $l$ with $p_l^u(n) \neq 0$. Using an inverted index, the set of parameters to update 
can be retrieved with a time complexity $O(|\bm w|_0+\sum_n \sum_{l:p^u_l(n)\neq 0} 1)$ instead of $O(L|\bm w|_0)$.\\
\end{description}

A basic sparse posterior inference algorithm computes 
$p'^u(\bm w)$, updates this for all $l$ with $p'(l)$ and $p'^u_l(\bm w)$, and normalizes to obtain the posterior 
$p(l|\bm w)$. This has the time complexity $O(L+|\bm w|_0 + \sum_n \sum_{l:p^u_l(n)\neq 0}1)$.
Using the precomputed values, we can compute joint probabilities as $p(\bm w, l) \propto
\exp(p'(l)+\sum_{n} w_n (p'^u(n) + \alpha'_l) +\sum_{n:p^u_l(n)\neq0} w_n p'^u_l(n)))$.
The posterior can be computed by normalizing $p(l|\bm w)= p(\bm w, l)/ \sum_{l'}p(\bm w, l')$. 
A key algorithmic idea here is that the conditional probabilities $p(\bm w| l)$ decompose into
separately computed terms for the smoothing distribution, the smoothing weight, and the 
label-conditional: $p(\bm w| l) \propto \exp(\sum_{n:w_n \neq 0}p'^u(n)+ |\bm w|_1 \alpha'_l + \sum_{n:p^u(n) \neq 0} p'^u_l(n))$.\\
 
\begin{algorithm}[h]
\caption{Sparse Posterior Inference for MNB}
\begin{algorithmic}[1]
\State $smooth\_logprob= 0$
\ForAll{$n:w_n \neq 0$}
 \State $smooth\_logprob+= p'^u(n) *w_n$ 
\EndFor
\ForAll{$l$}
 \State $logprobs_l= p'(l)+|\bm w|_1 \alpha'_l + smooth\_logprob$ 
\EndFor
\ForAll{$n:w_n \neq 0$}
 \ForAll{$(l, p'^u_l(n)) \in \bm \zeta_n$}
   \State $logprobs_l+=  p'^u_l(n) *w_n$ 
 \EndFor
\EndFor
\State $normalizer= -1000000$
\ForAll{$l$}
 \State $normalizer= log(exp(normalizer)+exp(logprobs_l))$
\EndFor
\ForAll{$l$}
 \State $logprobs_l-= normalizer$
\EndFor
\Return $logprobs$
\end{algorithmic}
\label{mnb_spi}
\end{algorithm}

Pseudo-code for the resulting sparse posterior inference algorithm returning $p(l|\bm w)$ is given in Algorithm 
\ref{mnb_spi}. Although the sparse inference algorithm is described here for MNB posterior inference, it is applicable to any linear model. 
The inference algorithm with precompiled values corresponds to a sparsely computed dot product $y(\bm \theta, \bm w)= \theta_0 + \sum_{n=1}^{N} \theta_n w_n$,
which is used for all linear classifiers, including Centroid, Perceptron, Logistic Regression and 
Support Vector Machine classifiers. Moreover, it can be generalized into structured models such as hierarchical mixture models. 
Therefore the textbook statement \citep{Manning:08} on optimality of $O(|\bm w|_0 L)$ 
posterior inference is not only incorrect for MNB, but for machine learning methods in general.\\

\section{Extension to Joint Inference on Hierarchically Smoothed Sequence Models}
The sparse inference in Algorithm \ref{mnb_spi} can be extended to hierarchically smoothed models, while retaining the same benefits in
computational complexity. As discussed in Chapter 3, there are many cases where multinomials and MNB models are 
extended so that the probabilities in a sequence model back off to less context-dependent models.
A variety of hierarchically smoothed sequence models are used with text, such as interpolated n-gram models. The smoothing hierarchy 
can come from sources such as word clusters 
\citep{Zitouni:08}, the local word context \citep{Chen:99}, or collection structure \citep{McCallum:99b,Zhang:02, Krikon:11}. Any combination of 
hierarchies can equally be used for backing-off. For example, a label-conditional passage bigram could first back-off to a label-conditional passage unigram, 
then to a label-conditional document unigram, and finally to a collection unigram model.\\

Let $M$ denote the depth of hierarchical smoothing for a sequence extension of MNB, so that the label-conditional probabilities for the model
are smoothed by the $M-1$ back-off layers in the hierarchy, where each node for layer $m$ is a categorical distribution that is used in the 
smoothing. An example of this would be a label-conditional n-gram model, where $M$ is the n-gram order. Node for layer $m=1$ corresponds to 
the root node of the back-off hierarchy, $m=M$ to the leaf node to be smoothed. The joint probability of a sequence $\ul{\bm w}$ and 
label $l$ for the model becomes:

\begin{align*}
p(\ul{\bm w}, l) &= p(l) \prod_j p_l(\ul{w}_j | \ul{\bm w})\\
 				   &= p(l) \sum_{\ul{\bm k}} \prod_j p_l(\ul{k}_j) \: p^u_{l\ul{k}_j}(\ul{w}_j | \ul{\bm w})\\
   				   &= p(l) \prod_j \sum_m p_l(m) p^u_{lm}(\ul{w}_j | \ul{\bm w})\\
				   &= p(l) \prod_j \sum_m (\prod_{m'=m+1}^{M} \alpha_{lm'} -\prod_{m'=m}^{M} \alpha_{lm'}) p^u_{lm}(\ul{w}_j | \ul{\bm w}),
\numberthis \label{hierarchy_expand}\\
\end{align*}
where $p_l(\ul{w}_j | \ul{\bm w})$ is the smoothed label-conditional probability, $p^u_{lm}(\ul{w}_j | \ul{\bm w})$ the component-conditional probabilities, 
and $\alpha_{lm}$ the back-off weight of component $m$. The conditioning variables $l$, $m$ and $\ul{\bm w}$ for the probabilities can 
be extended to include more variables, or can be tied to reduce the number of variables. For a label-conditional n-gram these would be tied as 
$p^u_{lm}(\ul{w}_j | \ul{\bm w})= p^u_{lm}(\ul{w}_j | \ul{w}_{j-m+1}...\ul{w}_{j-1})$.\\

Sparsity can be utilized by storing the non-zero parameters for each node in a precomputed form, by first 
smoothing and then precomputing the parameters for each node. Smoothed parameters are computed first for the root node $m=1$ : 
$p''^u_{lm}(\ul{w}_j | \ul{\bm w})= p^u_{lm}(\ul{w}_j | \ul{\bm w})$, and then for each $m>1$ up to to the leaf nodes:
$p''^u_{lm}(\ul{w}_j | \ul{\bm w})= \alpha_{lm} p''^u_{l(m-1)}(\ul{w}_j | \ul{\bm w}) + (1-\alpha_{lm}) p^u_{lm}(\ul{w}_j | \ul{\bm w})$.
Precomputed smoothed log-parameters are computed starting from the leafs $p'^u_{lm}(\ul{w}_j | \ul{\bm w})= \log(p''^u_{lm}(\ul{w}_j | \ul{\bm w}))
- \log(\alpha_{lm} \: p''^u_{l(m-1)}(\ul{w}_j | \ul{\bm w}))$ for the nodes $m>1$, and finally for the root-node 
$m=1$ : $p'^u_{lm}(\ul{w}_j | \ul{\bm w})= \log(p''^u_{lm}(\ul{w}_j | \ul{\bm w}))$. The smoothing weights and label priors can be
precomputed: $\alpha'_{lm}=\log(\alpha_{lm})$ and $p'(l)= \log(p(l))$.\\

With the precomputed values, 
joint probabilities of sequences $p(l|\ul{\bm w}, l)$ can be computed scalably by utilizing
sparsity. The joint probability in Equation \ref{hierarchy_expand} can be expressed in a factorized form as:
\begin{align*}
p(\ul{\bm w}, l) &= p(l) \prod_j \sum_m (\prod_{m'=m+1}^{M} \alpha_{lm'} -\prod_{m'=m}^{M} \alpha_{lm'}) p^u_{lm}(\ul{w}_j | \ul{\bm w})\\
				    &= p(l) \prod_m \xi(l,m)\\
				    &= \exp(p'(l)+\sum_m \xi'(l,m))\\
\xi'(l,m)&=\begin{cases}
        \sum_{j:p^u_{l{m}}(\ul{w}_j | \ul{\bm w})\neq 0} p'^u_{l{m}}(\ul{w}_j | \ul{\bm w}), & \text{if $m=1$}\\
       |\ul{\bm w}| \: \alpha'_{lm} + \sum_{j:p^u_{l{m}}(\ul{w}_j | \ul{\bm w})\neq 0} p'^u_{l{m}}(\ul{w}_j | \ul{\bm w}), & \text{otherwise,}\\
        \end{cases}
\numberthis \label{hierarchy_expand2}
\end{align*}
where $\xi(l,m)$ are factors for nodes $(l,m)$ explained in the following, and $\xi'(l,m)$= $\log(\xi(l,m))$.\\

Equation \ref{hierarchy_expand2} can be solved directly using dynamic programming. The factors $\xi(l,m)$ provide updates to computing 
$p(\ul{\bm w}| l)$ given its $m-1$ ancestor nodes. 
Starting from the root node, $\xi(l,m=1)$ computes $\log p(\ul{\bm w}| l)$ assuming that no descendant nodes exist. This is then updated 
iteratively by the descendant nodes $m>1$. The complexity of inference is reduced, because for each node only the non-zero unsmoothed parameters 
need to be considered, and these can be stored in postings lists retrieved from an 
inverted index. The time complexity is reduced from the dense $O(LMJ)$ to the sparse 
$O(L \: M + |\bm w|_0 + \sum_l \sum_m \sum_{j:p^u_{l{m}}(\ul{w}_j | \ul{\bm w})\neq 0} 1)$.\\

By using the shared hierarchical components between $l$ and the other conditioning variables, this complexity can be further reduced.
Since in a hierarchy the auxiliary variable values $\xi(l,m)$ are the same for all children of a node, $\xi(l,m)$ needs to be
computed only once for each shared node $h= (l,m)$ in the hierarchy, and updated according to the children. Exact hierarchical computation reduces
the complexity of computing $p(\ul{\bm w}, l)$ for all $l$ from 
$O(L \: M + |\bm w|_0 + \sum_l \sum_m \sum_{j:p^u_{l{m}}(\ul{w}_j | \ul{\bm w})\neq 0} 1)$ to
$O(L + |\bm w|_0+ \sum_{h}(1+\sum_{j:p^u_{h}(\ul{w}_j | \ul{\bm w})\neq 0} 1))$.\\

The hierarchical complexity can be further reduced in the case of constrained models or approximation. Assume that $p(l)$ are uniform in the following, and 
only the highest probability label is needed. If a node has no words that match the word sequence 
$\forall j:\ul{w}_j=0 \lor p^u_{lm}(\ul{w}_j | \ul{\bm w})=0$, then its update will be $\xi'(l,m)= |\ul{\bm w}| \: \alpha'_{lm}$. These nodes 
can be called the \emph{no-match nodes} for a given word sequence. If Jelinek-Mercer smoothing is used for estimating $\alpha_{lm}$, the back-off weights are 
the same for all children and $|\ul{\bm w}| \: \alpha'_{lm}$ can be 
added directly to the parent node $\xi'(l,m-1)$. Assuming matching nodes have no no-match ancestors, 
the sum $\sum_m \xi'(l,m))$ can be done only over the matching nodes $m$. Otherwise, a gap in the sum can be filled by computing $\xi(l,m-1)$ iteratively down 
to the first matching node. With the assumptions of no gaps in the hierarchy, Jelinek-Mercer for smoothing, and uniform $p(l)$, it suffices to compute 
the maximum probability $l$ by $\argmax_l p(\ul{\bm w}, l)= \argmax_l \sum_m \xi(l, m)$, where the sums $\sum_m \xi(l, m)$ can be done dynamically over 
the shared nodes $h$. The resulting time complexity is reduced to $O(|\bm w|_0+ \sum_h \sum_{j:p^u_{h}(\ul{w}_j | \ul{\bm w})\neq 0} 1)$, the
sum of non-zero features and matching nodes.\\

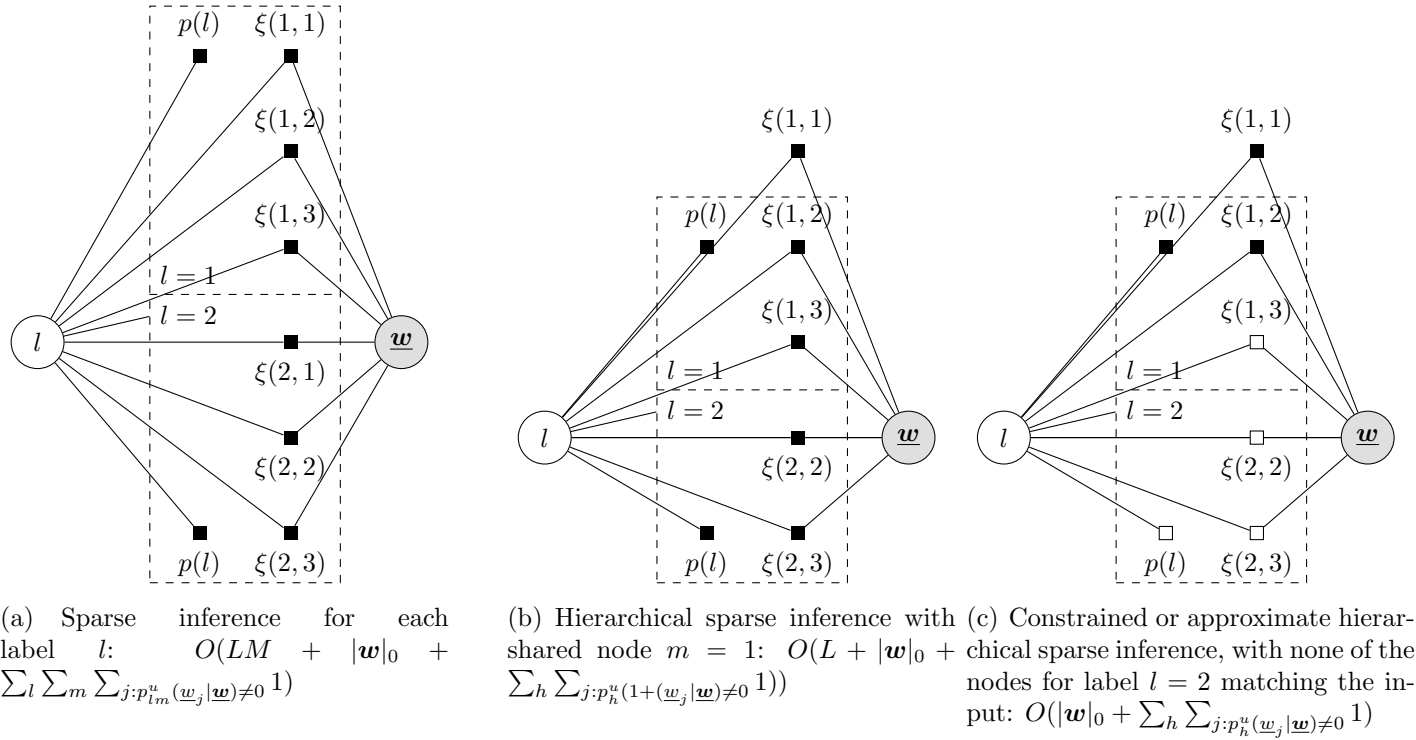
\begin{figure}[htp]
\centering
\subfigure[Sparse inference for each label $l$: $O(L M + |\bm w|_0 + \sum_l \sum_m \sum_{j:p^u_{l{m}}(\ul{w}_j | \ul{\bm w})\neq 0} 1)$]{
%
%
%
%


\begin{tikzpicture}[x=1.7cm,y=1.8cm]
  \node[obs]                       (w)           {$\ul{\bm w}$} ; %
  \factor[factor,left= of w]      {X21}     {below:$\xi(2,1)$}{}{}  ; %
  \factor[factor,below=of X21]    {X22}     {below:$\xi(2,2)$} {} {} ; %
  \factor[factor,below=of X22]    {X23}     {below:$\xi(2,3)$}{}{}  ; %
  \factor[factor,above=of X21]    {X13}     {$\xi(1,3)$} {} {} ; %
  \factor[factor,above=of X13]    {X12}     {$\xi(1,2)$}{}{}  ; %
  \factor[factor,above=of X12]    {X11}     {$\xi(1,1)$}{}{}  ; %
  \factor[factor,left=of X11]     {pl1}     {$p(l)$}{}{}  ; %
  \factor[factor,left= of X23]    {pl2}     {below:$p(l)$}{}{}  ; %
  \factor[space,left=of X21]     {s1}     {}{}{}  ; %
  \node[latent, left=of s1]      (l)        {$l$} ; %
  \factoredge {l} {X11} {} ; %
  \factoredge {l} {X12} {} ; %
  \factoredge {l} {X13} {} ; %
  \factoredge {l} {X21} {} ; %
  \factoredge {l} {X22} {} ; %
  \factoredge {l} {X23} {} ; %
  \factoredge {X11} {w} {} ; %
  \factoredge {X12} {w} {} ; %
  \factoredge {X13} {w} {} ; %
  \factoredge {X21} {w} {} ; %
  \factoredge {X22} {w} {} ; %
  \factoredge {X23} {w} {} ; %
  \factoredge {l} {pl1} {} ; %
  \factoredge {l} {pl2} {} ; %
  \hgate {gate1} {(X11)(X12)(X13)(pl1)} {$l=1$}%
  {(X21)(X22)(X23)(pl2)} {$l=2$}%
  {}; %
  \factoredge {l} {gate1} {} ; %
 
  



\end{tikzpicture}

}
\hspace{0.5cm}
\subfigure[Hierarchical sparse inference with shared node $m=1$: $O(L +|\bm w|_0+\sum_{h} \sum_{j:p^u_{h}(1+(\ul{w}_j | \ul{\bm w})\neq 0} 1))$]{
%
%
%
%


\begin{tikzpicture}[x=1.7cm,y=1.8cm]
  \node[obs]                       (w)           {$\ul{\bm w}$} ; %
  \factor[factor,left= of w]      {X22}     {below:$\xi(2,2)$}{}{}  ; %
  \factor[factor,below=of X22]    {X23}     {below:$\xi(2,3)$}{}{}  ; %
  \factor[factor,above=of X21]    {X13}     {$\xi(1,3)$} {} {} ; %
  \factor[factor,above=of X13]    {X12}     {$\xi(1,2)$}{}{}  ; %
  \factor[factor,above=of X12]    {X11}     {$\xi(1,1)$}{}{}  ; %
  \factor[factor,left=of X12]     {pl1}     {$p(l)$}{}{}  ; %
  \factor[factor,left= of X23]    {pl2}     {below:$p(l)$}{}{}  ; %
  \factor[space,left=of X22]     {s1}     {}{}{}  ; %
  \node[latent, left=of s1]      (l)        {$l$} ; %
  \factoredge {l} {X11} {} ; %
  \factoredge {l} {X12} {} ; %
  \factoredge {l} {X13} {} ; %
  \factoredge {l} {X21} {} ; %
  \factoredge {l} {X22} {} ; %
  \factoredge {l} {X23} {} ; %
  \factoredge {X11} {w} {} ; %
  \factoredge {X12} {w} {} ; %
  \factoredge {X13} {w} {} ; %
  \factoredge {X21} {w} {} ; %
  \factoredge {X22} {w} {} ; %
  \factoredge {X23} {w} {} ; %
  \factoredge {l} {pl1} {} ; %
  \factoredge {l} {pl2} {} ; %
  \hgate {gate1} {(X12)(X13)(pl1)} {$l=1$}%
  {(X21)(X22)(X23)(pl2)} {$l=2$}%
  {}; %
  \factoredge {l} {gate1} {} ; %
 
  



\end{tikzpicture}

}
\subfigure[Constrained or approximate hierarchical sparse inference, with none of the nodes for label $l=2$ matching the input: $O(|\bm w|_0+ \sum_h \sum_{j:p^u_{h}(\ul{w}_j | \ul{\bm w})\neq 0} 1)$]{
%
%
%
%


\begin{tikzpicture}[x=1.7cm,y=1.8cm]
  \node[obs]                       (w)           {$\ul{\bm w}$} ; %
  \factor[nofactor,left= of w]      {X22}     {below:$\xi(2,2)$}{}{}  ; %
  \factor[nofactor,below=of X22]    {X23}     {below:$\xi(2,3)$}{}{}  ; %
  \factor[nofactor,above=of X21]    {X13}     {$\xi(1,3)$} {} {} ; %
  \factor[factor,above=of X13]    {X12}     {$\xi(1,2)$}{}{}  ; %
  \factor[factor,above=of X12]    {X11}     {$\xi(1,1)$}{}{}  ; %
  \factor[factor,left=of X12]     {pl1}     {$p(l)$}{}{}  ; %
  \factor[nofactor,left= of X23]    {pl2}     {below:$p(l)$}{}{}  ; %
  \factor[space,left=of X22]     {s1}     {}{}{}  ; %
  \node[latent, left=of s1]      (l)        {$l$} ; %
  \factoredge {l} {X11} {} ; %
  \factoredge {l} {X12} {} ; %
  \factoredge {l} {X13} {} ; %
  \factoredge {l} {X21} {} ; %
  \factoredge {l} {X22} {} ; %
  \factoredge {l} {X23} {} ; %
  \factoredge {X11} {w} {} ; %
  \factoredge {X12} {w} {} ; %
  \factoredge {X13} {w} {} ; %
  \factoredge {X21} {w} {} ; %
  \factoredge {X22} {w} {} ; %
  \factoredge {X23} {w} {} ; %
  \factoredge {l} {pl1} {} ; %
  \factoredge {l} {pl2} {} ; %
  \hgate {gate1} {(X12)(X13)(pl1)} {$l=1$}%
  {(X21)(X22)(X23)(pl2)} {$l=2$}%
  {}; %
  \factoredge {l} {gate1} {} ; %
 
  



\end{tikzpicture}

}
\caption{Factor graph models visualizing the sparse inference complexity reductions, for the same input word sequence $\ul{w}$ and model with $L=2$, $M=3$. 
White factors correspond to no-match nodes for the word sequence and do not need to be considered in inference. The complexity of inference within 
each factor further reduces complexity, and is not illustrated.}
\label{sparse_infer_graphs}
\end{figure}

Figure \ref{sparse_infer_graphs} illustrates the resulting three types of sparse inference algorithms using factor graph notation.
There are several other cases where the hierarchical time complexity can be reduced using constraints and approximation. If Dirichet prior or discounting 
methods are used for estimating $\alpha_{lm}$, the back-off weights differ for the children. In this case the probabilities for the 
no-match children can be either computed exactly or approximated. A simple approximation is to precompute the mean back-off weight of the children for each 
node, and compute the probabilities for the no-match children using this mean weight. The priors $p(l)$ can be approximated by 0 for the labels that have 
only no-match nodes in the smoothing mixture. Using these two weaker approximations for $\alpha_{lm}$ and $p(l)$, the same complexity reduction can be 
obtained by computing $\argmax_l p(\ul{\bm w}, l)= \argmax_l \sum_m \xi(l, m)$ as before over the shared nodes. Other similar cases exist
where the required computation is reduced by the number of  no-match nodes. One practical case is ranked retrieval, where a set of top ranked
labels for documents can be inferred given the input word sequence for a a query.\\

\section{Extension to Joint Inference on Mixtures of Sequence Models}
The sparse inference discussed so far has mainly dealt with the scalable computation of hierarchically smoothed sequence models. It can be extended to cases 
where mixtures are defined over the sequence models, as well as to cases where mixtures are used both within and over sequences. A basic implementation for
inference with mixtures over sequence models would multiply the inference time complexity by the number of components, but the inference complexity can 
be reduced again using sparsity, constraints and approximation.\\

Similar to the mixture model view of MNB, we can consider the previous example of Equation \ref{hierarchy_expand} as a mixture model over the 
label-conditional hierarchically smoothed sequence models:
\begin{align*}
p(\ul{\bm w})    &= \sum_l p(\ul{\bm w}, l) \\
p(\ul{\bm w}, l) &= p(l) \prod_j \sum_m (\prod_{m'=m+1}^{M} \alpha_{lm'} -\prod_{m'=m}^{M} \alpha_{lm'}) p^u_{lm}(\ul{w}_j | \ul{\bm w})\\
				    &= \exp(p'(l)+\sum_m \xi'(l,m))
\numberthis \label{hierarchy_expand3}\\
\end{align*}

Lets assume a simple two-state HMM case of $M=2$, where $m=2$ is a label-independent background model and Jelinek-Mercer smoothing is used for 
selecting $\alpha'_{lM}$. The factor for the smoothing background model becomes shared: $\forall_l: \: \xi'(l,m=2)= \xi'(1,m=2)$.  
Given a model where most of the leaf nodes $p^u_{lM}(\ul{w}_j | \ul{\bm w})$ are sparse, many of the labels $l$ will have 
no-match leaf nodes $(l, M)$, so that 
$p(\ul{\bm w}| l)= \exp(\xi'(l=1,m=2) + |\ul{\bm w}| \: \alpha'_{lM})$. Let $L'(\ul{\bm w})$ indicate the number of no-match leaf nodes in Equation 
\ref{hierarchy_expand3}, $l'$ labels with leaf nodes matching the word sequence $\exists j:\ul{w}_J>0 \land p^u_{lm}(\ul{w}_j | \ul{\bm w})>0$, and 
$\alpha''$ the back-off weight for the leaf nodes. The marginalization can be computed as:

\begin{align*}
p(\ul{\bm w})= L'(\ul{\bm w}) (1-\sum_{l'}p(l')) \exp(\xi'(l=1,m=2) + |\ul{\bm w}| \alpha'') + \sum_{l'} p(\ul{\bm w}, l),
\numberthis \label{hierarchy_expand4}
\end{align*}
reducing marginalization time complexity from $O(L)$ to $O(L-L'(\ul{\bm w}))$.\\

Figure \ref{sparse_marginalization} illustrates the sparse marginalization using factor graphs.
The complexity of marginalizations can be reduced in more complex cases as well, such as: different smoothing methods, deeper smoothing hierarchies, and 
multi-layer mixtures over the labels variables. If smoothing other than Jelinek-Mercer is used, the back-off weights of the no-match children can be 
approximated by a mean value, or grouped in bins to reduce the approximation. If deeper smoothing hierarchies are used, the marginalizations can be 
conducted iteratively for each layer $m$ from $M$ to $1$. If multi-layer mixtures over the label variables are used, the marginalizations can be conducted similarly
by iterating from the leaf nodes to the root.\\

\begin{figure}[ht]
\centering\makebox[\textwidth]{
%
%
%
%

\begin{tikzpicture}
  \node[latent]        (l)        {$l$} ; %
   
   \matrix[row sep=0.0cm, column sep=0.0cm, right=of l,
      nodes={text width=1.2cm, text height=1.3cm,inner sep=0cm},
    ] (gates) {
	\node[space2]     (p1)     {}; & \node[space2]     (x12)     {}  ;\\
	\node[space2]     (p2)     {}; & \node[space2]     (x22)     {}  ; \\
	\node[space2]     (p3)     {}; & \node[space2]     (x32)     {}  ; \\
	\node[space2]     (p4)     {}; & \node[space2]     (x42)     {}  ; \\
	\node[space2]     (p5)     {}; & \node[space2]     (x52)     {}  ; \\
	\node[space2]     (p6)     {}; & \node[space2]     (x62)     {}  ; \\
    };
	\node[below right] at (p1.north west) {\scriptsize{l=1}};
	\node[below right] at (p2.north west) {\scriptsize{l=2}};
	\node[below right] at (p3.north west) {\scriptsize{l=3}};
	\node[below right] at (p4.north west) {\scriptsize{l=4}};
	\node[below right] at (p5.north west) {\scriptsize{l=5}};
	\node[below right] at (p6.north west) {\scriptsize{l=6}};
	\draw [dashed] (gates.north east) -- (gates.south east);
	\draw [dashed] (gates.north west) -- (gates.south west);
	\draw [dashed] (gates.north west) -- (gates.north east);
	\draw [dashed] (gates.south east) -- (gates.south west);
	\draw [dashed] (p1.south west) -- (x12.south east);
	\draw [dashed] (p2.south west) -- (x22.south east);
	\draw [dashed] (p3.south west) -- (x32.south east);
	\draw [dashed] (p4.south west) -- (x42.south east);
	\draw [dashed] (p5.south west) -- (x52.south east);
	\factor[factor,below of=x12]    {X12}     {below:$\xi(1,1)$}{}{}; %
	\factor[factor,below of=x22]    {X22}     {below:$\xi(2,1)$}{}{}; %
	\factor[nofactor,below of=x32]    {X32}     {below:$\xi(3,1)$}{}{}; %
	\factor[nofactor,below of=x42]    {X42}     {below:$\xi(4,1)$}{}{}; %
	\factor[nofactor,below of=x52]    {X52}     {below:$\xi(5,1)$}{}{}; %
	\factor[nofactor,below of=x62]    {X62}     {below:$\xi(6,1)$}{}{}; %
	\factor[factor,below of=p1]    {P1}     {below:$p(l=1)$}{}{}; %
	\factor[factor,below of=p2]    {P2}     {below:$p(l=2)$}{}{}; %
	\factor[nofactor,below of=p3]    {P3}     {below:$p(l=3)$}{}{}; %
	\factor[nofactor,below of=p4]    {P4}     {below:$p(l=4)$}{}{}; %
	\factor[nofactor,below of=p5]    {P5}     {below:$p(l=5)$}{}{}; %
	\factor[nofactor,below of=p6]    {P6}     {below:$p(l=6)$}{}{}; %
	\factor[factor,left= of p1]    {X11}     {below:$\xi(1,2)$}{}{}; %
	\node[obs, above right=of x12]                       (w)           {$\ul{\bm w}$} ; %
	\factoredge {l} {gates} {} ; %
	\factoredge {l} {X12} {} ; %
	\factoredge {l} {X11} {} ; %
	\factoredge {l} {P1} {} ; %
	\factoredge {w} {X12} {} ; %
	\factoredge {w} {X11} {} ; %
	\draw [decorate,decoration={brace,amplitude=10pt,mirror,raise=4pt},yshift=0pt]
(x12.south east) -- (x12.north east) node [black,midway,xshift=2.8cm] {\footnotesize $=p(l=1) \prod_m \xi(l=1,m)$};
	\draw [decorate,decoration={brace,amplitude=10pt,mirror,raise=4pt},yshift=0pt]
(x22.south east) -- (x22.north east) node [black,midway,xshift=2.8cm] {\footnotesize $=p(l=2) \prod_m \xi(l=2,m)$};
	\draw [decorate,decoration={brace,amplitude=10pt,mirror,raise=4pt},yshift=0pt]
(x62.south east) -- (x32.north east) node [black,midway,xshift=4.3cm] {\footnotesize $=(1-\sum_{l'}p(l')) \exp(\xi'(l=1,m=2) + |\ul{\bm w}| \alpha'') $};
\end{tikzpicture}

}
\caption{Sparse marginalization over $l$ illustrated using a factor graph for the hierarchically smoothed sequence model $p(\ul{\bm w})= \sum_l p(l) \prod_j \sum_m  p_l(m) p^u_{lm}(\ul{w}_j | \ul{\bm w})= \sum_l p(l) \prod_m \xi(l,m)$, with $L=6$, $M=2$ and component $m=1$ shared between labels. Leaf nodes for labels $l>2$ are no-match nodes and the white colored factors for these labels can be computed sparsely. For readability, factor edges for labels $l>1$ are 
not shown.}
\label{sparse_marginalization}
\end{figure}
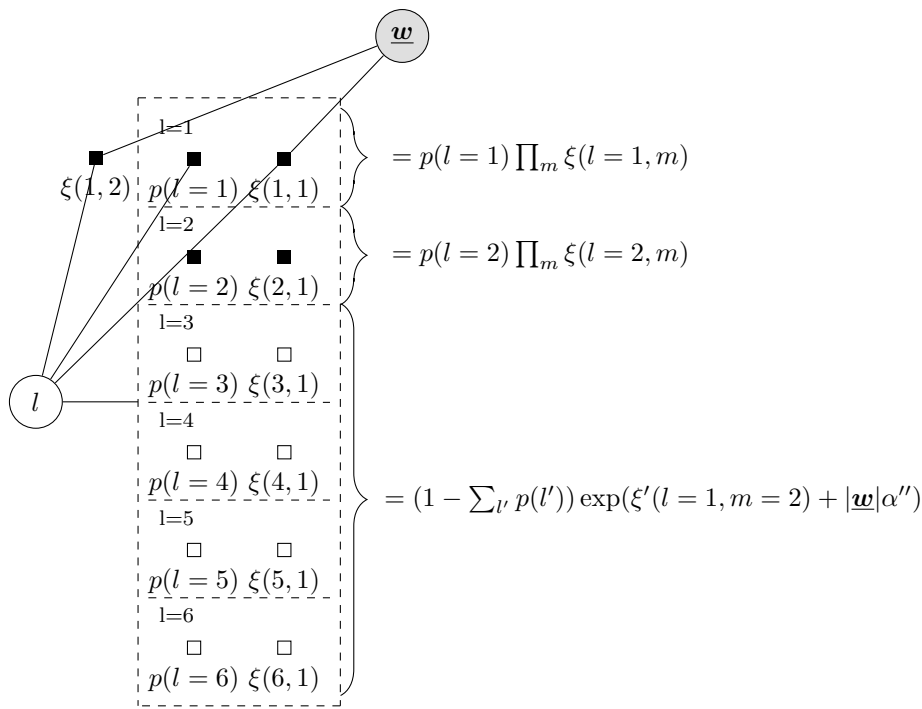

\section{Further Specialized Efficiency Improvements for Sparse Inference}
Sparse inference can be made substantially more efficient for many uses. Combination with parallelization and stream processing is
trivial, as subsets of the precomputed parameters $p'(l)$ and $\xi'(l,m)$ in Equation \ref{hierarchy_expand2} can be stored and processed by separate 
computing nodes, each node containing a shard of the full inverted index for the parameters. Aside from the generic methods for improving efficiency 
discussed earlier in Chapter 2, methods more specific to text mining can be applied. For ranking or classification only a subset of the labels is required, 
and therefore a number of further efficiency improvements are possible. These can be categorized as within-node pruning, between-node pruning, and 
search network minimization:

\begin{description}
\item[Within-node Pruning]
Computing a factor score $\xi(l, m)$ can be halted if it is unlikely to affect the result.
If classification or ranking with top-scoring labels is required, a ranked list of the top-scoring labels and their scores can be 
maintained, and evaluation of each label can be terminated once it cannot reach the top labels. In information retrieval, 
heuristics using top-ranked lists form one of the main techniques for improving search efficiency, one notable
algorithm being the MaxScore query evaluation algorithm \citep{Strohman:05}. The postings lists can be sorted in an order
such as decreasing magnitude of the parameters, enabling pruning of labels as early as possible. The inverted index can likewise be split into several indices, so 
that the apriori most likely labels are fully evaluated first; the labels in the following indices can be evaluated with much tighter pruning bounds.\\

\item[Between-node Pruning]
Computing the factor score $\xi(l, m)$ can be avoided altogether, if it is likely (or bound) to have no effect on the result.
If top-scoring labels are requested instead of the full posterior, the search can be organized hierarchically. 
Tree-based searches are commonly used with instance-based learning algorithms such as k-nearest neighbour classifiers. 
In tree-based search a hierarchy is constructed based on clustering label similarities, and a branch-and-bound search is used to retrieve
the top labels \citep{Ram:12}. Algorithm \ref{mnb_spi} can be extended into tree-based search by splitting the index
into indices for each layer in a clustered label hierarchy, and skipping a branch in the evaluation if their higher-level node 
does not reach the current score bound. This performs between-node pruning, since nodes are discarded even before their 
scoring is considered. Hierarchical top-ranking could be combined with hierarchical smoothing either separately or together, so 
that the same hierarchy would be used both for smoothing models and for scoring the labels efficiently.\\

\item[Search Network Minimization]
Search network minimization is used in complex graph search problems such as speech recognition systems \citep{Aubert:02}. 
For example, Finite State Machines and Hidden Markov Models are minimized by combining nodes where possible. Examples 
of this would be merging all nodes with a single child with the child node, or approximating similar nodes with a single clustered
node. Different types of minimizations of search networks can be used, if a graph of inverted indices is used for structured sparse 
inference.\\
\end{description}

\section{Tied Document Mixture: A Sparse Generative Model}

The sparse inference algorithms described in this chapter provide improved scalability for a variety of structured models. Mixture
variable nodes can be introduced at any structural level of a model, while incurring only modest increases in computational  
requirements. The nodes can be organized in layers, or form any type of a directed acyclic graph. The nodes can correspond 
to linguistic units, document structure, or any available metadata. Layers of nodes can include: subword units, phrases, 
sentences, passages, sections, fields, documents, labels, and labelsets.\\

Considering MNB and multinomial models for text, the most practical inclusion for most modeling purposes is 
a document node. These models are commonly used on datasets stored in the form of word vectors, where 
other metadata is not available, has been discarded, or varies between tasks and datasets. Document identifiers, in
contrast, are available in every application of these models. MNB and multinomial models are usually estimated by either averaging 
the document counts, or treating all documents as a single large document. Both types of modeling are unnatural in the sense
that they assume the documents to be identically distributed for a given label. This is a strong assumption, and one that does 
not generally hold with text data \citep{Puurula:13}. Explicitly modeling the documents avoids this problem.\\

Introducing a document node to MNB between the label and label conditional probabilities would in a general case 
increase the computational requirements considerably. Document mixture models \citep{Novovicova:03, Nigam:06} introduce a 
mixture over documents and learn the soft assignments of documents to mixture components using 
the EM algorithm. This requires the estimation of the number of components, the component weights and the component-conditional
probabilities. In general, optimizing the number of components has to be done by evaluation on development data, while optimizing
the other introduced parameters with EM will produce a local maximum of the model likelihood. The overall estimation is therefore
approximate, and the time complexity is multiplied by the number of restarts used to find the number of components and to avoid local
minima.\\

The use of approximate estimation with iterative algorithms can be avoided by constraining the mixture over documents in a suitable
way. One such way is to use a mixture with a component assigned for each document. This constrains the number of components to
the number of documents for the label, the component assignments to the documents, and the component-conditional probabilities
to document-conditional probabilities. This type of extension of MNB was proposed as Tied Document Mixture (TDM) \citep{Puurula:13}. 
In addition to these modeling choices, TDM smoothes the document models hierarchically, and uses a uniform distribution over the
document component weights.\\ 

The original version of TDM presented used simple hierarchical Jelinek-Mercer smoothing
\citep{Puurula:13}. With the theory of smoothing presented in Chapter 4, more refined smoothing methods 
can be attempted. The earlier version was also presented in word vector form, whereas
here it can be presented in the word sequence form together with the corresponding directed graphical model. Formally the TDM model
takes the form:

\begin{align*}
p(\ul{\bm w}, l) &= p(l) \sum_{i \in {I_l}} p_{l}(i) \prod_j p_{li}(\ul{w}_j)\\
&= p(l) \frac{1}{|I_l|}\sum_{i \in {I_l}} \prod_j p_{li}(\ul{w}_j)\\
&= p(l) \frac{1}{|I_l|}\sum_{i \in {I_l}} \prod_j \sum_m p_{li}(m) p_{lim}(\ul{w}_j)\\
&= p(l) \frac{1}{|I_l|}\sum_{i \in {I_l}} \prod_n \sum_m p_{li}(m) p_{lim}(n)^{w_n},
\numberthis \label{TDM}
\end{align*}
where $p(l)$, $p_{l}(i)$, $p_{li}(m)$ and $p_{lim}(\ul{w}_j)$ are all categoricals, and $I_l$ indicates the set of documents corresponding to label $l$ in the collection. Since a single document exists for each document indicator $i$, $p_{l}(i)= \frac{1}{|I_l|}$.\\

The document models $p_{li}(\ul{w}_j)= \sum_m p_{li}(m) p_{lim}(\ul{w}_j)$ are hierarchically tied for smoothing with four levels of nodes: $m=1$ uniform 
background model, $m=2$ collection model, $m=3$ label-conditional categoricals, and $m=4$ document-conditional categoricals. The original TDM 
used hierarchically Jelinek-Mercer smoothed document models, with a uniform distribution for the root-level smoothing \citep{Puurula:13}. An extension can 
be attempted so that both the document- and label-conditional models are smoothed with any of the methods described in Chapter 4. The document 
models are defined as:

\begin{align*}
p_{li}(\ul{w}_j)=& \sum_m p_{li}(m) p^u_{lim}(\ul{w}_j)\\
			    =& p_{li}(m=4) p^u_{li}(\ul{w}_j) + p_l(m=3) p^u_{l}(\ul{w}_j)\\
			    &+ p(m=2) p^u(\ul{w}_j) + p(m=1) U(\ul{w}_j)
\numberthis \label{} \\
\end{align*}

The label-conditional and collection models are estimated by discounting and normalizing averaged normalized document counts:
\begin{align*}
p^u_l(\ul{w}_j=n)\propto \max(0, (\sum_{i \in {I_l}} \frac {w^{(i)}_n} {|\bm w^{(i)}|_1}) - D(l, n)) \numberthis \\
p^u(\ul{w}_j=n)\propto \sum_l \max(0, (\sum_{i} \frac {w^{(i)}_n} {|\bm w^{(i)}|_1}) - D(l, n)), \numberthis 
\label{}
\end{align*}
where $D(l, n)$ is the given by the chosen discounting method for the hierarchy level, if any.\\

Normalized counts $E(C(l,n))= \sum_{i} \frac {w^{(i)}_n} {|\bm w^{(i)}|_1}$ are treated
as expected fractional counts for determining the smoothing and discounting values. Length normalization is used for the background models, 
to reduce the effect of untypically long documents on mean statistics.
The weights for the mixture components are computed dynamically from backoff-weights $\alpha$ for the different levels in the 
hierarchy: $p_{li}(m)= \prod_{m'=m+1}^{M} \alpha_{lim'} -\prod_{m'=m}^{M} \alpha_{lim'}$, where each $\alpha_{lim'}$ is produced 
by the smoothing method, as described in Chapter 4.\\

\begin{figure}[ht]
\centering\makebox[\textwidth]{
%
%
%
%


\begin{tikzpicture}[x=1.7cm,y=1.8cm]


  \node[obs]                   (w_j)      {$\ul w_j$} ; %
  \node[latent, above=of w_j]    (k_j)      {$k_j$} ; %
  \node[obs, above=of k_j]    (i)      {$i$} ; %
  \node[obs, above=of i]    (l)  {$l$}; %
  \edge {k_j} {w_j} ; %
  \edge {i} {k_j} ; %
  \edge {l} {i} ; %
  \plate {doc} {(w_j) (k_j)} {$j= 1..J$}; %
  \plate {docs} {(w_j) (k_j) (i) (doc)} {$\forall i\in I_l$} %
  \plate {classes} {(w_j) (k_j) (i) (l) (doc) (docs)} {$l= 1..L$}; %
  



\end{tikzpicture}

}
\caption{Directed graph for the Tied Document Mixture}
\label{TDM_graph}
\end{figure}
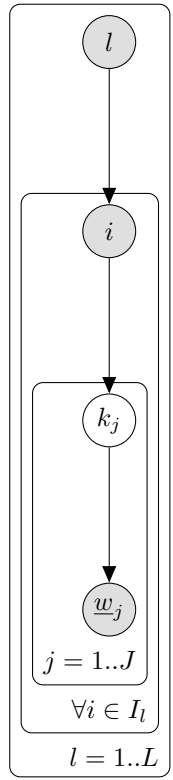

Figure \ref{TDM_graph} shows a directed graph for TDM. Estimating the smoothed model parameters has $O(\sum_i |\bm w^{(i)}|_0)$ time and 
space complexity. Using sparse inference, time and space complexity of approximate hierarchical inference for classification is 
$O(|\bm w|_0 + \sum_l \sum_{i \in {I_l}} \sum_{n:p^u_{li}(n)\neq 0} 1)$. Complexities of exact inference can be increased a little by the smoothing and type 
of inference. Computing the prior probabilities $p(l)$ exactly adds a $L$ term to the inference complexities, and using discounting or Dirichlet prior
smoothing exactly adds a $L$ term for the label nodes and $I$ for the document nodes.\\

The uniform distribution over document components can be seen as a type 
of kernel density model, where hierarchically smoothed multinomials are used instead of a Parzen kernel density using 
Gaussians \citep{Parzen:62}. The hierarchical smoothing can be formalized as a mean-shift \citep{Fukunaga:75} or data sharpening 
\citep{Choi:99} method, which shifts the document-conditional models closer to label-conditional models. Under this view, the classifier 
is a form of a Kernel Density Classifier \citep{Specht:88, John:95, Perez:09}. However, this kernel density formalization is complicated. 
The smoothed multinomial kernels are discrete, asymmetric, multivariate, bounded kernel functions, as well as local for each class. 
Each of these properties is treated as a deviation to a standard Parzen kernel density, and there is no known prior work on multinomial or
class-smoothed local kernel densities.
\chapter{Experiments}
This chapter presents a large-scale evaluation of the methods developed in the thesis, showing improvements in both efficiency and scalability.
A unified experimental framework is proposed for evaluation of classification and ranked retrieval, as well as optimization of model parameters with 
Gaussian random searches on development data. The performance measures, baseline methods, and statistical significance measurement across collections are discussed.
The set of text collections is described, along the chosen preprocessing, segmentation and statistics. Five sets of experiments are conducted, demonstrating 
considerable improvements from the methods developed in the thesis.\\

\section{Methodology}
\subsection{Experimental Framework}
Text mining (TM) applications are commonly decomposed into tasks solved using the methods of machine learning and statistics. The 
general task types are classification, ranking, clustering, regression and sequence labeling, of which the first two have been most extensively researched. 
Classification applications include spam classification, sentiment analysis, web page classification, and classification of documents into ontologies. Ranking is 
mostly applied for information retrieval (IR), where linear models implemented with inverted indices form the basis of modern web search engines. This 
chapter empirically explores a number of hypotheses drawn from the theory developed in Chapters 4 and 5. Experiments are conducted on ranking and 
classification datasets using the standard linear models for these tasks as baselines.\\

The experiments explore the following five research questions: 
\begin{description}
\item[Common formalization for smoothing] Chapter 4 formalized the various smoothing methods for generative models 
of text. Does this common formalization improve the effectiveness the models used in the basic TM tasks?
\item[Feature weighting and Extended MNB] Chapter 4 formalized weighting features for Multinomial Naive Bayes (MNB) and an extension of 
MNB that includes scaling of the prior and length modeling. Do these types of modifications provide improvements in effectiveness?
\item[Structured generative models] Chapter 5 introduced models that add constrained mixture modeling structure into MNB.
Do models of this type improve over MNB and Extended MNB models considerably?
\item[Strong linear models baselines] Chapters 4 and 5 propose generative models in a common framework for TM tasks. Do the proposed 
models improve over strong baselines for these tasks?
\item[Scalability of sparse inference] Chapter 5 introduced the idea of performing scalable probabilistic inference with generative models using inverted indices. 
How scalable are generative models utilizing sparse inference, compared to discriminative linear models and generative models implemented without 
inverted indices?\\
\end{description}

\begin{figure*}
\centering
 \includegraphics[scale=0.5, trim=100 110 100 90, clip=true]{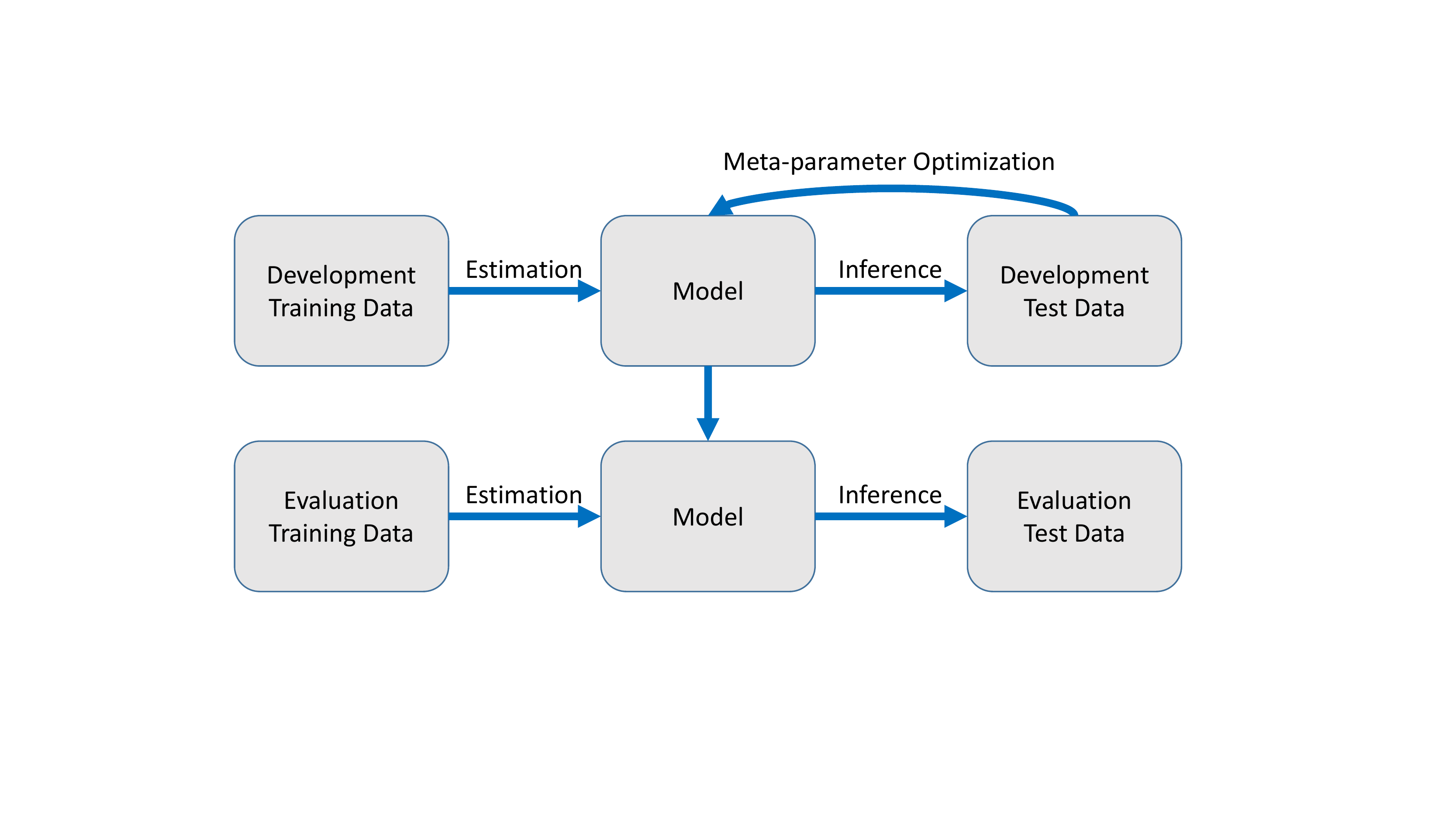}
\caption{The common framework for optimization and evaluation used for the experiments.}
\label{experiment_framework}
\end{figure*}

A common machine learning framework is applied to ranked retrieval and classification, illustrated in Figure \ref{experiment_framework}. For both cases 
datasets are segmented into a development training set $D^d$, a 
development test set $D^{d*}$, an evaluation training set $D^e$, and an evaluation test set $D^{e*}$. The development sets are used for random search 
optimization of the meta-parameters, such as the smoothing parameters. The evaluation sets are used to produce performance measures for the evaluated
models, which are then tested for statistically significant differences between the models across the datasets. \\

Within this framework, text classification and ad-hoc ranked retrieval tasks have fundamental differences only in how the datasets are organized. Classification 
operates on documents where both the training $\bm w$ and test set $\bm w^*$ documents are distributed in a similar fashion. Ad-hoc ranked retrieval refers 
to test set documents as queries, and to training set documents simply as documents. The queries often form word vectors much shorter than the retrieved 
documents, consisting of only a few keywords, a sentence, or a title indicating the search intent. The labels for classification datasets are distributed 
similarly between the train and test documents. The labels for ranked retrieval are document identifier variables, each training document corresponding to a single 
unique identifier, whereas queries correspond to multiple document labels that have been judged relevant to the query. The labels for multi-label classification 
can be described as binary indicator vectors $\bm c$, $\forall l: c_l \in (0, 1)$, constrained to $\sum_l c_l= 1$ for multi-class classification, and further $L= 2$ 
for binary-label classification. The labels for ranked retrieval can be described as either binary indicator vectors, or integer vectors when graded judgments 
are available for the queries. Aside from these fundamental differences in terminology and organization of data, ranked retrieval and text 
classification can be considered in the same experimental framework.\\

Multi-label classification tasks are converted into multi-class problems by using the Label Powerset method \citep{Boutell:04}. This maps each unique label
vector seen in training data into a categorical labelset variable, and maps the labelset variables back into label vectors after classification. This simplifies
model learning, but also increases the number of possible label variables in learning. Learning and optimizing meta-parameters for discriminative models for 
the large-scale multi-label datasets used in the experiments is not computationally feasible within practical times with Label Powerset or other basic
transformations of multi-label learning, and the results for these have not been computed.\\


\subsection{Performance Measures}

There are several commonly used performance measures for both ranking and classification. Classification measures need to 
consider the unbalanced label distributions common with text data: most labels have few associated documents, while most documents are labeled with one of 
the most common labels. Ranking measures need to consider the priority of ranking the top ranked labels accurately, compared to ranking all labels 
accurately. For the experiments conducted in this thesis, Micro-averaged F-score (Micro-F1) is used for evaluating classification, and Mean Average Precision
(MAP) and Normalized Discounted Cumulative Gain of top 20 documents (NDCG@20) are used to evaluate ranking. These measures are described
in the following.\\

For many classification tasks, F1-scores form the basis of the common evaluation measures. The F1-score is the harmonic mean of the precision and recall. Given 
binary label vectors of reference labels $\bm c$ and predicted labels $\bm{\hat{c}}$, precision is defined as the number of true positives TP divided by
the number of predicted positives PP: $\text{Precision}= \text{TP}/\text{PP}$, where $\text{TP}= \sum_{l:c_l= 1 \land \hat{c}_l= 1} 1$ and 
$\text{PP}= \sum_l \hat{c}_l$. Recall is defined as TP divided by the number of reference positives RP: $\text{Recall}= \text{TP}/\text{RP}$, where 
$\text{RP}= \sum_l c_l$. With these definitions, the F1-score for a single test document can be defined as:

\begin{align*}
\text{F1}= 2\: \frac{\text{Recall} \cdot \text{Precision}} { \text{Recall} + \text{Precision}}
\numberthis \label{KL}\\
\end{align*}

The Recall, Precision and F1-score measures have been developed in the context of IR, where the measures are commonly averaged across queries to
produce corresponding mean measures. Although Mean-F1 averaged across test label vectors $\bm c$ can be used for classification, the label imbalance in 
classification has made other types of averaged measures popular \citep{Tsoumakas:10}. With \emph{macro-averaging} a mean F1 is first computed for each label
independently, and these are averaged to produce the Macro-F1 measure. With \emph{micro-averaging}, the statistics of Recall and Precision are computed 
across the labels, and the Micro-F1 measure is then computed from the micro-averaged Recall and Precision statistics:

\begin{align*}
\text{Micro-F1}=&  2 \: \frac{\text{Micro-recall} \cdot \text{Micro-precision}} { \text{Micro-recall} + \text{Micro-precision}} \numberthis \\
\text{Micro-precision}=& \frac{\sum_{(\bm c, \bm{\hat{c}})} \sum_{l:c_l= 1 \land \hat{c}_l= 1} 1} {\sum_{(\bm c, \bm{\hat{c})}} \sum_l  \hat{c}_l} \numberthis \\
\text{Micro-recall}=& \frac{\sum_{(\bm c, \bm{\hat{c}})} \sum_{l:c_l= 1 \land \hat{c}_l= 1} 1} { \sum_{(\bm c, \bm{\hat{c}})} \sum_l  c_l} \numberthis
\label{microf1}\\
\end{align*}

Micro-F1 is one of the most commonly used measures for multi-label text classification. Unlike Micro-F1, Macro-F1 is strongly affected by 
label imbalance: a label occurring a thousand times has the same weight as one that occurs a single time. Micro-F1 for binary-label 
and multi-class tasks is also equivalent to mean Accuracy: the proportion of correct classifications \citep{Manning:08}. This makes Micro-F1 scores for these 
tasks directly comparable with much of the earlier research literature.\\

Precision forms the basis for MAP, the most common measure used for evaluating ranking in IR systems. MAP is computed as a mean of averaged Precision values
over possible levels of Recall. Let $\bm{\hat{y}}$ denote a vector of binary relevance decisions from the ranked and sorted scores for predicted labels 
$l$, and $\bm y$ a matching vector of binary reference scores. MAP is computed as:

\begin{align*}
\text{MAP}=& \frac{1}{|D^*|} \sum_{(\bm y, \bm{\hat{y}})} \frac{1}{\sum_k y_{k}} \sum_{k=1}^{L} \hat{y}_k \text{Precision}(k, \bm y) \numberthis \\
\text{Precision}(k, \bm y)=& \frac{\sum_{k'=1}^k y_{k'}}{k} \numberthis \\
\end{align*}

It can be seen that MAP ignores the actual predicted scores $\bm{\hat{y}}$, and uses only binary reference labels. 
MAP has since 
been supplemented with more sophisticated measures, such as the NDCG \citep{Jarvelin:02, Wang:13}. For NDCG the sorted reference scores $\bm y$ can take 
graded values. Standard NDCG@k is computed as:

\begin{align*}
\text{NDCG}(k)=& \frac{1}{|D^*|} \sum_{(\bm y, \bm{\hat{y}})} Z(k) \sum^k_{k'} \frac{2^{y_{k'}}-1} {\log_2(k'+1)} \numberthis \\
Z(k)=& \max_{\bm y'}(\sum^k_{k'} \frac{2^{y'_{k'}}-1} {\log_2(k'+1)}) \numberthis \\
\end{align*}

The normalizer $Z(k)$ is the maximum of possible Discounted Cumulative Gains (DCG), giving the DCG for the best possible ranking with $k$ ranked labels, and 
normalizing NDCG for each test document to a maximum score of $1$. Smaller values of $k$ are more effective in discriminating against ranking functions,
but become less robust \citep{Wang:13}. Larger values of $k$ are less brittle, but become less effective in discriminating ranking function performance. The value 
$k=20$ is used in the ranking experiments of this thesis.\\

Absolute differences in basic measures such as Micro-F1, NDCG@20 and MAP are affected by both the baseline model and the dataset. An improvement over a 
weak baseline or an easy dataset gives larger absolute differences than improvements over strong baselines and hard datasets. In text classification 
Micro-F1 is close to the perfect score of 1.0 for some tasks, whereas in text retrieval the reverse is the case, with MAP and NDCG results usually below 0.5.\\

The differences in Micro-F1 are further explored using Relative Error Reduction (RER), which is used in the evaluation of multinomial language models (LM)
in the context of speech recognition and machine translation \citep{Olive:11}. The RER can be defined as 
$1-(f^{max} -f^{s_2})/( f^{max}-f^{s_1})$, where $f^{max}$ is the maximum score, $f^{s_1}$ the score for baseline model, and $f^{s_2}$ the score for the
new model. Differences in NDCG@20 and MAP are further explored using Relative Improvement (RI), defined as $1-f^{s_2}/f^{s_1}$.
The derived measure RI is often used implicitly for describing the differences in IR ranking performance \citep{Zaragoza:07, Metzler:09}. 
Compared to absolute differences, relative differences are often more stable across varying baselines and datasets, and 
can give more intuitive measures of improvement.\\

\subsection{Baseline Methods}
Many of the leading solutions for TM tasks are instances of linear models, including those for classification and ranking tasks. For classification, linear classifiers
such as MNB, Support Vector Machines (SVM), and Logistic Regression (LR) have become the most common solutions. For ranking, the linear models 
Vector Space Model (VSM), Best Match 25 (BM25), and LM are the standard 
methods for ranked text retrieval in IR. The connections between linear models for TM were discussed in Chapter 2, and Chapters 3 and 4 examined 
the MNB model in detail, showing that a common framework for generative models covers the LMs used for ranking as a special case of MNB. The 
experiments conducted in this chapter use SVM, LR and MNB as baselines for classification, and BM25, VSM and MNB as baselines for ranking.\\

As discussed in Chapter 2, the multi-class linear scoring function takes the form:
\begin{align*}
y(\bm \theta_l, \bm w)= \theta_{l0} + \sum_{n=1}^{N} \theta_{ln} w_n
\numberthis \label{linear_score2}\\
\end{align*}

The following gives the parameter estimates for each of the baseline models. For most models, results with and without Term Frequency - Inverse Document
Frequency (TF-IDF) feature weighting are provided. TF-IDF modifies the training and test documents by applying the generalized TF-IDF presented in Chapter 4: 
$w_n = \log (1+\frac{w'_n} {|\bm w'|_0^{\phi}}) /|\bm w|_0^{1-\phi} \log(\max(1, \upsilon + \frac {I} {I_n}))$, where $w'_n$ are the original 
counts, $\phi$ is the parameter for length scaling and $\upsilon$ is the parameter for IDF lifting. Depending on the model, these parameters are fixed or
optimized. With fixed parameters, $\phi=0$, and length normalization is done after log normalization of counts. The fixed value $\upsilon= 0$ is used
for classification and ranking, producing Robertson-Walker IDF, whereas $\upsilon=-1$ is used for the scalability experiments, producing unsmoothed
Croft-Harper IDF and sparsifying the feature vectors to improve scalability.\\

MNB and LM are generative probabilistic models based on a multinomial or first-order Markov chain distribution of words conditional on each label variable.
An unsmoothed MNB or LM baseline model would have the parameter estimates $\theta_{ln}= \log \frac{\sum_{i:l^{i}= l} w_n^{(i)}}{\sum_i 
\sum_n w_n^{(i)}}$, while the bias is the log prior probability $\theta_{l0}= \log p(l)$ for MNB classification and uniform in the case of LM retrieval. 
The Dirichlet prior and Jelinek-Mercer smoothed MNB/LM models are used as basic baselines for classification
and ranking. These modify the unsmoothed log label-conditional probability parameters $\theta_{ln}$ as described in Chapter 4.
For both classification and ranking these baselines are compared with the proposed combinations of smoothing methods and extended generative models.\\
 
\label{LRSVM}
LR and SVM are discriminative linear classifiers, estimated with iterative algorithms that optimize a chosen loss function. With LR the loss function is
derived from an underlying discriminative probabilistic model, whereas with SVM the loss function is non-probabilistic.
L1 and L2-regularization with L2-SVM and LR are used as baselines for classification. Regularized LR and SVM models estimate the parameters 
$\theta_{ln}$ by optimizing $\theta_{ln}= \min_{\bm \theta} R(\bm \theta) + C \sum_i L(\bm \theta, D^{(i)})$, where $R(\bm \theta)$ is the 
regularization, $L(\bm \theta, D^{(i)})$ is the cost function, and $C$ is the regularization parameter. The loss function $L(\bm \theta, D^{(i)})$ for L2-SVM is 
$\max(0, 1 - l^{(i)}\bm \theta^T \bm w^{(i)})^2$, and $\log(1+ \epsilon^{-l^{(i)}\bm \theta^T \bm w^{(i)}})$ for LR. L1-regularization adds the term 
$R(\bm \theta)= |\bm \theta|_1$, while L2-regularization adds the term $R(\bm \theta)= \frac{1}{2} |\bm \theta|_2^2$.
Here L2-SVM is optimized using a coordinate descent algorithm \citep{Hsieh:08}, and LR is optimized using a trust region Newton method \citep{Lin:08}.\\

The VSM and BM25 methods used for ranking have a number of variants. VSM was initially defined as a cosine distance on word vector spaces 
\citep{Rocchio:71}, and 
this was later improved by applying TF-IDF feature transforms to training and test documents. Both of these basic models continue to be used in 
addition to more developed versions. The VSM models have the parameter estimates 
$\theta_{ln}= \frac{\sum_{i:l^{i}= l} w_n^{(i)}} {\sqrt{\sum_{n'}^N (\sum_{i:l^{i}= l} w_{n'}^{(i)})^2}}$, and the test document vector is 
L2-normalized: $w_n= \frac{w'_n}{|\bm w'|_2}$, where $\bm w'$ is the un-normalized vector. 
BM25 has been derived as an approximation to a probabilistic model, combined with a soft length normalization of counts \citep{Robertson:09}. The 
BM25 parameter estimates are $\theta_{ln}= IDF(n) \frac{(k_1 +1) w^{(i=l)}_n} {LN(i)+ w^{(i=l)}_n}$, and the test document counts are normalized 
$w_n= \frac{(k_3+1)w_n'}{k_3+w_n'}$, where $w'_n$ are the original un-normalized counts \citep{Manning:08, Robertson:09}. The training document 
length normalization is given by $LN(i)= k_1((1-b)+b|\bm w^{(i=l)}|_1/ A)$, with the average length $A=\sum_i |\bm w^{(i)}|_1/I$. The IDF 
for this standard BM25 is given by the smoothed Croft-Harper $IDF(n)=(I - I_n + 0.5)/(I_n + 0.5)$ \citep{Manning:08}. BM25 does not benefit from a further 
TF-IDF feature transform, as it includes an IDF function and has implicit count and document length normalization.\\

\subsection{Parameter Optimization}

The meta-parameters required by the models and TF-IDF are optimized on a development set for each dataset. A common practice is to either use a grid search 
of parameter estimates, or heuristic values \citep{Robertson:09}. Neither of these is guaranteed to provide optimal 
performance, and the results produced by unoptimized models can be misleading. The experiments shown in this thesis use random search
optimization of the meta-parameters, an approach that makes few assumptions about the optimized function and is efficient for the small-dimensional 
optimization problems encountered when optimizing TM linear model meta-parameters.\\

Grid search works by defining a grid of permissible parameter ranges ($\text{min}_q$, $\text{max}_q$) with small constant steps $\Delta_q$ for each 
meta-parameter $q$, such as increments of $0.1$ 
from $0.0$ to $1.0$. There are two main problems with this. First, the number of points to sample in the parameter space is an exponential function 
the dimension $Q$ of the parameter vector. With $Q=2$ parameters, a grid with range from $0.0$ to $1.0$ and increments of $0.1$ would require evaluation 
of $11^2$ points, while with $Q=5$ parameters the number of points would increase to $11^5= 161051$. This makes grid search efficient only when there are
few parameters. Second, if the steps for any of the parameters do not cover the optimum value, the optimization fails. With new models 
and data the permissible ranges and steps are not well known, and grid search can miss the optimal values.\\

Direct search optimization \citep{Powell:98}, also known as metaheuristics \citep{Luke:09} and black-box optimization, is a more complex method of 
parameter optimization. This seeks to optimize a function $f$ using a limited number of point evaluations
$f(\bm a)$, when very little is known about the properties of the function, such as smoothness, unimodality or convexity. Direct search problems of different types 
are encountered in a number of scientific disciplines, and as a result hundreds of methods have been extensively investigated. Some commonly known cases
are genetic algorithms and simulated annealing \citep{Luke:09}.\\

Random search \citep{Favreau:58,White:71} offers a type of direct search algorithm that is well suited for the small-dimensional, non-smooth and 
multi-modal functions encountered with the linear models in TM. A random search operates by improving the currently best point $\bm a$ by randomly 
generating new points $\bm d=\bm a+ \bm \Delta$ with steps $\bm \Delta$, bounding the points within the permissible ranges $\text{max}_q$ and 
$\text{min}_q$, and replacing $\bm a$ by $\bm d$ if the new point is good or better, i.e. $f(\bm d) \ge f(\bm a)$. Generating the steps by a Gaussian 
distribution produces the Random Optimization algorithm \citep{Matyas:65}, which can be improved by several commonly used heuristics:

\begin{itemize}
\item Decreasing step sizes\\
The step sizes can be gradually reduced by modifying the variance of the Gaussian distribution. The variance for each parameter can be initialized to be half the 
permissible range $0.5\cdot(\text{max}_q-\text{min}_q)$, and multiplied by $0.9$ after each iteration. This produces a log-curve decrease in step sizes, 
sampling most of the permissible ranges initially and searching locally later

\item Multiple parallel steps\\
The point evaluation can be parallelized, evaluating a subiteration of $Q$ points simultaneously and choosing the best point $\bm d_q= \bm a+ \bm \Delta_q$ 
as the new best point. This enables direct use of multiple processors for optimization, without any parallelization overhead

\item Multiple best points\\
In case of a tie, the best $X$ points from each subiteration can be used to replace the current best point, and sampling can be done uniformly from the 
this set $\bm a_x$: $\bm d_q= \bm a_{q\%X} + \bm \Delta_q$. This enables the search to spread out to $X$ different locations, in 
case a plateau is reached\\
\end{itemize}

\begin{figure*}
 \centering
 \includegraphics[scale=1.0, trim=283 140 280 135, clip=true]{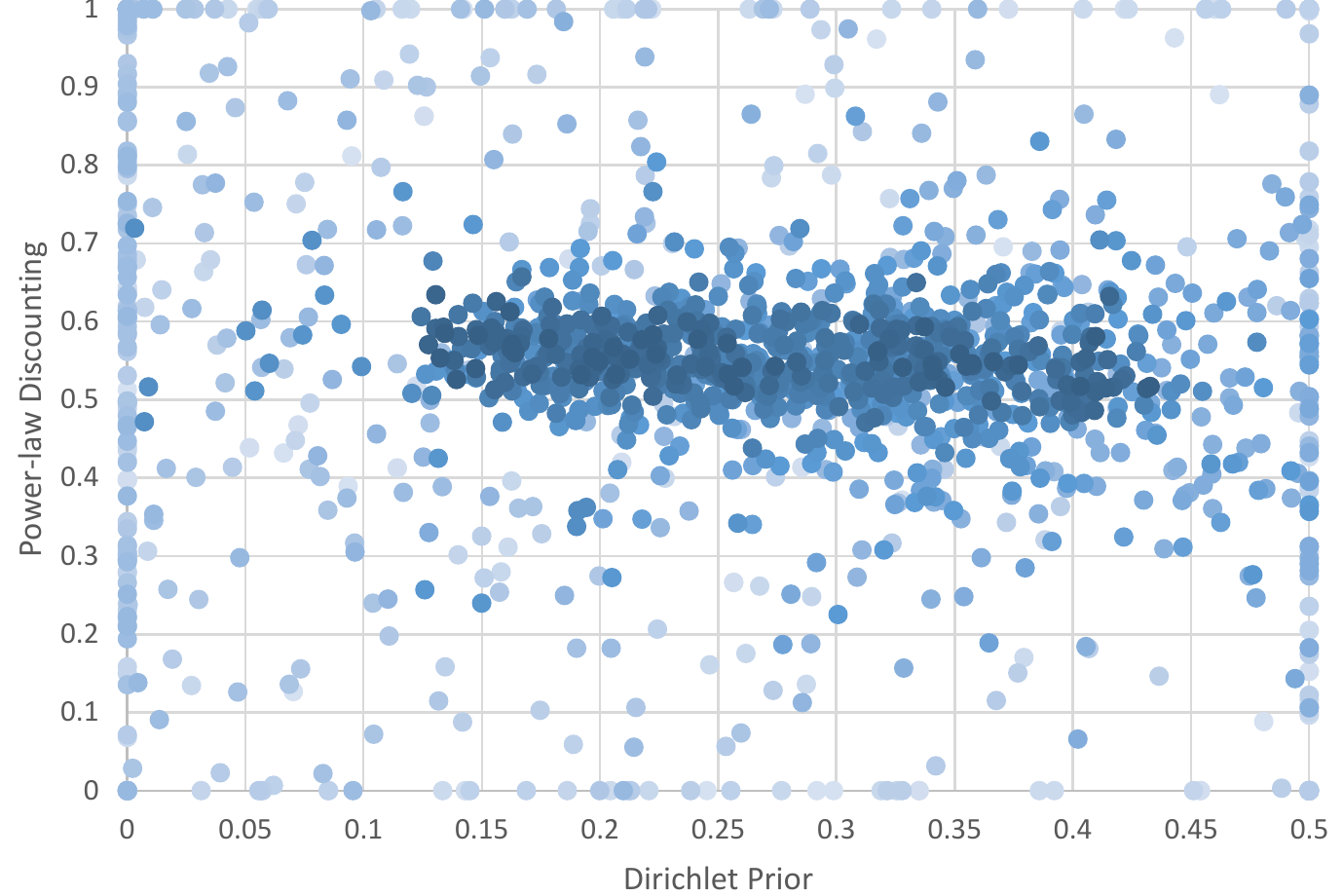}
\caption{Evaluated points for a 40x40 random search, optimizing the Dirichlet prior and power-law discounting parameter for a MNB model on the webkb text
classification dataset. Points color-coded from light to dark in the order of iterations. Light points show the global sampling done by the initial iterations, dark points 
in the center show the local search done by the last iterations. Dirichlet prior parameter is normalized by $L$ for optimization, adding up to $0.5$ to each count}
\label{randomsearch}
\end{figure*}

All of the above modifications are common variations to random searches, and generally improve search efficiency without introducing additional flaws into random 
search, such as vulnerability to local optima. A parallelized random search of 50 iterations and 8 subiterations can be denoted a 50x8 random search. 
Figure \ref{randomsearch} visualizes the evaluated points for a 40x40 search using this Gaussian random search. The models for retrieval were optimized
with 50x8 searches, the models for classification with 40x40 searches. LR and SVM models were optimized with 40x8 searches, due to longer estimation times
and the use of only up to three meta-parameters. For all model types, the random searches were iterated a few times with different permissible ranges.\\

\subsection{Significance Tests}

The parameters optimized on the development sets are used for measuring performance on the evaluation sets. Comparison of performance can be done within 
a dataset or between datasets. Within-dataset comparisons are more common when a limited number of datasets are available, as used to be the case in 
IR \citep{Hull:93, Sanderson:05, Smucker:07b, Cormack:07, Smucker:09}. Across-dataset comparisons are more common in fields where standardized datasets 
are publicly available, as is the case in machine learning \citep{Dietterich:98, Demsar:06}. These have the important advantage of measuring 
performance on a group of datasets, avoiding problems encountered when using folds of the same dataset for testing significance \citep{Demsar:06}. If the 
chosen datasets are distributed in the same way as a typical dataset for the task, then the discovered effects will hold on new datasets of the same task. 
Here a number of datasets have been segmented for both classification and ranking, and the methods are compared across the datasets.\\

Statistical significance of the evaluation set performance measures is assessed by performing one-tailed paired t-tests on the absolute 
between-dataset differences. The paired Student's t-test \citep{Gosset:08} is a basic test for comparing differences, recently advocated for evaluation of IR results \citep{Sanderson:05, Smucker:07b, Cormack:07, Smucker:09}. It compares the means and variances of two groups of observations, and assumes 
both groups have a Gaussian distribution. The null-hypothesis for a t-test posits that the difference between the means of the two groups is the result of the Gaussian variance. If the difference exceeds what the variance allows, it is considered to be statistically significant.  
Significance is computed by comparing the t-value of the test to a t-distribution, and discarding the null-hypothesis if the t-statistic deviates too far from the t-distribution for the chosen p-value of statistical significance.\\

A paired t-test compares observations from matched pairs, so that instead of the difference between means, the mean of the paired differences is compared. 
This makes the test more powerful, because the variance caused by datasets is subtracted from the comparison. A one-tailed t-test compares only the difference 
in the t-statistic in a single direction, instead of deviation in both directions. This makes the test twice as powerful in p-value, with the prior assumption that 
variance in one direction will not be significant. Paired one-tailed t-tests are conducted with two significance levels on the absolute differences in the measures:  
0.005 ($\dagger$) and 0.05 ($\ddagger$).\\

A problem with large-scale statistical testing is that the risk of false positive results is multiplied by the number of conducted comparisons. Modifying 
test statistics to penalize for the number of comparisons in turn weakens the significances of the individual comparisons. The strategy adopted here is to
constrain significance tests only on the differences that attempt to answer the research questions set out in advance, at the beginning of the chapter. 
This both simplifies the 
description of findings, and reduces the risk of false positives. Nevertheless, the datasets in question have been used to iteratively develop the models, and
a prior literature exists on the performance of the baseline linear models on these datasets. For these reasons the experiments and significance tests are
\emph{exploratory}, and \emph{confirmatory} evaluation of the discovered effects is left for future research on new datasets. For further exploration and 
confirmation of the findings, the full evaluation set results are provided as tables in Appendix A.\\

\section{Datasets}
\label{datasets_section}
\subsection{Dataset Overview}
A total of 13 datasets are used for ranking and 16 for classification. Additional 
experiments are conducted on a large Wikipedia dataset, which allows the scalability of linear classifiers to be assessed when the 
numbers of documents, features and labels are each scaled up to a million. Aside from the TREC1-8 datasets for ranking, all datasets are publicly available 
for research use, and the pre-processing scripts and pre-processed datasets are made available\footnote{http://www.cs.waikato.ac.nz/\~{}asp12/}.\\

The ranking datasets consist of the TREC 1-8\footnote{http://trec.nist.gov/data/test\_coll.html} collections split according to data source into 11 datasets, 
OHSU-TREC\footnote{http://trec.nist.gov/data/t9\_filtering.html} \citep{Hersh:94, Robertson:01} and FIRE 2008-2011\footnote{http://www.isical.ac.in/\~{}clia/} 
datasets. TREC1-8 \citep{Voorhees:99} contains the ad-hoc retrieval collections that were used in the 1990s to establish modern ranking functions and performance 
measures, most notably the BM25 ranking function \citep{Robertson:96} and the MAP evaluation measure \citep{Voorhees:99}. TREC was followed by other 
programs of for evaluating IR technology, such as NTCIR, INEX and FIRE. While the TREC ad-hoc track was suspended in 1999 in favor of more diverse tasks, some 
of these can be considered collections for ad-hoc ranked retrieval. The OHSU-TREC collection is a publicly available dataset of medical articles from PubMed, used 
for TREC9. The FIRE 2008-2011 English collections were constructed to evaluate IR in the major languages spoken in India, following the TREC ad-hoc 
evaluation paradigm.\\

Six binary-label, five multi-class, and five multi-label datasets are used. The binary-label datasets are 
TREC06\footnote{http://plg.uwaterloo.ca/\~{}gvcormac/treccorpus06/} \citep{Cormack:06}, ECUE1 and 
ECUE2\footnote{http://www.dit.ie/computing/staff/sjdelany/Dataset.htm} \citep{Delany:06} for spam classification, and  
ACL-IMDB\footnote{http://ai.stanford.edu/\~{}amaas/data/sentiment/} \citep{Maas:11}, 
TripAdvisor12\footnote{http://www.cs.virginia.edu/yanjun/paperA14/ecml12-cikm11-deepSC.htm \label{virginia_source}} \citep{Bespalov:12},
and Amazon12\textsuperscript{\ref{virginia_source}} \citep{Blitzer:07} for sentiment analysis. 
These have been made
available in the last 10 years, during which period spam classification and sentiment analysis became popular topics.
The multi-class datasets are R8, R52, WebKb, 20Ng and Cade\footnote{http://web.ist.utl.pt/\~{}acardoso/datasets/} \citep{Cardoso:07}.
These are older datasets, and versions of the first four provided the benchmarks used for early comparisons of text classification 
algorithms \citep{Lewis:92, Joachims:98, McCallum:98b}. The multi-label datasets, which are more recent and large-scale, are  
RCV1-v2-Ind\footnote{http://www.daviddlewis.com/resources/testcollections/rcv1/},
EUR-Lex\footnote{http://www.ke.tu-darmstadt.de/resources/eurlex} \citep{Mencia:10}, 
OHSU-TREC\footnote{http://trec.nist.gov/data/t9\_filtering.html} \citep{Robertson:01}, 
DMOZ2\footnote{http://lshtc.iit.demokritos.gr \label{lshtc_source}} and WikipMed2\textsuperscript{\ref{lshtc_source}}.\\

The large-scale Wikipedia dataset WikipLarge\textsuperscript{\ref{lshtc_source}} was used in the Large Scale Hierarchical Text Classification (LSHTC) evaluation 
of scalable multi-label classification of articles into the Wikipedia categorization hierarchy. It has word vector features with millions of words, documents and labels. 
The scalable probabilistic models developed in this thesis were used to win the 2014 LSHTC competition \citep{Puurula:14} on this data. For the systematic comparisons of 
scalability, the original dataset is pruned in terms of features, documents and labels, as described in the following subsections.\\


\subsection{Preprocessing}

Datasets were pre-processed from the original form into word vectors; however, some datasets are provided in word vector forms, making it impossible to perform 
identical processing steps. In all cases text was first lowercased. Next, stop-words were removed on most datasets, as well as short words ($< 3$ characters) and 
long words ($> 20$ characters). This was followed by Porter-stemming \citep{Porter:80} of the remaining words. The scripts used for both pre-processing 
and segmentation are publicly available\footnote{http://www.cs.waikato.ac.nz/\~{}asp12/}.\\

The binary-label datasets ecue1 and ecue2 are provided as pre-processed integer word vectors. Stop-word removal or stemming is not performed on the 
text \citep{Delany:06}, but documentation for any other pre-processing is not available. The trec06 dataset is provided as raw emails, including the 
metadata header. The header was removed, the remaining text was Porter-stemmed, and stop-words, short words, long words, non-words, and numbers were
removed. The aclimdb dataset is provided in pre-processed form, in lower-case with numbers removed, but without stemming or removal 
of stop-words or non-words \citep{Delany:06}. Opinion grades 1-4 and 1-7 were mapped to negative and positive labels, respectively \citep{Maas:11}.
The tripa12 and amazon12 datasets are provided in pre-processed form, with numbers replaced by ``NUMBER'', but without stemming, stop-word, 
short-word or non-word removal. Opinion grades 1-2 were mapped to negative label, 4-5 to positive \citep{Bespalov:12}.\\

The single-label datasets 20ng, cade, r52, r8, and webkb are provided as pre-processed word vectors. These use Porter-stemming, removal of
524 SMART stop-words, removal of short and truncation of long words \citep{Cardoso:07}. Where available, titles of documents are 
concatenated to text bodies. Here no further processing was done aside from format conversion of the files.\\

The multi-label dataset rcv1 is provided as pre-processed word vectors, with punctuation removal, Porter-stemming and SMART stop word 
removal \citep{Lewis:04}. The eurlex dataset is provided with lowercasing, Porter-stemming, stop-word removal and number removal \citep{Mencia:10}. 
The ohsu-trec
dataset is provided in the original OHSU-MED format \citep{Hersh:94}. Here we use the MEDLINE subject field as labels, and concatenate the title and description 
fields to form the word vectors, with Porter-stemming and short word removal as pre-processing. The DMOZ2, wikip\_med2 and wikip\_large datasets are provided
in pre-processed word vector forms, and no further pre-processing was made.\\

The TREC1-8 collections ``trec\_*'' for ranked retrieval were pre-processed by stop-word, xml-tag, non-word, number, and short word removal, followed by 
Porter-stemming. For queries the description fields of queries were used to form the query word vectors, and relevance judgements were 
converted into binary label vectors of relevant document identifiers. The fire\_en dataset was processed identically to the TREC datasets. For ohsu\_trec the 
pre-processing was performed identically, but the queries were concatenated from title and description fields, and the provided graded relevance judgments were 
preserved as integer-weighted label vectors with relevance judgement grades 0, 1 and 2.\\

\subsection{Segmentation}

The segmentation scripts first partitioned the datasets into a development set for optimizing parameters and an evaluation set for computing the
performance measures of the optimized models. Dataset-dependent segmentation choices were made to make the parameter optimization reliable, and 
keep processing complexity within practical bounds. Pre-existing dataset partitions were used where available. Otherwise random sampling was used to 
further segment the data. All datasets were mapped into 
the same framework of segmentation, with the development sets used for optimizing parameters and the evaluation sets used for conducting the 
evaluated experiment results. Both sets were further divided into a training sets ($D^d$, $D^e$) for learning the linear models, and test sets 
($D^{d*}$, $D^{e*}$) for measuring performance.\\

Many of the classification datasets are provided with existing development and evaluation partitions. The original partitions for ohsu-trec 
and rcv1 were swapped, as this provided more data for learning models. The evaluation training set for all classification datasets was concatenated from
the development training and test sets, while the evaluation test set composed of a held-out portion. The single-label and binary-label datasets 
ecue1 and ecue2 had insufficient data to form reliable development sets. For these datasets, 5-fold cross-validation over the development partition was used,
with 200 documents preserved for each fold as the development test set, and the rest as the development train set. The classification datasets amazon12, rcv1, 
ohsu-trec, wikip\_med2, and DMOZ2 had the lowest and highest document frequency words pruned, to reduce the total number of counts to 8 million per 
dataset, enabling efficient experimentation with less memory use.\\

The TREC datasets are commonly divided by combinations of data source, year, TIPSTER disk, and query number to form smaller segments for experiments.
Here the datasets were segmented by data source to form the 11 trec\_* datasets. The trec\_* and fire\_en datasets were further segmented according to
queries to form the development and evaluation sets, so that the first 20 queries from each year were concatenated to form the development
test set, and the remaining 30 queries from each year were concatenated to form the evaluation test set. The ohsu\_trec dataset was segmented according to
the existing document partition, reserving ohsumed.87 for development and ohsumed.88-91 for evaluation. For all of the retrieval datasets, documents
not given a relevance judgement for any of the queries in the test set were removed from the training set, greatly improving the efficiency for performing
experiments.\\

The wikip\_large dataset was segmented different from the general framework. The original training dataset provided for LSHTC4 was segmented by
random sampling to reserve $1\%$ of the data as an evaluation test set, and the remaining $99\%$ was used as the evaluation training set. These evaluation
sets were then further pruned by documents, features and labelsets so that each of these dimensions scaled up to a million. Documents were pruned
in the order they occurred in a shuffled training dataset, with the number of preserved documents varied with the thresholds
(10, 100, 1000, 10000, 100000, 1000000). Features were pruned to preserve the most frequent words, with the number or preserved features varied with the 
thresholds (10, 100, 1000, 10000, 100000, 1000000). Labelsets were similarly pruned to preserve the most common labelsets with the thresholds 
(1, 10, 1000, 1000000). Overall, these pruning choices resulted in 144 pruned versions of wikip\_large for testing scalability of the models. The scalability 
experiments were then conducted on the evaluation sets using fixed parameters, and no development sets were constructed for optimization.\\

\subsection{Dataset Statistics}
\begin{table}
\footnotesize
\centering
\caption{Basic statistics of the pre-processed and segmented experiment datasets. $D^d$ is the development training set,
$D^{d*}$ the development test set, $D^e$ the evaluation training set and $D^{e*}$ the evaluation test set. Statistics
for the number of label variables $L$ and the number of word features $N$ are denoted correspondingly}
\begin{tabular}{|l|c|c|c|c|c|c|c|c|c|} \hline
dataset & $|D^d|$ & $|D^{d*}|$ & $|D^e|$ & $|D^{e*}|$ & $L^d$ & $L^e$ & $N^d$ & $N^e$ \\ \hline \hline
20ng & 11093 & 200 & 11293 & 7528 & 20 & 20 & 54112 & 54580 \\ \hline
cade & 27122 & 200 & 27322 & 13661 & 12 & 12 & 156751 & 157483 \\ \hline
r52 & 6332 & 200 & 6532 & 2568 & 52 & 52 & 15882 & 16145 \\ \hline
r8 & 5285 & 200 & 5485 & 2189 & 8 & 8 & 14334 & 14575 \\ \hline
webkb & 2585 & 200 & 2785 & 1396 & 4 & 4 & 7287 & 7287 \\ \hline
ecue1 & 9778 & 200 & 9978 & 1000 & 2 & 2 & 100000 & 100000 \\ \hline
ecue2 & 10665 & 200 & 10865 & 1000 & 2 & 2 & 159579 & 161155 \\ \hline
trec06 & 34039 & 1000 & 34039 & 2783 & 2 & 2 & 797772 & 797772 \\ \hline
tripa12 & 55299 & 4999 & 55299 & 10077 & 2 & 2 & 76364 & 76364 \\ \hline
aclimdb & 45000 & 2000 & 45000 & 3000 & 2 & 2 & 89527 & 89527 \\ \hline
amazon12 & 257877 & 9998 & 257877 & 100556 & 2 & 2 & 86914 & 86914 \\ \hline
rcv1 & 342117 & 1000 & 342117 & 8644 & 350 & 350 & 160281 & 160281 \\ \hline
eurlex & 16381 & 1000 & 16381 & 1933 & 3828 & 3828 & 172928 & 172928 \\ \hline
ohsu-trec & 196555 & 1000 & 196555 & 35890 & 14373 & 14373 & 290117 & 290117 \\ \hline
DMOZ2 & 390809 & 2000 & 390809 & 1947 & 27874 & 27874 & 111939 & 111939 \\ \hline
wikip\_med2 & 452318 & 2000 & 452318 & 2568 & 36463 & 36463 & 47021 & 47021 \\ \hline
wikip\_large & NA & NA & 2341782 & 23654 & NA & 324634 & NA & 1608946 \\ \hline \hline
fire\_en & 21919 & 90 & 16075 & 60 & 21919 & 16075 & 103551 & 91089 \\ \hline
ohsu\_trec & 36890 & 63 & 196555 & 63 & 36890 & 196555 & 77994 & 220256 \\ \hline
trec\_ap & 47172 & 150 & 33474 & 100 & 47172 & 33474 & 201591 & 162284 \\ \hline
trec\_cr & 5063 & 60 & 4006 & 40 & 5063 & 4006 & 198170 & 188513 \\ \hline
trec\_doe & 10053 & 89 & 7717 & 59 & 10053 & 7717 & 32352 & 28569 \\ \hline
trec\_fbis & 23207 & 90 & 17315 & 60 & 23207 & 17315 & 202033 & 175660 \\ \hline
trec\_fr & 25185 & 240 & 20581 & 160 & 25185 & 20581 & 252577 & 242648 \\ \hline
trec\_ft & 41452 & 120 & 30549 & 80 & 41452 & 30549 & 228547 & 187797 \\ \hline
trec\_la & 25944 & 90 & 17834 & 60 & 25944 & 17834 & 162531 & 129299 \\ \hline
trec\_pt & 1635 & 30 & 1792 & 20 & 1635 & 1792 & 111883 & 106147 \\ \hline
trec\_sjmn & 9160 & 30 & 6469 & 20 & 9160 & 6469 & 74447 & 59992 \\ \hline
trec\_wsj & 57117 & 150 & 45078 & 100 & 57117 & 45078 & 247771 & 215497 \\ \hline
trec\_zf & 19901 & 150 & 13763 & 99 & 19901 & 13763 & 192489 & 158042 \\ \hline
\end{tabular}
\label{basic_dataset_stats}
\end{table}

The common framework for classification and ranking tasks enables direct comparison of the dataset properties. Table \ref{basic_dataset_stats} shows the
basic dataset statistics of numbers of documents, features and labels for the development and evaluation sets. Table \ref{mean_dataset_stats} shows the
mean numbers of features and labels per document. For the datasets that use 5-fold cross-validation for development, the first fold is used to compute the
statistics for the tables. For the retrieval datasets, label variables for labeled non-relevant documents are not included for showing the number of document labels
per query, as these are assumed to be as relevant as non-labeled documents.\\

\begin{table}
\footnotesize
\centering
\caption{Mean statistics of the pre-processed and segmented experiment datasets. $|\bm w^d|_0$ is the mean number of unique words and $|\bm c^d|_0$ the 
mean number of labels per document in the training set. The corresponding statistics are given for the development test set ($\bm w^{d*}$, $\bm c^{d*}$), 
evaluation training set ($\bm w^e$, $\bm c^e$) and evaluation test set ($\bm w^{e*}$, $\bm c^{e*}$)}
\begin{tabular}{|l|c|c|c|c|c|c|c|c|c|} \hline
dataset & $|\bm w^d|_0$ & $|\bm w^{d*}|_0$ & $|\bm w^e|_0$ & $|\bm w^{e*}|_0$ & $|\bm c^d|_0$ & $|\bm c^{d*}|_0$ & $|\bm c^e|_0$ & $|\bm c^{e*}|_0$ \\ \hline \hline
20ng & 84.20 & 90.62 & 84.32 & 83.14 & 1.00 & 1.00 & 1.00 & 1.00 \\ \hline
cade & 62.24 & 74.68 & 62.33 & 59.83 & 1.00 & 1.00 & 1.00 & 1.00 \\ \hline
r52 & 43.08 & 44.08 & 43.11 & 39.71 & 1.00 & 1.00 & 1.00 & 1.00 \\ \hline
r8 & 41.34 & 37.49 & 41.20 & 37.28 & 1.00 & 1.00 & 1.00 & 1.00 \\ \hline
webkb & 76.86 & 79.79 & 77.07 & 79.03 & 1.00 & 1.00 & 1.00 & 1.00 \\ \hline
ecue1 & 186.54 & 165.13 & 186.11 & 211.28 & 1.00 & 1.00 & 1.00 & 1.00 \\ \hline
ecue2 & 144.02 & 145.66 & 144.05 & 132.91 & 1.00 & 1.00 & 1.00 & 1.00 \\ \hline
trec06 & 107.21 & 85.01 & 107.21 & 93.97 & 1.00 & 1.00 & 1.00 & 1.00 \\ \hline
tripa12 & 105.25 & 106.85 & 105.25 & 104.61 & 1.00 & 1.00 & 1.00 & 1.00 \\ \hline
aclimdb & 136.51 & 129.71 & 136.51 & 131.53 & 1.00 & 1.00 & 1.00 & 1.00 \\ \hline
amazon12 & 30.69 & 30.20 & 30.69 & 30.63 & 1.00 & 1.00 & 1.00 & 1.00 \\ \hline
rcv1 & 22.44 & 21.02 & 22.44 & 21.84 & 1.60 & 1.58 & 1.60 & 1.57 \\ \hline
eurlex & 271.42 & 252.74 & 271.42 & 271.84 & 5.32 & 5.29 & 5.32 & 5.32 \\ \hline
ohsu-trec & 40.12 & 37.00 & 40.12 & 37.35 & 12.39 & 11.66 & 12.39 & 11.93 \\ \hline
DMOZ2 & 20.44 & 20.03 & 20.44 & 23.72 & 1.03 & 1.02 & 1.03 & 1.03 \\ \hline
wikip\_med2 & 17.27 & 18.38 & 17.27 & 15.52 & 1.84 & 2.01 & 1.84 & 1.67 \\ \hline
wikip\_large & NA & NA & 42.54 & 42.27 & NA & NA & 3.26 & 3.27 \\ \hline \hline
fire\_en & 145.84 & 6.92 & 148.56 & 6.57 & 1.00 & 52.48 & 1.00 & 41.02 \\ \hline
ohsu\_trec & 50.87 & 6.41 & 53.25 & 6.41 & 1.00 & 10.63 & 1.00 & 50.87 \\ \hline
trec\_ap & 164.77 & 8.71 & 165.26 & 8.62 & 1.00 & 82.78 & 1.00 & 81.66 \\ \hline
trec\_cr & 473.82 & 7.33 & 531.94 & 8.50 & 1.00 & 14.75 & 1.00 & 7.00 \\ \hline
trec\_doe & 43.65 & 9.62 & 45.36 & 9.97 & 1.00 & 9.63 & 1.00 & 22.93 \\ \hline
trec\_fbis & 205.15 & 6.78 & 216.20 & 7.97 & 1.00 & 30.07 & 1.00 & 28.40 \\ \hline
trec\_fr & 233.84 & 7.98 & 253.04 & 8.38 & 1.00 & 6.50 & 1.00 & 7.38 \\ \hline
trec\_ft & 155.63 & 6.91 & 158.34 & 7.71 & 1.00 & 29.03 & 1.00 & 37.53 \\ \hline
trec\_la & 202.05 & 6.78 & 205.21 & 7.97 & 1.00 & 23.04 & 1.00 & 24.35 \\ \hline
trec\_pt & 563.39 & 7.30 & 517.50 & 6.50 & 1.00 & 0.83 & 1.00 & 0.75 \\ \hline
trec\_sjmn & 178.74 & 7.30 & 179.12 & 6.50 & 1.00 & 28.53 & 1.00 & 22.05 \\ \hline
trec\_wsj & 188.03 & 8.71 & 194.51 & 8.62 & 1.00 & 66.42 & 1.00 & 69.81 \\ \hline
trec\_zf & 216.07 & 8.71 & 219.00 & 8.66 & 1.00 & 17.65 & 1.00 & 18.35 \\ \hline
\end{tabular}
\label{mean_dataset_stats}
\end{table}

When compared under the same framework, it can be seen from Table \ref{mean_dataset_stats} that the training and test set documents in retrieval have 
very different properties. The test set documents are queries, which are often 20 times shorter than the training documents, and have a large number of labels; 
whereas the retrieved documents have each one document identifier label. In stark contrast, text classification training and test set documents are generally 
drawn from the same type of data. Both types of tasks can have significant differences between development and evaluation conditions, although generally 
these are defined to be similar enough for meta-parameter optimization to be possible.\\ 

\section{Experiments and Results}

\subsection{Evaluated Linear Model Modifications} 
\begin{table}
\footnotesize
\centering
\caption{Modification affixes and reference pages in the thesis}
\begin{tabular}{|l|c|c|} \hline
affix & modification & reference pages\\ \hline
u & uniform background model & \pageref{background_models} \\
c & collection background model & \pageref{background_models} \\
uc & uniform-smoothed collection background model & \pageref{background_models} \\
dp & Dirichlet prior smoothing & \pageref{mnb_smoothing} \\
jm & Jelinek-Mercer smoothing & \pageref{mnb_smoothing} \\
ad & absolute discounting & \pageref{mnb_smoothing} \\
pd & power-law discounting & \pageref{mnb_smoothing} \\
kdp & kernel Dirichlet prior smoothing & \pageref{mnb_smoothing},  \pageref{TDM}\\
kjm & kernel Jelinek-Mercer smoothing & \pageref{mnb_smoothing},  \pageref{TDM} \\
kpd & kernel power-law discounting & \pageref{mnb_smoothing},  \pageref{TDM} \\
po & Poisson document length modeling & \pageref{extended_MNB} \\
ps & prior scaling &  \pageref{extended_MNB} \\
qidf & query IDF weighting & \pageref{tfidf2} \\
qidfX & query IDF weighting with IDF lifting & \pageref{tfidf2} \\
ti & TF-IDF weighting & \pageref{tfidf2} \\
tXi & TF-IDF weighting with length scaling & \pageref{tfidf2} \\
tiX & TF-IDF weighting with IDF lifting & \pageref{tfidf2} \\
tXiX & TF-IDF weighting with length scaling and IDF lifting & \pageref{tfidf2} \\
l1r & L1 regularization for LR/SVM & \pageref{LRSVM} \\
l2r & L2 regularization for LR/SVM & \pageref{LRSVM} \\ \hline
\end{tabular}
\label{modification_table}
\end{table}

The experiments compare a number of models across datasets and measures. Modifications to models are denoted by acronym affixes separated by underscores, for example ``jm'' for Jelinek-Mercer smoothed models, and ``u\_dp'' for uniform-background Dirichlet prior smoothed models. Throughout the experiments only a subset of possible modifications is attempted, because there are 39 combinations for the basic smoothing methods, and far more when feature weighting and structural models are included. Table \ref{modification_table} summarizes the modifications used in the experiments, with references to descriptions.
None of the MNB or TDM modifications increase the time or space complexities of estimation or inference, or introduce considerable additional constants to 
processing requirements. IDF lifting can decrease the complexities, sparsifying documents further by weighting frequently occurring words to $0$.\\ 

\subsection{Smoothing Methods} 

The first set of experiments evaluates the usefulness of the common framework for multinomial models of text presented in Chapter 4. Four methods are used
for discounting and smoothing: Dirichlet prior (dp), Jelinek-Mercer (jm), absolute discounting (ad), and power-law discounting (pd). These are combined with three
choices for background distribution: uniform (u), collection (c), and uniform-smoothed collection (uc). Absolute discounting proved early in the 
experiments to be inferior to power-law discounting, and further combinations with the other methods are not shown. This was expected from the 
literature \citep{Huang:10}. Combinations for the other models were explored based on the initial performance of the collection-smoothed models.\\

\begin{figure*}
\centering
 \includegraphics[scale=1.0, trim=285 120 285 120, clip=true]{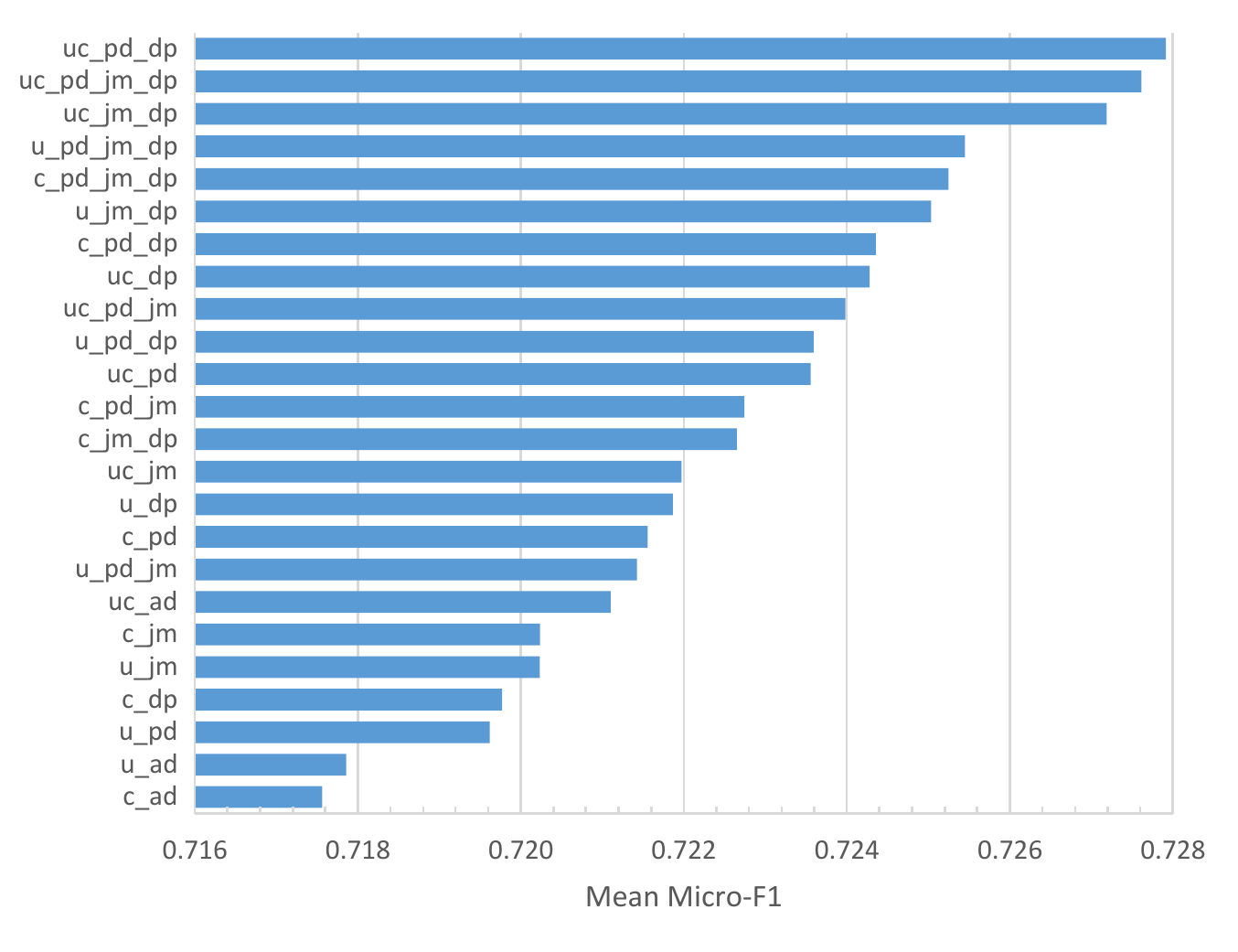}
\caption{Mean Micro-F1 across the text classification datasets with the different multinomial smoothing methods}
\label{smooth_fig1}
\end{figure*}

\begin{figure*}
\centering
 \includegraphics[scale=1.0, trim=285 120 285 120, clip=true]{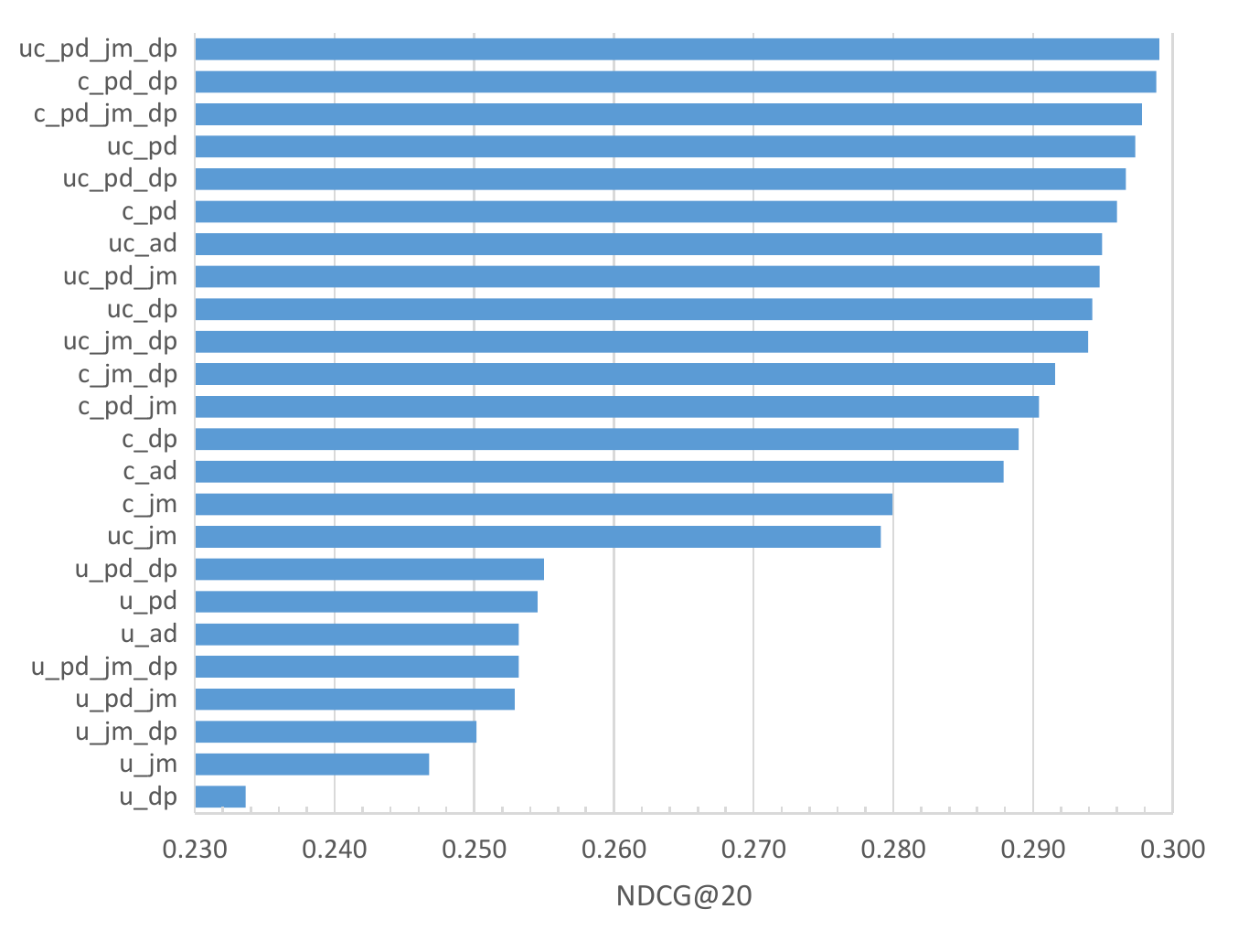}
\caption{Mean NDCG@20 across the text retrieval datasets with the different multinomial smoothing methods}
\label{smooth_fig2}
\end{figure*}

\begin{figure*}
\centering
 \includegraphics[scale=1.0, trim=285 120 285 120, clip=true]{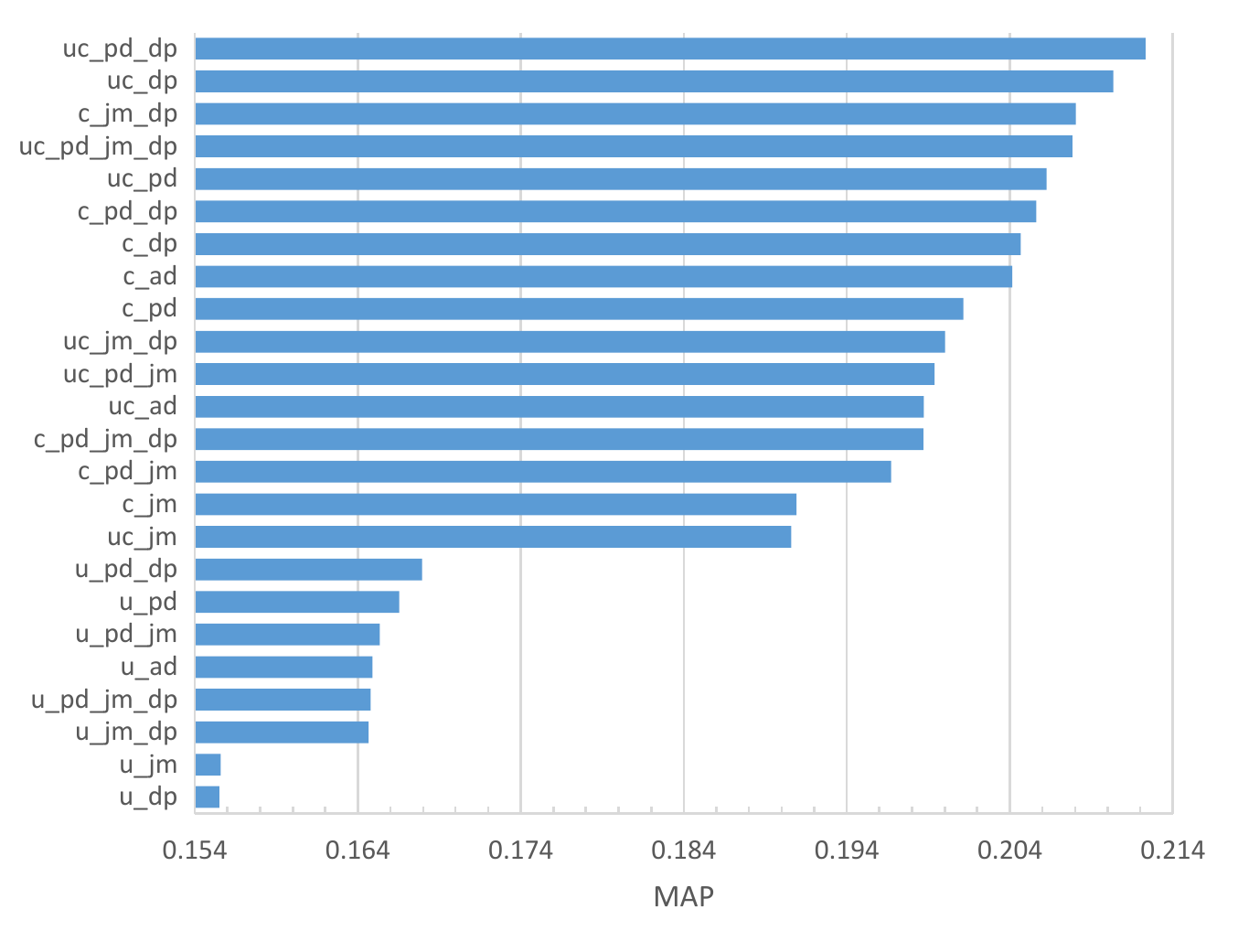}
\caption{Mean MAP across the text retrieval datasets with the different multinomial smoothing methods}
\label{smooth_fig3}
\end{figure*}

Figure \ref{smooth_fig1} shows the mean Micro-F1 across the text classification datasets for the different smoothing methods. Figures \ref{smooth_fig2} 
and \ref{smooth_fig3} shows the mean NDCG@20 and MAP across the text retrieval datasets for the different smoothing methods, respectively. The first
visible effect is the overall improvement from the combinations, compared to the individual smoothing methods in standard use, such as c\_dp
and c\_jm. Overall, the differences are small but consistent.\\

The main hypothesis to test is whether the generalized smoothing method (uc\_pd\_jm\_dp) improves over the baselines Dirichlet prior (c\_dp) and 
Pitman-Yor process smoothing (c\_pd\_dp).
Comparing the combined smoothing method uc\_pd\_jm\_dp to the baseline c\_dp, a relative reduction of 2.80\%$\dagger$ in Micro-F1 and relative 
improvements of 3.37\% in NDCG@20 and 1.53\% in MAP are seen. Comparing to the stronger baseline of c\_pd\_dp, the corresponding reductions are 
1.03\%$\ddagger$, 0.03\%, and 0.28\%.\\

\subsection{Feature Weighting and the Extended MNB}

The second set of experiments compares feature weighting and the Extended MNB model that adds document length modeling and scaling 
of the label prior. Feature weighting in text retrieval has been proposed in the form of query weighting \citep{Smucker:06, Momtazi:10}, whereas in text 
classification both training and test set documents are weighted similarly \citep{Rennie:03}. Document length modeling and prior scaling are applicable for the 
text classification experiments, where test documents can be assumed to have the same length distributions as training documents, and the prior probabilities 
of labels have varying degrees of usefulness for prediction.\\ 

The experiments first compared the usefulness of Poisson length modeling (po) and prior scaling (ps). Parameters for both were allowed to vary within 
the permissible range from 0.0 to 2.0. Prior scaling was noticed to improve classification considerably on most datasets. Poisson length modeling gave no 
significant improvement on average, and no additional gain was observed in combination with prior scaling. The following experiments on text classification 
used prior scaling, but not length modeling.\\

Feature weighting was attempted with both query idf weighting (qidf) and TF-IDF training and test document weighting (ti). Parameterized versions used
IDF lifting (tiX, qidfX) in the range -1.0 to 50.0, length scaling (tXi) in the range -1.0 to 2.0, or both (tXiX). A limited selection of the best-performing smoothing
models were chosen for these experiments, with uniform or uniform-smoothed collection distributions for background models, and prior scaling for the text 
classification datasets.\\

\begin{figure*}
\centering
 \includegraphics[scale=1.0, trim=80 110 80 110, clip=true]{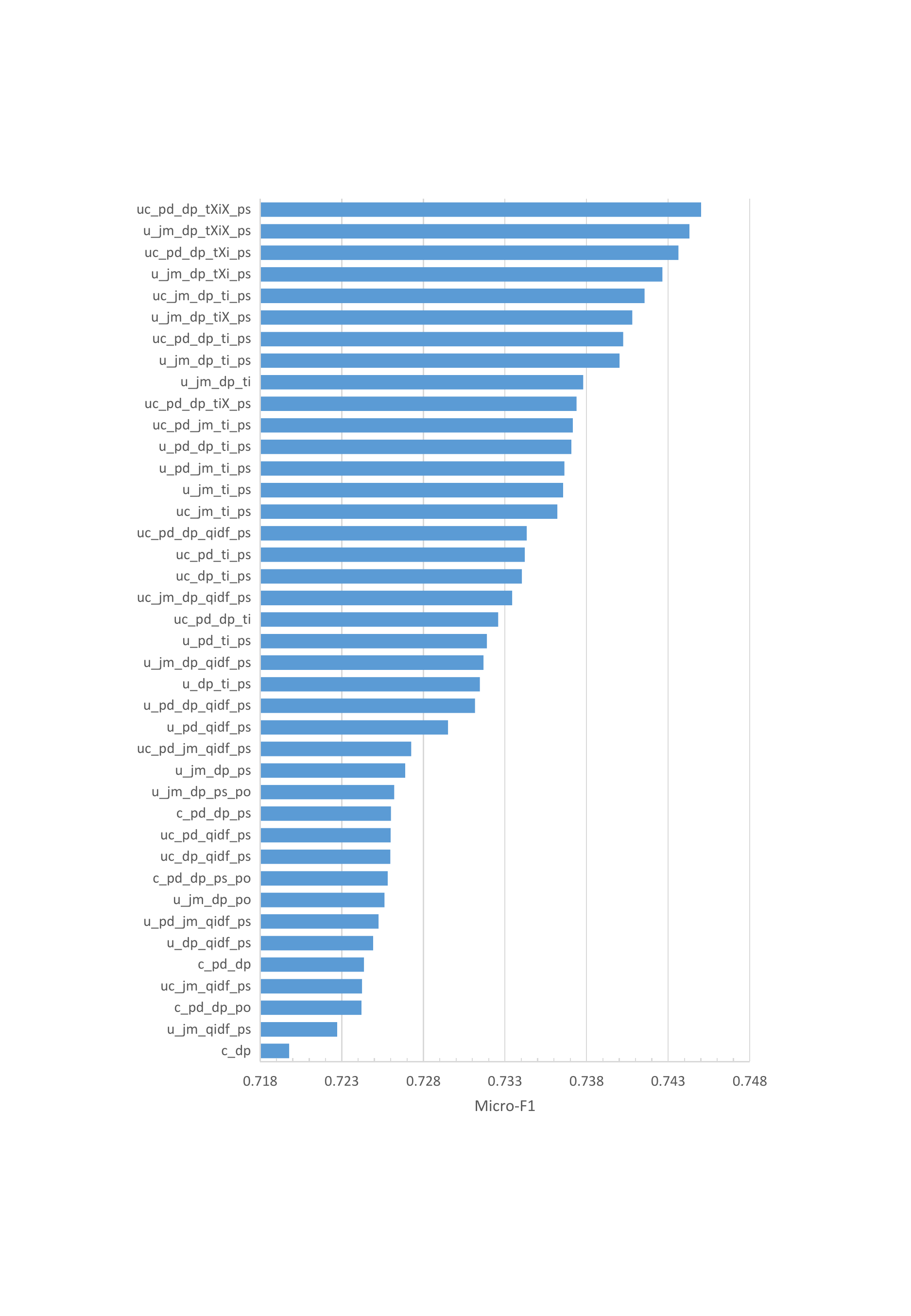}
\caption{Mean Micro-F1 across the text classification datasets with the Extended MNB models. Baseline models c\_dp and c\_pd\_dp included for comparison}
\label{extend_fig1}
\end{figure*}

\begin{figure*}
\centering
 \includegraphics[scale=1.0, trim=80 150 80 150, clip=true]{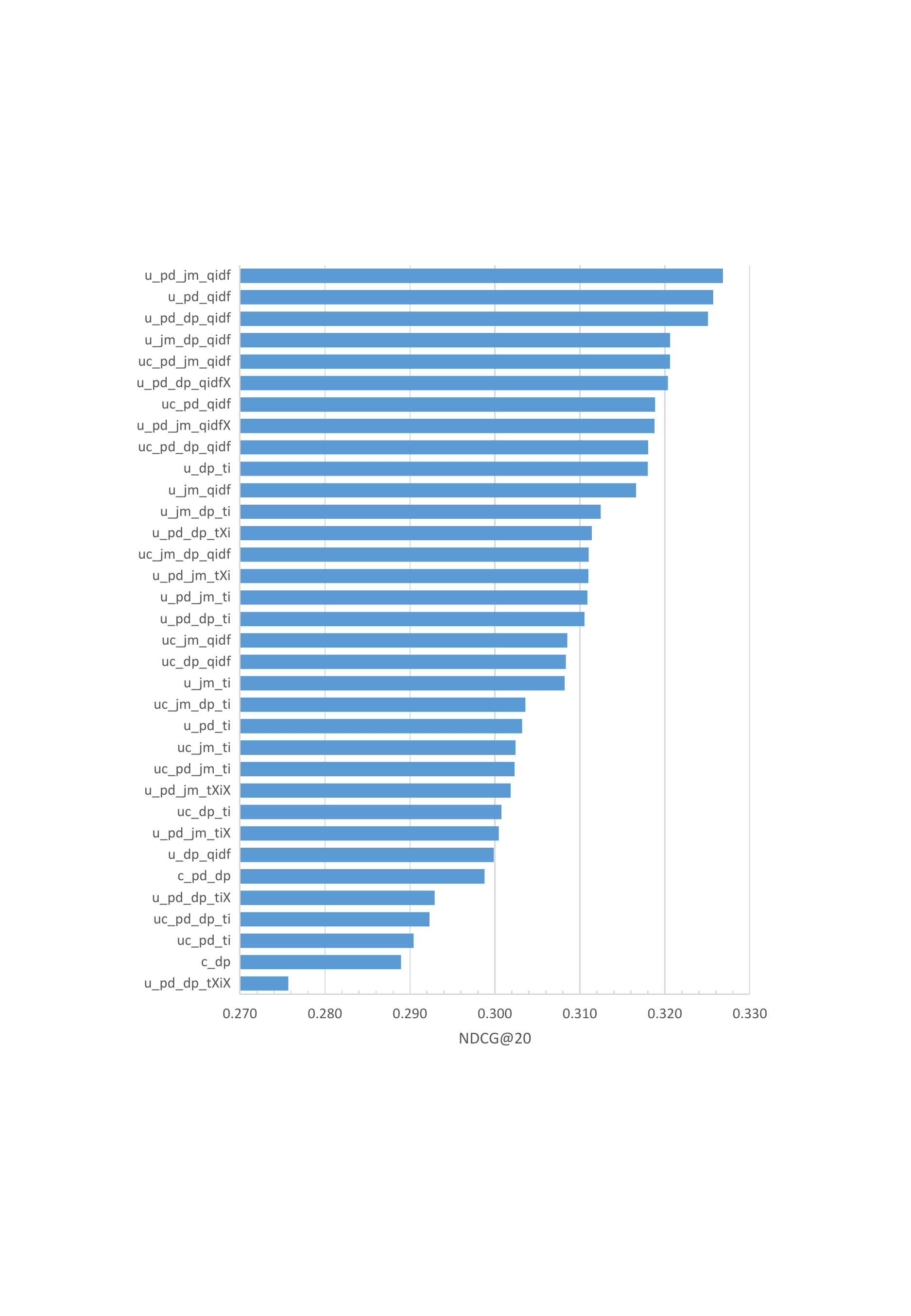}
\caption{Mean NDCG@20 across the text retrieval datasets with the Extended MNB models. Baseline models c\_dp and c\_pd\_dp included for comparison}
\label{extend_fig2}
\end{figure*}

\begin{figure*}
\centering
 \includegraphics[scale=1.0, trim=80 150 80 150, clip=true]{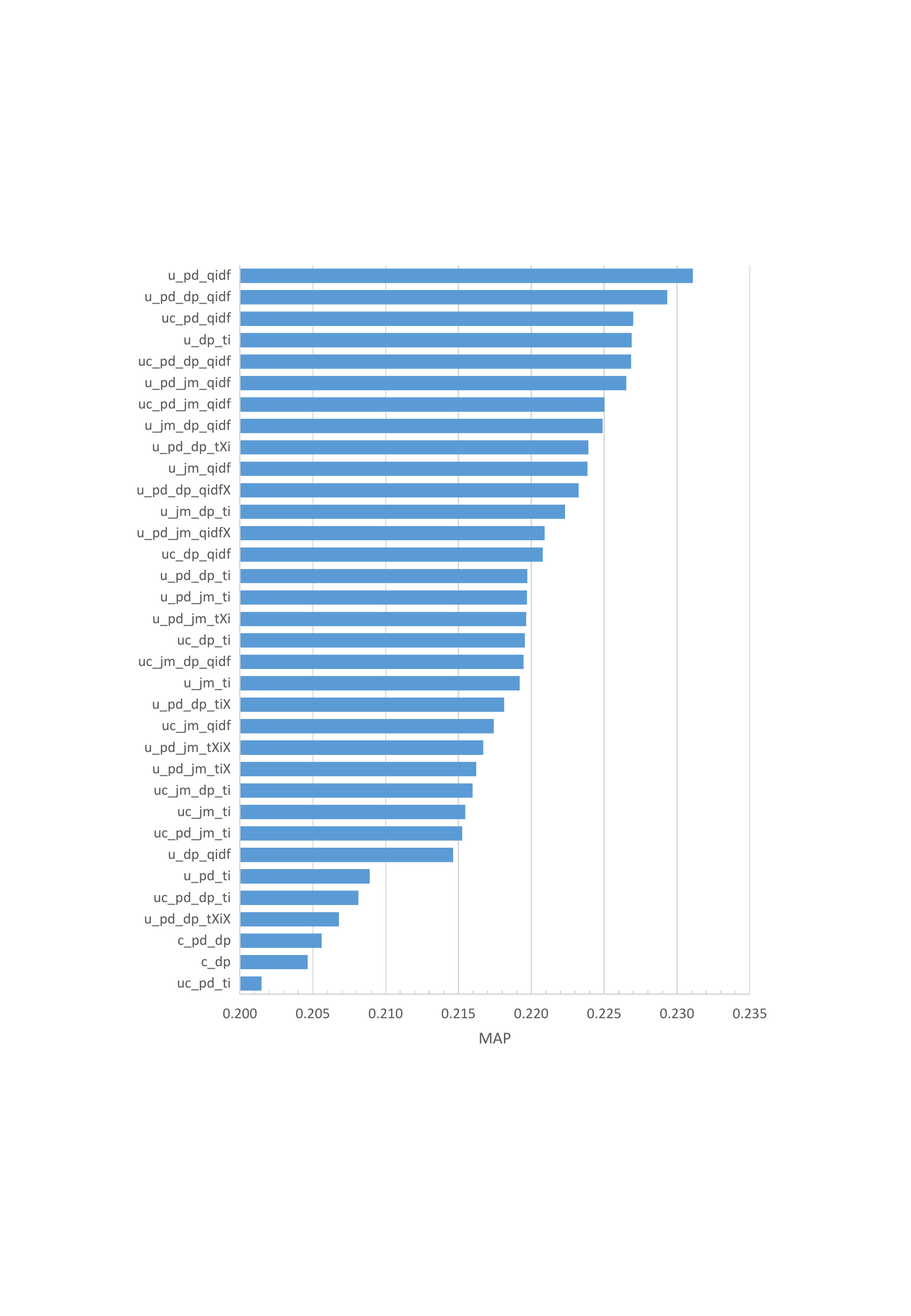}
\caption{Mean MAP across the text retrieval datasets with the Extended MNB models. Baseline models c\_dp and c\_pd\_dp included for comparison}
\label{extend_fig3}
\end{figure*}

Figures \ref{extend_fig1}, \ref{extend_fig2} and \ref{extend_fig3} show the results for the second set of experiments, averaged across the datasets.
For reference, the baselines c\_dp and c\_pd\_dp are included in the figures. Compared to the smoothing method variants, the improvements from
feature weighting and prior scaling are substantial. Averaged over the datasets, uc\_pd\_dp\_tXiX\_ps produces a relative error reduction of 7.50\%$\dagger$
in Micro-F1 over c\_pd\_dp in text classification, and u\_pd\_qidf produces relative improvements over c\_pd\_dp of 8.26\%$\ddagger$ in NDCG@20 and 
11.03\%$\ddagger$ in MAP.\\

\subsection{Tied Document Mixture}

The third set of experiments explored the Tied Document Mixture (TDM) model proposed in Chapter 5 for text classification. Smoothing the TDM kernel 
densities was done with Jelinek-Mercer (kjm), power-law discounting (kpd), and Dirichlet prior (kdp). Due to longer processing times on the largest datasets, as 
well as a much larger number of possible combinations for smoothing, a small number of combinations successful on single-label datasets were chosen for
a full set of experiments. For simplifying the comparisons, the feature weighting and prior scaling combinations were also excluded from the TDM experiments,
and left for future experimentation.\\

\begin{figure*}
\centering
 \includegraphics[scale=1.0, trim=285 175 285 175, clip=true]{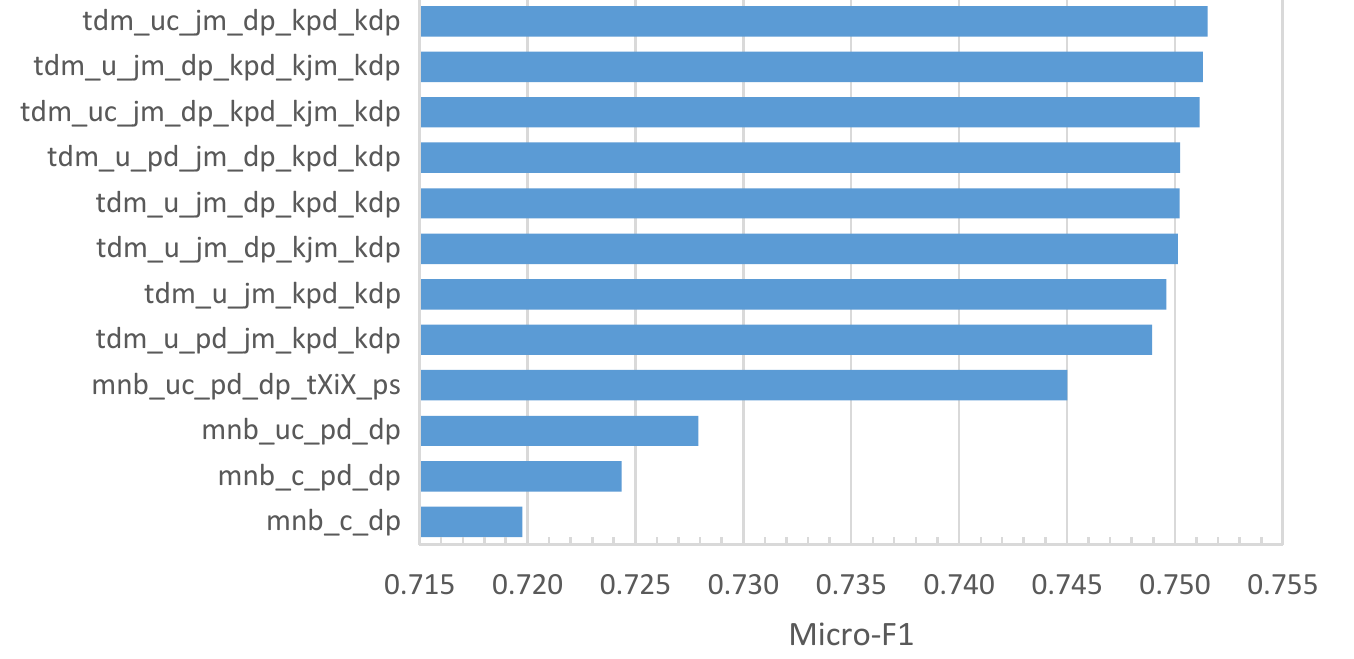}
\caption{Mean Micro-F1 across the text classification datasets with different TDM models. MNB baseline models included for comparison}
\label{structure_fig1}
\end{figure*}

Figure \ref{structure_fig1} shows the Micro-F1 results averaged across the text classification datasets. For comparison, MNB baselines from the previous sets of experiments have been included, and results for the two models are separated by the ``tdm'' and ``mnb'' affixes. The best performing model 
tdm\_uc\_jm\_dp\_kpd\_kdp produces a relative Micro-F1 improvement of 2.55\% over mnb\_uc\_pd\_dp\_tXiX\_ps and 8.67\%$\dagger$ 
over mnb\_uc\_pd\_dp.\\

\subsection{Comparison with Strong Linear Model Baselines}
The fourth set of experiments compared strong linear model baselines to the results from the first three sets of experiments. For ad-hoc text retrieval
a strong baseline is the BM25 model (bm25), while results from the earlier VSMs can be included for comparison. For text classification tasks the strong baselines 
are LR (lr) and l2-SVM (l2svm) models, combined with the parameterized TF-IDF feature weighting used with the Extended MNB models.\\

\begin{figure*}
\centering
 \includegraphics[scale=1.0, trim=285 175 285 175, clip=true]{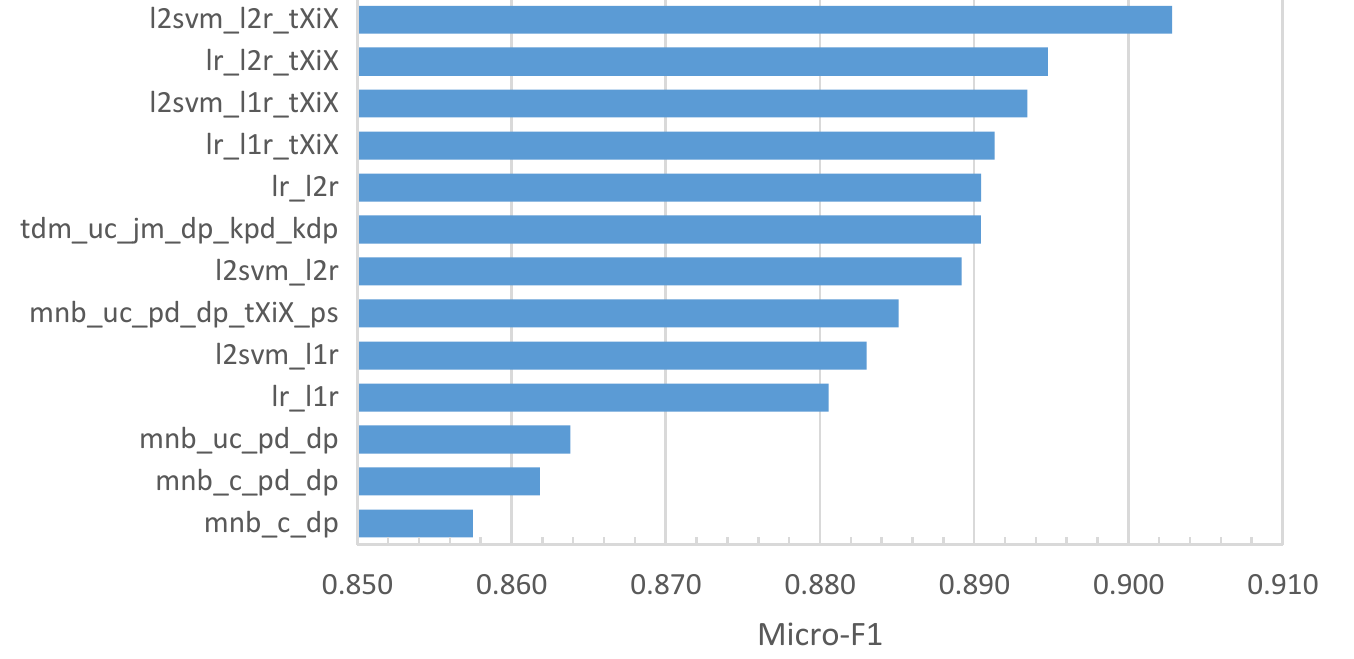}
\caption{Mean Micro-F1 across the binary-label and multi-class text classification datasets for the baseline LR and SVM models, compared to MNB and TDM models}
\label{comparison_fig1}
\end{figure*}

Figure \ref{comparison_fig1} summarizes the Micro-F1 results for the LR and SVM baselines compared to MNB and TDM, averaged across the binary-label 
and multi-class datasets. Overall, the feature weighted SVM and LR models seem to outperform the generative models, l2svm\_l2r\_tXiX in particular showing
exceptionally high performance. The difference of l2svm\_l2r\_tXiX is 11.32\% to tdm\_uc\_jm\_dp\_kpd\_kdp, 15.43\%$\dagger$ to 
mnb\_uc\_pd\_dp\_tXiX\_ps, and 28.66\%$\dagger$ to mnb\_uc\_pd\_dp. Despite the difference in mean scores, none of the LR and SVM 
models significantly improve over TDM, when tested across datasets.\\

\begin{figure*}
\centering
 \includegraphics[scale=1.0, trim=285 145 285 145, clip=true]{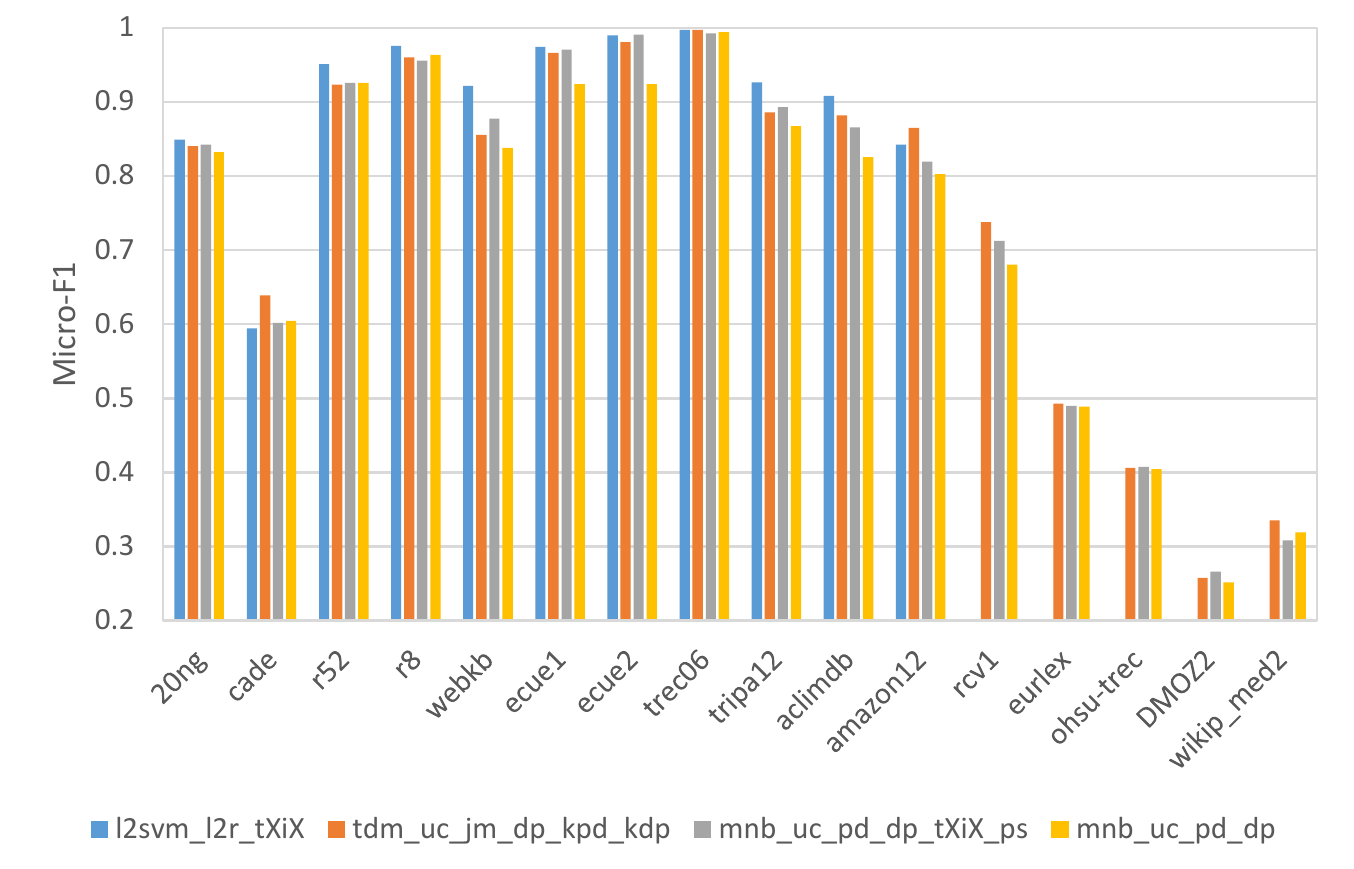}
\caption{Micro-F1 on each text classification dataset. TDM outperforms the L2-regularized SVM with parameterized TF-IDF on cade and amazon12}
\label{comparison_fig1b}
\end{figure*}

The differences within each dataset can be further examined. Figure \ref{comparison_fig1b} shows the differences within each dataset. On both
cade and amazon12, tdm\_uc\_jm\_dp\_kpd\_kdp outperforms l2svm\_l2r\_tXiX, while mnb\_uc\_pd\_dp\_tXiX\_ps never outperforms l2svm\_l2r\_tXiX.
The most likely reason for this is that both feature weighted MNB and SVM models are linear models, and the parameter estimation for learned linear
models provides more accurate classifiers in linearly separable classification problems. Datasets such as cade and amazon12 are possibly more non-linear,
and TDM outperforms in these cases. With large-scale multi-label datasets SVM and LR are not directly usable in reasonable processing time, and on rcv1 
and wikip\_med2 TDM improves on MNB by a small margin.\\

\begin{figure*}
\centering
 \includegraphics[scale=1.0, trim=285 200 285 200, clip=true]{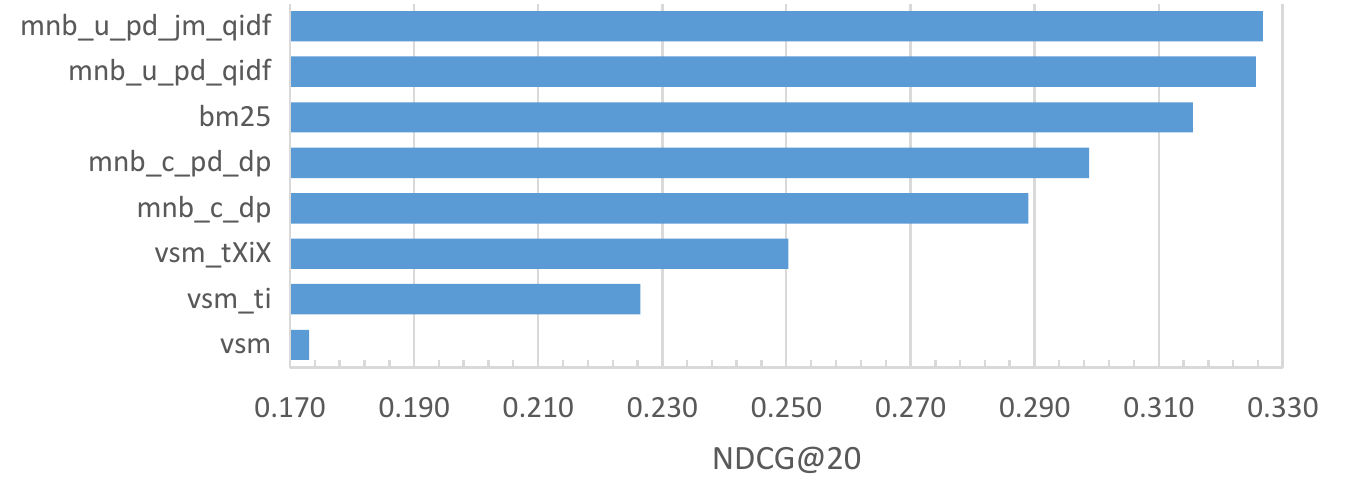}
\caption{Mean NDCG@20 across the text retrieval datasets with the baseline VSM and BM25 models, compared to MNB models}
\label{comparison_fig2}
\end{figure*}

\begin{figure*}
\centering
 \includegraphics[scale=1.0, trim=285 200 285 200, clip=true]{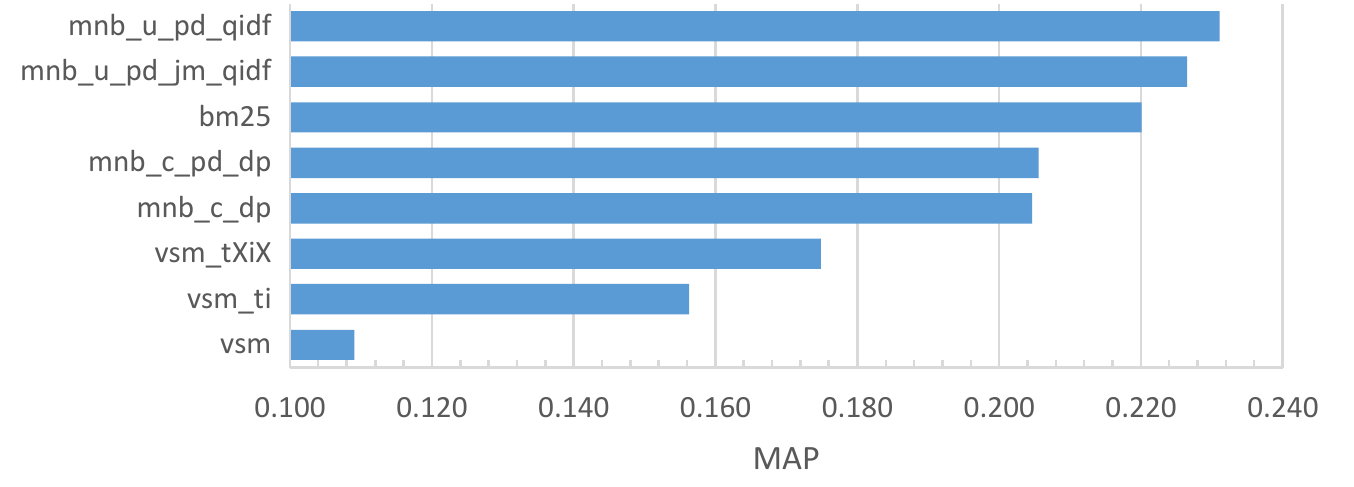}
\caption{Mean MAP across the text retrieval datasets with the baseline VSM and BM25 models, compared to MNB models}
\label{comparison_fig3}
\end{figure*}

\begin{figure*}
\centering
 \includegraphics[scale=1.0, trim=285 145 285 145, clip=true]{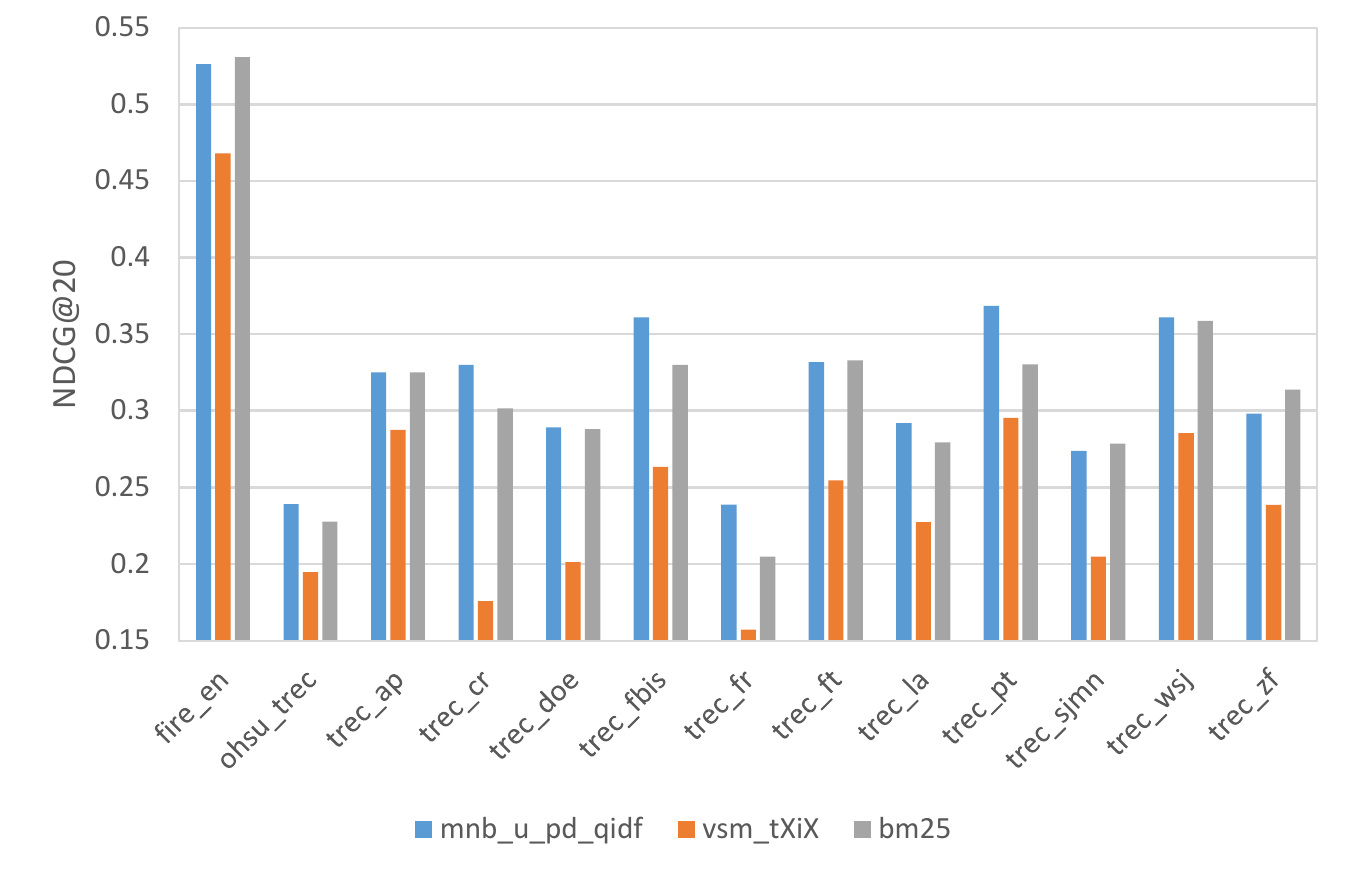}
\caption{NDCG@20 on each text retrieval dataset, comparing MNB with VSM and BM25}
\label{comparison_fig2b}
\end{figure*}

\begin{figure*}
\centering
 \includegraphics[scale=1.0, trim=285 145 285 145, clip=true]{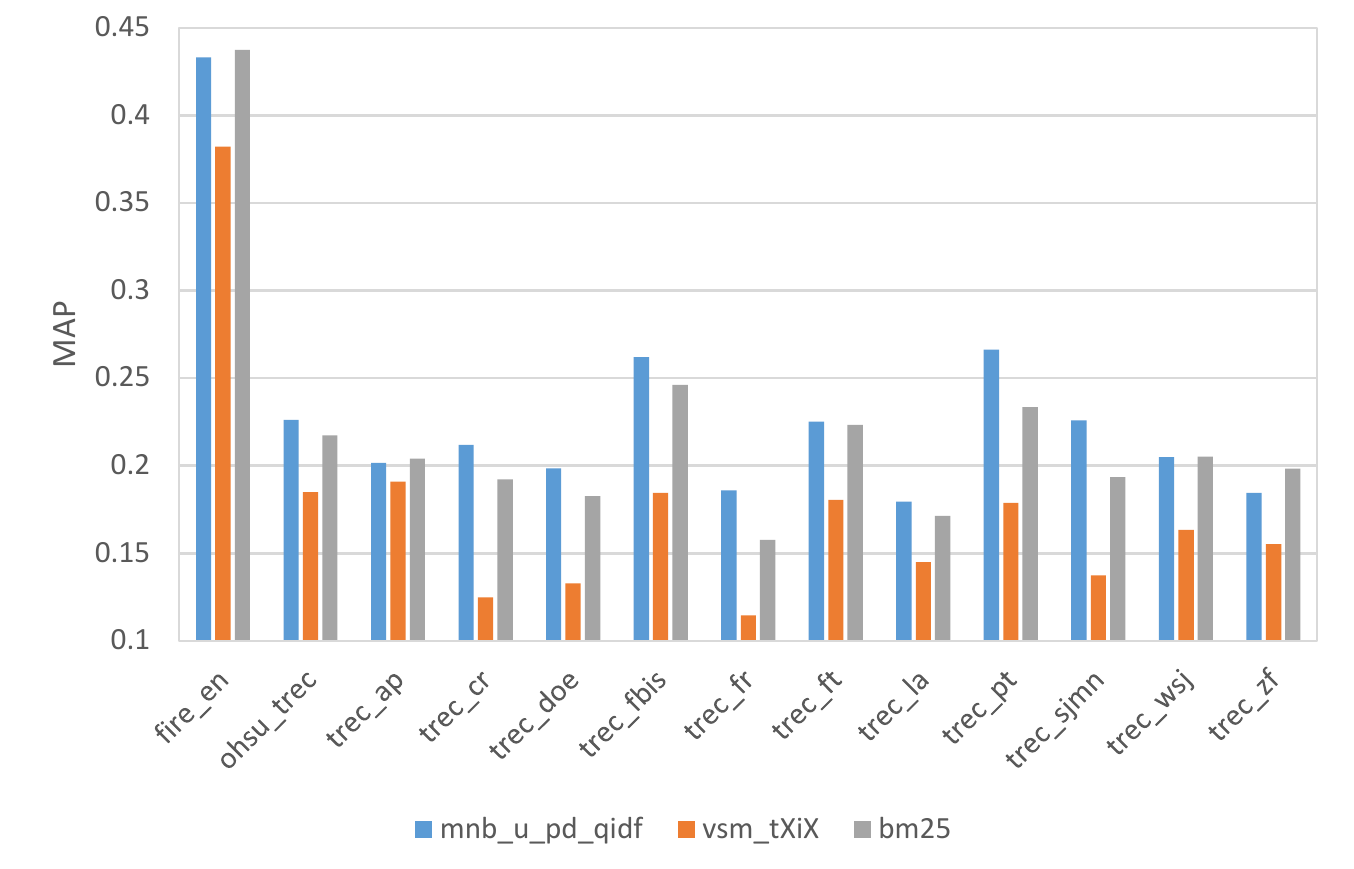}
\caption{MAP on each text retrieval dataset, comparing MNB with VSM and BM25}
\label{comparison_fig3b}
\end{figure*}

Figures \ref{comparison_fig2} and \ref{comparison_fig3} show comparisons of the MNB models to VSM and BM25 models in mean NDCG@20 and MAP
across the datasets. The results are nearly identical under both measures. VSM models fall behind MNB models, while BM25 outperforms basic MNB models,
but not the MNB models using query weighting. The improvement of mnb\_u\_pd\_qidf over vsm\_tXiX is 23.42\%$\dagger$ in NDCG@20 and 24.33\%$\dagger$ in
MAP. The improvement over bm25 is 3.12\%$\ddagger$ in NDCG@20 and 4.76\%$\ddagger$ in MAP.\\

The NDCG@20 and MAP differences within each retrieval dataset are illustrated in Figures \ref{comparison_fig2b} and \ref{comparison_fig3b}. 
While mnb\_u\_pd\_qidf and bm25 have similar performance, on several datasets mnb\_u\_pd\_qidf outperforms bm25 by a margin on both measures, resulting
in the significant differences across the datasets.\\

\subsection{Scalability and Efficiency}

The fifth set of experiments explored the scalability of linear models for classification in estimation and inference. The large Wikipedia dataset from the LSHTC4
competition was pruned to scale the number documents, features, and labelsets, each up to a million. The Label Powerset method \citep{Boutell:04} was used to 
map the multi-label learning problems into a multi-class problem directly learnable by the LR and SVM models. For improving scalability, all evaluated models 
used TF-IDF feature weighting with the unsmoothed Croft-Harper IDF, further pruning the words occurring in more than half of the training set documents. Each
model configuration was allowed to run for four hours on a 3.40GHz i7-2600 processor with 16GB of RAM memory, and runs taking longer were terminated.
The learned models lr\_l2r\_tXiX, l2svm\_l2r\_tXiX, l2svm\_l1r\_tXiX, lr\_l1r\_tXiX were evaluated, as well as mnb\_u\_jm\_tXiX and tdm\_u\_jm\_kjm\_tXiX. 
Hash table implementations of the generative models instead of an inverted index were tested to evaluate the significance of the 
sparse posterior inference. These perform MNB inference by updating the conditional probabilities for each label, for each word in the test document. 
This gives the commonly considered ``optimal'' time complexity for MNB \citep{Manning:08}.\\

\begin{figure*}
\centering
\subfigure[L2-regularized L2-Support Vector Machines, labelsets pruned to $1$]{
 \includegraphics[scale=1.0, trim=287 210 280 205, clip=true]{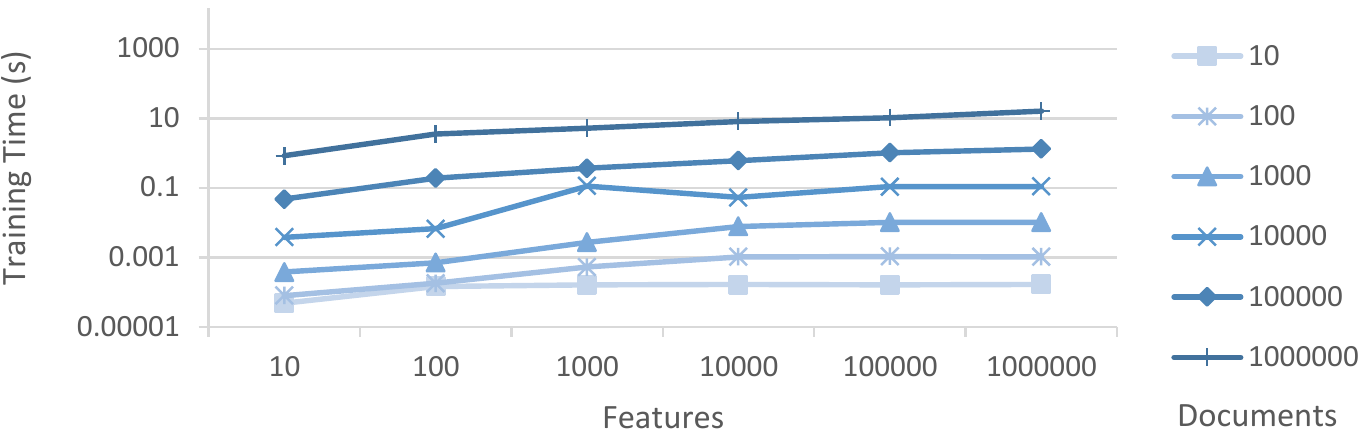}
}
\subfigure[L2-regularized L2-Support Vector Machines, labelsets pruned to $10$]{
 \includegraphics[scale=1.0, trim=287 210 280 205, clip=true]{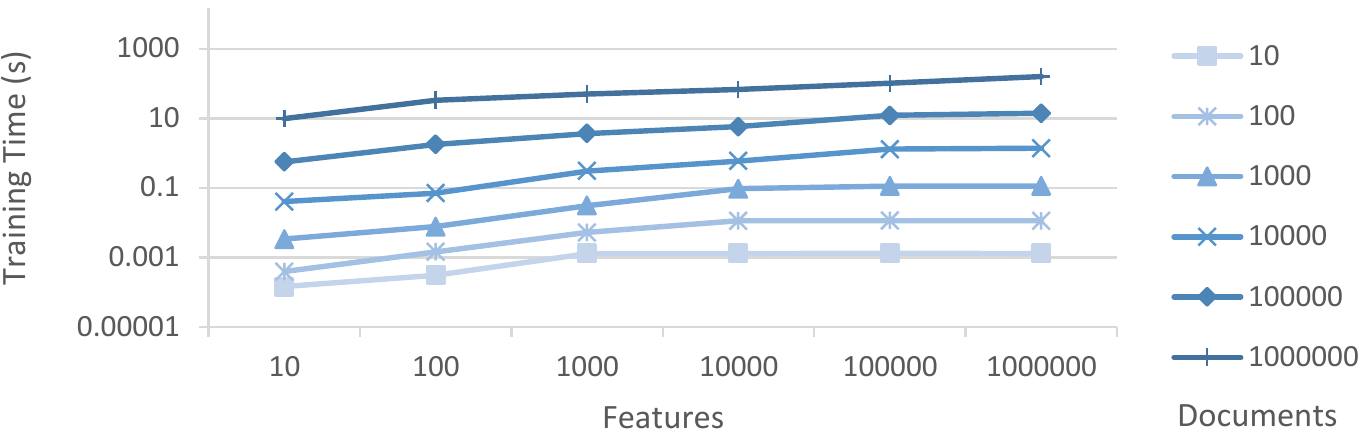}
}
\subfigure[L2-regularized L2-Support Vector Machines, labelsets pruned to $1000$]{
 \includegraphics[scale=1.0, trim=287 210 280 205, clip=true]{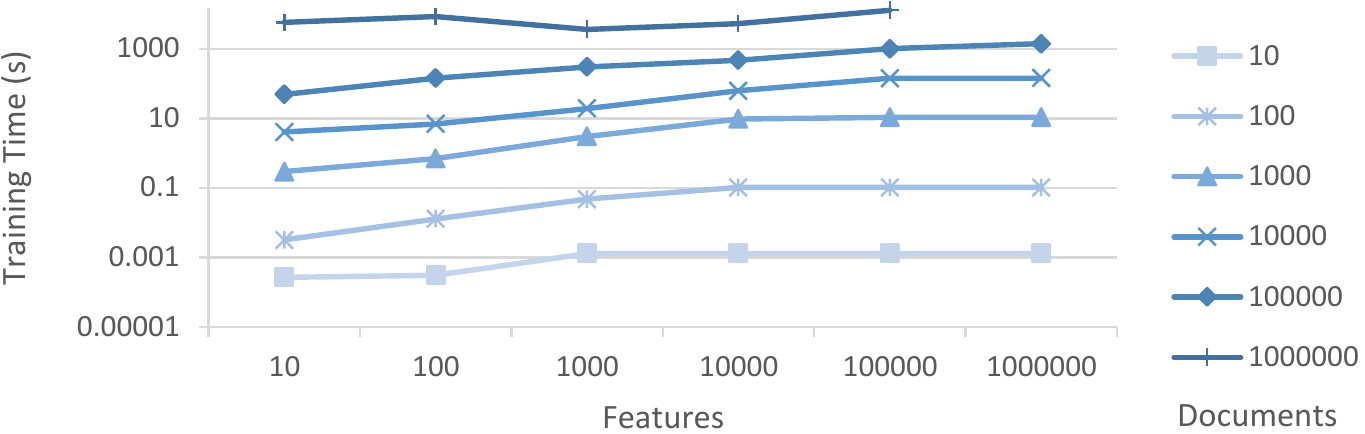}
}
\subfigure[L2-regularized L2-Support Vector Machines, labelsets pruned to $1000000$]{
 \includegraphics[scale=1.0, trim=287 210 280 205, clip=true]{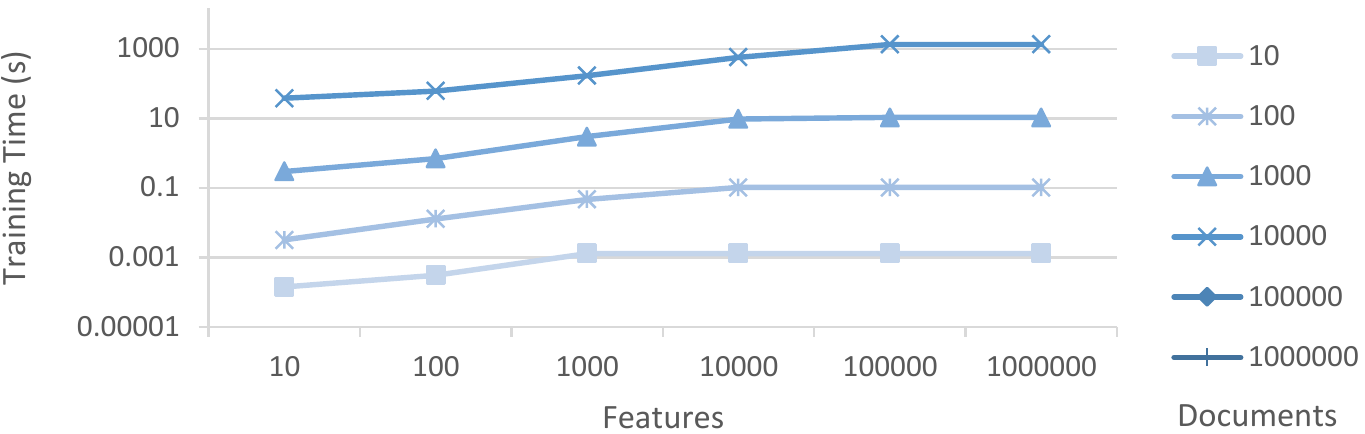}
}
\caption{Estimation times for L2-regularized L2-Support Vector Machines on wikip\_large, with different pruning of documents, features and labelsets}
\label{l2svm_l2r_train}
\end{figure*}

\begin{figure*}
\centering
\subfigure[Multinomial Naive Bayes with an inverted index, labelsets pruned to $1$]{
 \includegraphics[scale=1.0, trim=287 210 280 205, clip=true]{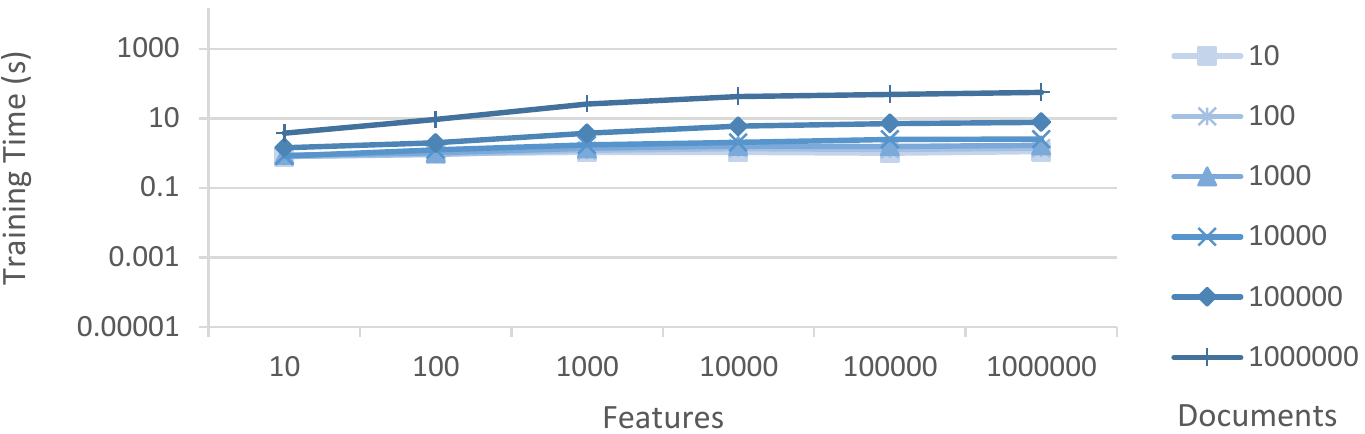}
}
\subfigure[Multinomial Naive Bayes with an inverted index, labelsets pruned to $10$]{
 \includegraphics[scale=1.0, trim=287 210 280 205, clip=true]{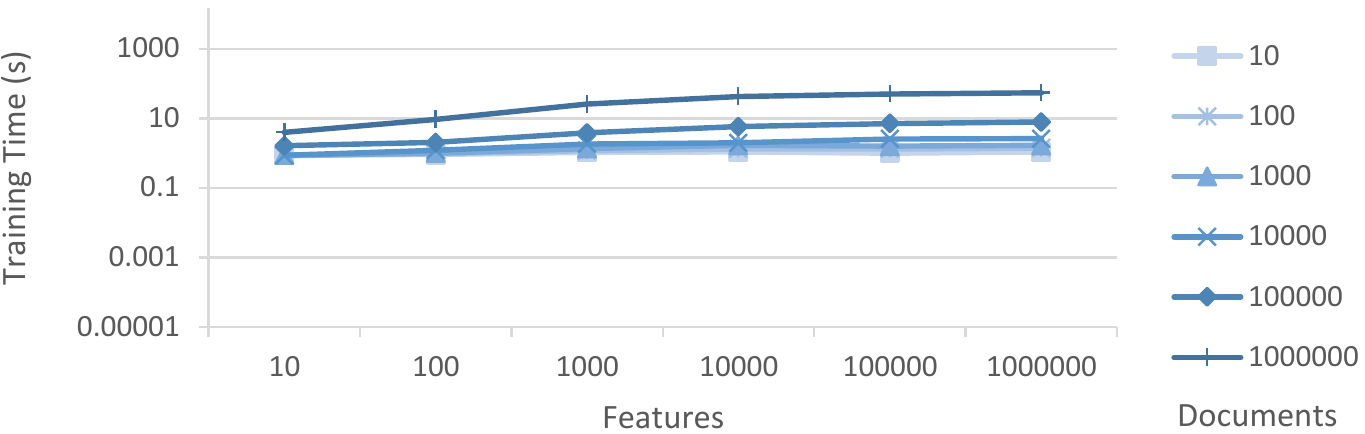}
}
\subfigure[Multinomial Naive Bayes with an inverted index, labelsets pruned to $1000$]{
 \includegraphics[scale=1.0, trim=287 210 280 205, clip=true]{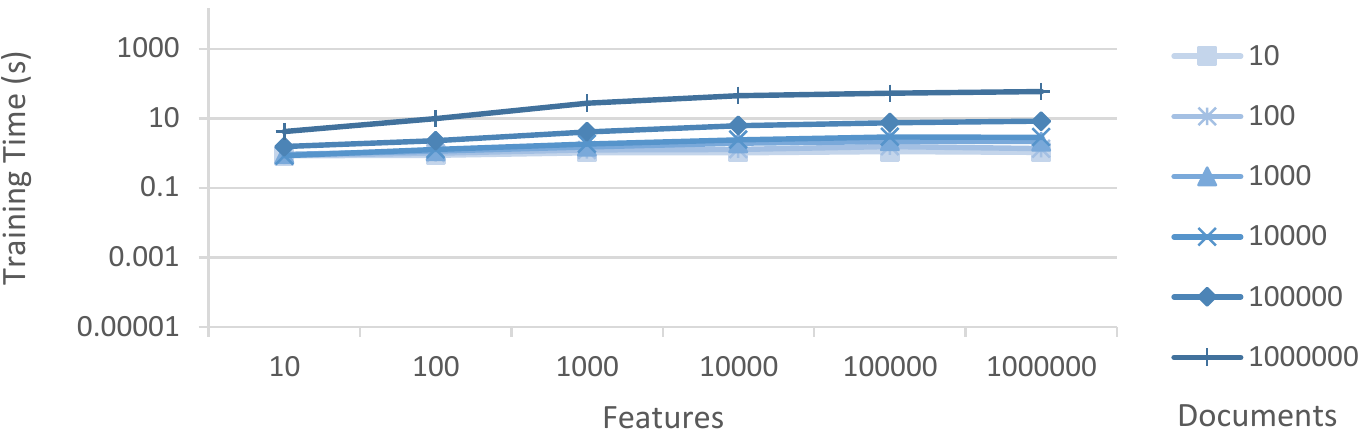}
}
\subfigure[Multinomial Naive Bayes with an inverted index, labelsets pruned to $1000000$]{
 \includegraphics[scale=1.0, trim=287 210 280 205, clip=true]{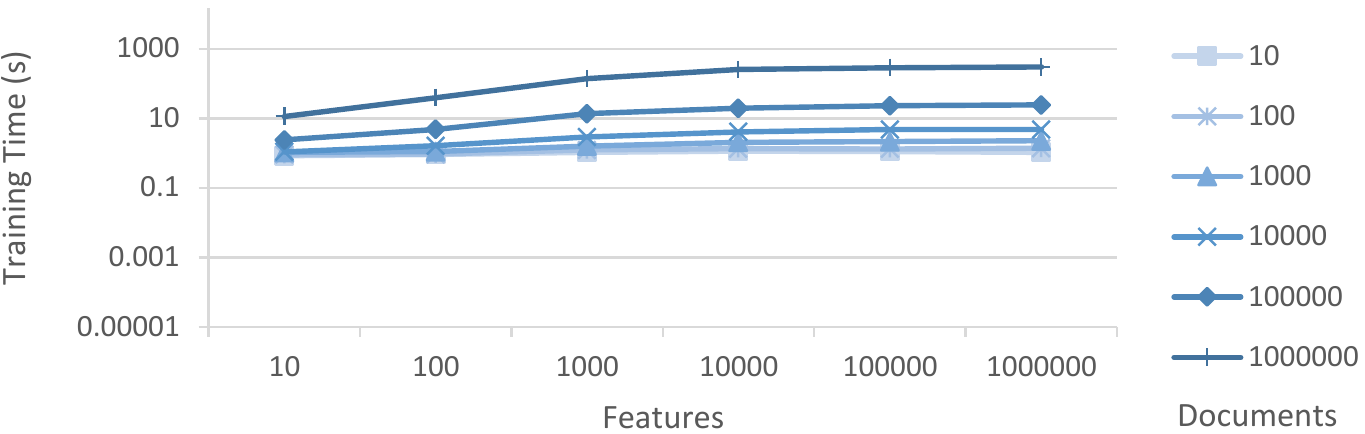}
}
\caption{Estimation times for Multinomial Naive Bayes with an inverted index on wikip\_large, with different pruning of documents, features and labelsets}
\label{mnb_train}
\end{figure*}

Figure \ref{l2svm_l2r_train} shows the estimation times for l2svm\_l2r\_tXiX and Figure \ref{mnb_train} for mnb\_u\_jm\_tXiX. 
Figure \ref{l2svm_l2r_train} for the SVM model shows exponentially scaling estimation times, with the times increasing rapidly with more documents and
labelsets. Figure \ref{mnb_train} for the MNB model shows linear estimation times, unaffected by the number of labels, and only marginally affected by the 
number of features. The SVM model does not complete the task in the allowed time when the allowed documents and labelsets both number over 10000, while
the MNB model completes in all but few of the largest configurations. The estimation times of the other learner linear models behave similarly to 
l2svm\_l2r\_tXiX, while estimation for tdm\_u\_jm\_kjm\_tXiX behaves similarly to mnb\_u\_jm\_tXiX. The constant difference in small numbers of documents 
in favor of SVM is due to a pre-processing difference: the LR and SVM models were implemented in C with LibLinear, with Python feature reading times 
subtracted from the training times, while the generative models were implemented in Java with SGMWeka and the training times include a constant from 
reading the feature files.\\

\begin{figure*}
\centering
\subfigure[Multinomial Naive Bayes with an inverted index]{
 \includegraphics[scale=1.0, trim=287 210 280 200, clip=true]{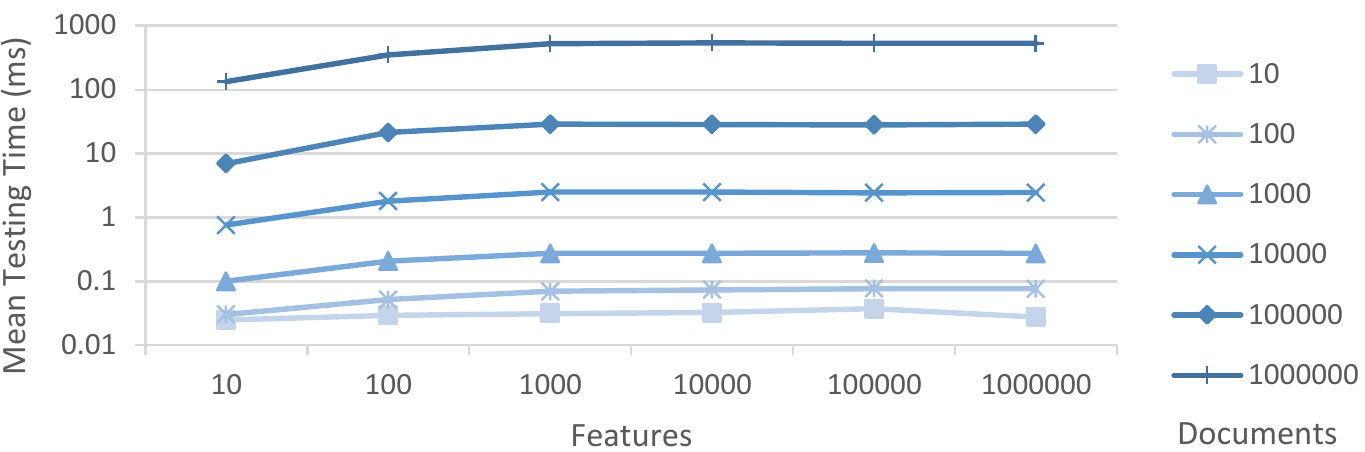}
}
\subfigure[Multinomial Naive Bayes with a hash table]{
 \includegraphics[scale=1.0, trim=287 210 280 200, clip=true]{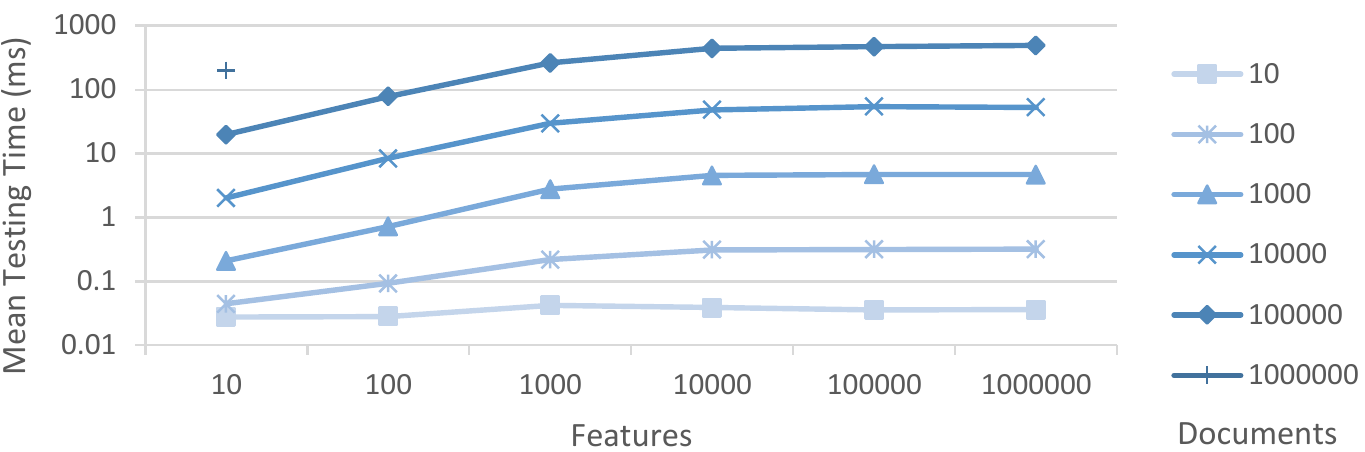}
}
\subfigure[Tied Document Mixture with an inverted index]{
 \includegraphics[scale=1.0, trim=287 210 280 200, clip=true]{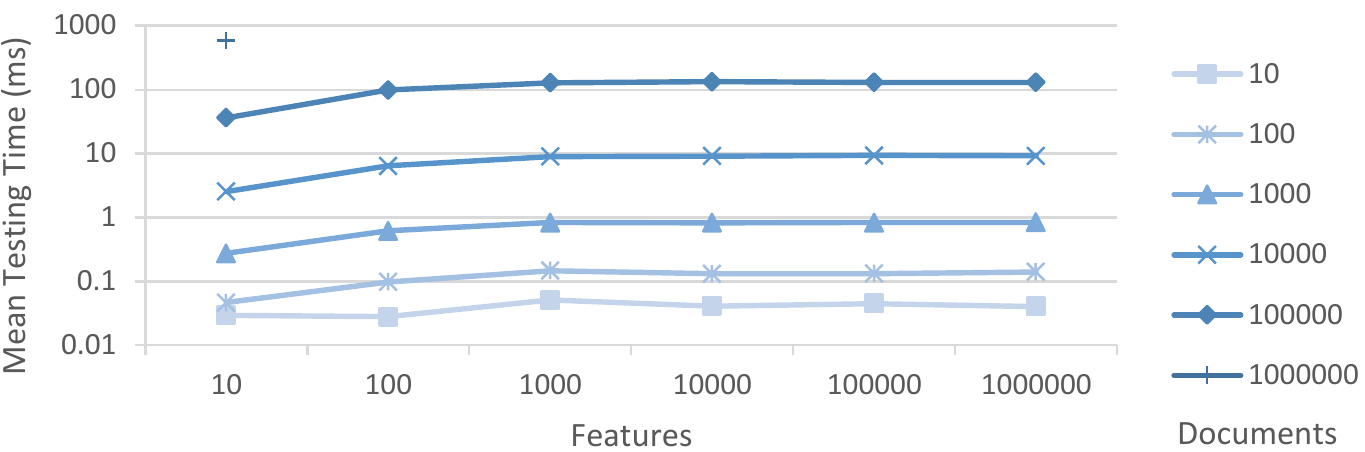}
}
\subfigure[Tied Document Mixture with a hash table]{
 \includegraphics[scale=1.0, trim=287 210 280 200, clip=true]{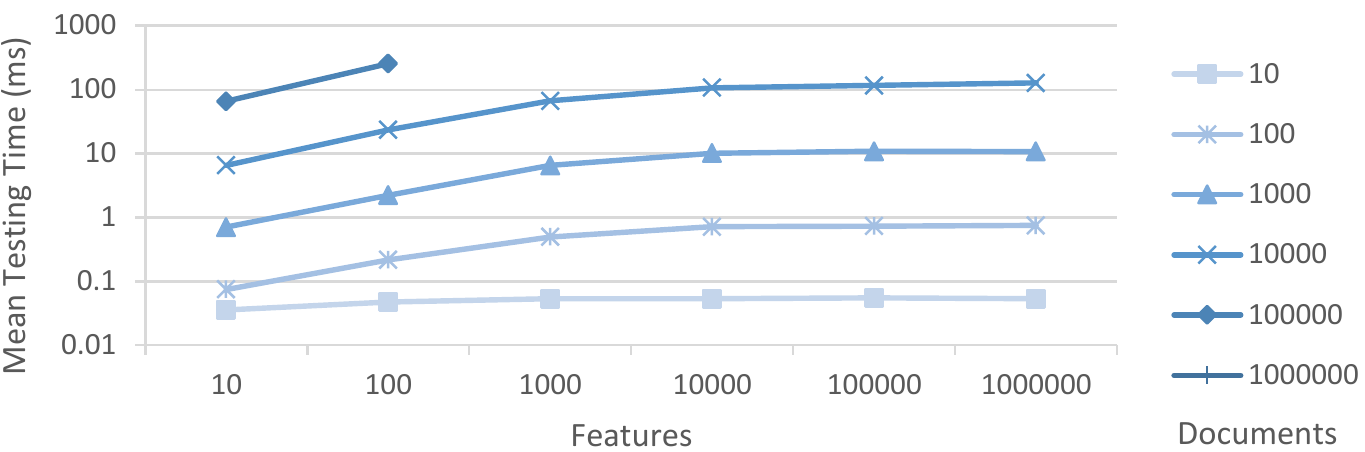}
}
\caption{Mean inference times per document on wikip\_large with 1000000 labelsets. Multinomial Naive Bayes and Tied Document Mixture compared, with inverted index and hash table 
implementations}
\label{index_hash_infer}
\end{figure*}

The inference times for the models depend on both sparsity of the parameters and their representation. 
Application of sparse inference reduces the inference complexities for both linear and non-linear 
models according to sparsity of the parameters. Figure \ref{index_hash_infer} compares the inference times for mnb\_u\_jm\_tXiX and 
tdm\_u\_jm\_kjm\_tXiX using an inverted index and a hash table, with labelsets pruned to 1000000.\\

Figure \ref{index_hash_infer} exhibits several overlapping effects. With a hash table, inference for MNB and TDM has similar complexity, producing nearly
identical scaling. With an inverted index, the inference becomes more scalable for both models. While most of the configurations with over 100000 
training documents do not complete
in time with a hash table, the inverted index implementations scale more easily to 1000000 training documents. For both MNB and TDM, the sparse
inference times scale linearly with the number of training documents, since this also increases the number of seen labelsets closer to $1000000$. For both models,
with high numbers of documents the number of features induces an exponential growth with the hash table, whereas with inverted index the growth becomes less
exponential. Figure \ref{index_hash_infer2} shows this effect in more detail. The highest-dimensional TDM model (10000 features, 10000 documents) to complete 
the task within four hours with the hash table implementation took 106.8 ms per classification. The corresponding inverted index implementation took 
9.1 ms per classification, an order of magnitude reduction in mean classification times. The gap between a common hash table implementation and the 
sparse inference with an inverted index will only increase from this in higher dimensional tasks.\\

\begin{figure*}
\centering
\includegraphics[scale=1.0, trim=285 210 280 200, clip=true]{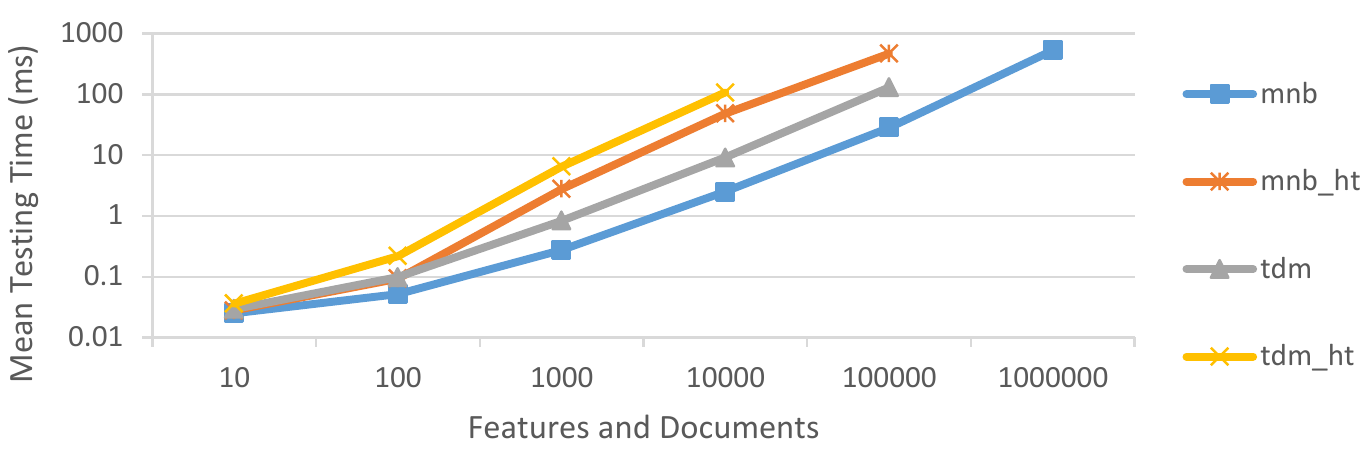}
\caption{Mean inference times per document on wikip\_large with 1000000 labelsets. Inverted index and hash table (\_ht) implementations compared, with both the
number of features and documents increased}
\label{index_hash_infer2}
\end{figure*}

\chapter{Conclusion}
This chapter concludes the thesis with a discussion. First a summary of the thesis results and implications of the findings
are discussed. The thesis statement is revisited, arguing that models extending Multinomial Naive Bayes offer both a versatile
solution in terms of both effectiveness and scalability. Limitations of the thesis and future work are discussed, considering
current developments related to text mining.\\

\section{Summary of Results}
This thesis proposed generative models of text using sparse computation as a general solution for text mining applications.
The problems of fragmentation of research and scalability of models were identified as central problems for text mining.
A solution based on modified generative multinomial models of text combined with a novel type of exact sparse inference
was proposed as a solution for a variety of text mining tasks.\\

Building on overviews of both text mining and generative multinomial models for text, the thesis showed the connection of
the Multinomial Naive Bayes (MNB) models to linear models and directed generative graphical models. Modifications and
extensions to MNB such as smoothing and feature weighting were formalized as constrained graphical models estimated
with the maximum likelihood principle.\\

Inference using inverted indices was shown to reduce the complexity of inference with linear models according to the sparsity 
of the model representation. This sparse inference was shown to be equally applicable to structured extensions
of linear models. A hierarchical extension of MNB called Tied Document Mixture (TDM) was proposed as a basic extension
of MNB with document-level nodes. Empirical evaluation of the TDM and modified MNB models showed that the models offer
highly competitive performance across text classification and ranking tasks, and sparse inference reduced the
inference times by an order of magnitude in the largest considered experiments that completed within the allowed time.\\

\section{Implications of Findings}

Current research in machine learning considers generative Bayes models for classification generally inferior 
to discriminative models. The formalization of model modifications and the experiment results show that the Bayes 
model framework is considerably more flexible and effective than thought. Moreover, these findings directly extend to other 
types of text mining tasks, such as ranked retrieval of documents. Structured generative models such as the proposed TDM 
were shown to further improve modeling effectiveness, leading to new types of scalable generative models for text mining.\\

The generalized smoothing function presents virtually all of the commonly used smoothing functions for multinomial 
and n-gram models of text in a common mixture model framework. The methods differ only in the chosen discounting method, and how 
the smoothing coefficient is chosen. This describes the decades of statistical language modeling research in a concise form, 
as well as simplifies the development and analysis of new smoothing methods. Formalizing all the smoothing methods as 
approximate maximum likelihood estimation on a Hidden Markov Model re-establishes a probabilistic formulation for the functions. 
The formalization of feature transforms and weighted words as inference over probabilistic data similarly re-establishes 
the use of these methods in a probabilistic framework. Feature transforms were shown to greatly improved MNB 
performance for classification and ranking, and has potential implications for other types of generative text models.\\

Scalability limits the range of applications for probabilistic models. The naive inference used here as baseline is widely considered 
to be optimal: \emph{``Because we have to look at the data at least once, NB can be said to have optimal time complexity.''} 
\citep{Manning:08}. The presented sparse inference enables improved scaling of linear models and structured 
extensions to different types of tasks. Unlike parallelization or approximation, sparse inference reduces the total 
required computation non-linearly and provides exact results. The inference can be further combined with parallelization 
and many other efficiency improvements used in information retrieval and machine learning. Especially the application of  
structured sparse models becomes more scalable compared to naive inference.\\

The experimental results show that the commonly used generative models for text classification and ranking are comparatively 
weak baselines, whereas the modified and extended generative models have performance on par with strong task-specific methods. 
Among the possible models, the combination of TF-IDF weighting with uniform Jelinek-Mercer smoothing is one single-parameter 
option that performs well in both types of tasks. With more parameters and optimization for the specific dataset, a number of 
stronger models can be learned. The results show that the models developed in the thesis provide improved solutions for a 
variety of common applications, such as classification of sentiment, spam, e-mails, news, web-pages and Wikipedia articles, and different 
types of ranked retrieval of text documents. The obtained improvements should extend naturally to other tasks that process text with 
generative models, as well as to future text mining applications.\\

\section{Revisiting the Thesis Statement}
The thesis statement was posed in Chapter 1 as:
\emph{Generative models of text combined with inference using inverted indices provide sparse generative models for text mining that are both 
versatile and scalable, providing state-of-the-art effectiveness and high scalability for various text mining tasks.}\\

The theory and experiment results provide strong support to the thesis statement. Modifications and extensions to MNB models 
of text were formalized as well-defined graphical models. The experiments showed high effectiveness 
of the developed models for a variety of text classification and clustering tasks, and the obtained improvements should hold in many current and future
applications of generative text modeling. The idea of Naive Bayes models as ``punching bags of machine learning'' should therefore be reconsidered. 
The theory for sparse inference was developed to produce scalability as a function of model sparsity to linear models and their structural extensions. 
In practice this reduced the processing times of the largest completed experiments by an order of magnitude. Given the theory and results of the 
thesis, the ``curse of dimensionality'' of high-dimensional sparse text data should perhaps be considered a useful property.\\

\section{Limitations of the Thesis}
The experiments in the thesis were restricted to the main applications of text classification and ad-hoc text retrieval, where performance of 
modified MNB models were shown to be competitive with high-performing task-specific solutions. These are by far the most 
successful applications of text mining, but do not cover all of the possible types of current and future text mining tasks. Some applications 
are less suited for the assumptions of generative multinomial models than others. Generative models estimated to maximize likelihood are neither 
guaranteed to give sufficient performance, if the perfomance measure is very different from maximum likelihood. For example, discriminative 
models directly optimizing posterior probabilities can be a more suitable choice, if high precision of posterior probabilities is required.\\

Over the last years a number of research directions have been proposed as widely applicable solutions for machine learning. Deep learning 
combining modern parallel computing hardware with developments in optimization has brought a resurgence of interest in multi-layer neural 
networks \citep{Bengio:09, Poon:11, Mikolov:11, Collobert:11}. Probabilistic programming combined with factor graphs is enabling flexible development 
of complex graphical model architectures \citep{McCallum:09b, Minka:12, Andres:12}. Gaussian processes are developing into a high-performing 
solution to a wide variety of applications \citep{Rasmussen:03, Pietro:13}. Connections to these research directions is outside the scope of the 
thesis. Research in these topics will continue, but none of these frameworks currently form a paradigm for performing a variety of text processing 
mining tasks with high scalability.\\

Scalability has become an increasingly prominent topic in machine learning during the time used for conducting the thesis research. Particularly, 
parallelized and online stream learning algorithms have become popular in the context of the ``Big Data'' and data science movement. Online algorithms 
such as adaptive stochastic gradient descent \citep{Duchi:11, McMahan:13} are used to learn deep neural networks, and parallelization frameworks such as
MapReduce \citep{Dean:08} combined with cloud computing are enabling new types of applications. Both types of improvements are increasingly
popular for text mining. Linear models combining parallelized online learning with approximations \citep{Li:13, McMahan:13, Agarwal:14} have been proposed as
one highly scalable solution. Extensive comparison to these developments is outside the scope of this thesis. As discussed in Chapters 4 and 5, 
both parallelization and stream processing can be trivially combined in estimation and inference with the algorithms presented in this thesis.
Furthermore, unlike the sparse inference developed in this thesis, these developments do not reduce the computational complexity of scaling to 
large numbers of labels and other latent variable nodes.\\

\section{Future Work}
Experiments in text mining tasks other than ranking and classification should prove the models developed in this thesis useful across text mining applications.
Clustering uses the posterior probabilities from generative models in the same way as classification and ranking, and there is no reason to doubt that
the shown performance improvements extend to clustering. Regression with Bayes models \citep{Frank:98} is likely improved substantially, since 
TDM enables both modeling of each continuous variable value using a document-conditional distribution, and hierarchical smoothing of the value-conditional
models.\\

One highly useful application of the models is in n-gram language models that extend the multinomial models with context variables. These
have a variety of uses such as speech recognition, machine translation, text compression, text prediction and optical character recognition. The generalized
smoothing and feature weighting combined with random search for metaparameters could provide superior models to basic n-gram models using
Kneser-Ney smoothing. Sparse inference could be applied to language model decoding, although the most common operation with language models
is the query of individual conditional probabilities of words given the context, not the computation of marginals for the Bayes rule.\\

The use of weighted words from feature transforms was shown to improve classification and ranking performance, but the use of arbitrary weighted words 
for generative models can have other uses. Topic models are commonly estimated from word count or sequence data, but the derivation presented in
Chapter 5 allows the use of fractional counts for these models as well. Alternatively, the posterior probabilities of a topic model or outputs of a 
Non-Negative Matrix Factorization \citep{Paatero:94} could be used as features for a generative model. Non-probabilistic topic models such as Latent
Semantic Analysis \citep{Deerwester:88} have traditionally used TF-IDF to improve topic separation, and the same should work for generative topic 
models.\\

Sparse inference as well as the modifications to generative models were provided for the case of MNB. However, sparse inference is
applicable whenever the parameters of any linear model can be represented with sparse vectors for each label and less sparse vectors for back-off nodes, with 
the most basic case being a single background distribution node. Inference of kernel densities over sparse data is particularly scalable. As shown with structured 
extensions of MNB, many types of graphs for the back-off nodes can be efficiently computed. It would therefore be valuable 
to know which types of linear and non-linear models can be represented in sparse forms, and how much the scalability of inference is improved.\\

Sparse inference algorithms reduce the complexity of inference with large numbers of labels, and more generally with large numbers of 
hidden variables in structured models. The TDM model examined in this thesis was an elementary expansion of a linear model with a mixture over
document-conditional models. Extension of these findings into more structured cases can yield rich models that both have higher performance and
offer the possibility to make different types of inference using the joint probabilities of hidden variables. For example, adding layers of topic 
variables could enable highly scalable topic modeling.\\

Combining the sparse inference with other improvements for scalability was outlined, but empirical experimentation was left for future work. The combination 
with parallelization should provide further linear improvements in processing speeds, but in practice parallelization of tasks across networks introduces many
complications. Combining the algorithm with the other efficiency improvements for text mining mentioned in Chapter 5 can provide more than linear reductions,
but the exact scale of these improvements depend on the data and structure of the model. Tree-based searches can provide considerable reductions, if the 
node-conditional models separate into different clusters \citep{Ram:12}. Search network minimization can produce considerably smaller models, if nodes can
be merged or removed with minimal loss in performance \citep{Aubert:02}. A considerable literature exists in information retrieval to improve the efficiency  
of inverted indices, that can be directly leveraged for sparse inference \citep{Zobel:06}.\\

Extensions of the TDM model were used by the author for the 2014 Kaggle competitions LHSTC4\footnote{http://www.kaggle.com/c/lshtc/} \citep{Puurula:14} 
and WISE\footnote{http://www.kaggle.com/c/wise-2014} \citep{Tsoumakas:14}. Both competitions were large-scale multi-label text classification tasks with 
over a hundred competing teams. The submission to LSHTC4 won the competition, while the submission for WISE placed narrowly second. TDM extended with 
label hierarchy nodes proved to be the most useful model, and an ensemble of sparse generative models proved to be a high-performing solution. Future work 
will apply models such as TDM to other cases where ground-breaking effectiveness and scalability is required.\\

\begin{appendices}
\chapter{Tables of Results}
This appendix contains the tables of results for the experiments summarized in Chapter 6. Tables \ref{smooth1} to \ref{smooth9}
show the results for the MNB smoothing methods. Tables \ref{extend1} to \ref{extend12} show the results for the Extended MNB models.
Table \ref{structure1} shows the results for the TDM model. Tables \ref{comparison1} to \ref{comparison2} shows the results for the 
strong linear model baselines. Tables \ref{train_times1} to \ref{train_times8} show the training times for the scalability experiments,
and tables \ref{test_times1} to \ref{test_times8} show the corresponding testing times. In tables \ref{train_times1} to  \ref{test_times8}
the modifier appendix "ht" indicates a hash table implementation for inference instead of an inverted index.

\clearpage 

\begin{table}
\footnotesize
\centering
\caption{Evaluation set Micro-F1 results of smoothed MNB models on text classification datasets, optimized with 40x40 Gaussian random searches on the development set}
\begin{tabular}{|l|c|c|c|c|c|c|c|c|c|} \hline
Model & \multicolumn{8}{c|}{MNB} \\ \hline
Background & \multicolumn{8}{c|}{c} \\ \hline
Smoothing &  jm & dp & ad & pd & jm\_dp & pd\_dp & pd\_jm & pd\_jm\_dp \\
\hline \hline
20ng & 83.04 & 82.90 & 82.93 & $\bm{83.22}$ & 83.06 & $\bm{83.22}$ & $\bm{83.22}$ & $\bm{83.22}$ \\ \hline
cade & 60.70 & 59.90 & 60.44 & 60.45 & $\bm{60.74}$ & 60.46 & 60.68 & $\bm{60.74}$ \\ \hline
r52 & 91.51 & 90.69 & 89.68 & 93.14 & 91.19 & 93.18 & 93.18 & $\bm{93.26}$ \\ \hline
r8 & 96.25 & 94.83 & 95.43 & 95.75 & 96.29 & 95.93 & $\bm{96.84}$ & $\bm{96.84}$ \\ \hline
webkb & 83.16 & 83.09 & 83.30 & 83.09 & 83.16 & 83.09 & 83.16 & $\bm{83.38}$ \\
\hline
\end{tabular}
\label{smooth1}
\end{table}

\begin{table}
\footnotesize
\centering
\caption{Evaluation set Micro-F1 results of smoothed MNB models on text classification datasets, optimized with 40x40 Gaussian random searches on development sets}
\begin{tabular}{|l|c|c|c|c|c|c|c|c|c|} \hline
Model & \multicolumn{8}{c|}{MNB} \\ \hline
Background & \multicolumn{8}{c|}{c} \\ \hline
Smoothing &  jm & dp & ad & pd & jm\_dp & pd\_dp & pd\_jm & pd\_jm\_dp \\
\hline \hline
20ng & 82.08 & 82.66 & 82.50 & 82.46 & $\bm{82.67}$ & $\bm{82.67}$ & 82.05 & 82.66 \\ \hline
cade & $\bm{59.90}$ & 59.76 & 59.83 & 59.83 & 59.74 & 59.84 & 59.83 & 59.83 \\ \hline
r52 & $\bm{91.90}$ & 91.08 & 91.54 & 91.78 & 91.54 & 91.78 & 91.86 & $\bm{91.90}$ \\ \hline
r8 & $\bm{96.57}$ & 96.43 & 96.43 & 96.02 & 96.39 & 96.39 & 96.07 & 96.48 \\ \hline
webkb & 84.02 & 84.02 & 83.73 & 83.73 & $\bm{84.09}$ & 84.02 & 83.81 & $\bm{84.09}$ \\ \hline
ecue1 & 90.80 & 92.40 & 91.80 & $\bm{92.60}$ & 92.40 & 92.40 & 92.40 & 92.40 \\ \hline
ecue2 & $\bm{92.40}$ & $\bm{92.40}$ & $\bm{92.40}$ & $\bm{92.40}$ & $\bm{92.40}$ & $\bm{92.40}$ & $\bm{92.40}$ & $\bm{92.40}$ \\ \hline
trec06 & 99.28 & $\bm{99.46}$ & 99.42 & 99.42 & $\bm{99.46}$ & $\bm{99.46}$ & 99.38 & 99.28 \\ \hline
tripa12 & 86.76 & 86.79 & 86.78 & 86.78 & 86.78 & $\bm{86.80}$ & 86.78 & 86.79 \\ \hline
aclimdb & 82.50 & 82.56 & $\bm{82.60}$ & 82.56 & 82.53 & 82.56 & 82.56 & 82.56 \\ \hline
amazon12 & $\bm{80.28}$ & 80.21 & 80.21 & 80.24 & $\bm{80.28}$ & 80.27 & 80.26 & 80.27 \\ \hline
rcv1 & 64.99 & 67.95 & 64.61 & 64.61 & 67.86 & $\bm{67.96}$ & 65.03 & 67.95 \\ \hline
eurlex & 48.04 & 45.59 & 45.41 & 45.39 & 48.03 & 45.39 & 47.99 & $\bm{48.06}$ \\ \hline
ohsu-trec & 40.19 & 39.85 & 39.54 & 40.06 & 40.19 & 40.06 & $\bm{40.35}$ & 40.04 \\ \hline
DMOZ2 & 22.28 & 24.62 & 21.97 & 22.54 & 24.64 & $\bm{24.80}$ & 22.54 & 24.05 \\ \hline
wikip\_med2 & 30.35 & 29.14 & 29.73 & 30.89 & 30.99 & 30.89 & 30.91 & $\bm{31.90}$ \\ \hline \hline
mean & 72.02 & 72.18 & 71.78 & 71.96 & 72.50 & 72.35 & 72.14 & $\bm{72.54}$ \\ \hline
\end{tabular}
\label{smooth2}
\end{table} \clearpage 

\begin{table}
\footnotesize
\centering
\caption{Evaluation set Micro-F1 results of smoothed MNB models on text classification datasets, optimized with 40x40 Gaussian random searches on development sets}
\begin{tabular}{|l|c|c|c|c|c|c|c|c|c|} \hline
Model & \multicolumn{8}{c|}{MNB} \\ \hline
Background & \multicolumn{8}{c|}{uc} \\ \hline
Smoothing &  jm & dp & ad & pd & jm\_dp & pd\_dp & pd\_jm & pd\_jm\_dp \\
\hline \hline
20ng & 83.06 & 83.03 & 82.79 & 83.22 & $\bm{83.39}$ & 83.22 & 83.22 & 83.22 \\ \hline
cade & $\bm{60.77}$ & 59.71 & 60.45 & 60.43 & 60.70 & 60.46 & 60.70 & 60.74 \\ \hline
r52 & 91.66 & 91.31 & 91.47 & 92.60 & $\bm{92.79}$ & 92.56 & 92.56 & 92.67 \\ \hline
r8 & $\bm{96.57}$ & 96.43 & 96.39 & 96.29 & 96.48 & 96.34 & $\bm{96.57}$ & 96.16 \\ \hline
webkb & 83.66 & $\bm{84.02}$ & 83.45 & 83.73 & 83.88 & 83.81 & 83.95 & 83.95 \\ \hline
ecue1 & 91.10 & 92.40 & 91.90 & $\bm{92.60}$ & 91.60 & 92.40 & 92.40 & 92.40 \\ \hline
ecue2 & 92.20 & $\bm{92.40}$ & 92.20 & 92.20 & $\bm{92.40}$ & $\bm{92.40}$ & 92.20 & $\bm{92.40}$ \\ \hline
trec06 & 99.24 & $\bm{99.42}$ & 99.38 & 99.38 & 99.24 & $\bm{99.42}$ & 99.17 & 99.20 \\ \hline
tripa12 & 86.75 & 86.79 & $\bm{86.80}$ & 86.74 & 86.77 & 86.73 & 86.74 & $\bm{86.80}$ \\ \hline
aclimdb & 82.43 & 82.53 & $\bm{82.60}$ & 82.56 & 82.53 & 82.56 & 82.56 & 82.53 \\ \hline
amazon12 & $\bm{80.27}$ & 80.22 & 80.20 & 80.25 & $\bm{80.27}$ & 80.25 & 80.25 & $\bm{80.27}$ \\ \hline
rcv1 & 65.10 & $\bm{68.28}$ & 64.61 & 64.60 & 68.03 & 68.07 & 65.00 & 68.12 \\ \hline
eurlex & 48.93 & 48.80 & 48.88 & 49.12 & 49.08 & 48.89 & $\bm{49.27}$ & 49.26 \\ \hline
ohsu-trec & $\bm{40.60}$ & 40.34 & 40.10 & 40.42 & 40.59 & 40.42 & 40.19 & 40.44 \\ \hline
DMOZ2 & 22.40 & 23.93 & 22.05 & 22.50 & 24.52 & $\bm{25.16}$ & 22.56 & 25.02 \\ \hline
wikip\_med2 & 30.35 & 29.18 & 30.43 & 30.98 & 31.17 & $\bm{31.91}$ & 30.97 & 30.93 \\ \hline \hline
mean & 72.19 & 72.42 & 72.10 & 72.35 & 72.71 & $\bm{72.79}$ & 72.39 & 72.76 \\ \hline
\end{tabular}
\label{smooth3}
\end{table} 

\begin{table}
\footnotesize
\centering
\caption{Evaluation set NDCG@20 results of smoothed MNB models on text retrieval datasets, optimized with 50x8 Gaussian random searches on development sets}
\begin{tabular}{|l|c|c|c|c|c|c|c|c|c|} \hline
Model & \multicolumn{8}{c|}{MNB} \\ \hline
Background & \multicolumn{8}{c|}{c} \\ \hline
Smoothing & jm & dp & ad & pd & jm\_dp & pd\_dp & pd\_jm & pd\_jm\_dp \\
\hline \hline
fire\_en & 49.06 & 50.43 & 50.47 & 50.11 & 50.27 & 50.71 & 50.10 & $\bm{50.90}$ \\ \hline
ohsu\_trec & 23.18 & 22.98 & $\bm{23.75}$ & 23.57 & 23.03 & 22.79 & 23.26 & 22.95 \\ \hline
trec\_ap & 30.38 & 29.86 & 29.79 & 30.79 & 30.67 & $\bm{31.48}$ & 30.90 & 30.97 \\ \hline
trec\_cr & 26.80 & 26.52 & 26.73 & 26.02 & 27.61 & 26.56 & 25.93 & $\bm{27.76}$ \\ \hline
trec\_doe & 26.84 & 28.66 & 28.24 & 29.16 & 28.95 & 28.84 & 28.88 & $\bm{30.38}$ \\ \hline
trec\_fbis & 31.90 & $\bm{33.96}$ & 31.64 & 33.17 & 31.63 & 33.17 & 32.05 & 32.95 \\ \hline
trec\_fr & 19.68 & 22.43 & 21.68 & 23.16 & 22.88 & 23.41 & 23.09 & $\bm{24.25}$ \\ \hline
trec\_ft & 27.74 & 33.18 & 31.87 & 31.62 & 32.87 & 32.86 & 31.42 & $\bm{33.97}$ \\ \hline
trec\_la & 25.99 & 25.45 & $\bm{27.74}$ & 26.67 & 25.83 & 26.67 & 26.61 & 25.64 \\ \hline
trec\_pt & 15.65 & 14.24 & 18.24 & $\bm{23.23}$ & 17.30 & $\bm{23.23}$ & 17.33 & 17.30 \\ \hline
trec\_sjmn & 27.66 & $\bm{29.53}$ & 26.81 & 25.97 & 27.66 & 26.14 & 26.65 & 25.98 \\ \hline
trec\_wsj & 32.31 & 33.94 & 33.08 & 33.62 & 34.18 & 34.88 & 33.59 & $\bm{35.15}$ \\ \hline
trec\_zf & 26.70 & 24.42 & 24.12 & 27.64 & 26.09 & 27.64 & 27.64 & $\bm{28.84}$ \\ \hline \hline
mean & 27.99 & 28.89 & 28.78 & 29.59 & 29.15 & $\bm{29.88}$ & 29.04 & 29.77 \\ \hline
\end{tabular}
\label{smooth4}
\end{table} \clearpage 

\begin{table}
\footnotesize
\centering
\caption{Evaluation set MAP results of smoothed MNB models on text retrieval datasets, optimized with 50x8 Gaussian random searches on development sets}
\begin{tabular}{|l|c|c|c|c|c|c|c|c|c|} \hline
Model & \multicolumn{8}{c|}{MNB} \\ \hline
Background & \multicolumn{8}{c|}{c} \\ \hline
Smoothing & jm & dp & ad & pd & jm\_dp & pd\_dp & pd\_jm & pd\_jm\_dp \\
\hline \hline
fire\_en & 40.72 & 41.95 & 41.45 & 41.55 & $\bm{42.05}$ & 41.55 & 41.55 & 41.54 \\ \hline
ohsu\_trec & 20.95 & 20.40 & 21.54 & $\bm{21.74}$ & 21.41 & 21.60 & $\bm{21.74}$ & 20.41 \\ \hline
trec\_ap & 19.19 & 18.40 & 18.36 & 19.22 & 19.24 & 19.25 & 19.15 & $\bm{19.26}$ \\ \hline
trec\_cr & $\bm{17.39}$ & 16.07 & 17.17 & 17.30 & 16.49 & 17.30 & 17.30 & 17.32 \\ \hline
trec\_doe & 17.24 & 20.19 & 19.37 & 20.33 & 20.05 & $\bm{20.57}$ & 17.29 & 17.21 \\ \hline
trec\_fbis & 22.47 & 25.93 & 22.92 & 23.09 & $\bm{26.08}$ & 25.86 & 22.72 & 23.58 \\ \hline
trec\_fr & 14.06 & 17.17 & 12.26 & 16.59 & 17.17 & $\bm{18.05}$ & 16.54 & 17.09 \\ \hline
trec\_ft & 19.95 & 22.78 & 20.96 & 21.78 & $\bm{22.88}$ & 22.59 & 21.73 & 21.89 \\ \hline
trec\_la & 16.41 & 16.13 & 16.49 & 16.37 & 16.63 & 16.51 & 16.37 & $\bm{16.99}$ \\ \hline
trec\_pt & 11.06 & 11.96 & $\bm{18.44}$ & 11.00 & 11.07 & 11.00 & 11.06 & 11.08 \\ \hline
trec\_sjmn & 14.72 & 20.68 & $\bm{23.16}$ & 16.73 & 20.86 & 16.73 & 14.48 & 14.41 \\ \hline
trec\_wsj & 18.55 & 19.33 & 18.27 & 19.06 & 19.58 & 19.52 & 19.07 & $\bm{19.72}$ \\ \hline
trec\_zf & 15.42 & 15.00 & 14.93 & 16.69 & 16.87 & 16.69 & 16.70 & $\bm{17.75}$ \\ \hline \hline
mean & 19.09 & 20.46 & 20.41 & 20.11 & $\bm{20.80}$ & 20.56 & 19.67 & 19.87 \\ \hline
\end{tabular}
\label{smooth5}
\end{table}

\begin{table}
\footnotesize
\centering
\caption{Evaluation set NDCG@20 results of smoothed MNB models on text retrieval datasets, optimized with 50x8 Gaussian random searches on development sets}
\begin{tabular}{|l|c|c|c|c|c|c|c|c|c|} \hline
Model & \multicolumn{8}{c|}{MNB} \\ \hline
Background & \multicolumn{8}{c|}{u} \\ \hline
Smoothing & jm & dp & ad & pd & jm\_dp & pd\_dp & pd\_jm & pd\_jm\_dp \\
\hline \hline
fire\_en & 47.84 & 46.55 & 48.64 & 48.48 & 49.17 & $\bm{49.25}$ & 48.54 & 48.61 \\ \hline
ohsu\_trec & 20.88 & 21.01 & $\bm{22.25}$ & 21.10 & 21.15 & 21.37 & 21.13 & 21.18 \\ \hline
trec\_ap & 25.23 & 24.67 & 24.72 & $\bm{26.56}$ & 25.26 & $\bm{26.56}$ & 26.07 & 26.53 \\ \hline
trec\_cr & 26.54 & 24.62 & 27.54 & $\bm{27.77}$ & 26.56 & 27.27 & 25.91 & 25.91 \\ \hline
trec\_doe & 22.62 & 22.22 & 23.22 & $\bm{23.40}$ & 22.62 & 23.14 & 22.62 & 22.81 \\ \hline
trec\_fbis & 20.34 & 24.02 & $\bm{25.54}$ & 24.98 & 21.74 & 24.99 & 23.56 & 24.48 \\ \hline
trec\_fr & 12.97 & 12.68 & $\bm{14.18}$ & 12.63 & 13.17 & 12.63 & 12.63 & 12.62 \\ \hline
trec\_ft & 26.66 & 24.42 & 27.32 & 28.05 & 27.31 & 27.33 & $\bm{28.09}$ & 28.04 \\ \hline
trec\_la & 23.45 & 20.82 & 22.51 & 24.56 & 24.71 & 24.62 & $\bm{24.94}$ & 24.64 \\ \hline
trec\_pt & 25.50 & 22.72 & $\bm{26.40}$ & 24.66 & 23.01 & 24.66 & 24.66 & 25.62 \\ \hline
trec\_sjmn & 26.78 & 26.80 & $\bm{28.74}$ & 28.52 & 27.80 & 28.52 & 28.45 & 28.48 \\ \hline
trec\_wsj & 27.42 & 22.75 & 26.83 & $\bm{28.43}$ & 28.26 & $\bm{28.43}$ & 28.08 & 27.61 \\ \hline
trec\_zf & $\bm{14.52}$ & 10.36 & 11.20 & 11.67 & 14.38 & 12.65 & 14.02 & 12.55 \\ \hline \hline
mean & 24.67 & 23.36 & 25.31 & 25.45 & 25.01 & $\bm{25.49}$ & 25.28 & 25.31 \\ \hline
\end{tabular}
\label{smooth6}
\end{table} \clearpage 

\begin{table}
\footnotesize
\centering
\caption{Evaluation set MAP results of smoothed MNB models on text retrieval datasets, optimized with 50x8 Gaussian random searches on development sets}
\begin{tabular}{|l|c|c|c|c|c|c|c|c|c|} \hline
Model & \multicolumn{8}{c|}{MNB} \\ \hline
Background & \multicolumn{8}{c|}{u} \\ \hline
Smoothing & jm & dp & ad & pd & jm\_dp & pd\_dp & pd\_jm & pd\_jm\_dp \\
\hline \hline
fire\_en & 36.71 & 37.29 & 37.97 & 38.06 & $\bm{38.60}$ & 38.03 & 38.02 & 38.05 \\ \hline
ohsu\_trec & 17.13 & 17.26 & $\bm{18.47}$ & 17.77 & 17.44 & 17.85 & 17.79 & 17.85 \\ \hline
trec\_ap & 14.70 & 13.11 & 13.78 & 14.87 & 14.70 & 14.88 & 14.91 & $\bm{14.92}$ \\ \hline
trec\_cr & 15.25 & 14.53 & 16.36 & 16.97 & 15.25 & $\bm{16.99}$ & 15.62 & 15.34 \\ \hline
trec\_doe & 14.72 & 13.27 & 14.62 & $\bm{15.59}$ & 14.72 & $\bm{15.59}$ & 14.75 & 14.76 \\ \hline
trec\_fbis & 11.87 & $\bm{16.75}$ & 16.37 & 16.51 & 13.47 & 16.45 & 16.49 & 15.04 \\ \hline
trec\_fr & 9.01 & 8.90 & $\bm{10.12}$ & 8.60 & 9.55 & 8.62 & 8.60 & 9.88 \\ \hline
trec\_ft & 16.52 & 14.18 & 16.09 & 16.84 & 16.80 & 16.84 & 16.89 & $\bm{16.95}$ \\ \hline
trec\_la & 13.46 & 11.90 & 12.82 & 14.16 & 14.01 & 14.17 & 14.20 & $\bm{14.22}$ \\ \hline
trec\_pt & 8.84 & 15.49 & $\bm{15.93}$ & 13.90 & 15.51 & 15.71 & 13.90 & 14.07 \\ \hline
trec\_sjmn & 20.37 & 20.88 & 20.95 & $\bm{21.20}$ & 20.37 & $\bm{21.20}$ & $\bm{21.20}$ & $\bm{21.20}$ \\ \hline
trec\_wsj & 14.08 & 11.76 & 13.40 & 14.63 & $\bm{14.82}$ & 14.63 & 14.69 & 14.57 \\ \hline
trec\_zf & $\bm{9.53}$ & 6.77 & 7.45 & 7.31 & 8.74 & 7.30 & 7.81 & 7.30 \\ \hline \hline
mean & 15.55 & 15.55 & 16.49 & 16.65 & 16.46 & $\bm{16.79}$ & 16.53 & 16.47 \\ \hline
\end{tabular}
\label{smooth7}
\end{table} 

\begin{table}
\footnotesize
\centering
\caption{Evaluation set NDCG@20 results of smoothed MNB models on text retrieval datasets, optimized with 50x8 Gaussian random searches on development sets}
\begin{tabular}{|l|c|c|c|c|c|c|c|c|c|} \hline
Model & \multicolumn{8}{c|}{MNB} \\ \hline
Background & \multicolumn{8}{c|}{uc} \\ \hline
Smoothing & jm & dp & ad & pd & jm\_dp & pd\_dp & pd\_jm & pd\_jm\_dp \\
\hline \hline
fire\_en & 49.14 & 50.83 & 50.68 & 50.91 & 51.80 & $\bm{51.89}$ & 50.96 & 51.64 \\ \hline
ohsu\_trec & 23.08 & 23.32 & $\bm{23.71}$ & 23.48 & 23.13 & 23.09 & 23.17 & 23.33 \\ \hline
trec\_ap & 30.42 & 29.70 & 29.75 & 30.92 & $\bm{31.02}$ & 30.79 & 30.87 & 30.86 \\ \hline
trec\_cr & 26.80 & 26.52 & 26.71 & 26.04 & 26.45 & $\bm{28.22}$ & 25.88 & 26.39 \\ \hline
trec\_doe & 26.84 & 28.84 & 28.42 & 29.16 & 27.73 & $\bm{29.45}$ & 28.88 & 28.64 \\ \hline
trec\_fbis & 31.93 & $\bm{33.77}$ & 31.67 & 33.17 & 31.90 & 33.17 & 32.05 & 32.71 \\ \hline
trec\_fr & 20.26 & 22.35 & 21.64 & $\bm{23.15}$ & 22.43 & 22.47 & 23.10 & 22.94 \\ \hline
trec\_ft & 26.31 & 32.89 & 31.86 & 31.29 & 33.01 & $\bm{33.71}$ & 30.28 & 32.99 \\ \hline
trec\_la & 25.94 & 25.37 & $\bm{27.74}$ & 26.62 & 25.83 & 26.41 & 26.30 & 25.70 \\ \hline
trec\_pt & 15.65 & 19.06 & $\bm{26.54}$ & 22.19 & 22.34 & 15.47 & 22.31 & 22.22 \\ \hline
trec\_sjmn & 27.67 & $\bm{29.51}$ & 27.55 & 26.93 & 22.68 & 26.83 & 27.42 & 28.13 \\ \hline
trec\_wsj & 32.31 & 34.10 & 33.07 & 33.53 & 34.33 & $\bm{35.27}$ & 33.62 & 34.38 \\ \hline
trec\_zf & 26.41 & 26.19 & 24.03 & 29.07 & $\bm{29.39}$ & 28.80 & 28.27 & 28.76 \\ \hline \hline
mean & 27.90 & 29.42 & 29.49 & 29.73 & 29.39 & 29.66 & 29.47 & $\bm{29.90}$ \\ \hline
\end{tabular}
\label{smooth8}
\end{table} \clearpage 

\begin{table}
\footnotesize
\centering
\caption{Evaluation set MAP results of smoothed MNB models on text retrieval datasets, optimized with 50x8 Gaussian random searches on development sets}
\begin{tabular}{|l|c|c|c|c|c|c|c|c|c|} \hline
Model & \multicolumn{8}{c|}{MNB} \\ \hline
Background & \multicolumn{8}{c|}{uc} \\ \hline
Smoothing & jm & dp & ad & pd & jm\_dp & pd\_dp & pd\_jm & pd\_jm\_dp \\
\hline \hline
fire\_en & 40.81 & 41.93 & 41.41 & 41.89 & $\bm{42.25}$ & 41.84 & 41.84 & 42.06 \\ \hline
ohsu\_trec & 20.91 & 21.52 & 21.54 & $\bm{21.74}$ & 20.48 & 21.15 & 21.61 & 20.44 \\ \hline
trec\_ap & 19.18 & 18.98 & 18.23 & 18.97 & 19.16 & $\bm{19.22}$ & 19.06 & 19.03 \\ \hline
trec\_cr & $\bm{17.34}$ & 16.80 & 17.17 & 17.30 & 16.56 & 17.30 & 17.12 & 17.30 \\ \hline
trec\_doe & 17.23 & 20.46 & 19.38 & 20.33 & 20.37 & $\bm{20.58}$ & 17.87 & 20.45 \\ \hline
trec\_fbis & 22.45 & 25.61 & 22.91 & 23.10 & 24.01 & 25.98 & 22.58 & $\bm{26.17}$ \\ \hline
trec\_fr & 13.98 & 17.62 & 12.26 & 16.65 & 17.16 & 17.34 & 16.69 & $\bm{17.63}$ \\ \hline
trec\_ft & 19.54 & 22.79 & 20.95 & 21.78 & 22.43 & 22.74 & 21.16 & $\bm{22.85}$ \\ \hline
trec\_la & 16.38 & 16.13 & 16.50 & 16.36 & 16.66 & 16.18 & 16.38 & $\bm{16.82}$ \\ \hline
trec\_pt & 11.12 & 15.11 & $\bm{15.72}$ & 13.25 & 11.08 & 15.69 & 12.64 & 15.70 \\ \hline
trec\_sjmn & 14.69 & 20.76 & 19.07 & 21.12 & 14.69 & $\bm{21.74}$ & 16.60 & 15.72 \\ \hline
trec\_wsj & 18.61 & 19.30 & 18.27 & 19.08 & 19.24 & $\bm{19.69}$ & 18.96 & 19.22 \\ \hline
trec\_zf & 15.46 & 16.39 & 14.86 & 16.52 & 15.87 & 16.52 & 16.62 & $\bm{16.76}$ \\ \hline \hline
mean & 19.05 & 21.03 & 19.87 & 20.62 & 20.00 & $\bm{21.23}$ & 19.93 & 20.78 \\ \hline
\end{tabular}
\label{smooth9}
\end{table}

\begin{table}
\footnotesize
\centering
\caption{Evaluation set Micro-F1 results of Extended MNB models and MNB baselines on text classification datasets, 
optimized with 40x40 Gaussian random searches on development sets}
\begin{tabular}{|l|c|c|c|c|c|c|c|c|c|} \hline
Model & \multicolumn{8}{c|}{MNB} \\ \hline
Background & \multicolumn{4}{c|}{c} & \multicolumn{4}{c|}{u}\\ \hline
Smoothing & \multicolumn{4}{c|}{pd\_dp} & \multicolumn{4}{c|}{jm\_dp}\\ \hline
Factors & po & ps & po\_ps & - & po & ps & po\_ps & - \\
\hline \hline
20ng & $\bm{83.23}$ & 83.22 & 83.20 & 83.22 & 82.71 & 82.67 & 82.75 & 82.67 \\ \hline
cade & 60.46 & 60.85 & $\bm{60.88}$ & 60.46 & 60.44 & 59.68 & 60.44 & 59.74 \\ \hline
r52 & $\bm{93.18}$ & 92.71 & 92.67 & $\bm{93.18}$ & 91.90 & 91.90 & 91.62 & 91.54 \\ \hline
r8 & 95.70 & 96.29 & 95.84 & 95.93 & $\bm{96.61}$ & 96.57 & $\bm{96.61}$ & 96.39 \\ \hline
webkb & 83.23 & 83.09 & 83.38 & 83.09 & 84.02 & $\bm{84.09}$ & 84.02 & $\bm{84.09}$ \\ \hline
ecue1 & 91.10 & 91.50 & 91.50 & 91.40 & $\bm{92.70}$ & 92.40 & 91.70 & 92.40 \\ \hline
ecue2 & 92.20 & 92.20 & 92.20 & 92.20 & $\bm{92.40}$ & $\bm{92.40}$ & $\bm{92.40}$ & $\bm{92.40}$ \\ \hline
trec06 & 99.38 & 99.38 & 99.38 & 99.38 & $\bm{99.46}$ & 99.42 & 99.20 & $\bm{99.46}$ \\ \hline
tripa12 & 86.68 & 86.69 & 86.69 & 86.69 & $\bm{86.80}$ & 86.79 & 86.79 & 86.78 \\ \hline
aclimdb & 82.40 & 82.40 & 82.30 & 82.33 & $\bm{82.56}$ & 82.53 & 82.53 & 82.53 \\ \hline
amazon12 & 80.11 & 80.17 & 80.16 & 80.11 & $\bm{80.33}$ & 80.28 & $\bm{80.33}$ & 80.28 \\ \hline
rcv1 & 67.35 & 67.66 & 67.77 & 67.35 & 68.70 & 68.03 & $\bm{68.71}$ & 67.86 \\ \hline
eurlex & 48.86 & $\bm{49.06}$ & 48.94 & 48.99 & 48.10 & 47.97 & 47.83 & 48.03 \\ \hline
ohsu-trec & 40.36 & 40.36 & 40.36 & 40.36 & $\bm{40.43}$ & $\bm{40.43}$ & 40.07 & 40.19 \\ \hline
DMOZ2 & 24.85 & $\bm{25.44}$ & $\bm{25.44}$ & 24.83 & 23.87 & 24.77 & 25.21 & 24.64 \\ \hline
wikip\_med2 & 29.56 & 30.52 & 30.51 & 29.39 & $\bm{31.92}$ & 30.98 & 31.64 & 30.99 \\ \hline
mean & 72.42 & 72.60 & 72.58 & 72.43 & $\bm{72.68}$ & 72.56 & 72.62 & 72.50 \\ \hline
\end{tabular}
\label{extend1}
\end{table}  \clearpage 

\begin{table}
\footnotesize
\centering
\caption{Evaluation set Micro-F1 results of Extended MNB models on text classification datasets, optimized with 40x40 Gaussian random searches on development sets}
\begin{tabular}{|l|c|c|c|c|c|c|c|c|c|} \hline
Model & \multicolumn{8}{c|}{MNB} \\ \hline
Background & \multicolumn{2}{c|}{uc} & \multicolumn{2}{c|}{u} & \multicolumn{2}{c|}{uc} & \multicolumn{2}{c|}{u} \\ \hline
Smoothing & \multicolumn{4}{c|}{pd\_dp} & \multicolumn{4}{c|}{jm}\\ \hline
Weighting & qidf & ti & qidf & ti & qidf & ti & qidf & ti \\ \hline
Factors & \multicolumn{8}{c|}{ps} \\ \hline
\hline \hline
20ng & 82.53 & 82.81 & 82.59 & 82.30 & 83.06 & $\bm{83.62}$ & 82.77 & 83.32 \\ \hline
cade & $\bm{61.12}$ & 59.26 & 59.05 & 57.10 & 60.10 & 58.26 & 58.71 & 57.50 \\ \hline
r52 & $\bm{93.06}$ & 92.60 & 92.17 & 91.51 & 90.14 & 92.60 & 90.10 & 92.60 \\
\hline
\end{tabular}
\label{extend1b}
\end{table}

\begin{table}
\footnotesize
\centering
\caption{Evaluation set Micro-F1 results of Extended MNB models on text classification datasets, optimized with 40x40 Gaussian random searches on development sets}
\begin{tabular}{|l|c|c|c|c|c|c|c|c|c|} \hline
Model & \multicolumn{8}{c|}{MNB} \\ \hline
Background & \multicolumn{2}{c|}{uc} & \multicolumn{2}{c|}{u} & \multicolumn{2}{c|}{uc} & \multicolumn{2}{c|}{u} \\ \hline
Smoothing & \multicolumn{4}{c|}{jm\_dp} & \multicolumn{4}{c|}{pd}\\ \hline
Weighting & qidf & ti & qidf & ti & qidf & ti & qidf & ti \\ \hline
Factors & \multicolumn{8}{c|}{ps} \\ \hline
\hline \hline
20ng & 83.03 & $\bm{83.51}$ & 82.65 & 83.35 & 82.57 & 82.26 & 82.46 & 82.27 \\ \hline
cade & 59.74 & 59.05 & 59.62 & 57.52 & $\bm{61.40}$ & 58.92 & 58.70 & 57.08 \\ \hline
r52 & 92.87 & 93.02 & 91.08 & 92.44 & $\bm{93.18}$ & 92.52 & 92.91 & 91.51 \\
\hline
\end{tabular}
\label{extend2}
\end{table} \clearpage 

\begin{table}
\footnotesize
\centering
\caption{Evaluation set Micro-F1 results of Extended MNB models on text classification datasets, optimized with 40x40 Gaussian random searches on development sets}
\begin{tabular}{|l|c|c|c|c|c|c|c|c|c|} \hline
Model & \multicolumn{8}{c|}{MNB} \\ \hline
Background & \multicolumn{2}{c|}{uc} & \multicolumn{2}{c|}{u} & \multicolumn{2}{c|}{uc} & \multicolumn{2}{c|}{u} \\ \hline
Smoothing & \multicolumn{4}{c|}{pd} & \multicolumn{4}{c|}{dp}\\ \hline
Weighting & qidf & ti & qidf & ti & qidf & ti & qidf & ti \\ \hline
Factors & \multicolumn{8}{c|}{ps} \\ \hline
\hline \hline
20ng & 82.99 & $\bm{83.67}$ & 82.63 & 83.35 & 82.66 & 83.18 & 82.94 & 83.24 \\ \hline
cade & $\bm{61.15}$ & 58.72 & 58.71 & 58.14 & 59.46 & 59.70 & 59.21 & 57.71 \\ \hline
r52 & $\bm{93.14}$ & 92.56 & 92.95 & 91.51 & 88.51 & 88.74 & 88.31 & 87.73 \\
\hline
\end{tabular}
\label{extend3}
\end{table}

\begin{table}
\footnotesize
\centering
\caption{Evaluation set Micro-F1 results of Extended MNB models on text classification datasets, optimized with 40x40 Gaussian random searches on development sets}
\begin{tabular}{|l|c|c|c|c|c|c|c|c|c|} \hline
Model & \multicolumn{8}{c|}{MNB} \\ \hline
Background & uc & u & uc & u & uc & u & uc & u \\ \hline
Smoothing & pd\_dp & jm\_dp & pd\_dp & jm\_dp & pd\_dp & jm\_dp & pd\_dp & jm\_dp \\ \hline
Weighting & \multicolumn{2}{c|}{ti} & \multicolumn{2}{c|}{tXi} & \multicolumn{2}{c|}{tiX} & \multicolumn{2}{c|}{tXiX}\\ \hline
Factors & \multicolumn{8}{c|}{ps} \\ \hline
\hline \hline
20ng & 83.04 & 83.26 & 83.87 & 83.74 & 83.31 & 83.42 & $\bm{84.23}$ & 83.55 \\ \hline
cade & 59.25 & 57.55 & 59.62 & 57.56 & 59.55 & 57.96 & $\bm{60.17}$ & 57.78 \\ \hline
r52 & 88.43 & 90.84 & 91.90 & $\bm{92.83}$ & 87.92 & 92.64 & 92.60 & 92.56 \\
\hline
\end{tabular}
\label{extend4}
\end{table} \clearpage 

\begin{table}
\footnotesize
\centering
\caption{Evaluation set NDCG@20 results of Extended MNB models on text retrieval datasets, optimized with 50x8 Gaussian random searches on development sets}
\begin{tabular}{|l|c|c|c|c|c|c|c|c|c|} \hline
Model & \multicolumn{8}{c|}{MNB} \\ \hline
Background & \multicolumn{2}{c|}{uc} & \multicolumn{2}{c|}{u} & \multicolumn{2}{c|}{uc} & \multicolumn{2}{c|}{u} \\ \hline
Smoothing & \multicolumn{4}{c|}{pd\_dp} & \multicolumn{4}{c|}{jm} \\ \hline
Weighting & qidf & ti & qidf & ti & qidf & ti & qidf & ti \\ \hline
\hline \hline
fire\_en & $\bm{53.91}$ & 49.37 & 52.22 & 51.67 & 51.30 & 50.60 & 51.30 & 50.66 \\ \hline
ohsu\_trec & 23.75 & 23.76 & 23.76 & $\bm{23.78}$ & 23.21 & $\bm{23.78}$ & 23.28 & $\bm{23.78}$ \\ \hline
trec\_ap & 31.68 & 29.64 & $\bm{32.50}$ & 31.95 & 31.71 & 30.74 & 31.39 & 31.33 \\ \hline
trec\_cr & 27.15 & 26.02 & $\bm{32.98}$ & 27.54 & 27.91 & 26.47 & 32.86 & 29.69 \\ \hline
trec\_doe & 29.28 & 25.91 & 28.90 & 27.08 & $\bm{29.53}$ & 25.18 & 28.44 & 26.34 \\ \hline
trec\_fbis & 34.41 & 26.21 & $\bm{36.51}$ & 30.00 & 33.34 & 32.73 & 32.39 & 33.29 \\ \hline
trec\_fr & 23.54 & 23.10 & 23.88 & $\bm{23.93}$ & 23.22 & 21.96 & 22.17 & 21.51 \\ \hline
trec\_ft & 32.69 & 28.25 & $\bm{33.17}$ & 32.50 & 29.87 & 30.09 & 30.99 & 31.94 \\ \hline
trec\_la & 28.31 & 26.12 & $\bm{28.75}$ & 28.55 & 28.26 & 27.79 & 28.48 & 27.80 \\ \hline
trec\_pt & 36.79 & 31.09 & $\bm{36.85}$ & 32.94 & 33.25 & 35.95 & 36.65 & 35.95 \\ \hline
trec\_sjmn & 26.52 & 24.45 & 27.59 & 27.57 & 26.78 & 27.44 & $\bm{30.34}$ & 26.87 \\ \hline
trec\_wsj & 35.51 & 35.33 & $\bm{35.72}$ & 35.43 & 34.52 & 33.44 & 34.52 & 33.52 \\ \hline
trec\_zf & 29.82 & 30.67 & 29.71 & $\bm{30.70}$ & 28.11 & 26.92 & 28.75 & 27.93 \\ \hline \hline
mean & 31.80 & 29.23 & $\bm{32.50}$ & 31.05 & 30.85 & 30.24 & 31.66 & 30.82 \\ \hline
\end{tabular}
\label{extend5}
\end{table}

\begin{table}
\footnotesize
\centering
\caption{Evaluation set MAP results of Extended MNB models on text retrieval datasets, optimized with 50x8 Gaussian random searches on development sets}
\begin{tabular}{|l|c|c|c|c|c|c|c|c|c|} \hline
Model & \multicolumn{8}{c|}{MNB} \\ \hline
Background & \multicolumn{2}{c|}{uc} & \multicolumn{2}{c|}{u} & \multicolumn{2}{c|}{uc} & \multicolumn{2}{c|}{u} \\ \hline
Smoothing & \multicolumn{4}{c|}{pd\_dp} & \multicolumn{4}{c|}{jm} \\ \hline
Weighting & qidf & ti & qidf & ti & qidf & ti & qidf & ti \\ \hline
\hline \hline
fire\_en & 43.27 & 42.15 & $\bm{43.28}$ & 42.98 & 42.05 & 42.45 & 42.42 & 42.45 \\ \hline
ohsu\_trec & $\bm{22.73}$ & 22.63 & 22.70 & 22.44 & 22.05 & 22.39 & 22.08 & 22.31 \\ \hline
trec\_ap & $\bm{20.55}$ & 18.93 & 20.20 & 20.40 & 20.15 & 20.30 & 19.90 & 20.29 \\ \hline
trec\_cr & 17.31 & 15.88 & $\bm{21.17}$ & 18.90 & 17.02 & 16.61 & 19.70 & 19.18 \\ \hline
trec\_doe & 19.39 & 17.41 & $\bm{19.84}$ & 18.47 & 18.99 & 16.82 & 19.30 & 18.36 \\ \hline
trec\_fbis & 26.21 & 22.46 & 26.18 & $\bm{26.48}$ & 24.97 & 24.38 & 23.38 & 24.21 \\ \hline
trec\_fr & $\bm{19.33}$ & 17.01 & 17.48 & 17.64 & 17.35 & 16.68 & 16.44 & 16.00 \\ \hline
trec\_ft & 23.06 & 20.50 & 22.50 & $\bm{23.53}$ & 22.90 & 22.68 & 21.84 & 22.89 \\ \hline
trec\_la & 17.92 & 15.56 & 17.90 & $\bm{17.98}$ & 17.19 & 17.09 & 17.62 & 17.32 \\ \hline
trec\_pt & 25.59 & 23.54 & 25.22 & 22.43 & 25.51 & $\bm{26.81}$ & 26.70 & $\bm{26.81}$ \\ \hline
trec\_sjmn & 18.42 & 16.27 & 22.57 & 18.45 & 16.97 & 17.45 & $\bm{23.44}$ & 17.27 \\ \hline
trec\_wsj & $\bm{20.94}$ & 20.63 & 20.51 & 20.52 & 19.78 & 19.71 & 19.91 & 19.70 \\ \hline
trec\_zf & $\bm{20.11}$ & 17.54 & 18.53 & 15.37 & 17.66 & 16.67 & 18.22 & 18.11 \\ \hline \hline
mean & 22.68 & 20.81 & $\bm{22.93}$ & 21.97 & 21.74 & 21.54 & 22.38 & 21.92 \\ \hline
\end{tabular}
\label{extend6}
\end{table} \clearpage 

\begin{table}
\footnotesize
\centering
\caption{Evaluation set NDCG@20 results of Extended MNB models on text retrieval datasets, optimized with 50x8 Gaussian random searches on development sets}
\begin{tabular}{|l|c|c|c|c|c|c|c|c|c|} \hline
Model & \multicolumn{8}{c|}{MNB} \\ \hline
Background & \multicolumn{2}{c|}{uc} & \multicolumn{2}{c|}{u} & \multicolumn{2}{c|}{uc} & \multicolumn{2}{c|}{u} \\ \hline
Smoothing & \multicolumn{4}{c|}{jm\_dp} & \multicolumn{4}{c|}{pd} \\ \hline
Weighting & qidf & ti & qidf & ti & qidf & ti & qidf & ti \\ \hline
\hline \hline
fire\_en & 51.96 & 50.55 & $\bm{53.12}$ & 50.59 & 52.88 & 49.44 & 52.63 & 49.48 \\ \hline
ohsu\_trec & 23.91 & 23.78 & 23.77 & 23.78 & 23.75 & $\bm{23.99}$ & 23.90 & 23.75 \\ \hline
trec\_ap & 31.32 & 30.79 & 32.49 & 31.33 & 32.16 & 29.86 & $\bm{32.50}$ & 30.07 \\ \hline
trec\_cr & 27.91 & 26.51 & 32.22 & 29.56 & 26.34 & 25.89 & $\bm{32.98}$ & 27.69 \\ \hline
trec\_doe & $\bm{29.47}$ & 25.18 & 28.26 & 28.23 & 29.15 & 28.87 & 28.90 & 29.27 \\ \hline
trec\_fbis & 35.09 & 34.71 & 32.39 & 33.18 & 34.42 & 26.25 & $\bm{36.08}$ & 29.98 \\ \hline
trec\_fr & 22.96 & 23.40 & 22.17 & 21.17 & $\bm{25.32}$ & 22.33 & 23.88 & 22.42 \\ \hline
trec\_ft & 32.63 & 29.02 & $\bm{33.41}$ & 32.25 & 31.90 & 27.61 & 33.17 & 29.38 \\ \hline
trec\_la & 26.17 & 26.70 & 28.08 & 27.62 & 28.07 & 25.90 & $\bm{29.18}$ & 25.71 \\ \hline
trec\_pt & 34.09 & 35.95 & $\bm{38.78}$ & 35.95 & 36.92 & 31.13 & 36.85 & 35.18 \\ \hline
trec\_sjmn & 26.13 & 27.62 & 26.99 & 27.07 & $\bm{27.84}$ & 24.46 & 27.38 & 27.28 \\ \hline
trec\_wsj & 34.54 & 33.52 & $\bm{36.26}$ & 35.30 & 35.93 & 31.09 & 36.09 & 32.46 \\ \hline
trec\_zf & 28.11 & 26.85 & 28.81 & 30.11 & 29.76 & 30.71 & 29.80 & $\bm{31.41}$ \\ \hline \hline
mean & 31.10 & 30.35 & 32.06 & 31.24 & 31.88 & 29.04 & $\bm{32.56}$ & 30.32 \\ \hline
\end{tabular}
\label{extend7}
\end{table}

\begin{table}
\footnotesize
\centering
\caption{Evaluation set MAP results of Extended MNB models on text retrieval datasets, optimized with 50x8 Gaussian random searches on development sets}
\begin{tabular}{|l|c|c|c|c|c|c|c|c|c|} \hline
Model & \multicolumn{8}{c|}{MNB} \\ \hline
Background & \multicolumn{2}{c|}{uc} & \multicolumn{2}{c|}{u} & \multicolumn{2}{c|}{uc} & \multicolumn{2}{c|}{u} \\ \hline
Smoothing & \multicolumn{4}{c|}{jm\_dp} & \multicolumn{4}{c|}{pd} \\ \hline
Weighting & qidf & ti & qidf & ti & qidf & ti & qidf & ti \\ \hline
\hline \hline
fire\_en & 42.60 & 42.46 & $\bm{43.93}$ & 42.45 & 43.29 & 42.57 & 43.31 & 42.66 \\ \hline
ohsu\_trec & $\bm{22.69}$ & 22.38 & 22.43 & 22.31 & 22.64 & 22.54 & 22.60 & 22.22 \\ \hline
trec\_ap & 20.12 & 20.22 & 20.20 & 20.30 & $\bm{20.53}$ & 17.70 & 20.15 & 18.63 \\ \hline
trec\_cr & 17.01 & 16.59 & 19.35 & 19.18 & 17.32 & 15.63 & $\bm{21.17}$ & 17.60 \\ \hline
trec\_doe & 19.65 & 16.82 & 19.30 & 19.14 & 19.41 & 18.79 & $\bm{19.84}$ & 19.11 \\ \hline
trec\_fbis & 26.21 & 24.42 & 23.39 & 24.21 & $\bm{26.39}$ & 18.40 & 26.18 & 20.13 \\ \hline
trec\_fr & 17.36 & 17.49 & 17.58 & 18.17 & $\bm{19.12}$ & 16.91 & 18.57 & 16.61 \\ \hline
trec\_ft & $\bm{23.50}$ & 22.74 & 21.83 & 22.78 & 22.95 & 21.10 & 22.50 & 21.39 \\ \hline
trec\_la & 17.47 & 17.09 & 17.37 & 17.32 & 17.78 & 15.61 & $\bm{17.93}$ & 15.91 \\ \hline
trec\_pt & 24.81 & $\bm{26.81}$ & 24.81 & $\bm{26.81}$ & 26.61 & 23.37 & 26.61 & 24.73 \\ \hline
trec\_sjmn & 16.61 & 15.72 & $\bm{23.44}$ & 18.45 & 18.42 & 15.50 & 22.57 & 18.00 \\ \hline
trec\_wsj & 19.57 & 19.70 & 20.22 & 19.70 & $\bm{20.79}$ & 17.52 & 20.48 & 18.44 \\ \hline
trec\_zf & 17.66 & 18.26 & 18.45 & 18.12 & $\bm{19.78}$ & 16.25 & 18.44 & 16.08 \\ \hline \hline
mean & 21.94 & 21.59 & 22.49 & 22.23 & 22.70 & 20.14 & $\bm{23.10}$ & 20.89 \\ \hline
\end{tabular}
\label{extend8}
\end{table} \clearpage 

\begin{table}
\footnotesize
\centering
\caption{Evaluation set NDCG@20 results of Extended MNB models on text retrieval datasets, optimized with 50x8 Gaussian random searches on development sets}
\begin{tabular}{|l|c|c|c|c|c|c|c|c|c|} \hline
Model & \multicolumn{8}{c|}{MNB} \\ \hline
Background & \multicolumn{2}{c|}{uc} & \multicolumn{2}{c|}{u} & \multicolumn{2}{c|}{uc} & \multicolumn{2}{c|}{u} \\ \hline
Smoothing & \multicolumn{4}{c|}{pd\_jm} & \multicolumn{4}{c|}{dp} \\ \hline
Weighting & qidf & ti & qidf & ti & qidf & ti & qidf & ti \\ \hline
\hline \hline
fire\_en & 52.58 & 50.57 & $\bm{52.73}$ & 50.67 & 51.62 & 51.62 & 51.63 & 51.62 \\ \hline
ohsu\_trec & 23.60 & 23.84 & $\bm{23.86}$ & 23.72 & 23.27 & 23.02 & 23.24 & 23.38 \\ \hline
trec\_ap & 32.00 & 30.79 & $\bm{32.53}$ & 31.02 & 31.16 & 31.92 & 30.79 & 31.90 \\ \hline
trec\_cr & $\bm{34.24}$ & 28.50 & 34.23 & 29.69 & 28.28 & 16.00 & 27.69 & 28.44 \\ \hline
trec\_doe & 29.04 & 25.97 & 28.90 & $\bm{29.58}$ & 27.12 & 24.52 & 26.29 & 28.03 \\ \hline
trec\_fbis & 34.34 & 33.21 & 35.59 & 32.86 & $\bm{36.27}$ & 35.58 & 34.92 & 34.99 \\ \hline
trec\_fr & 24.00 & 21.86 & 23.63 & 21.80 & 21.66 & $\bm{24.46}$ & 21.17 & 24.17 \\ \hline
trec\_ft & 31.81 & 28.25 & $\bm{33.21}$ & 31.98 & 32.05 & 32.42 & 30.28 & 32.43 \\ \hline
trec\_la & 28.39 & 27.80 & $\bm{29.26}$ & 27.79 & 25.20 & 28.69 & 24.28 & 28.55 \\ \hline
trec\_pt & 34.09 & 35.95 & $\bm{38.48}$ & 35.99 & 36.87 & 30.04 & 37.33 & 36.29 \\ \hline
trec\_sjmn & 27.38 & 25.88 & 27.39 & 26.88 & 26.39 & 26.94 & $\bm{30.67}$ & 27.29 \\ \hline
trec\_wsj & $\bm{36.08}$ & 33.41 & 35.11 & 33.41 & 32.30 & 35.32 & 30.18 & 35.47 \\ \hline
trec\_zf & 29.16 & 26.92 & 29.92 & 28.70 & 28.61 & 30.40 & 21.32 & $\bm{30.77}$ \\ \hline \hline
mean & 32.06 & 30.23 & $\bm{32.68}$ & 31.08 & 30.83 & 30.07 & 29.98 & 31.80 \\ \hline
\end{tabular}
\label{extend9}
\end{table}

\begin{table}
\footnotesize
\centering
\caption{Evaluation set MAP results of Extended MNB models on text retrieval datasets, optimized with 50x8 Gaussian random searches on development sets}
\begin{tabular}{|l|c|c|c|c|c|c|c|c|c|} \hline
Model & \multicolumn{8}{c|}{MNB} \\ \hline
Background & \multicolumn{2}{c|}{uc} & \multicolumn{2}{c|}{u} & \multicolumn{2}{c|}{uc} & \multicolumn{2}{c|}{u} \\ \hline
Smoothing & \multicolumn{4}{c|}{pd\_jm} & \multicolumn{4}{c|}{dp} \\ \hline
Weighting & qidf & ti & qidf & ti & qidf & ti & qidf & ti \\ \hline
\hline \hline
fire\_en & $\bm{43.34}$ & 42.45 & 43.26 & 42.46 & 42.25 & 43.00 & 42.25 & 43.04 \\ \hline
ohsu\_trec & 22.64 & 22.64 & 22.58 & 22.44 & $\bm{22.65}$ & 22.25 & 22.38 & 22.27 \\ \hline
trec\_ap & 20.32 & 20.30 & 20.20 & 20.31 & 19.05 & 20.36 & 18.41 & $\bm{20.38}$ \\ \hline
trec\_cr & 17.23 & 16.72 & $\bm{21.42}$ & 19.18 & 17.58 & 16.12 & 17.20 & 18.16 \\ \hline
trec\_doe & 19.31 & 16.89 & 19.84 & 18.97 & $\bm{20.30}$ & 19.55 & 18.96 & 19.13 \\ \hline
trec\_fbis & 26.34 & 24.42 & 25.63 & 24.21 & 27.01 & 21.98 & 26.81 & $\bm{27.74}$ \\ \hline
trec\_fr & $\bm{19.43}$ & 17.01 & 18.57 & 16.00 & 18.32 & 18.16 & 16.30 & 17.71 \\ \hline
trec\_ft & 22.95 & 22.76 & 22.36 & 22.78 & 22.62 & $\bm{23.42}$ & 20.89 & 23.38 \\ \hline
trec\_la & $\bm{17.97}$ & 17.10 & 17.91 & 17.31 & 17.24 & 17.89 & 15.14 & 17.96 \\ \hline
trec\_pt & 25.07 & 26.80 & 25.45 & 26.81 & 25.29 & 25.88 & 25.29 & $\bm{26.94}$ \\ \hline
trec\_sjmn & 17.88 & 16.69 & 18.22 & 17.28 & 17.08 & 17.06 & $\bm{24.38}$ & 18.45 \\ \hline
trec\_wsj & $\bm{20.82}$ & 19.71 & 20.51 & 19.70 & 19.37 & 20.61 & 16.32 & 20.67 \\ \hline
trec\_zf & $\bm{19.16}$ & 16.28 & 18.47 & 18.11 & 18.20 & 19.08 & 14.62 & 19.08 \\ \hline \hline
mean & 22.50 & 21.52 & 22.65 & 21.97 & 22.07 & 21.95 & 21.46 & $\bm{22.68}$ \\ \hline
\end{tabular}
\label{extend10}
\end{table} \clearpage 

\begin{table}
\footnotesize
\centering
\caption{Evaluation set NDCG@20 results of Extended MNB models on text retrieval datasets, optimized with 50x8 Gaussian random searches on development sets}
\begin{tabular}{|l|c|c|c|c|c|c|c|c|c|} \hline
Model & \multicolumn{8}{c|}{MNB} \\ \hline
Background & \multicolumn{8}{c|}{u}\\ \hline
Smoothing & pd\_dp & pd\_jm & pd\_dp & pd\_jm & pd\_dp & pd\_jm & pd\_dp & pd\_jm\\ \hline
Weighting & \multicolumn{2}{c|}{qidfX} & \multicolumn{2}{c|}{tXi} & \multicolumn{2}{c|}{tiX} & \multicolumn{2}{c|}{tXiX}\\ \hline
\hline \hline
fire\_en & 51.57 & $\bm{52.35}$ & 51.76 & 50.68 & 49.82 & 49.61 & 42.63 & 50.06 \\ \hline
ohsu\_trec & 23.77 & 23.97 & 23.89 & 23.96 & 23.66 & $\bm{24.16}$ & 23.74 & 23.64 \\ \hline
trec\_ap & $\bm{32.87}$ & 32.84 & 31.87 & 31.04 & 31.79 & 31.12 & 29.46 & 30.05 \\ \hline
trec\_cr & 30.73 & 31.08 & 28.99 & 29.86 & 30.01 & 30.91 & 28.73 & $\bm{31.51}$ \\ \hline
trec\_doe & 28.90 & $\bm{28.96}$ & 27.55 & 26.69 & 28.77 & 26.96 & 27.31 & 26.63 \\ \hline
trec\_fbis & $\bm{36.23}$ & 36.04 & 30.24 & 33.18 & 30.19 & 31.30 & 36.16 & 32.16 \\ \hline
trec\_fr & 22.75 & $\bm{24.35}$ & 24.01 & 21.63 & 24.02 & 22.42 & 11.21 & 22.12 \\ \hline
trec\_ft & 32.90 & $\bm{32.93}$ & 31.45 & 32.05 & 30.86 & 30.14 & 30.99 & 30.35 \\ \hline
trec\_la & 29.12 & $\bm{29.50}$ & 28.56 & 27.87 & 26.52 & 28.30 & 26.79 & 28.04 \\ \hline
trec\_pt & 34.18 & 28.95 & 32.99 & $\bm{38.68}$ & 14.42 & 27.50 & 14.71 & 28.41 \\ \hline
trec\_sjmn & 27.40 & 27.55 & 27.24 & 26.91 & 28.01 & 26.02 & $\bm{28.19}$ & 26.75 \\ \hline
trec\_wsj & 35.24 & 35.11 & $\bm{35.48}$ & 33.40 & 32.31 & 34.33 & 32.48 & 34.25 \\ \hline
trec\_zf & 30.74 & $\bm{30.75}$ & 30.72 & 28.28 & 30.35 & 27.77 & 25.95 & 28.39 \\ \hline \hline
mean & $\bm{32.03}$ & 31.88 & 31.13 & 31.09 & 29.29 & 30.04 & 27.56 & 30.18 \\ \hline
\end{tabular}
\label{extend11}
\end{table} 

\begin{table}
\footnotesize
\centering
\caption{Evaluation set MAP results of Extended MNB models on text retrieval datasets, optimized with 50x8 Gaussian random searches on development sets}
\begin{tabular}{|l|c|c|c|c|c|c|c|c|c|} \hline
Model & \multicolumn{8}{c|}{MNB} \\ \hline
Background & \multicolumn{8}{c|}{u}\\ \hline
Smoothing & pd\_dp & pd\_jm & pd\_dp & pd\_jm & pd\_dp & pd\_jm & pd\_dp & pd\_jm\\ \hline
Weighting & \multicolumn{2}{c|}{qidfX} & \multicolumn{2}{c|}{tXi} & \multicolumn{2}{c|}{tiX} & \multicolumn{2}{c|}{tXiX}\\ \hline
\hline \hline
fire\_en & 51.57 & $\bm{52.35}$ & 51.76 & 50.68 & 49.82 & 49.61 & 42.63 & 50.06 \\ \hline
ohsu\_trec & 23.77 & 23.97 & 23.89 & 23.96 & 23.66 & $\bm{24.16}$ & 23.74 & 23.64 \\ \hline
trec\_ap & $\bm{32.87}$ & 32.84 & 31.87 & 31.04 & 31.79 & 31.12 & 29.46 & 30.05 \\ \hline
trec\_cr & 30.73 & 31.08 & 28.99 & 29.86 & 30.01 & 30.91 & 28.73 & $\bm{31.51}$ \\ \hline
trec\_doe & 28.90 & $\bm{28.96}$ & 27.55 & 26.69 & 28.77 & 26.96 & 27.31 & 26.63 \\ \hline
trec\_fbis & $\bm{36.23}$ & 36.04 & 30.24 & 33.18 & 30.19 & 31.30 & 36.16 & 32.16 \\ \hline
trec\_fr & 22.75 & $\bm{24.35}$ & 24.01 & 21.63 & 24.02 & 22.42 & 11.21 & 22.12 \\ \hline
trec\_ft & 32.90 & $\bm{32.93}$ & 31.45 & 32.05 & 30.86 & 30.14 & 30.99 & 30.35 \\ \hline
trec\_la & 29.12 & $\bm{29.50}$ & 28.56 & 27.87 & 26.52 & 28.30 & 26.79 & 28.04 \\ \hline
trec\_pt & 34.18 & 28.95 & 32.99 & $\bm{38.68}$ & 14.42 & 27.50 & 14.71 & 28.41 \\ \hline
trec\_sjmn & 27.40 & 27.55 & 27.24 & 26.91 & 28.01 & 26.02 & $\bm{28.19}$ & 26.75 \\ \hline
trec\_wsj & 35.24 & 35.11 & $\bm{35.48}$ & 33.40 & 32.31 & 34.33 & 32.48 & 34.25 \\ \hline
trec\_zf & 30.74 & $\bm{30.75}$ & 30.72 & 28.28 & 30.35 & 27.77 & 25.95 & 28.39 \\ \hline \hline
mean & $\bm{32.03}$ & 31.88 & 31.13 & 31.09 & 29.29 & 30.04 & 27.56 & 30.18 \\ \hline
\end{tabular}
\label{extend12}
\end{table}\clearpage 

\begin{table}
\footnotesize
\centering
\caption{Evaluation set Micro-F1 results of TDM models on text classification datasets, optimized with 40x40 Gaussian random searches on development sets}
\begin{tabular}{|l|c|c|c|c|c|c|c|c|c|} \hline
Model & \multicolumn{8}{c|}{TDM} \\ \hline
Background & u & u & u & uc  & u & u & u & uc \\ \hline
\multirow{3}{*}{Smoothing} & jm & jm & jm & dp & jm & pd & pd & jm \\ 
& dp & dp & dp & dp & & pd & jm & dp \\
& & & & & & & dp & \\ \hline
\multirow{3}{*}{Kernel} & pd & jm & pd & pd & pd & pd & pd & pd \\
& dp & dp & jm & jm & dp & dp & dp & dp \\
& & & dp & dp & & & & \\
\hline
\end{tabular}
\label{structure1}
\end{table}\clearpage 

\begin{table}
\footnotesize
\centering
\caption{Evaluation set Micro-F1 results of LR and SVM models on text classification datasets, optimized with 40x8 Gaussian random searches on development sets}
\begin{tabular}{|l|c|c|c|c|c|c|c|c|c|} \hline
Model & lr & l2svm & l2svm & lr & lr & l2svm & l2svm & lr \\ \hline
Regularizer & l2r & l2r & l1r & l1r & l2r & l2r & l1r & l1r \\ \hline
Weighting & \multicolumn{4}{c|}{-} & \multicolumn{4}{c|}{tXiX} \\ \hline \hline
20ng & 80.28 & 80.73 & 76.58 & 78.87 & 83.79 & $\bm{84.92}$ & 81.30 & 80.67 \\ \hline
cade & 57.86 & 57.79 & 56.89 & 51.05 & $\bm{59.74}$ & 59.46 & 59.21 & 58.89 \\ \hline
r52 & 93.06 & 93.22 & 93.38 & 93.30 & 92.28 & $\bm{95.13}$ & 94.19 & 94.19 \\ \hline
r8 & 97.03 & 97.03 & 96.93 & 96.84 & 97.03 & $\bm{97.57}$ & 96.80 & 96.80 \\ \hline
webkb & 91.33 & 91.26 & 89.18 & 90.40 & 91.97 & $\bm{92.19}$ & 91.83 & 91.69 \\ \hline
ecue1 & 95.20 & 95.30 & 96.20 & 95.50 & 95.50 & $\bm{97.40}$ & 97.10 & 96.20 \\ \hline
ecue2 & 98.50 & 97.50 & 97.40 & 97.60 & 97.60 & $\bm{99.00}$ & 96.90 & 96.50 \\ \hline
trec06 & 99.42 & 99.38 & 99.64 & 99.60 & 99.49 & $\bm{99.71}$ & 99.64 & 99.56 \\ \hline
tripa12 & 92.10 & 91.91 & 91.31 & 91.70 & 92.51 & $\bm{92.62}$ & 91.98 & 91.91 \\ \hline
aclimdb & 90.36 & 89.93 & 89.50 & 89.23 & 89.93 & $\bm{90.83}$ & 89.80 & 89.86 \\ \hline
amazon12 & 84.32 & 84.01 & 84.26 & $\bm{84.47}$ & 84.35 & 84.26 & 84.00 & 84.14 \\ \hline
rcv1 & NA & NA & NA & NA & NA & NA & NA & NA \\ \hline
eurlex & NA & NA & NA & NA & NA & NA & NA & NA \\ \hline
ohsu-trec & NA & NA & NA & NA & NA & NA & NA & NA \\ \hline
DMOZ2 & NA & NA & NA & NA & NA & NA & NA & NA \\ \hline
wikip\_med2 & NA & NA & NA & NA & NA & NA & NA & NA \\ \hline \hline
mean & 89.04 & 88.91 & 88.30 & 88.05 & 89.47 & $\bm{90.28}$ & 89.34 & 89.13 \\ \hline
\end{tabular}
\label{comparison1}
\end{table}

\begin{table}
\footnotesize
\centering
\caption{Evaluation set NDCG@20 and MAP results of VSM and BM25 modes on text classification datasets, optimized with 50x8 Gaussian random searches on development sets}
\begin{tabular}{|l|c|c|c|c|c|c|c|c|c|} \hline
Measure & \multicolumn{4}{c|}{NDCG@20} & \multicolumn{4}{c|}{MAP} \\ \hline
Model & \multicolumn{3}{c|}{vsm} & bm25 & \multicolumn{3}{c|}{vsm} & bm25 \\ \hline
Weighting & - & ti & tXiX & - & - & ti & tXiX & - \\ \hline \hline
fire\_en & 39.42 & 45.59 & 46.80 & $\bm{53.08}$ & 28.84 & 36.08 & 38.22 & $\bm{43.73}$ \\ \hline
ohsu\_trec & 13.87 & 18.04 & 19.46 & $\bm{22.76}$ & 9.76 & 15.00 & 18.48 & $\bm{21.71}$ \\ \hline
trec\_ap & 20.50 & 24.27 & 28.74 & $\bm{32.50}$ & 11.56 & 15.89 & 19.07 & $\bm{20.39}$ \\ \hline
trec\_cr & 14.35 & 18.51 & 17.56 & $\bm{30.16}$ & 9.53 & 12.75 & 12.45 & $\bm{19.20}$ \\ \hline
trec\_doe & 18.67 & 21.58 & 20.14 & $\bm{28.79}$ & 11.23 & 12.40 & 13.27 & $\bm{18.26}$ \\ \hline
trec\_fbis & 13.95 & 24.14 & 26.33 & $\bm{32.99}$ & 8.46 & 17.38 & 18.43 & $\bm{24.60}$ \\ \hline
trec\_fr & 8.25 & 14.88 & 15.72 & $\bm{20.49}$ & 4.72 & 9.74 & 11.44 & $\bm{15.75}$ \\ \hline
trec\_ft & 18.59 & 23.79 & 25.46 & $\bm{33.29}$ & 11.55 & 16.58 & 18.04 & $\bm{22.32}$ \\ \hline
trec\_la & 16.86 & 18.22 & 22.73 & $\bm{27.94}$ & 10.76 & 12.75 & 14.49 & $\bm{17.12}$ \\ \hline
trec\_pt & 12.19 & 20.65 & 29.55 & $\bm{33.01}$ & 6.91 & 14.73 & 17.87 & $\bm{23.33}$ \\ \hline
trec\_sjmn & 19.03 & 14.78 & 20.49 & $\bm{27.85}$ & 12.03 & 11.14 & 13.73 & $\bm{19.34}$ \\ \hline
trec\_wsj & 19.15 & 25.39 & 28.53 & $\bm{35.88}$ & 9.50 & 14.03 & 16.31 & $\bm{20.50}$ \\ \hline
trec\_zf & 10.09 & 24.50 & 23.84 & $\bm{31.39}$ & 6.94 & 14.67 & 15.50 & $\bm{19.81}$ \\ \hline \hline
mean & 17.30 & 22.64 & 25.03 & $\bm{31.55}$ & 10.91 & 15.63 & 17.48 & $\bm{22.00}$ \\ \hline
\end{tabular}
\label{comparison2}
\end{table}\clearpage 

\begin{table}
\footnotesize
\centering
\caption{Training times in seconds for lr\_l2r\_tXiX}
\begin{tabular}{|l|l|c|c|c|c|c|c|} \hline
Labelsets & Documents & \multicolumn{6}{c|}{Features} \\ \cline{3-8}
& & 10 & 100 & 1000 & 10000 & 100000 & 1000000 \\ \hline
\multirow{5}{*}{1} & 10 & 0.0001 & 0.0001 & 0.0001 & 0.0001 & 0.0001 & 0.0001 \\ \cline{2-8}
& 100 & 0.0001 & 0.0002 & 0.0004 & 0.0008 & 0.0008 & 0.0008 \\ \cline{2-8}
& 1000 & 0.0009 & 0.0013 & 0.0020 & 0.0040 & 0.0052 & 0.0053 \\ \cline{2-8}
& 10000 & 0.0117 & 0.0159 & 0.0281 & 0.0509 & 0.0801 & 0.1044 \\ \cline{2-8}
& 100000 & 0.1683 & 0.4092 & 0.7425 & 0.8779 & 0.9760 & 1.0540 \\ \cline{2-8}
& 1000000 & 2.0355 & 6.2743 & 9.0974 & 9.4415 & 11.883 & 14.129 \\
\hline
\end{tabular}
\label{train_times1}
\end{table}

\begin{table}
\footnotesize
\centering
\caption{Training times in seconds for l2svm\_l2r\_tXiX}
\begin{tabular}{|l|l|c|c|c|c|c|c|} \hline
Labelsets & Documents & \multicolumn{6}{c|}{Features} \\ \cline{3-8}
& & 10 & 100 & 1000 & 10000 & 100000 & 1000000 \\ \hline
\multirow{5}{*}{1} & 10 & 0.0000 & 0.0001 & 0.0002 & 0.0002 & 0.0002 & 0.0002 \\ \cline{2-8}
& 100 & 0.0001 & 0.0002 & 0.0005 & 0.0011 & 0.0011 & 0.0011 \\ \cline{2-8}
& 1000 & 0.0004 & 0.0007 & 0.0028 & 0.0079 & 0.0104 & 0.0104 \\ \cline{2-8}
& 10000 & 0.0038 & 0.0068 & 0.1135 & 0.0531 & 0.1095 & 0.1101 \\ \cline{2-8}
& 100000 & 0.0478 & 0.1920 & 0.3682 & 0.6134 & 1.0333 & 1.3279 \\ \cline{2-8}
& 1000000 & 0.8341 & 3.5295 & 5.2274 & 8.1350 & 10.373 & 16.083 \\
\hline
\end{tabular}
\label{train_times2}
\end{table} \clearpage

\begin{table}
\footnotesize
\centering
\caption{Training times in seconds for l2svm\_l1r\_tXiX}
\begin{tabular}{|l|l|c|c|c|c|c|c|} \hline
Labelsets & Documents & \multicolumn{6}{c|}{Features} \\ \cline{3-8}
& & 10 & 100 & 1000 & 10000 & 100000 & 1000000 \\ \hline
\multirow{5}{*}{1} & 10 & 0.0000 & 0.0000 & 0.0001 & 0.0001 & 0.0001 & 0.0001 \\ \cline{2-8}
& 100 & 0.0001 & 0.0005 & 0.0003 & 0.0005 & 0.0006 & 0.0006 \\ \cline{2-8}
& 1000 & 0.0002 & 0.0026 & 0.0077 & 0.0099 & 0.0106 & 0.0103 \\ \cline{2-8}
& 10000 & 0.0016 & 0.0064 & 0.0857 & 0.1177 & 0.1315 & 0.1333 \\ \cline{2-8}
& 100000 & 0.0110 & 0.0966 & 0.4699 & 0.7891 & 1.5672 & 0.9556 \\ \cline{2-8}
& 1000000 & 0.2419 & 1.3368 & 5.6083 & 16.567 & 19.312 & 21.434 \\
\hline
\end{tabular}
\label{train_times3}
\end{table}

\begin{table}
\footnotesize
\centering
\caption{Training times in seconds for lr\_l1r\_tXiX}
\begin{tabular}{|l|l|c|c|c|c|c|c|} \hline
Labelsets & Documents & \multicolumn{6}{c|}{Features} \\ \cline{3-8}
& & 10 & 100 & 1000 & 10000 & 100000 & 1000000 \\ \hline
\multirow{5}{*}{1} & 10 & 0.0000 & 0.0001 & 0.0001 & 0.0001 & 0.0001 & 0.0001 \\ \cline{2-8}
& 100 & 0.0001 & 0.0002 & 0.0001 & 0.0002 & 0.0002 & 0.0002 \\ \cline{2-8}
& 1000 & 0.0006 & 0.0010 & 0.0019 & 0.0053 & 0.0070 & 0.0069 \\ \cline{2-8}
& 10000 & 0.0065 & 0.0124 & 0.1104 & 0.0414 & 0.0665 & 0.0896 \\ \cline{2-8}
& 100000 & 0.0667 & 0.1451 & 0.3758 & 0.4872 & 0.6406 & 0.7311 \\ \cline{2-8}
& 1000000 & 0.9081 & 2.8916 & 11.386 & 16.454 & 15.465 & 17.160 \\
\hline
\end{tabular}
\label{train_times4}
\end{table} \clearpage

\begin{table}
\footnotesize
\centering
\caption{Training times in seconds for mnb\_c\_jm\_tXiX}
\begin{tabular}{|l|l|c|c|c|c|c|c|} \hline
Labelsets & Documents & \multicolumn{6}{c|}{Features} \\ \cline{3-8}
& & 10 & 100 & 1000 & 10000 & 100000 & 1000000 \\ \hline
\multirow{5}{*}{1} & 10 & 0.7960 & 0.9570 & 1.0570 & 1.0500 & 0.9890 & 1.1000 \\ \cline{2-8}
& 100 & 0.7980 & 0.9110 & 1.1990 & 1.3600 & 1.2840 & 1.3700 \\ \cline{2-8}
& 1000 & 0.8730 & 1.0230 & 1.3720 & 1.5950 & 1.5670 & 1.6630 \\ \cline{2-8}
& 10000 & 0.8240 & 1.2430 & 1.7280 & 2.0370 & 2.5070 & 2.5170 \\ \cline{2-8}
& 100000 & 1.4380 & 1.9760 & 3.7140 & 5.9010 & 7.0570 & 7.6850 \\ \cline{2-8}
& 1000000 & 3.7440 & 9.1760 & 25.912 & 42.388 & 49.062 & 56.350 \\
\hline
\end{tabular}
\label{train_times5}
\end{table}

\begin{table}
\footnotesize
\centering
\caption{Training times in seconds for mnb\_ht\_c\_jm\_tXiX}
\begin{tabular}{|l|l|c|c|c|c|c|c|} \hline
Labelsets & Documents & \multicolumn{6}{c|}{Features} \\ \cline{3-8}
& & 10 & 100 & 1000 & 10000 & 100000 & 1000000 \\ \hline
\multirow{5}{*}{1} & 10 & 0.7960 & 0.9570 & 1.0570 & 1.0500 & 0.9890 & 1.1000 \\ \cline{2-8}
& 100 & 0.7980 & 0.9110 & 1.1990 & 1.3600 & 1.2840 & 1.3700 \\ \cline{2-8}
& 1000 & 0.8730 & 1.0230 & 1.3720 & 1.5950 & 1.5670 & 1.6630 \\ \cline{2-8}
& 10000 & 0.8240 & 1.2430 & 1.7280 & 2.0370 & 2.5070 & 2.5170 \\ \cline{2-8}
& 100000 & 1.4380 & 1.9760 & 3.7140 & 5.9010 & 7.0570 & 7.6850 \\ \cline{2-8}
& 1000000 & 3.7440 & 9.1760 & 25.912 & 42.388 & 49.062 & 56.350 \\
\hline
\end{tabular}
\label{train_times6}
\end{table} \clearpage

\begin{table}
\footnotesize
\centering
\caption{Training times in seconds for tdm\_c\_jm\_tXiX}
\begin{tabular}{|l|l|c|c|c|c|c|c|} \hline
Labelsets & Documents & \multicolumn{6}{c|}{Features} \\ \cline{3-8}
& & 10 & 100 & 1000 & 10000 & 100000 & 1000000 \\ \hline
\multirow{5}{*}{1} & 10 & 0.8710 & 0.9900 & 1.1090 & 1.1300 & 1.1230 & 1.1400 \\ \cline{2-8}
& 100 & 0.8790 & 1.1550 & 1.2540 & 1.5090 & 1.3910 & 1.4310 \\ \cline{2-8}
& 1000 & 0.9560 & 1.1320 & 1.7020 & 2.2030 & 2.3380 & 2.4050 \\ \cline{2-8}
& 10000 & 1.2020 & 1.8080 & 3.3090 & 4.7660 & 5.2520 & 5.3160 \\ \cline{2-8}
& 100000 & 2.8890 & 6.0580 & 16.262 & 23.959 & 30.482 & 29.778 \\ \cline{2-8}
& 1000000 & 12.928 & NA & NA & NA & NA & NA \\
\hline
\end{tabular}
\label{train_times7}
\end{table}

\begin{table}
\footnotesize
\centering
\caption{Training times in seconds for tdm\_ht\_c\_jm\_tXiX}
\begin{tabular}{|l|l|c|c|c|c|c|c|} \hline
Labelsets & Documents & \multicolumn{6}{c|}{Features} \\ \cline{3-8}
& & 10 & 100 & 1000 & 10000 & 100000 & 1000000 \\ \hline
\multirow{5}{*}{1} & 10 & 0.9170 & 1.0090 & 1.1980 & 1.2200 & 1.1280 & 1.1960 \\ \cline{2-8}
& 100 & 0.9260 & 1.1380 & 1.4110 & 1.5050 & 1.5130 & 1.4470 \\ \cline{2-8}
& 1000 & 0.9450 & 1.2260 & 1.5930 & 2.1850 & 2.3790 & 2.1340 \\ \cline{2-8}
& 10000 & 1.1440 & 1.7140 & 3.3050 & 4.9150 & 5.1990 & 5.1930 \\ \cline{2-8}
& 100000 & 2.8150 & 5.6730 & 14.9470 & 23.4950 & NA & NA  \\ \cline{2-8}
& 1000000 & 12.9140 & NA & NA & NA & NA & NA \\ \hline
\multirow{5}{*}{10} & 10 & 0.9280 & 1.0530 & 1.1830 & 1.1660 & 1.2070 & 1.1500 \\ \cline{2-8}
& 100 & 0.9980 & 1.1130 & 1.3230 & 1.4630 & 1.4720 & 1.5410 \\ \cline{2-8}
& 1000 & 1.0040 & 1.2290 & 1.6500 & 2.0790 & 2.4190 & 2.5340 \\ \cline{2-8}
& 10000 & 1.1520 & 1.7460 & 3.4040 & 4.7840 & 5.2020 & 5.2730 \\ \cline{2-8}
& 100000 & 2.7420 & 5.7550 & 14.9800 & 23.5480 & NA & NA \\ \cline{2-8}
& 1000000 & 12.8880 & NA & NA & NA & NA & NA \\ \hline
\multirow{5}{*}{1000} & 10 & 0.9290 & 0.9800 & 1.1970 & 1.2270 & 1.2490 & 1.1910 \\ \cline{2-8}
& 100 & 1.0780 & 1.1240 & 1.3380 & 1.4720 & 1.4470 & 1.4830 \\ \cline{2-8}
& 1000 & 1.0600 & 1.1980 & 1.7890 & 2.4960 & 2.2760 & 2.2940 \\ \cline{2-8}
& 10000 & 1.1910 & 1.9680 & 3.4280 & 5.2450 & 5.4840 & 5.4810 \\ \cline{2-8}
& 100000 & 2.8110 & 6.2980 & 15.3180 & 23.9250 & NA & NA  \\ \cline{2-8}
& 1000000 & 11.7400 & NA & NA & NA & NA & NA \\ \hline
\multirow{5}{*}{1000000} & 10 & 1.0010 & 1.0590 & 1.1250 & 1.1250 & 1.1090 & 1.1940 \\ \cline{2-8}
& 100 & 1.0120 & 1.0660 & 1.3480 & 1.5150 & 1.5530 & 1.5380 \\ \cline{2-8}
& 1000 & 1.1350 & 1.3500 & 1.7670 & 2.3450 & 2.2720 & 2.2760 \\ \cline{2-8}
& 10000 & 1.4920 & 2.4200 & 4.4770 & 6.2660 & 6.9170 & 7.2160 \\ \cline{2-8}
& 100000 & 3.6570 & 8.6960 & NA & NA & NA & NA \\ \cline{2-8}
& 1000000 & NA & NA & NA & NA & NA & NA \\ \hline
\end{tabular}
\label{train_times8}
\end{table} \clearpage

\begin{table}
\footnotesize
\centering
\caption{Mean test times in milliseconds for lr\_l2r\_tXiX}
\begin{tabular}{|l|l|c|c|c|c|c|c|} \hline
Labelsets & Documents & \multicolumn{6}{c|}{Features} \\ \cline{3-8}
& & 10 & 100 & 1000 & 10000 & 100000 & 1000000 \\ \hline
\multirow{5}{*}{1} & 10 & 0.0201 & 0.0347 & 0.0694 & 0.0705 & 0.0698 & 0.0709 \\ \cline{2-8}
& 100 & 0.0218 & 0.0475 & 0.1031 & 0.1439 & 0.1461 & 0.1460 \\ \cline{2-8}
& 1000 & 0.0262 & 0.0474 & 0.1189 & 0.1862 & 0.1965 & 0.1927 \\ \cline{2-8}
& 10000 & 0.0238 & 0.0492 & 0.1239 & 0.2089 & 0.2235 & 0.2190 \\ \cline{2-8}
& 100000 & 0.0268 & 0.0552 & 0.1334 & 0.2069 & 0.2306 & 0.2356 \\ \cline{2-8}
& 1000000 & 0.0536 & 0.0881 & 0.1744 & 0.2772 & 0.3131 & 0.3229 \\
\hline
\end{tabular}
\label{test_times1}
\end{table}

\begin{table}
\footnotesize
\centering
\caption{Mean test times in milliseconds for l2svm\_l2r\_tXiX}
\begin{tabular}{|l|l|c|c|c|c|c|c|} \hline
Labelsets & Documents & \multicolumn{6}{c|}{Features} \\ \cline{3-8}
& & 10 & 100 & 1000 & 10000 & 100000 & 1000000 \\ \hline
\multirow{5}{*}{1} & 10 & 0.0225 & 0.0354 & 0.0698 & 0.0718 & 0.0695 & 0.0733 \\ \cline{2-8}
& 100 & 0.0240 & 0.0431 & 0.1046 & 0.1460 & 0.1462 & 0.1484 \\ \cline{2-8}
& 1000 & 0.0239 & 0.0496 & 0.1207 & 0.1868 & 0.1966 & 0.1945 \\ \cline{2-8}
& 10000 & 0.0239 & 0.0491 & 0.1252 & 0.2051 & 0.2195 & 0.2248 \\ \cline{2-8}
& 100000 & 0.0250 & 0.0521 & 0.1304 & 0.2072 & 0.2313 & 0.2343 \\ \cline{2-8}
& 1000000 & 0.0532 & 0.0836 & 0.1798 & 0.2707 & 0.3067 & 0.3162 \\
\hline
\end{tabular}
\label{test_times2}
\end{table} \clearpage

\begin{table}
\footnotesize
\centering
\caption{Mean test times in milliseconds for l2svm\_l1r\_tXiX}
\begin{tabular}{|l|l|c|c|c|c|c|c|} \hline
Labelsets & Documents & \multicolumn{6}{c|}{Features} \\ \cline{3-8}
& & 10 & 100 & 1000 & 10000 & 100000 & 1000000 \\ \hline
\multirow{5}{*}{1} & 10 & 0.0204 & 0.0351 & 0.0704 & 0.0724 & 0.0701 & 0.0708 \\ \cline{2-8}
& 100 & 0.0246 & 0.0435 & 0.1053 & 0.1450 & 0.1441 & 0.1457 \\ \cline{2-8}
& 1000 & 0.0238 & 0.0470 & 0.1202 & 0.1844 & 0.1946 & 0.1928 \\ \cline{2-8}
& 10000 & 0.0248 & 0.0495 & 0.1268 & 0.1994 & 0.2243 & 0.2254 \\ \cline{2-8}
& 100000 & 0.0257 & 0.0529 & 0.1286 & 0.2109 & 0.2322 & 0.2360 \\ \cline{2-8}
& 1000000 & 0.0517 & 0.0815 & 0.1749 & 0.2831 & 0.3148 & 0.3291 \\
\hline
\end{tabular}
\label{test_times3}
\end{table}

\begin{table}
\footnotesize
\centering
\caption{Mean test times in milliseconds for lr\_l1r\_tXiX}
\begin{tabular}{|l|l|c|c|c|c|c|c|} \hline
Labelsets & Documents & \multicolumn{6}{c|}{Features} \\ \cline{3-8}
& & 10 & 100 & 1000 & 10000 & 100000 & 1000000 \\ \hline
\multirow{5}{*}{1} & 10 & 0.0204 & 0.0389 & 0.0699 & 0.0728 & 0.0703 & 0.0707 \\ \cline{2-8}
& 100 & 0.0256 & 0.0442 & 0.1051 & 0.1482 & 0.1464 & 0.1449 \\ \cline{2-8}
& 1000 & 0.0268 & 0.0474 & 0.1187 & 0.1863 & 0.1918 & 0.1940 \\ \cline{2-8}
& 10000 & 0.0259 & 0.0486 & 0.1264 & 0.2031 & 0.2238 & 0.2213 \\ \cline{2-8}
& 100000 & 0.0275 & 0.0527 & 0.1299 & 0.2032 & 0.2323 & 0.2371 \\ \cline{2-8}
& 1000000 & 0.0528 & 0.0829 & 0.1764 & 0.2795 & 0.3139 & 0.3277 \\
\hline
\end{tabular}
\label{test_times4}
\end{table} \clearpage

\begin{table}
\footnotesize
\centering
\caption{Mean test times in milliseconds for mnb\_c\_jm\_tXiX}
\begin{tabular}{|l|l|c|c|c|c|c|c|} \hline
Labelsets & Documents & \multicolumn{6}{c|}{Features} \\ \cline{3-8}
& & 10 & 100 & 1000 & 10000 & 100000 & 1000000 \\ \hline
\multirow{5}{*}{1} & 10 & 0.0269 & 0.0309 & 0.0320 & 0.0259 & 0.0280 & 0.0274 \\ \cline{2-8}
& 100 & 0.0273 & 0.0265 & 0.0330 & 0.0430 & 0.0457 & 0.0411 \\ \cline{2-8}
& 1000 & 0.0264 & 0.0285 & 0.0384 & 0.0362 & 0.0402 & 0.0394 \\ \cline{2-8}
& 10000 & 0.0278 & 0.0264 & 0.0380 & 0.0405 & 0.0433 & 0.0427 \\ \cline{2-8}
& 100000 & 0.0322 & 0.0332 & 0.0433 & 0.0377 & 0.0459 & 0.0493 \\ \cline{2-8}
& 1000000 & 0.0264 & 0.0329 & 0.0373 & 0.0409 & 0.0479 & 0.0493 \\
\hline
\end{tabular}
\label{test_times5}
\end{table}

\begin{table}
\footnotesize
\centering
\caption{Mean test times in milliseconds for mnb\_ht\_c\_jm\_tXiX}
\begin{tabular}{|l|l|c|c|c|c|c|c|} \hline
Labelsets & Documents & \multicolumn{6}{c|}{Features} \\ \cline{3-8}
& & 10 & 100 & 1000 & 10000 & 100000 & 1000000 \\ \hline
\multirow{5}{*}{1} & 10 & 0.0262 & 0.0310 & 0.0265 & 0.0271 & 0.0275 & 0.0267 \\ \cline{2-8}
& 100 & 0.0258 & 0.0256 & 0.0390 & 0.0380 & 0.0413 & 0.0397 \\ \cline{2-8}
& 1000 & 0.0265 & 0.0249 & 0.0386 & 0.0389 & 0.0435 & 0.0459 \\ \cline{2-8}
& 10000 & 0.0295 & 0.0256 & 0.0436 & 0.0385 & 0.0461 & 0.0460 \\ \cline{2-8}
& 100000 & 0.0362 & 0.0252 & 0.0450 & 0.0392 & 0.0430 & 0.0433 \\ \cline{2-8}
& 1000000 & 0.0360 & 0.0364 & 0.0361 & 0.0399 & 0.0464 & 0.0483 \\
\hline
\end{tabular}
\label{test_times6}
\end{table} \clearpage

\begin{table}
\footnotesize
\centering
\caption{Mean test times in milliseconds for tdm\_c\_jm\_tXiX}
\begin{tabular}{|l|l|c|c|c|c|c|c|} \hline
Labelsets & Documents & \multicolumn{6}{c|}{Features} \\ \cline{3-8}
& & 10 & 100 & 1000 & 10000 & 100000 & 1000000 \\ \hline
\multirow{5}{*}{1} & 10 & 0.0274 & 0.0332 & 0.0479 & 0.0488 & 0.0483 & 0.0498 \\ \cline{2-8}
& 100 & 0.0395 & 0.0602 & 0.0962 & 0.1059 & 0.0975 & 0.1035 \\ \cline{2-8}
& 1000 & 0.1741 & 0.3373 & 0.4816 & 0.4866 & 0.4907 & 0.4951 \\ \cline{2-8}
& 10000 & 1.4231 & 3.2992 & 4.6781 & 4.6911 & 4.5519 & 4.6349 \\ \cline{2-8}
& 100000 & 19.237 & 46.185 & 62.657 & 64.243 & 63.606 & 62.989 \\ \cline{2-8}
& 1000000 & 267.21 & NA & NA & NA & NA & NA \\
\hline
\end{tabular}
\label{test_times7}
\end{table}

\begin{table}
\footnotesize
\centering
\caption{Mean test times in milliseconds for tdm\_ht\_c\_jm\_tXiX}
%
\label{test_times8}
\end{table}
\chapter{Kaggle LSHTC4 Winning Solution}
The appended document describes the winning solution to Kaggle LSHTC4\footnote{http://www.kaggle.com/c/lshtc/} competition organized in 2014, that had
119 participating teams. The solution was based on the methods presented in the thesis, including the sparse inference, the modified and extended MNB models, the
TDM model, and the random search development framework. Some additional techniques such as model-based feedback, label thresholding, and 
ensemble learning were developed for the competition, that fall outside the scope of the thesis. The developed ensemble learning is a natural 
continuation of the ideas presented in the thesis. It combines an ensemble of sparse generative model base-classifiers using a mixture model, where
the weight of each component is dynamically predicted using a large number of meta-features, and the outputs of the base-classifiers are
restricted to predicting a single most likely output for each input. The LSHTC ensemble solution was later extended for the Kaggle 
WISE2014\footnote{http://www.kaggle.com/c/wise-2014/} competition, where it came second out of 120 competing teams.\\ \clearpage 
\includepdf[pages=-,pagecommand=, fitpaper= false, offset= 0mm 10mm]{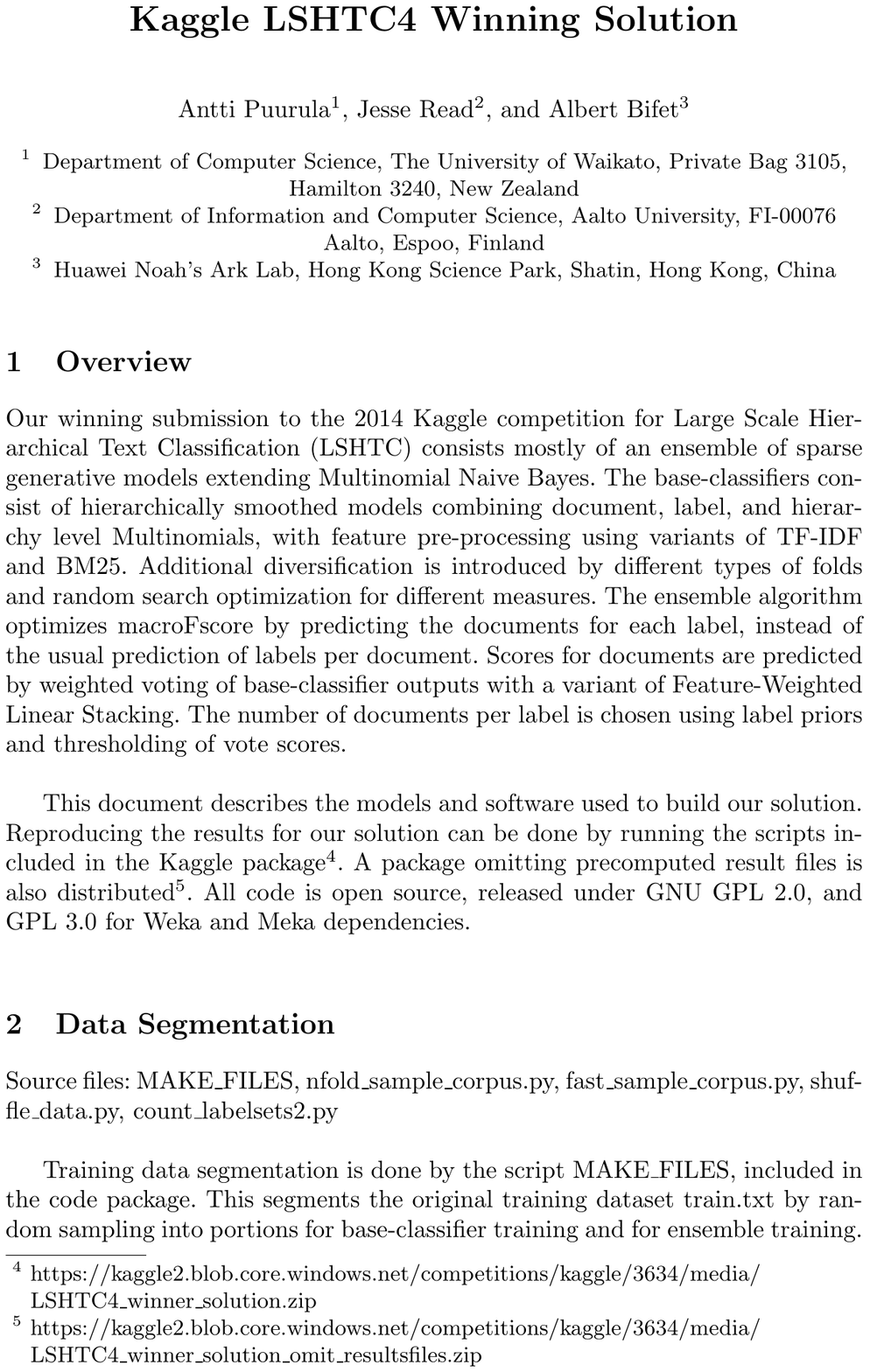}
\end{appendices}

\bibliographystyle{plainnat}
\bibliography{thesis} 
\end{document}